\documentclass[fleqn,onecolumn,showpacs,preprintnumbers,showkeys]{revtex4}
\usepackage{graphicx}
\usepackage{amssymb}
\usepackage{amsmath}\allowdisplaybreaks
\usepackage{bm}
\usepackage[flushleft]{caption2}
\usepackage[chatter]{rotating}

\begin{document}
\setcounter{page}{1}
\title{Some helicities in electromagnetic and gravitational fields}
\author{Zi-Hua Weng}
\email{xmuwzh@xmu.edu.cn.}
\affiliation{School of Physics and Mechanical \& Electrical Engineering, Xiamen University, Xiamen 361005, China}

\begin{abstract}
The paper investigates the influences of the helicity on the gravitational mass density, the field source, the charge continuity equation, and the mass continuity equation etc in the electromagnetic field and gravitational field. By means of the algebra of octonions, the magnetic helicity, the current helicity, the cross helicity, the kinetic helicity, the field energy, the enstrophy, and some new helicity terms can be derived from the octonion definitions of the linear mentum and the force in some field descriptions with different operators. The study claims that the gravitational mass density, the field source, the charge continuity equation, and the mass continuity equation will be impacted by the helicity, the field strength, and the vorticity of the rotational objects and of the spinning charged objects in the gravitational field and the electromagnetic field with their adjoint fields.
\end{abstract}



\maketitle

%--1--%

\section{INTRODUCTION}

The concept of the helicity \cite{seehafer, mitra} was originated by C. F. Gauss's studying of the orbits of the comet and the asteroid in the astronomy. Recently it is extended to the researches regarding the double helix structure of DNA molecular chain \cite{malyshev}, the solar electromagnetic field, the fluid dynamics, the meteorology, and the topology etc. In the astronomy the helicity can be used to investigate the helical structure of the solar magnetic field \cite{reinard, gibbon}, including the magnetic helicity, the current helicity, the cross helicity \cite{podesta}, and the kinetic helicity. In the fluid dynamics \cite{trueba} the helicity can be used to illustrate the vorticity fields, the kinetic helicity, and the enstrophy. It is found that the helicity will impact the gravitational mass density, the charge continuity equation, and the mass continuity equation.

The algebra of quaternions was invented in 1843 by W. R. Hamilton \cite{hamilton}, and was first used by J. C. Maxwell \cite{maxwell} to describe the physical features of the electromagnetic field. Later the algebra of quaternions is used to represent the physical properties of the gravitational field. According to the 'Spacetime equality postulation' \cite{weng1}, the quaternion space for the gravitational field is independent to that for the electromagnetic field \cite{weng2}. These two quaternion spaces are perpendicular to each other, and are combined together to become one octonion space, which is related to the algebra of octonions. The latter was invented by A. Cayley \cite{cayley} etc. Consequently the algebra of octonions can be used to depict the features of the gravitational field and the electromagnetic field simultaneously, including the mass continuity equation, the charge continuity equation, the gravitational mass, and the helicity etc.

In terms of the algebra of octonions, the helicity of the rotational objects and the spinning charged objects in the electromagnetic field and the gravitational field \cite{newton} can be found in the definition of the field source and the definition of the force-power in the octonion space. The helicities cover all known helicity terms in the classical theories, including the magnetic helicity, the current helicity \cite{yeates}, the cross helicity, the kinetic helicity, the field energy, the enstrophy, and some new helicity terms.

%--2--%

\section{The fields regarding the operator ($\lozenge, \mathbb{B}$) }

In the electromagnetic field and gravitational field \cite{heaviside}, the octonion operator $(\lozenge + k_b \mathbb{B})$ can draw out directly the physical properties of two fields, including the octonion field potential, the octonion field strength, the octonion field source, the octonion linear momentum, the octonion angular momentum, the energy, the torque, the power, the force, and some helicities \cite{malanushenko} of the rotational objects and the spinning charged objects \cite{uhlenbeck} etc. It is found that the helicity, the field strength, and the field source will impact the gravitational mass, the charge continuity equation, and the mass continuity equation etc in the octonion space with the operator $(\lozenge + k_b \mathbb{B})$.

\subsection{Field source}

In the octonion space, we can define the source of the electromagnetic field and the source of the gravitational field. In the quaternion space for the gravitational field, the basis vector is $\mathbb{E}_g$ = ($\emph{\textbf{i}}_0$,  $\emph{\textbf{i}}_1$, $\emph{\textbf{i}}_2$, $\emph{\textbf{i}}_3$), the radius vector is $\mathbb{R}_g$ = ($r_0$, $r_1$, $r_2$, $r_3$), and the velocity is $\mathbb{V}_g$ = ($v_0^\delta$, $v_1$, $v_2$, $v_3$). In the quaternion space for the electromagnetic field, the basis vector is $\mathbb{E}_e$ = ($\emph{\textbf{I}}_0$, $\emph{\textbf{I}}_1$, $\emph{\textbf{I}}_2$, $\emph{\textbf{I}}_3$), the radius vector is $\mathbb{R}_e$ = ($R_0$, $R_1$, $R_2$, $R_3$), and the velocity is $\mathbb{V}_e$ = ($V_0$, $V_1$, $V_2$, $V_3$), with $\mathbb{E}_e$ = $\mathbb{E}_g$ $\circ$ $\emph{\textbf{I}}_0$ . The $\mathbb{E}_e$ is independent of the $\mathbb{E}_g$ , and that they can combine together to become the basis vector of octonion space, $ \mathbb{E} = (1, \emph{\textbf{i}}_1, \emph{\textbf{i}}_2, \emph{\textbf{i}}_3, \emph{\textbf{I}}_0, \emph{\textbf{I}}_1, \emph{\textbf{I}}_2, \emph{\textbf{I}}_3) $. The octonion radius vectors $\mathbb{R}_g$ and $\mathbb{R}_e$ can be combined together to become the octonion radius vector, $\mathbb{R} = \Sigma ( \emph{\textbf{i}}_i r_i + k_{eg} \emph{\textbf{I}}_i R_i)$. And then the octonion velocity is $\mathbb{V} = \Sigma ( \emph{\textbf{i}}_i v_i + k_{eg} \emph{\textbf{I}}_i V_i)$. Herein $\emph{\textbf{i}}_0 = 1$; $r_0 = v_0 t$, $t$ is the time; $v_0$ and $k_{eg}$ are the coefficients for the dimensional homogeneity, and $v_0$ is the speed of light in comparison with the classical theory; the symbol $\circ$ denotes the octonion multiplication.

The gravitational potential $\mathbb{A}_g = \Sigma ( a_i \emph{\textbf{i}}_i)$ is combined with the electromagnetic potential $\mathbb{A}_e = \Sigma ( A_i \emph{\textbf{I}}_i)$ to become the octonion field potential, $\mathbb{A} = \mathbb{A}_g + k_{eg} \mathbb{A}_e $. While the octonion field strength $\mathbb{B} = \lozenge \circ \mathbb{A} = \mathbb{B}_g + k_{eg} \mathbb{B}_e$ consists of the gravitational strength, $\mathbb{B}_g = \Sigma ( h_i \emph{\textbf{i}}_i)$, and the electromagnetic strength, $\mathbb{B}_e = \Sigma (H_i \emph{\textbf{I}}_i)$. The gauge equations are $h_0 = 0$ and $H_0 = 0$. The gravitational strength $\mathbb{B}_g$ includes two components, $\textbf{g} / v_0 = \partial_0 \textbf{a} + \nabla a_0 $ and $\textbf{b} =  \nabla \times \textbf{a}$, and the electromagnetic strength $\mathbb{B}_e$ involves two parts, $\textbf{E} / v_0 = \partial_0 \textbf{A} + \nabla \circ \textbf{A}_0 $ and $\textbf{B} = \nabla \times \textbf{A}$. Herein $\textbf{a} = \Sigma ( a_j \emph{\textbf{i}}_j)$, $\textbf{A} = \Sigma ( A_j \emph{\textbf{I}}_j)$, $\textbf{A}_0 = A_0 \emph{\textbf{I}}_0$; $\lozenge = \Sigma (\emph{\textbf{i}}_i \partial_i)$, with $\partial_i = \partial / \partial r_i$. $i = 0, 1, 2, 3$. $\nabla = \Sigma (\emph{\textbf{i}}_j \partial_j)$. $j = 1, 2, 3$.

The linear momentum density $\mathbb{S}_g = m \mathbb{V}_g$ is the source of the gravitational field, while the electric current density $\mathbb{S}_e = q \mathbb{V}_e$ is that of the electromagnetic field. The octonion field source $\mathbb{S}$ satisfies
\begin{eqnarray}
\mu \mathbb{S} = - (\lozenge + k_b \mathbb{B})^* \circ \mathbb{B} = \mu_g \mathbb{S}_g + k_{eg} \mu_e \mathbb{S}_e - k_b \mathbb{B}^*
\circ \mathbb{B}~,
\end{eqnarray}
where $-\mu_g \mathbb{S}_g = \lozenge^* \circ \mathbb{B}_g$, $-\mu_e \mathbb{S}_e = \lozenge^* \circ \mathbb{B}_e$; $k_b = 1 / v_0$; $\mathbb{B}^* \circ \mathbb{B}/ \mu_g = \mathbb{B}_g^* \circ \mathbb{B}_g / \mu_g + \mathbb{B}_e^* \circ \mathbb{B}_e / \mu_e$; $k_{eg}^2 = \mu_g /\mu_e$; $\mu_g$ and $\mu_e$ are the gravitational constant and the electromagnetic constant respectively; $q$ is the electric charge density; $m$ is the inertial mass density; $*$ denotes the conjugate of the octonion.

\begin{table}[h]
\caption{\label{tab:table1}The octonion multiplication table.}
\centering
\begin{tabular}{c|cccc|cccc}
\hline\hline
$ $ & $1$ & $\emph{\textbf{i}}_1$  & $\emph{\textbf{i}}_2$ &
$\emph{\textbf{i}}_3$  & $\emph{\textbf{I}}_0$  &
$\emph{\textbf{I}}_1$
& $\emph{\textbf{I}}_2$  & $\emph{\textbf{I}}_3$  \\
\hline
$1$ & $1$ & $\emph{\textbf{i}}_1$  & $\emph{\textbf{i}}_2$ &
$\emph{\textbf{i}}_3$  & $\emph{\textbf{I}}_0$  &
$\emph{\textbf{I}}_1$
& $\emph{\textbf{I}}_2$  & $\emph{\textbf{I}}_3$  \\
$\emph{\textbf{i}}_1$ & $\emph{\textbf{i}}_1$ & $-1$ &
$\emph{\textbf{i}}_3$  & $-\emph{\textbf{i}}_2$ &
$\emph{\textbf{I}}_1$
& $-\emph{\textbf{I}}_0$ & $-\emph{\textbf{I}}_3$ & $\emph{\textbf{I}}_2$  \\
$\emph{\textbf{i}}_2$ & $\emph{\textbf{i}}_2$ &
$-\emph{\textbf{i}}_3$ & $-1$ & $\emph{\textbf{i}}_1$  &
$\emph{\textbf{I}}_2$  & $\emph{\textbf{I}}_3$
& $-\emph{\textbf{I}}_0$ & $-\emph{\textbf{I}}_1$ \\
$\emph{\textbf{i}}_3$ & $\emph{\textbf{i}}_3$ &
$\emph{\textbf{i}}_2$ & $-\emph{\textbf{i}}_1$ & $-1$ &
$\emph{\textbf{I}}_3$  & $-\emph{\textbf{I}}_2$
& $\emph{\textbf{I}}_1$  & $-\emph{\textbf{I}}_0$ \\
\hline
$\emph{\textbf{I}}_0$ & $\emph{\textbf{I}}_0$ &
$-\emph{\textbf{I}}_1$ & $-\emph{\textbf{I}}_2$ &
$-\emph{\textbf{I}}_3$ & $-1$ & $\emph{\textbf{i}}_1$
& $\emph{\textbf{i}}_2$  & $\emph{\textbf{i}}_3$  \\
$\emph{\textbf{I}}_1$ & $\emph{\textbf{I}}_1$ &
$\emph{\textbf{I}}_0$ & $-\emph{\textbf{I}}_3$ &
$\emph{\textbf{I}}_2$  & $-\emph{\textbf{i}}_1$
& $-1$ & $-\emph{\textbf{i}}_3$ & $\emph{\textbf{i}}_2$  \\
$\emph{\textbf{I}}_2$ & $\emph{\textbf{I}}_2$ &
$\emph{\textbf{I}}_3$ & $\emph{\textbf{I}}_0$  &
$-\emph{\textbf{I}}_1$ & $-\emph{\textbf{i}}_2$
& $\emph{\textbf{i}}_3$  & $-1$ & $-\emph{\textbf{i}}_1$ \\
$\emph{\textbf{I}}_3$ & $\emph{\textbf{I}}_3$ &
$-\emph{\textbf{I}}_2$ & $\emph{\textbf{I}}_1$  &
$\emph{\textbf{I}}_0$  & $-\emph{\textbf{i}}_3$
& $-\emph{\textbf{i}}_2$ & $\emph{\textbf{i}}_1$  & $-1$ \\
\hline\hline
\end{tabular}
\end{table}

\subsection{Torque and force}

In the presence of electromagnetic and gravitational fields, the octonion angular momentum density $\mathbb{L}$ can be defined from the octonion linear momentum density $\mathbb{P} = \mu \mathbb{S} / \mu_g = \Sigma (p_i \emph{\textbf{i}}_i ) + \Sigma (P_i \emph{\textbf{I}}_i)$, the octonion radius vector $\mathbb{R}$, and the octonion physics quantity $\mathbb{X}$ ,
\begin{eqnarray}
\mathbb{L} = (\mathbb{R} + k_{rx} \mathbb{X} ) \circ \mathbb{P}~,
\end{eqnarray}
where $\mathbb{L} = \Sigma (l_i \emph{\textbf{i}}_i ) + \Sigma (L_i \emph{\textbf{I}}_i)$; $\mathbb{X} = \Sigma (x_i \emph{\textbf{i}}_i) + k_{eg} \Sigma (X_i \emph{\textbf{I}}_i)$. $ \textbf{p} = \Sigma (p_j \emph{\textbf{i}}_j)$; $\textbf{P}_0 = P_0 \emph{\textbf{I}}_0 $; $\textbf{P} = \Sigma (P_j \emph{\textbf{I}}_j )$. $\textbf{l} = \Sigma (l_j \emph{\textbf{i}}_j)$, $\textbf{L}_0 = L_0 \emph{\textbf{I}}_0$, $\textbf{L} = \Sigma (L_j \emph{\textbf{I}}_j)$. $p_0 = \widehat{m} v_0$, $v_0^\delta = v_0 \lozenge \cdot \mathbb{R}$, $\widehat{m} = \triangle m + m (v_0^\delta / v_0) $, $\triangle m = - k_b \mathbb{B}^* \circ \mathbb{B}/ ( \mu_g v_0)$. $\widehat{m}$ is the gravitational mass density. $\lozenge \cdot \mathbb{R}$ is the scalar part of $\lozenge \circ \mathbb{R}$. $k_{rx} = 1 / v_0$ is a coefficient for the dimensional homogeneity. The derivation of octonion quantity $\mathbb{X}$ will yield the gravitational and electromagnetic potentials simultaneously.

In the electromagnetic field \cite{faraday} and the gravitational field, the octonion torque-energy density $\mathbb{W}$ is defined from the octonion angular momentum density $\mathbb{L}$ and the octonion field strength $\mathbb{B}$,
\begin{eqnarray}
\mathbb{W} = v_0 (\lozenge + k_b \mathbb{B}) \circ \mathbb{L}~,
\end{eqnarray}
where $\mathbb{W} = \Sigma (w_i \emph{\textbf{i}}_i ) + \Sigma (W_i \emph{\textbf{I}}_i )$ ; the $-w_0/2$ is the energy density, the $\textbf{w}/2 = \Sigma (w_j \emph{\textbf{i}}_j )/2$ is the torque density. The scalar $w_0$ is written as,
\begin{eqnarray}
w_0 = v_0 \partial_0 l_0 + v_0 \nabla \cdot \textbf{l} + \textbf{h} \cdot \textbf{l} + k_{eg} \textbf{H} \cdot \textbf{L}~,
\end{eqnarray}
where $-w_0/2$ includes the kinetic energy, gravitational potential energy, field energy, work, electric potential energy, magnetic potential energy, the interacting energy between dipole moment with electromagnetic strength, and some new terms. $a_0 / v_0 = \partial_0 x_0 + \nabla \cdot \textbf{x}$ and $\textbf{a} = \partial_0 \textbf{x} + \nabla x_0 + \nabla \times \textbf{x}$ are the scalar and vectorial potential of the gravitational field respectively. $\textbf{A}_0 / v_0 = \partial_0 \textbf{X}_0 + \nabla \cdot \textbf{X}$ and $\textbf{A} = \partial_0 \textbf{X} + \nabla \circ \textbf{X}_0 + \nabla \times \textbf{X}$ are the scalar and vectorial potential of the electromagnetic field respectively. $\textbf{h} = \Sigma (h_j \emph{\textbf{i}}_j)$, $\textbf{H}_0 = H_0 \emph{\textbf{I}}_0$, $\textbf{H} = \Sigma (H_j \emph{\textbf{I}}_j)$. $\textbf{h} = \textbf{g} / v_0 + \textbf{b}$, $\textbf{H} = \textbf{E} / v_0 + \textbf{B}$.

In a similar way, expressing the torque density $\textbf{w}$ as
\begin{eqnarray}
\textbf{w} = v_0 \partial_0 \textbf{l} + v_0 \nabla l_0 + v_0 \nabla \times \textbf{l} + l_0 \textbf{h} + \textbf{h} \times \textbf{l}
+ k_{eg} \textbf{H} \times \textbf{L} + k_{eg} \textbf{H} \circ \textbf{L}_0~,
\end{eqnarray}
where the above covers the torque density caused by the inertial force, the electromagnetic force, the gravitational force, and other force terms etc.

The octonion force-power density $\mathbb{N} = v_0 (\lozenge + k_b \mathbb{B})^* \circ \mathbb{W} $ can be defined from the torque-energy density $\mathbb{W}$. In the octonion space, the force-power density is written as $\mathbb{N} = \Sigma (n_i \emph{\textbf{i}}_i) + \Sigma (N_i \emph{\textbf{I}}_i)$. The power density is $f_0 = - n_0/(2 v_0)$, while the force density is $\textbf{f} = - \textbf{n} / (2 v_0)$, with the vectorial part $\textbf{n} = \Sigma (n_j \emph{\textbf{i}}_j )$ in the space $\mathbb{E}_g$. The other two vectorial parts, $\textbf{N}_0 = N_0 \emph{\textbf{I}}_0$ and $\textbf{N} = \Sigma (N_j \emph{\textbf{I}}_j)$, are in the space $\mathbb{E}_e$, and may not be detected directly.

Further expressing the scalar $n_0$ of the octonion force-power density $\mathbb{N}$ as
\begin{eqnarray}
n_0 = v_0 \partial_0 w_0 + v_0 \nabla^* \cdot \textbf{w} + \textbf{h}^* \cdot \textbf{w} + k_{eg} \textbf{H}^* \cdot \textbf{W}~.
\end{eqnarray}

The force density $\textbf{f}$ in the gravitational field and the electromagnetic field can be defined from the vectorial part $\textbf{n}$ of the octonion force-power density $\mathbb{N}$,
\begin{eqnarray}
- 2 \textbf{f} = \partial_0 \textbf{w} + \nabla^* w_0 + \textbf{h}^* \times \textbf{w}/v_0 + \nabla^* \times \textbf{w} + w_0 \textbf{h}^*/v_0
+ k_{eg} \textbf{H}^* \times \textbf{W}/v_0 + k_{eg} \textbf{H}^* \circ \textbf{W}_0/v_0~,
\end{eqnarray}
where the force density $\textbf{f}$ includes that of the inertial force, gravity, Lorentz force, gradient of energy, and interacting force between
dipole moment with magnetic strength etc. This force definition is much more complicated than that in the classical field theory, and encompasses more new force terms regarding the gradient of energy etc.

\begin{table}[h]
\caption{The operator and multiplication of the physical quantity in the octonion space.}
\label{tab:table3}
\centering
\begin{tabular}{ll}
\hline\hline
definitions                 &  meanings                                               \\
\hline
$\nabla \cdot \textbf{a}$   &  $-(\partial_1 a_1 + \partial_2 a_2 + \partial_3 a_3)$  \\
$\nabla \times \textbf{a}$  &  $\emph{\textbf{i}}_1 ( \partial_2 a_3
                                 - \partial_3 a_2 ) + \emph{\textbf{i}}_2 ( \partial_3 a_1
                                 - \partial_1 a_3 )
                                 + \emph{\textbf{i}}_3 ( \partial_1 a_2
                                 - \partial_2 a_1 )$                                  \\
$\nabla a_0$                &  $\emph{\textbf{i}}_1 \partial_1 a_0
                                 + \emph{\textbf{i}}_2 \partial_2 a_0
                                 + \emph{\textbf{i}}_3 \partial_3 a_0  $              \\
$\partial_0 \textbf{a}$     &  $\emph{\textbf{i}}_1 \partial_0 a_1
                                 + \emph{\textbf{i}}_2 \partial_0 a_2
                                 + \emph{\textbf{i}}_3 \partial_0 a_3  $              \\
\hline
$\nabla \cdot \textbf{P}$   &  $-(\partial_1 P_1 + \partial_2 P_2 + \partial_3 P_3) \emph{\textbf{I}}_0 $  \\
$\nabla \times \textbf{P}$  &  $-\emph{\textbf{I}}_1 ( \partial_2
                                 P_3 - \partial_3 P_2 ) - \emph{\textbf{I}}_2 ( \partial_3 P_1
                                 - \partial_1 P_3 )
                                 - \emph{\textbf{I}}_3 ( \partial_1 P_2 - \partial_2 P_1 )$    \\
$\nabla \circ \textbf{P}_0$ &  $\emph{\textbf{I}}_1 \partial_1 P_0
                                 + \emph{\textbf{I}}_2 \partial_2 P_0
                                 + \emph{\textbf{I}}_3 \partial_3 P_0  $             \\
$\partial_0 \textbf{P}$     &  $\emph{\textbf{I}}_1 \partial_0 P_1
                                 + \emph{\textbf{I}}_2 \partial_0 P_2
                                 + \emph{\textbf{I}}_3 \partial_0 P_3  $             \\
\hline\hline
\end{tabular}
\end{table}

\subsection{Helicity}

The part force density $\mathbb{F}$ is one part of the octonion force-power density $\mathbb{N}$, and is defined from the octonion linear momentum density $\mathbb{P} = \mu \mathbb{S} / \mu_g$ ,
\begin{eqnarray}
\mathbb{F} = v_0 (\lozenge + k_b \mathbb{B})^* \circ \mathbb{P}~,
\end{eqnarray}
where the part force density includes that of the gravitational force, Lorentz force, inertial force, and the interacting force between electromagnetic strength with magnetic moment etc.

The scalar $f_0$ of $\mathbb{F}$ is written as,
\begin{eqnarray}
f_0 / v_0 = \partial_0 p_0 + \nabla^* \cdot \textbf{p} + \textbf{h}^* \cdot \textbf{p} / v_0 + k_{eg} \textbf{H}^* \cdot \textbf{P} /v_0~,
\end{eqnarray}
where $k_{eg} \textbf{H}^* \cdot \textbf{P}$ is the field source helicity of the electromagnetic field, which involves the electric current helicity $k_{eg} \textbf{B}^* \cdot \textbf{P}$ and the helicity term $k_{eg} \textbf{E}^* \cdot \textbf{P} / v_0$; the $\textbf{h}^* \cdot \textbf{p}$ is the field source helicity of the gravitational field.

The scalar part of the octonion quantity is reserved in the octonion coordinate transformation. Therefore the above is the mass continuity equation in the electromagnetic field and the gravitational field when $f_0 = 0$, and is influenced by the current helicity terms \cite{su} of the electromagnetic field and of the gravitational field, although the impact of the field source helicity may be quite tiny.

A new physical quantity $\mathbb{F}_q$ can be defined from the part force density $\mathbb{F}$,
\begin{eqnarray}
\mathbb{F}_q = \mathbb{F} \circ \emph{\textbf{I}}_0^*~.
\end{eqnarray}

The scalar part $F_0$ of the $\mathbb{F}_q$ is written as,
\begin{eqnarray}
F_0 / v_0 = (\partial_0 \textbf{P}_0) \circ \emph{\textbf{I}}_0^* + (\nabla^* \cdot \textbf{P}) \circ \emph{\textbf{I}}_0^* + (\textbf{h}^* \cdot \textbf{P} / v_0 + k_{eg} \textbf{H}^* \cdot \textbf{p} /v_0) \circ \emph{\textbf{I}}_0^* ~,
\end{eqnarray}
where the $(\textbf{h}^* \cdot \textbf{P} + k_{eg} \textbf{H}^* \cdot \textbf{p}) \circ \emph{\textbf{I}}_0^*$ can be considered as one kind of the helicity.

The above is the charge continuity equation in the case for coexistence of the gravitational field and electromagnetic field when $F_0 = 0$. And the charge continuity equation is the invariant under the octonion coordinate transformation also. It states that the gravitational strength and electromagnetic strength have the influence on the charge continuity equation, although the $(\textbf{h}^* \cdot \textbf{P} + k_{eg} \textbf{H}^* \cdot \textbf{p}) \circ \emph{\textbf{I}}_0^*$ and $\triangle m$ both are usually small when the fields are weak.

\begin{table}[h]
\caption{Some physical quantities in the octonion spaces with the operator ($\lozenge + k_b \mathbb{B}$).}
\label{tab:table3}
\centering
\begin{tabular}{ll}
\hline\hline
$ definitions $                                                               & $ meanings $ \\
\hline
$\mathbb{X}$                                                                  & field quantity \\
$\mathbb{A} = \lozenge \circ \mathbb{X}$                                      & field potential \\
$\mathbb{B} = \lozenge \circ \mathbb{A}$                                      & field strength \\
$\mathbb{R}$                                                                  & radius vector \\
$\mathbb{V} = v_0 \lozenge \circ \mathbb{R}$                                  & velocity \\
$\mathbb{U} = \lozenge \circ \mathbb{V}$                                      & velocity curl \\
$\mu \mathbb{S} = - ( \lozenge + k_b \mathbb{B} )^* \circ \mathbb{B}$         & field source \\
$\mathbb{H}_b = k_b \mathbb{B}^* \cdot \mathbb{B}$                            & field strength helicity\\
$\mathbb{P} = \mu \mathbb{S} / \mu_g$                                         & linear momentum density \\
$\mathbb{\bar{R}} = \mathbb{R} + k_{rx} \mathbb{X}$                           & compounding radius vector \\
$\mathbb{L} = \mathbb{\bar{R}} \circ \mathbb{P}$                              & angular momentum density \\
$\mathbb{W} = v_0 ( \lozenge + k_b \mathbb{B} ) \circ \mathbb{L}$             & torque-energy densities \\
$\mathbb{N} = v_0 ( \lozenge + k_b \mathbb{B} )^* \circ \mathbb{W}$           & force-power density \\
$\mathbb{F} = - \mathbb{N} / (2v_0)$                                          & force density \\
$\mathbb{H}_s = k_b \mathbb{B}^* \cdot \mathbb{P}$                            & field source helicity\\
\hline\hline
\end{tabular}
\end{table}

%--3--%

\section{The fields regarding the operator ($\lozenge, \mathbb{\bar{B}}$) }

The features of the gravitational field can be described by the algebra of quaternions, including the field source and the mass continuity equation etc. The latter can be impacted by the gravitational strength and the linear momentum etc. The characteristics of the gravitational field and the electromagnetic field can be investigated simultaneously by the algebra of octonions, and the mass continuity equation will be influenced by the electromagnetic strength and the electric current directly, besides the gravitational strength and the linear momentum etc.

\subsection{Gravitational field}

In the quaternion space for electromagnetic fields, the quaternion radius vector is $\mathbb{R}_g = \Sigma (r_i \emph{\textbf{i}}_i)$, the quaternion velocity is $\mathbb{V}_g = v_0 \lozenge \circ \mathbb{R}_g = v_0^\delta + \Sigma (v_j \emph{\textbf{i}}_j)$, and the quaternion velocity curl is $\mathbb{U}_g = \lozenge \circ \mathbb{V}_g = \Sigma (u_i \emph{\textbf{i}}_i)$. The gravitational potential is $\mathbb{A}_g  = \lozenge \circ \mathbb{X}_g = \Sigma (a_i \emph{\textbf{i}}_i)$, and is defined from the quaternion quantity $\mathbb{X}_g$. While the gravitational strength is defined from the gravitational potential, that is $\mathbb{B}_g = \lozenge \circ \mathbb{A}_g = \Sigma (h_i \emph{\textbf{i}}_i)$.

The radius vector $\mathbb{R}_g$ and the quaternion $\mathbb{X}_g = \Sigma (x_i \emph{\textbf{i}}_i)$ can be combined together to become the compounding radius vector $\mathbb{\bar{R}}_g = \mathbb{R}_g + k_{rx} \mathbb{X}_g$, and the compounding quantity $\mathbb{\bar{X}}_g = \mathbb{X}_g + K_{rx} \mathbb{R}_g$. The related space is called as the quaternion compounding space, which is one kind of function space. In this compounding space, the coordinate is $\bar{r}_i = r_i + k_{rx} x_i$ for the basis vector $\emph{\textbf{i}}_i$, and the radius vector is $\mathbb{\bar{R}}_g = \Sigma (\emph{\textbf{i}}_i \bar{r}_i)$. We gain the compounding velocity $\mathbb{\bar{V}}_g = v_0 \lozenge \circ \mathbb{\bar{R}}_g = \mathbb{V}_g + v_0 k_{rx} \mathbb{A}_g$, the compounding velocity curl (or the vorticity) $\mathbb{\bar{U}}_g = \lozenge \circ \mathbb{\bar{V}}_g = \mathbb{U}_g + v_0 k_{rx} \mathbb{B}_g$, the compounding gravitational potential $\mathbb{\bar{A}}_g = \lozenge \circ \mathbb{\bar{X}}_g = \mathbb{A}_g + K_{rx} \mathbb{V}_g / v_0 $, and the compounding gravitational strength $\mathbb{\bar{B}}_g = \lozenge \circ \mathbb{\bar{A}}_g = \mathbb{B}_g + K_{rx} \mathbb{U}_g / v_0 $. Herein $K_{rx} = 1 / k_{rx}$, and $\bar{x}_i = x_i + K_{rx} r_i$.

\subsubsection{Torque and force}

The compounding source $\mathbb{\bar{S}}$ of the gravitational field in the quaternion compounding space comprises the linear momentum density $\mathbb{\bar{S}}_g = m \mathbb{\bar{V}}_g$ , and can be defined as,
\begin{eqnarray}
\mu \mathbb{\bar{S}} = - ( \lozenge + k_b \mathbb{\bar{B}}_g )^* \circ \mathbb{\bar{B}}_g = \mu_g \mathbb{\bar{S}}_g - \mathbb{\bar{B}}_g^* \circ \mathbb{\bar{B}}_g / v_0~,
\end{eqnarray}
where $\mathbb{\bar{B}}_g^* \circ \mathbb{\bar{B}}_g / (2\mu_g)$ is the field energy density of gravitational field. $\mathbb{\bar{V}}_g = \Sigma (\bar{r}_i \emph{\textbf{i}}_i  ) $ .

From the $\mathbb{\bar{R}}_g$ and compounding linear momentum density $\mathbb{\bar{P}}_g = \mu \mathbb{\bar{S}} / \mu_g$, the compounding angular momentum density is defined as $\mathbb{\bar{L}}_g = \mathbb{\bar{R}}_g \circ \mathbb{\bar{P}}_g$ . The compounding torque-energy density $\mathbb{\bar{W}}_g$ can be defined from the above compounding angular momentum density $\mathbb{\bar{L}}_g = \Sigma (\bar{l}_i \emph{\textbf{i}}_i) $ ,
\begin{eqnarray}
\mathbb{\bar{W}}_g  = v_0 ( \lozenge + k_b \mathbb{\bar{B}}_g )^* \circ \mathbb{\bar{L}}_g~,
\end{eqnarray}
where the compounding torque-energy includes the potential energy, the kinetic energy, the torque, and the work etc in the gravitational field. The $\mathbb{\bar{P}}_g = \Sigma (\bar{p}_i \emph{\textbf{i}}_i ) $ is the extension of the $\mathbb{\bar{S}}_g$ .

From the above, the compounding force-power $\mathbb{\bar{N}}_g$ density is defined as,
\begin{eqnarray}
\mathbb{\bar{N}}_g  = v_0 ( \lozenge + k_b \mathbb{\bar{B}}_g )^* \circ \mathbb{\bar{W}}_g~,
\end{eqnarray}
where the $\mathbb{\bar{N}}_g$ is the function of force density and power density in the gravitational field. And the compounding force density is $- \Sigma ( \emph{\textbf{i}}_j \bar{n}_j ) / ( 2 v_0 )$, with $\Sigma ( \emph{\textbf{i}}_j \bar{n}_j )$ being the vectorial part of $\mathbb{\bar{N}}_g$ .

\subsubsection{Field strength helicity}

In the quaternion compounding space, the compounding linear momentum density is defined from the quaternion compounding field source $\mathbb{\bar{S}}$ ,
\begin{eqnarray}
\mathbb{\bar{P}}_g = \widehat{m} v_0 + \Sigma (m \bar{v}_j \emph{\textbf{i}}_j)~,
\end{eqnarray}
where $\bar{p}_0 = \widehat{m} \bar{v}_0$, $\widehat{m} = \triangle m + m (\bar{v}_0^\delta / \bar{v}_0) $, $\bar{v}_0^\delta = v_0 \lozenge \cdot \mathbb{\bar{R}}$, $\triangle m = - k_b \mathbb{\bar{B}}_g^* \circ \mathbb{\bar{B}}_g / ( \mu_g \bar{v}_0 )$. $m$ is the inertial mass density, while the $\widehat{m}$ is the gravitational mass density. $\lozenge \cdot \mathbb{\bar{R}}$ is the scalar part of the $\lozenge \circ \mathbb{\bar{R}}$. And $\bar{r}_0$ is written as the $\bar{r}_0 = \bar{v}_0 t$, with the $\bar{v}_0$ being the coefficient for the dimensional homogeneity.

According to the quaternion feature, the gravitational mass $\widehat{m}$ is one reserved quantity, and is changed with the field strength $\mathbb{B}_g$, and the enstrophy $\textbf{u}^* \cdot \textbf{u} / 8$ etc, with the $\textbf{u}$ being the vectorial part of the vorticity $\mathbb{U}_g$ .

\subsubsection{Field source helicity}

As an especial and simple case, the $\mathbb{\bar{F}}_g$ is one part of the quaternion compounding force-power density $\mathbb{\bar{N}}_g$ , and is defined from the $\mathbb{\bar{P}}_g$ ,
\begin{eqnarray}
\mathbb{\bar{F}}_g = v_0 ( \lozenge + k_b \mathbb{\bar{B}}_g )^* \circ \mathbb{\bar{P}}_g~,
\end{eqnarray}
where the above includes the density of gravitational force, Lorentz force, and inertial force etc.

The scalar $\bar{f}_0$ of $\mathbb{\bar{F}}_g$ is written as,
\begin{eqnarray}
\bar{f}_0 / v_0 = \partial_0 \bar{p}_0 + \nabla^* \cdot \bar{\textbf{p}} + \bar{\textbf{h}}^* \cdot \bar{\textbf{p}} / v_0~,
\end{eqnarray}
where the term $\bar{\textbf{h}}^* \cdot \bar{\textbf{p}}$ is the field source helicity of the gravitational field.

The above is the mass continuity equation in the gravitational field when $\bar{f}_0 = 0$, and is influenced by the field source helicity of gravitational field, although the impact of helicity is quite weak.

\subsection{Gravitational and electromagnetic fields}

The gravitational field and electromagnetic field both can be depicted by the quaternion, and the quaternion spaces will be combined together to become the octonion space. In other words, the gravitational field and the electromagnetic field can be described with the octonion space simultaneously.

In the octonion space, the octonion basis vector is $\mathbb{E} = (\emph{\textbf{i}}_0, \emph{\textbf{i}}_1, \emph{\textbf{i}}_2, \emph{\textbf{i}}_3, \emph{\textbf{I}}_0, \emph{\textbf{I}}_1, \emph{\textbf{I}}_2, \emph{\textbf{I}}_3)$, the octonion radius vector is $\mathbb{R} = \Sigma (\emph{\textbf{i}}_i r_i + k_{eg} \emph{\textbf{I}}_i R_i )$, the octonion velocity is $\mathbb{V} = v_0 \lozenge \circ \mathbb{R} = \Sigma ( v_0^\delta + \emph{\textbf{i}}_j v_j + k_{eg} \emph{\textbf{I}}_i V_i )$, and the octonion vorticity is $\mathbb{U} = \lozenge \circ \mathbb{V} = \Sigma (\emph{\textbf{i}}_i u_i + k_{eg} \emph{\textbf{I}}_i U_i )$. The gravitational potential $\mathbb{A}_g$ and the electromagnetic potential $\mathbb{A}_e = \Sigma (\emph{\textbf{I}}_i A_i)$ are combined together to become the octonion field potential, $\mathbb{A} = \lozenge \circ \mathbb{X} = \mathbb{A}_g + k_{eg} \mathbb{A}_e $, and it can be defined from the octonion quantity $\mathbb{X} = \mathbb{X}_g + k_{eg} \mathbb{X}_e$. The octonion strength, $\mathbb{B} = \lozenge \circ \mathbb{A} = \mathbb{B}_g + k_{eg} \mathbb{B}_e$, consists of the gravitational strength $\mathbb{B}_g$ and the electromagnetic strength $\mathbb{B}_e = \lozenge \circ \mathbb{A}_e = \Sigma (H_i \emph{\textbf{I}}_i)$. Herein $\mathbb{X}_e = \Sigma (X_i \emph{\textbf{I}}_i)$.

The octonion radius vector $\mathbb{R}$ and the octonion physical quantity $\mathbb{X}$ in the octonion space can be combined together to become the compounding radius vector $\mathbb{\bar{R}} = \mathbb{R} + k_{rx} \mathbb{X} = \Sigma ( \bar{r}_i \emph{\textbf{i}}_i + k_{eg} \bar{R}_i \emph{\textbf{I}}_i )$, as well as the compounding quantity $\mathbb{\bar{X}} = \mathbb{X} + K_{rx} \mathbb{R} = \Sigma ( \bar{x}_i \emph{\textbf{i}}_i + k_{eg} \bar{X}_i \emph{\textbf{I}}_i )$ respectively. The related space is called as the octonion compounding space, which is one kind of function space. In this space, the octonion compounding field potential is $\mathbb{\bar{A}} = \lozenge \circ \mathbb{\bar{X}} = \mathbb{A} + K_{rx} \mathbb{V} / v_0 = \Sigma ( \bar{a}_i \emph{\textbf{i}}_i + k_{eg} \bar{A}_i \emph{\textbf{I}}_i )$, the octonion compounding field strength is $\mathbb{\bar{B}} = \lozenge \circ \mathbb{\bar{A}} = \mathbb{B} + K_{rx} \mathbb{U} / v_0 = \Sigma ( \bar{h}_i \emph{\textbf{i}}_i + k_{eg} \bar{H}_i \emph{\textbf{I}}_i )$, the octonion compounding velocity is $\mathbb{\bar{V}} = \mathbb{V} + v_0 k_{rx} \mathbb{A} = \Sigma ( \bar{v}_i \emph{\textbf{i}}_i + k_{eg} \bar{V}_i \emph{\textbf{I}}_i )$, as well as the octonion compounding velocity curl is $\mathbb{\bar{U}} = \mathbb{U} + v_0 k_{rx} \mathbb{B} = \Sigma ( \bar{u}_i \emph{\textbf{i}}_i + k_{eg} \bar{U}_i \emph{\textbf{I}}_i )$. Herein $\bar{R}_i = R_i + k_{rx} X_i$, and $\bar{X}_i = X_i + K_{rx} R_i$.

\begin{table}[h]
\caption{Comparison between the fields in the octonion space with that in the octonion compounding space.}
\label{tab:table3}
\centering
\begin{tabular}{lll}
\hline\hline
$ octonion~space $                                                                 &  $ octonion~compounding~space $                                & $ descriptions $ \\

\hline
$\nabla \cdot \textbf{b} = 0$                                                      &  $\nabla \cdot \bar{\textbf{b}} = 0$                                                        & (Gauss's law of gravitation) \\
$\partial_0 \textbf{b} + \nabla^* \times \textbf{g} / v_0 = 0$                     &  $\partial_0 \bar{\textbf{b}} + \nabla^* \times \bar{\textbf{g}} / v_0 = 0$                       & (Faraday's law of gravitation) \\
$\nabla^* \cdot \textbf{g} = - \widehat{m} / \varepsilon_g$                        &  $\nabla^* \cdot \bar{\textbf{g}} = - \widehat{m} / \varepsilon_g$                          & Newton's law of gravitation \\
$\partial_0 \textbf{g} / v_0 + \nabla^* \times \textbf{b} = - \mu_g \textbf{s}$    &  $\partial_0 \bar{\textbf{g}} / v_0 + \nabla^* \times \bar{\textbf{b}} = - \mu_g \bar{\textbf{s}}$       & (Ampere's law of gravitation) \\

\hline
$\nabla \cdot \textbf{B} = 0$                                                      &  $\nabla \cdot \bar{\textbf{B}} = 0$                                                        & Gauss's law of magnetism \\
$\partial_0 \textbf{B} + \nabla^* \times \textbf{E} / V_0 = 0$                     &  $\partial_0 \bar{\textbf{B}} + \nabla^* \times \bar{\textbf{E}} / V_0 = 0$                       & Faraday's law \\
$\nabla^* \cdot \textbf{E} = - (q / \varepsilon_e) \emph{\textbf{I}}_0$            &  $\nabla^* \cdot \bar{\textbf{E}} = - (q / \varepsilon_e) \emph{\textbf{I}}_0$              & Gauss's law \\
$\partial_0 \textbf{E} / V_0 + \nabla^* \times \textbf{B} = - \mu_e \textbf{S}$    &  $\partial_0 \bar{\textbf{E}} / V_0 + \nabla^* \times \bar{\textbf{B}} = - \mu_e \bar{\textbf{S}}$      & Ampere-Maxwell law \\

\hline
$\textbf{b} = 0$                                                                   &  $\bar{\textbf{b}} = 0$                      &  the extreme case of weak field   \\  $\textbf{g} = 0$                                                                   &  $\bar{\textbf{g}} = 0$                      &  the extreme case of weak field   \\  $\textbf{B} = 0$                                                                   &  $\bar{\textbf{B}} = 0$                      &  the extreme case of weak field   \\  $\textbf{E} = 0$                                                                   &  $\bar{\textbf{E}} = 0$                      &  the extreme case of weak field   \\

\hline\hline
\end{tabular}
\end{table}

\subsubsection{Torque and force}

In the coordinate system of the octonion compounding space, the coordinate is $(\bar{r}_i + k_{eg} \bar{R}_i \emph{\textbf{I}}_0)$ for the basis vector $\emph{\textbf{i}}_i$ , and the radius vector is written as $\mathbb{\bar{R}} = \Sigma \left\{\emph{\textbf{i}}_i \circ (\bar{r}_i + k_{eg} \bar{R}_i \emph{\textbf{I}}_0) \right\}$. In this octonion compounding space, the compounding source $\mathbb{\bar{S}}$ involves the linear momentum density $\mathbb{\bar{S}}_g$, and the electric current density $\mathbb{\bar{S}}_e = q \mathbb{\bar{V}}_e$. While the latter is the field source of the electromagnetic field. The octonion compounding source $\mathbb{\bar{S}}$ satisfies,
\begin{eqnarray}
\mu \mathbb{\bar{S}} = - ( \lozenge + k_b \mathbb{\bar{B}} )^* \circ \mathbb{\bar{B}} = \mu_g \mathbb{\bar{S}}_g + k_{eg} \mu_e \mathbb{\bar{S}}_e - \mathbb{\bar{B}}^* \circ \mathbb{\bar{B}} / v_0~,
\end{eqnarray}
where $\mathbb{\bar{B}} = \mathbb{\bar{B}}_g + k_{eg} \mathbb{\bar{B}}_e$; $\mathbb{\bar{B}}^* \circ \mathbb{\bar{B}}/ \mu_g = \mathbb{\bar{B}}_g^* \circ \mathbb{\bar{B}}_g / \mu_g + \mathbb{\bar{B}}_e^* \circ \mathbb{\bar{B}}_e / \mu_e$ is the field energy density.

The compounding angular momentum density is $\mathbb{\bar{L}} = \mathbb{\bar{R}} \circ \mathbb{\bar{P}}$ in the octonion compounding space. The compounding torque-energy density $\mathbb{\bar{W}}$ is defined from the compounding strength $\mathbb{\bar{B}}$ and the angular momentum density $\mathbb{\bar{L}}$ ,
\begin{eqnarray}
\mathbb{\bar{W}}  = v_0 ( \lozenge + k_b \mathbb{\bar{B}} ) \circ \mathbb{\bar{L}}~,
\end{eqnarray}
where the compounding torque-energy includes the potential energy, the kinetic energy, the torque, the gyroscopic torque, and the work etc in the gravitational field and the electromagnetic field.

The compounding force-power density $\mathbb{\bar{N}}  = v_0 ( \mathbb{\bar{B}} / v_0 + \lozenge)^* \circ \mathbb{\bar{W}}$ is defined from the compounding torque-energy density $\mathbb{\bar{W}}$. The $\mathbb{\bar{N}}$ is the function of the force density and the power density in the electromagnetic field and gravitational field. And the compounding force density is $- \Sigma ( \emph{\textbf{i}}_j \bar{n}_j ) / ( 2 v_0 )$, with the $\Sigma ( \emph{\textbf{i}}_j \bar{n}_j )$ being one vectorial part of $\mathbb{\bar{N}}$ .

\begin{table}[b]
\caption{Some physical quantities in the octonion compounding spaces with the operator ($\lozenge + k_b \mathbb{\bar{B}}$).}
\label{tab:table3}
\centering
\begin{tabular}{ll}
\hline\hline
$ definitions $                                                                                & $ meanings $ \\
\hline
$\mathbb{\bar{X}}$                                                                             & field quantity \\
$\mathbb{\bar{A}} = \lozenge \circ \mathbb{\bar{X}}$                                           & field potential \\
$\mathbb{\bar{B}} = \lozenge \circ \mathbb{\bar{A}}$                                           & field strength \\
$\mathbb{\bar{R}}$                                                                             & radius vector \\
$\mathbb{\bar{V}} = v_0 \lozenge \circ \mathbb{\bar{R}}$                                       & velocity \\
$\mathbb{\bar{U}} = \lozenge \circ \mathbb{\bar{V}}$                                           & velocity curl \\
$\mu \mathbb{\bar{S}} = - ( \lozenge + k_b \mathbb{\bar{B}} )^* \circ \mathbb{\bar{B}}$        & field source \\
$\mathbb{\bar{H}}_b = k_b \mathbb{\bar{B}}^* \cdot \mathbb{\bar{B}}$                           & field strength helicity\\
$\mathbb{\bar{P}} = \mu \mathbb{\bar{S}} / \mu_g$                                              & linear momentum density \\
$\mathbb{\bar{R}} = \mathbb{R} + k_{rx} \mathbb{X}$                                            & compounding radius vector \\
$\mathbb{\bar{L}} = \mathbb{\bar{R}} \circ \mathbb{\bar{P}}$                                   & angular momentum density \\
$\mathbb{\bar{W}} = v_0 ( \lozenge + k_b \mathbb{\bar{B}} ) \circ \mathbb{\bar{L}}$            & torque-energy densities \\
$\mathbb{\bar{N}} = v_0 ( \lozenge + k_b \mathbb{\bar{B}} )^* \circ \mathbb{\bar{W}}$          & force-power density \\
$\mathbb{\bar{F}} = - \mathbb{\bar{N}} / (2v_0)$                                               & force density \\
$\mathbb{\bar{H}}_s = k_b \mathbb{\bar{B}}^* \cdot \mathbb{\bar{P}}$                           & field source helicity\\
\hline\hline
\end{tabular}
\end{table}

\subsubsection{Field strength helicity}

In the octonion compounding space, the octonion compounding linear momentum density can be defined from the octonion compounding field source $\mathbb{\bar{S}}$ ,
\begin{eqnarray}
\mathbb{\bar{P}} = \widehat{m} v_0 + \Sigma (m \bar{v}_j \emph{\textbf{i}}_j) + \Sigma (M \bar{V}_i \emph{\textbf{I}}_i)~,
\end{eqnarray}
where $M = k_{eg} \mu_e q / \mu_g$. $\widehat{m} = \triangle m + m (\bar{v}_0^\delta / \bar{v}_0) $, $\bigtriangleup m = - k_b \mathbb{\bar{B}}^* \circ \mathbb{\bar{B}} / (\bar{v}_0 \mu_g )$; $m$ and $\widehat{m}$ are the inertial mass and the gravitational mass respectively. $\mathbb{\bar{B}}$ consists of the field strength $\mathbb{B}$ and the vorticity $\mathbb{U}$.

Making use of the octonion feature and the above, the gravitational mass $\widehat{m}$ is one reserved quantity similarly, and is changed with the gravitational strength $\mathbb{B}_g$, the electromagnetic strength $\mathbb{B}_e$, and the enstrophy $\textbf{u}^* \cdot \textbf{u} / 8$ etc, with the $\textbf{u}$ being the vectorial part of the vorticity $\mathbb{U}_g$ .

\subsubsection{Field source helicity}

In the octonion compounding space, the part force density $\mathbb{\bar{F}}$ is one part of the force-power density $\mathbb{\bar{N}}$, and is defined from the octonion linear momentum density $\mathbb{\bar{P}}$ ,
\begin{eqnarray}
\mathbb{\bar{F}} = v_0 (\lozenge + k_b \mathbb{\bar{B}})^* \circ \mathbb{\bar{P}}~,
\end{eqnarray}
where the above includes the density of the gravitational force, the Lorentz force, the inertial force, and the interacting force between electromagnetic strength with magnetic moment etc.

The scalar $\bar{f}_0$ of the $\mathbb{\bar{F}}$ is written as,
\begin{eqnarray}
\bar{f}_0 / v_0 = \partial_0 \bar{p}_0 + \nabla^* \cdot \bar{\textbf{p}} + \bar{\textbf{h}}^* \cdot \bar{\textbf{p}} / v_0 + k_{eg} \bar{\textbf{H}}^* \cdot \bar{\textbf{P}} /v_0~,
\end{eqnarray}
where $\mathbb{\bar{B}}_g = \Sigma (\bar{h}_i \emph{\textbf{i}}_i)$, $\mathbb{\bar{B}}_e = \Sigma (\bar{H}_i \emph{\textbf{I}}_i)$. $\bar{\textbf{h}} = \Sigma (\bar{h}_j \emph{\textbf{i}}_j) = \bar{\textbf{g}} / v_0 + \bar{\textbf{b}}$, $\bar{\textbf{H}} = \Sigma (\bar{H}_j \emph{\textbf{I}}_j) = \bar{\textbf{E}} / v_0 + \bar{\textbf{B}}$. The $\mathbb{\bar{B}}^* \cdot \mathbb{\bar{P}}$ is the scalar part of $\mathbb{\bar{B}}^* \circ \mathbb{\bar{P}}$. The term $k_{eg} \bar{\textbf{H}}^* \cdot \bar{\textbf{P}}$ is the field source helicity of the spinning charged objects \cite{dirac} in the electromagnetic fields, including the current helicity $k_{eg} \bar{\textbf{B}}^* \cdot \bar{\textbf{P}}$ and the helicity term $k_{eg} \bar{\textbf{E}}^* \cdot \bar{\textbf{P}} / v_0$; $\bar{\textbf{h}}^* \cdot \bar{\textbf{p}}$ is the field source helicity of the rotational objects in the gravitational fields.

The above is the mass continuity equation in the electromagnetic and gravitational fields, and is influenced by the current helicity and the enstrophy \cite{paret} etc of either the electromagnetic field or the gravitational field, although the impact of field source helicity may be quite tiny.

A new physical quantity $\mathbb{\bar{F}}_q$ can be defined from the part force density $\mathbb{\bar{F}}$,
\begin{eqnarray}
\mathbb{\bar{F}}_q = \mathbb{\bar{F}} \circ \emph{\textbf{I}}_0^*~.
\end{eqnarray}

The scalar part $\bar{F}_0$ of the octonion $\mathbb{\bar{F}}_q$ is written as,
\begin{eqnarray}
\bar{F}_0 / v_0 = (\partial_0 \bar{\textbf{P}}_0) \circ \emph{\textbf{I}}_0^* + (\nabla^* \cdot \bar{\textbf{P}}) \circ \emph{\textbf{I}}_0^* + (\bar{\textbf{h}}^* \cdot \bar{\textbf{P}} / v_0 + k_{eg} \bar{\textbf{H}}^* \cdot \bar{\textbf{p}} /v_0) \circ \emph{\textbf{I}}_0^* ~,
\end{eqnarray}
where the $(\bar{\textbf{h}}^* \cdot \bar{\textbf{P}} + k_{eg} \bar{\textbf{H}}^* \cdot \bar{\textbf{p}}) \circ \emph{\textbf{I}}_0^*$ can be considered as one kind of the helicity.

The above is the charge continuity equation in the presence of the gravitational field and the electromagnetic field when $\bar{F}_0 = 0$. And the charge continuity equation is the invariant under the octonion coordinate transformation. It means that the gravitational strength and the electromagnetic strength exert an influence on the charge continuity equation, although the $(\bar{\textbf{h}}^* \cdot \bar{\textbf{P}} + k_{eg} \bar{\textbf{H}}^* \cdot \bar{\textbf{p}}) \circ \emph{\textbf{I}}_0^*$ and $\triangle m$ both are usually slight when the fields are overlooked.

%--4--%

\section{The fields regarding the operator ($\lozenge, \mathbb{A}$) }

In the electromagnetic and gravitational fields, the operator ($\lozenge + k_b \mathbb{B}$) derives the field potential, field strength, field source, linear momentum, angular momentum, energy, torque, power, force, and helicity etc, except for the magnetic helicity. The existence of magnetic helicity reveals that there may exist other kinds of operators for some features of the electromagnetic and gravitational fields. In this section, the octonion operator ($\lozenge + k_b \mathbb{B}$) should be replaced by the new operator ($\lozenge + k_a \mathbb{A}$), with the $k_a$ being one coefficient for the dimensional homogeneity.

In terms of the new octonion operator ($\lozenge + k_a \mathbb{A}$) in the electromagnetic field \cite{lorentz} and the gravitational field, there is the field potential, the field strength, the field source, the linear momentum, the angular momentum, the energy, the torque, the power, the force, and the helicity, including the current helicity and the magnetic helicity etc.

\subsection{Field source and torque}

In the electromagnetic field and the gravitational field with the operator ($\lozenge + k_a \mathbb{A}$), the some physical quantities will remain the same, except for the gravitational mass density etc.

In the octonion space, the octonion basis vector is $ \mathbb{E} = (1, \emph{\textbf{i}}_1, \emph{\textbf{i}}_2, \emph{\textbf{i}}_3, \emph{\textbf{I}}_0, \emph{\textbf{I}}_1, \emph{\textbf{I}}_2, \emph{\textbf{I}}_3) $, the octonion radius vector is $\mathbb{R} = \Sigma ( \emph{\textbf{i}}_i r_i + k_{eg} \emph{\textbf{I}}_i R_i)$, and the octonion velocity is $\mathbb{V} = \Sigma ( \emph{\textbf{i}}_i v_i + k_{eg} \emph{\textbf{I}}_i V_i)$. Meanwhile the octonion field strength, $\mathbb{B} = (\lozenge + k_a \mathbb{A}) \circ \mathbb{A} = \mathbb{B}_g + k_{eg} \mathbb{B}_e$, is defined from the octonion field potential $\mathbb{A} = \mathbb{A}_g + k_{eg} \mathbb{A}_e$. The gravitational strength is $\mathbb{B}_g = \Sigma ( h_i \emph{\textbf{i}}_i) = \lozenge \circ \mathbb{A}_g + k_a ( \mathbb{A}_g \circ \mathbb{A}_g + k_{eg}^2 \mathbb{A}_e \circ \mathbb{A}_e )$ in the space $\mathbb{E}_g$, while the electromagnetic strength is $\mathbb{B}_e = \Sigma (H_i \emph{\textbf{I}}_i) = \lozenge \circ \mathbb{A}_e + k_a k_{eg} ( \mathbb{A}_e \circ \mathbb{A}_g + \mathbb{A}_g \circ \mathbb{A}_e)$ in the space $\mathbb{E}_e$. The gauge equations are $h_0 = 0$ and $H_0 = 0$.

The linear momentum density $\mathbb{S}_g$ is the source for the gravitational field, and the electric current density $\mathbb{S}_e$ is that for the electromagnetic field. The octonion field source $\mathbb{S}$ satisfies
\begin{eqnarray}
\mu \mathbb{S} = - (\lozenge + k_a \mathbb{A})^* \circ \mathbb{B} = \mu_g \mathbb{S}_g + k_{eg} \mu_e \mathbb{S}_e - k_a \mathbb{A}^* \circ \mathbb{B}~,
\end{eqnarray}
where the $k_a \mathbb{A}^* \cdot \mathbb{B}$ is the scalar part of the octonion $k_a \mathbb{A}^* \circ \mathbb{B}$, and is one kind of the helicity, while the $k_a \mathbb{A}^* \cdot \mathbb{B}/ (2 \mu_g)$ is one sort of the field energy density.

The octonion linear momentum density is $\mathbb{P} = \mu \mathbb{S} / \mu_g = \Sigma (p_i \emph{\textbf{i}}_i + P_i \emph{\textbf{I}}_i)$, and the octonion angular momentum density is $\mathbb{L} = (\mathbb{R} + k_{rx} \mathbb{X} ) \circ \mathbb{P} = \Sigma (l_i \emph{\textbf{i}}_i + L_i \emph{\textbf{I}}_i)$, with the octonion physical quantity $\mathbb{X}$. The octonion torque-energy density $\mathbb{W}$ is defined from the octonion angular momentum density $\mathbb{L}$ and the octonion field potential $\mathbb{A}$,
\begin{eqnarray}
\mathbb{W} = v_0 (\lozenge + k_a \mathbb{A}) \circ \mathbb{L}~,
\end{eqnarray}
where the octonion $\mathbb{W} = \Sigma (w_i \emph{\textbf{i}}_i ) + \Sigma (W_i \emph{\textbf{I}}_i )$; the $-w_0/2$ is the energy density, the $\textbf{w}/2 = \Sigma (w_j \emph{\textbf{i}}_j )/2$ is the torque density. $\textbf{p} = \Sigma (p_j \emph{\textbf{i}}_j)$; $\textbf{P}_0 = P_0 \emph{\textbf{I}}_0 $; $\textbf{P} = \Sigma (P_j \emph{\textbf{I}}_j )$.

The scalar $w_0$ of $\mathbb{W}$ is written as,
\begin{eqnarray}
w_0 / v_0 = \partial_0 l_0 + \nabla \cdot \textbf{l} + k_a \textbf{a} \cdot \textbf{l} + k_a a_0 l_0 + k_{eg} k_a ( \textbf{A} \cdot \textbf{L} + \textbf{A}_0 \circ \textbf{L}_0 )~,
\end{eqnarray}
where $-w_0/2$ includes the kinetic energy, the field energy, the work, the interacting energy between the dipole moment with the field potential, and some new terms. The octonion field potential is $\mathbb{A} = \Sigma ( a_i \emph{\textbf{i}}_i + k_{eg} A_i \emph{\textbf{I}}_i )$. $\textbf{a} = \Sigma (a_j \emph{\textbf{i}}_j)$, $\textbf{A}_0 = A_0 \emph{\textbf{I}}_0$, $\textbf{A} = \Sigma (A_j \emph{\textbf{I}}_j)$. $\textbf{l} = \Sigma (l_j \emph{\textbf{i}}_j)$, $\textbf{L}_0 = L_0 \emph{\textbf{I}}_0$, $\textbf{L} = \Sigma (L_j \emph{\textbf{I}}_j)$.

In a similar way, expressing the torque density $\textbf{w}$ of $\mathbb{W}$ as
\begin{eqnarray}
\textbf{w} / v_0 = \partial_0 \textbf{l} + \nabla l_0 + \nabla \times \textbf{l} + k_a \textbf{a} l_0 + k_a \textbf{a} \times \textbf{l} + k_a a_0 \textbf{l}
+ k_{eg} k_a ( \textbf{A} \times \textbf{L} + \textbf{A}_0 \circ \textbf{L} + \textbf{A} \circ \textbf{L}_0)~,
\end{eqnarray}
where the above encompasses some new terms of the torque density.

\subsection{Force}

In the octonion space with the operator $(\lozenge + k_a \mathbb{A})$, from the torque-energy density $\mathbb{W}$, the octonion force-power density, $\mathbb{N} = \Sigma (n_i \emph{\textbf{i}}_i) + \Sigma (N_i \emph{\textbf{I}}_i)$, is defined as follows,
\begin{eqnarray}
\mathbb{N} = v_0 (\lozenge + k_a \mathbb{A})^* \circ \mathbb{W}~,
\end{eqnarray}
where the power density is $f_0 = - n_0/(2 v_0)$, and the force density is $\textbf{f} = - \textbf{n} / (2 v_0)$. The vectorial parts are $\textbf{n} = \Sigma (n_j \emph{\textbf{i}}_j )$, $\textbf{N}_0 = N_0 \emph{\textbf{I}}_0$, and $\textbf{N} = \Sigma (N_j \emph{\textbf{I}}_j)$.

Further expressing the scalar $n_0$ of $\mathbb{N}$ as
\begin{eqnarray}
n_0 / v_0 = \partial_0 w_0 + \nabla^* \cdot \textbf{w} + k_a \textbf{a}^* \cdot \textbf{w} + k_a a_0 w_0 + k_{eg} k_a ( \textbf{A}^* \cdot \textbf{W} + \textbf{A}_0^* \circ \textbf{W}_0 )~.
\end{eqnarray}

The force density $\textbf{f}$ in the gravitational and electromagnetic fields with the operator $(\lozenge + k_a \mathbb{A})$ can be defined from the vectorial part $\textbf{n}$ of the octonion force-power density $\mathbb{N}$ ,
\begin{eqnarray}
- 2 \textbf{f} = \partial_0 \textbf{w} + \nabla^* w_0 + \nabla^* \times \textbf{w} + k_a ( \textbf{a}^* w_0 + \textbf{a}^* \times \textbf{w} + a_0 \textbf{w})
+ k_{eg} k_a ( \textbf{A}^* \times \textbf{W} + \textbf{A}_0^* \circ \textbf{W} + \textbf{A}^* \circ \textbf{W}_0)~,
\end{eqnarray}
where the force density $\textbf{f}$ includes that of the inertial force, gradient of energy, and interacting force between dipole moment with field potential etc. This force definition is a little different from that in the classical field theory, and encompasses more new force terms regarding the gradient of energy and the field potential etc. In case the field potential $\mathbb{A}$ is weak enough, the force terms about the field potential may be too tiny to be neglected.

\subsection{Helicity}

As the deduction of octonion fields with the operator $(\lozenge + k_a \mathbb{A})$, the current helicity and magnetic helicity \cite{charbonneau} may impact the gravitational mass density, the charge continuity equation, and the mass continuity equation etc.

\subsubsection{Field strength helicity}

The octonion linear momentum density $\mathbb{P} = \mu \mathbb{S} / \mu_g$ can be defined from the octonion field source $\mathbb{S}$ in the octonion space with the operator $(\lozenge + k_a \mathbb{A})$,
\begin{eqnarray}
\mathbb{P} = \widehat{m} v_0 + \Sigma (m v_j \emph{\textbf{i}}_j) + \Sigma (M V_i \emph{\textbf{I}}_i)~,
\end{eqnarray}
where $p_0 = \widehat{m} v_0$, $\widehat{m} = \triangle m + m (v_0^\delta / v_0) $, $\triangle m = - k_a \mathbb{A}^* \cdot \mathbb{B}/ ( \mu_g v_0)$. $\mathbb{A}^* \cdot \mathbb{B}$ is the scalar part of the octonion $\mathbb{A}^* \circ \mathbb{B}$.

According to the octonion features, the gravitational mass $\widehat{m}$ is one reserved scalar quantity in the above, and is changed with the field strength $\mathbb{B}_g$ and $\mathbb{B}_e$, the field potential $\mathbb{A}_g$ and $\mathbb{A}_e$, and the helicity $\mathbb{A}^* \cdot \mathbb{B}$ etc, including the magnetic helicity $ \textbf{A} \cdot \textbf{B} $.

\subsubsection{Field source helicity}

The part force density $\mathbb{F}$ is one part of the octonion force-power density $\mathbb{N}$, and is defined from the octonion linear momentum density $\mathbb{P}$ ,
\begin{eqnarray}
\mathbb{F} = v_0 (\lozenge + k_a \mathbb{A})^* \circ \mathbb{P}~,
\end{eqnarray}
where the part force density includes that of the inertial force and of the interacting force between the field potential with the linear momentum density etc.

The scalar $f_0$ of $\mathbb{F}$ is written as,
\begin{eqnarray}
f_0 / v_0= \partial_0 p_0 + \nabla^* \cdot \textbf{p} + k_a ( \textbf{a}^* \cdot \textbf{p} + a_0 p_0 ) + k_{eg} k_a ( \textbf{A}^* \cdot \textbf{P} + \textbf{A}_0^* \circ \textbf{P}_0 )~,
\end{eqnarray}
where the terms $( \textbf{a}^* \cdot \textbf{p} + a_0 p_0 )$ and $( \textbf{A}^* \cdot \textbf{P} + \textbf{A}_0^* \circ \textbf{P}_0 )$ can be considered as the helicity in the fields with the operator $(\lozenge + k_a \mathbb{A})$, as comparing to the fields with the operator $(\lozenge + k_b \mathbb{B})$.

The above is the mass continuity equation in the electromagnetic and gravitational fields with the octonion operator $(\lozenge + k_a \mathbb{A})$, and is influenced by the helicity of the electromagnetic and gravitational fields, although the impact of the helicity may be quite tiny. And the variation of the mass density with the time may not be continuous always.

A new physical quantity $\mathbb{F}_q$ can be defined from the part force density $\mathbb{F}$,
\begin{eqnarray}
\mathbb{F}_q = \mathbb{F} \circ \emph{\textbf{I}}_0^*~.
\end{eqnarray}

The scalar part $F_0$ of the $\mathbb{F}_q$ is written as,
\begin{eqnarray}
F_0 / v_0 = (\partial_0 \textbf{P}_0 + \nabla^* \cdot \textbf{P}) \circ \emph{\textbf{I}}_0^* + k_a ( \textbf{a}^* \cdot \textbf{P} + a_0 \textbf{P}_0 ) \circ \emph{\textbf{I}}_0^* + k_{eg} k_a ( \textbf{A}^* \cdot \textbf{p} + p_0 \textbf{A}_0^* ) \circ \emph{\textbf{I}}_0^*~,
\end{eqnarray}
where the $( \textbf{a}^* \cdot \textbf{P} + a_0 \textbf{P}_0 ) \circ \emph{\textbf{I}}_0^*$ and $( \textbf{A}^* \cdot \textbf{p} + p_0 \textbf{A}_0^* ) \circ \emph{\textbf{I}}_0^*$ can be considered as the helicity in the fields too.

The above is the charge continuity equation in the gravitational field and the electromagnetic field with the octonion operator $(\lozenge + k_a \mathbb{A})$ when $F_0 = 0$. And the charge continuity equation is the invariant under the octonion coordinate transformation. It states that the gravitational potential and the electromagnetic potential have the influence on the charge continuity equation, although the terms $( \textbf{a}^* \cdot \textbf{P} + a_0 \textbf{P}_0 ) \circ \emph{\textbf{I}}_0^*$, $( \textbf{A}^* \cdot \textbf{p} + p_0 \textbf{A}_0^* ) \circ \emph{\textbf{I}}_0^*$, and $\triangle m$ are usually trivial when the fields are faint. And the variation of the charge density with the time should not be continual forever.

\begin{table}[h]
\caption{Some physical quantities in the octonion spaces with the operator ($\lozenge + k_a \mathbb{A}$).}
\label{tab:table3}
\centering
\begin{tabular}{ll}
\hline\hline
$ definitions $                                                                    & $ meanings $ \\
\hline
$\mathbb{X}$                                                                       & field quantity \\
$\mathbb{A} = \lozenge \circ \mathbb{X}$                                           & field potential \\
$\mathbb{B} = ( \lozenge + k_a \mathbb{A} ) \circ \mathbb{A}$                      & field strength \\
$\mathbb{R}$                                                                       & radius vector \\
$\mathbb{V} = v_0 \lozenge \circ \mathbb{R}$                                       & velocity \\
$\mathbb{U} = \lozenge \circ \mathbb{V}$                                           & velocity curl \\
$\mu \mathbb{S} = - ( \lozenge + k_a \mathbb{A} )^* \circ \mathbb{B}$              & field source \\
$\mathbb{H}_b = k_a \mathbb{A}^* \cdot \mathbb{B}$                                 & field strength helicity\\
$\mathbb{P} = \mu \mathbb{S} / \mu_g$                                              & linear momentum density \\
$\mathbb{\bar{R}} = \mathbb{R} + k_{rx} \mathbb{X}$                                & compounding radius vector \\
$\mathbb{L} = \mathbb{\bar{R}} \circ \mathbb{P}$                                   & angular momentum density \\
$\mathbb{W} = v_0 ( \lozenge + k_a \mathbb{A} ) \circ \mathbb{L}$                  & torque-energy densities \\
$\mathbb{N} = v_0 ( \lozenge + k_a \mathbb{A} )^* \circ \mathbb{W}$                & force-power density \\
$\mathbb{F} = - \mathbb{N} / (2v_0)$                                               & force density \\
$\mathbb{H}_s = k_a \mathbb{A}^* \cdot \mathbb{P}$                                 & field source helicity\\
\hline\hline
\end{tabular}
\end{table}

%--5--%

\section{The fields regarding the operator ($\lozenge, \mathbb{A}, \mathbb{B}$) }

In the electromagnetic field and the gravitational field, the two fields with the octonion operator ($\lozenge + k_a \mathbb{A}$) can conclude some properties except for the current helicity etc, meanwhile the octonion operator ($\lozenge + k_b \mathbb{B}$) may deduce some other parts of features of two fields except for the magnetic helicity etc. Both of these two descriptions are not exact enough to describe the characteristics of the electromagnetic field and the gravitational field. In this section, the operator ($\lozenge + k_a \mathbb{A}$) and ($\lozenge + k_b \mathbb{B}$) should be replaced by the combined operator ($\lozenge + k_a \mathbb{A} + k_b \mathbb{B}$) to cover more characteristics of the electromagnetic field and the gravitational field simultaneously.

By means of the octonion operator ($\lozenge + k_a \mathbb{A} + k_b \mathbb{B}$) in the electromagnetic field and the gravitational field, we can depict the field strength, the field source, the linear momentum, the energy, the torque, the power, the force, and the helicity, including the current helicity and the magnetic helicity etc.

\subsection{Field source and torque}

In the electromagnetic field and gravitational field with the operator ($\lozenge + k_a \mathbb{A} + k_b \mathbb{B}$), the most of physical quantities will remain the same, except for the gravitational mass density and the helicity etc.

In the fields with the octonion operator ($\lozenge + k_a \mathbb{A} + k_b \mathbb{B}$), the octonion basis vector is $ \mathbb{E} = (1, \emph{\textbf{i}}_1, \emph{\textbf{i}}_2, \emph{\textbf{i}}_3, \emph{\textbf{I}}_0, \emph{\textbf{I}}_1, \emph{\textbf{I}}_2, \emph{\textbf{I}}_3) $, the octonion radius vector is $\mathbb{R} = \Sigma ( \emph{\textbf{i}}_i r_i + k_{eg} \emph{\textbf{I}}_i R_i)$, and the octonion velocity is $\mathbb{V} = \Sigma ( \emph{\textbf{i}}_i v_i + k_{eg} \emph{\textbf{I}}_i V_i)$. The octonion field strength, $\mathbb{B} = (\lozenge + k_a \mathbb{A}) \circ \mathbb{A} = \mathbb{B}_g + k_{eg} \mathbb{B}_e$, is defined from the octonion field potential $\mathbb{A} = \mathbb{A}_g + k_{eg} \mathbb{A}_e$ still. Herein the gravitational strength is $\mathbb{B}_g = \Sigma ( h_i \emph{\textbf{i}}_i) = \lozenge \circ \mathbb{A}_g + k_a ( \mathbb{A}_g \circ \mathbb{A}_g + k_{eg}^2 \mathbb{A}_e \circ \mathbb{A}_e )$, while the electromagnetic strength is $\mathbb{B}_e = \Sigma (H_i \emph{\textbf{I}}_i) = \lozenge \circ \mathbb{A}_e + k_a k_{eg} ( \mathbb{A}_e \circ \mathbb{A}_g + \mathbb{A}_g \circ \mathbb{A}_e)$. The gauge equations are $h_0 = 0$ and $H_0 = 0$.

The linear momentum density $\mathbb{S}_g$ is the source for the gravitational field, and the electric current density $\mathbb{S}_e$ is that for the electromagnetic field. The octonion field source $\mathbb{S}$ satisfies
\begin{eqnarray}
\mu \mathbb{S} = - (\lozenge + k_a \mathbb{A} + k_b \mathbb{B})^* \circ \mathbb{B} = \mu_g \mathbb{S}_g + k_{eg} \mu_e \mathbb{S}_e - k_a \mathbb{A}^* \circ \mathbb{B} - k_b \mathbb{B}^* \circ \mathbb{B} ~,
\end{eqnarray}
where the $(k_a \mathbb{A}^* \cdot \mathbb{B} + k_b \mathbb{B}^* \cdot \mathbb{B})$ is the field strength helicity, while the $(k_a \mathbb{A}^* \cdot \mathbb{B} + k_b \mathbb{B}^* \cdot \mathbb{B}) / (2 \mu_g)$ is one sort of the field energy density.

The octonion linear momentum density is $\mathbb{P} = \mu \mathbb{S} / \mu_g = \Sigma (p_i \emph{\textbf{i}}_i + P_i \emph{\textbf{I}}_i)$, the octonion angular momentum density is $\mathbb{L} = (\mathbb{R} + k_{rx} \mathbb{X} ) \circ \mathbb{P} = \Sigma (l_i \emph{\textbf{i}}_i + L_i \emph{\textbf{I}}_i)$, with the octonion physical quantity $\mathbb{X}$. The octonion torque-energy density $\mathbb{W}$ is defined from the octonion angular momentum density $\mathbb{L}$, the octonion field potential $\mathbb{A}$, and the octonion field strength $\mathbb{B}$,
\begin{eqnarray}
\mathbb{W} = v_0 (\lozenge + k_a \mathbb{A} + k_b \mathbb{B}) \circ \mathbb{L}~,
\end{eqnarray}
where the octonion $\mathbb{W} = \Sigma (w_i \emph{\textbf{i}}_i ) + \Sigma (W_i \emph{\textbf{I}}_i )$ ; the $-w_0/2$ is the energy density, the $\textbf{w}/2 = \Sigma (w_j \emph{\textbf{i}}_j )/2$ is the torque density. $\textbf{p} = \Sigma (p_j \emph{\textbf{i}}_j)$; $\textbf{P}_0 = P_0 \emph{\textbf{I}}_0 $; $\textbf{P} = \Sigma (P_j \emph{\textbf{I}}_j )$.

The scalar $w_0$ of $\mathbb{W}$ is written as,
\begin{eqnarray}
w_0 / v_0 = \partial_0 l_0 + \nabla \cdot \textbf{l} + k_a \textbf{a} \cdot \textbf{l} + k_a a_0 l_0 + k_{eg} k_a ( \textbf{A} \cdot \textbf{L} + \textbf{A}_0 \circ
\textbf{L}_0 ) + k_b ( \textbf{h} \cdot \textbf{l} + k_{eg} \textbf{H} \cdot \textbf{L} )~,
\end{eqnarray}
where $-w_0/2$ includes the kinetic energy, the gravitational potential energy, the electric potential energy, the magnetic potential energy, the field energy, the work, the interacting energy between the dipole moment with the fields, and some new energy terms. The octonion field potential is $\mathbb{A} = \Sigma ( a_i \emph{\textbf{i}}_i + k_{eg} A_i \emph{\textbf{I}}_i )$. $\mathbb{A}_g = \Sigma ( a_i \emph{\textbf{i}}_i )$, $\mathbb{A}_g = \Sigma ( A_i \emph{\textbf{I}}_i )$. $\textbf{a} = \Sigma (a_j \emph{\textbf{i}}_j)$, $\textbf{A}_0 = A_0 \emph{\textbf{I}}_0$, $\textbf{A} = \Sigma (A_j \emph{\textbf{I}}_j)$. $\textbf{l} = \Sigma (l_j \emph{\textbf{i}}_j)$, $\textbf{L}_0 = L_0 \emph{\textbf{I}}_0$, $\textbf{L} = \Sigma (L_j \emph{\textbf{I}}_j)$.

In a similar way, expressing the torque density $\textbf{w}$ of $\mathbb{W}$ as
\begin{align}
\textbf{w} / v_0 = & ~ \partial_0 \textbf{l} + \nabla l_0 + \nabla \times \textbf{l} + k_b l_0 \textbf{h} + k_b \textbf{h} \times \textbf{l} + k_{eg} k_b (\textbf{H} \times \textbf{L} + \textbf{H} \circ \textbf{L}_0)
\nonumber
\\
& + k_a \textbf{a} l_0 + k_a \textbf{a} \times \textbf{l} + k_a a_0 \textbf{l} + k_{eg} k_a ( \textbf{A} \times \textbf{L} + \textbf{A}_0 \circ \textbf{L} + \textbf{A} \circ \textbf{L}_0)~,
\end{align}
where the above encompasses some new terms of the torque density.

\subsection{Force}

In the octonion space with the operator $(\lozenge + k_a \mathbb{A} + k_b \mathbb{B})$, the octonion force-power density, $\mathbb{N} = \Sigma (n_i \emph{\textbf{i}}_i) + \Sigma (N_i \emph{\textbf{I}}_i)$, is defined from the torque-energy density $\mathbb{W}$,
\begin{eqnarray}
\mathbb{N} = v_0 (\lozenge + k_a \mathbb{A} + k_b \mathbb{B})^* \circ \mathbb{W}~,
\end{eqnarray}
where the power density is $f_0 = - n_0/(2 v_0)$, and the force density is $\textbf{f} = - \textbf{n} / (2 v_0)$. The vectorial parts are $\textbf{n} = \Sigma (n_j \emph{\textbf{i}}_j )$, $\textbf{N}_0 = N_0 \emph{\textbf{I}}_0$, and $\textbf{N} = \Sigma (N_j \emph{\textbf{I}}_j)$.

Further expressing the scalar $n_0$ of the octonion force-power density $\mathbb{N}$ as
\begin{eqnarray}
n_0 / v_0 = \partial_0 w_0 + \nabla^* \cdot \textbf{w} + k_a ( \textbf{a}^* \cdot \textbf{w} + a_0 w_0 )
 + k_{eg} k_a ( \textbf{A}^* \cdot \textbf{W} + \textbf{A}_0^* \circ \textbf{W}_0 ) + k_b ( \textbf{h}^* \cdot \textbf{w} + k_{eg} \textbf{H}^* \cdot \textbf{W} )~.
\end{eqnarray}

In the gravitational field and the electromagnetic field with the operator $(\lozenge + k_a \mathbb{A} + k_b \mathbb{B})$, the force density $\textbf{f}$ can be defined from the vectorial part $\textbf{n}$ of the octonion force-power density $\mathbb{N}$ ,
\begin{align}
- 2 \textbf{f} = & ~ \partial_0 \textbf{w} + \nabla^* w_0 + \nabla^* \times \textbf{w} + k_b w_0 \textbf{h}^* + k_b \textbf{h}^* \times \textbf{w} + k_{eg} k_b (\textbf{H}^* \times \textbf{W} + \textbf{H}^* \circ \textbf{W}_0)
\nonumber\\
& + k_a ( \textbf{a}^* w_0 + \textbf{a}^* \times \textbf{w} + a_0 \textbf{w})
+ k_{eg} k_a ( \textbf{A}^* \times \textbf{W} + \textbf{A}_0^* \circ \textbf{W} + \textbf{A}^* \circ \textbf{W}_0)~,
\end{align}
where the force density $\textbf{f}$ includes that of the inertial force, the gravitational force, the gradient of energy, Lorentz force, and the interacting force between the dipole moment with the two fields etc. This force definition is much more complex than that in the classical field theory, and includes more new force terms regarding the gradient of energy and the field potential etc. In case the field potential $\mathbb{A}$ is distinct enough, the force terms about the field potential will be detected.

\subsection{Helicity}

In the octonion space with the operator $(\lozenge + k_a \mathbb{A} + k_b \mathbb{B})$, the current helicity and magnetic helicity \cite{brandenburg} may impact the gravitational mass density, the charge continuity equation, and the mass continuity equation etc.

\subsubsection{Field strength helicity}

The octonion linear momentum density $\mathbb{P} = \mu \mathbb{S} / \mu_g$ can be defined from the octonion field source $\mathbb{S}$ in the octonion space with the operator $(\lozenge + k_a \mathbb{A} + k_b \mathbb{B})$,
\begin{eqnarray}
\mathbb{P} = \widehat{m} v_0 + \Sigma (m v_j \emph{\textbf{i}}_j) + \Sigma (M V_i \emph{\textbf{I}}_i)~,
\end{eqnarray}
where $\widehat{m} = m (v_0^\delta / v_0) + \bigtriangleup m$, $\bigtriangleup m = - (k_a \mathbb{A}^* \cdot \mathbb{B} + k_b \mathbb{B}^* \cdot \mathbb{B}) / (v_0 \mu_g )$.

According to the octonion features, the gravitational mass $\widehat{m}$ is one reserved scalar in the above, and is changed with the field strength $\mathbb{B}_g$ and $\mathbb{B}_e$, the field potential $\mathbb{A}_g$ and $\mathbb{A}_e$, and the helicity $(k_a \mathbb{A}^* \cdot \mathbb{B} + k_b \mathbb{B}^* \cdot \mathbb{B})$ etc, including the magnetic helicity $ \textbf{A} \cdot \textbf{B} $.

\subsubsection{Field source helicity}

The part force density $\mathbb{F}$ is one part of the octonion force-power density $\mathbb{N}$, and is defined from the octonion linear momentum density $\mathbb{P}$ ,
\begin{eqnarray}
\mathbb{F} = v_0 (\lozenge + k_a \mathbb{A} + k_b \mathbb{B})^* \circ \mathbb{P}~,
\end{eqnarray}
where the part force density includes that of the inertial force, gravitational force, Lorentz force, and the interacting force between the fields with the dipoles etc.

The scalar $f_0$ of $\mathbb{F}$ is written as,
\begin{eqnarray}
f_0 / v_0 =  \partial_0 p_0 + \nabla^* \cdot \textbf{p} + k_a ( \textbf{a}^* \cdot \textbf{p} + a_0 p_0 )  + k_{eg} k_a ( \textbf{A}^* \cdot \textbf{P} + \textbf{A}_0^* \circ \textbf{P}_0 )  + k_b ( \textbf{h}^* \cdot \textbf{p} + k_{eg} \textbf{H}^* \cdot \textbf{P} )~,
\end{eqnarray}
where the terms $( \textbf{a}^* \cdot \textbf{p} + a_0 p_0 )$, $( \textbf{A}^* \cdot \textbf{P} + \textbf{A}_0^* \circ \textbf{P}_0 )$, and $( \textbf{h}^* \cdot \textbf{p} + k_{eg} \textbf{H}^* \cdot \textbf{P} )$ are the helicity in the fields with the operator $(\lozenge + k_a \mathbb{A} + k_b \mathbb{B})$, including the current helicity $k_{eg} \textbf{B}^* \cdot \textbf{P} $.

The above is the mass continuity equation in the electromagnetic field and the gravitational field with the operator $(\lozenge + k_a \mathbb{A} + k_b \mathbb{B})$, and it is influenced by the helicity of the electromagnetic field and of the gravitational field. The impact of the helicity may be significant especially in the strong fields.

A new physical quantity $\mathbb{F}_q$ can be defined from the part force density $\mathbb{F}$,
\begin{eqnarray}
\mathbb{F}_q = \mathbb{F} \circ \emph{\textbf{I}}_0^*~.
\end{eqnarray}

The scalar part $F_0$ of the $\mathbb{F}_q$ is written as,
\begin{align}
F_0 / v_0 = & ~ (\partial_0 \textbf{P}_0 + \nabla^* \cdot \textbf{P}) \circ \emph{\textbf{I}}_0^* + k_a ( \textbf{a}^* \cdot \textbf{P} + a_0 \textbf{P}_0 ) \circ \emph{\textbf{I}}_0^*
\nonumber
\\
& + k_{eg} k_a ( \textbf{A}^* \cdot \textbf{p} + p_0 \textbf{A}_0^* ) \circ \emph{\textbf{I}}_0^*  + k_b ( \textbf{h}^* \cdot \textbf{P} + k_{eg} \textbf{H}^* \cdot \textbf{p} ) \circ \emph{\textbf{I}}_0^*
~,
\end{align}
where the last three terms are the helicity in the gravitational field and electromagnetic field.

The above is the charge continuity equation in the case for coexistence of the gravitational field and electromagnetic field with the octonion operator $(\lozenge + k_a \mathbb{A} + k_b \mathbb{B})$ when $F_0 = 0$. And this charge continuity equation is the invariant under the octonion coordinate transformation. It states that the potential and the strength of the gravitational and electromagnetic fields have the influence on the charge continuity equation, although the terms $( \textbf{a}^* \cdot \textbf{P} + a_0 \textbf{P}_0 ) \circ \emph{\textbf{I}}_0^*$, $( \textbf{A}^* \cdot \textbf{p} + p_0 \textbf{A}_0^* ) \circ \emph{\textbf{I}}_0^*$, $( \textbf{h}^* \cdot \textbf{P} + k_{eg} \textbf{H}^* \cdot \textbf{p} ) \circ \emph{\textbf{I}}_0^*$, and $\triangle m$ are usually trifling in the weak fields.

\begin{table}[h]
\caption{Some physical quantities in the octonion spaces with the operator ($\lozenge + k_a \mathbb{A} + k_b \mathbb{B}$).}
\label{tab:table3}
\centering
\begin{tabular}{ll}
\hline\hline
$ definitions $                                                                                                 & $ meanings $ \\
\hline
$\mathbb{X}$                                                                                                    & field quantity \\
$\mathbb{A} = \lozenge \circ \mathbb{X}$                                                                        & field potential \\
$\mathbb{B} = ( \lozenge + k_a \mathbb{A} ) \circ \mathbb{A}$                                                   & field strength \\
$\mathbb{R}$                                                                                                    & radius vector \\
$\mathbb{V} = v_0 \lozenge \circ \mathbb{R}$                                                                    & velocity \\
$\mathbb{U} = \lozenge \circ \mathbb{V}$                                                                        & velocity curl \\
$\mu \mathbb{S} = - ( \lozenge + k_a \mathbb{A} + k_b \mathbb{B} )^* \circ \mathbb{B}$                          & field source \\
$\mathbb{\mathbb{H}}_b = ( k_a \mathbb{A} + k_b \mathbb{B} )^* \cdot \mathbb{B}$                                & field strength helicity\\
$\mathbb{P} = \mu \mathbb{S} / \mu_g$                                                                           & linear momentum density \\
$\mathbb{\bar{R}} = \mathbb{R} + k_{rx} \mathbb{X}$                                                             & compounding radius vector \\
$\mathbb{L} = \mathbb{\bar{R}} \circ \mathbb{P}$                                                                & angular momentum density \\
$\mathbb{W} = v_0 ( \lozenge + k_a \mathbb{A} + k_b \mathbb{B} ) \circ \mathbb{L}$                              & torque-energy densities \\
$\mathbb{N} = v_0 ( \lozenge + k_a \mathbb{A} + k_b \mathbb{B} )^* \circ \mathbb{W}$                            & force-power density \\
$\mathbb{F} = - \mathbb{N} / (2v_0)$                                                                            & force density \\
$\mathbb{H}_s = ( k_a \mathbb{A} + k_b \mathbb{B} )^* \cdot \mathbb{P}$                                         & field source helicity\\
\hline\hline
\end{tabular}
\end{table}

%--6--%

\section{The fields regarding the operator ($\lozenge, \mathbb{\bar{A}}, \mathbb{\bar{B}}$)}

In the electromagnetic field and gravitational field, the operator ($\lozenge + k_a \mathbb{A} + k_b \mathbb{B}$) can deduce some properties of two fields, including the angular momentum, current helicity, and magnetic helicity etc. However this description can not depict the cross helicity and the kinetic helicity etc in the electromagnetic field and gravitational field. In this section, the operator ($\lozenge + k_a \mathbb{A} + k_b \mathbb{B}$) should be replaced by one new combined operator ($\lozenge + k_a \mathbb{\bar{A}} + k_b \mathbb{\bar{B}}$) to encompass more features of the electromagnetic field and gravitational field simultaneously.

By means of the operator ($\lozenge + k_a \mathbb{\bar{A}} + k_b \mathbb{\bar{B}}$) in the electromagnetic field and gravitational field, we can describe the field strength, field source, linear momentum, energy, torque, power, force, and helicity, including the current helicity, magnetic helicity, cross helicity, and kinetic helicity etc.

\subsection{Field source and torque}

In the electromagnetic field and gravitational field with the operator ($\lozenge + k_a \mathbb{\bar{A}} + k_b \mathbb{\bar{B}}$), some physical quantities will keep unchanged, except for the gravitational mass density and the helicity etc.

In the octonion compounding space with the octonion operator ($\lozenge + k_a \mathbb{\bar{A}} + k_b \mathbb{\bar{B}}$), the octonion basis vector is written as $ \mathbb{E} = (1, \emph{\textbf{i}}_1, \emph{\textbf{i}}_2, \emph{\textbf{i}}_3, \emph{\textbf{I}}_0, \emph{\textbf{I}}_1, \emph{\textbf{I}}_2, \emph{\textbf{I}}_3) $, the octonion radius vector is $\mathbb{\bar{R}} = \Sigma ( \emph{\textbf{i}}_i \bar{r}_i + k_{eg} \emph{\textbf{I}}_i \bar{R}_i)$, and the octonion velocity is $\mathbb{\bar{V}} = \Sigma ( \emph{\textbf{i}}_i \bar{v}_i + k_{eg} \emph{\textbf{I}}_i \bar{V}_i)$. The octonion field strength, $\mathbb{\bar{B}} = (\lozenge + k_a \mathbb{\bar{A}}) \circ \mathbb{\bar{A}} = \mathbb{\bar{B}}_g + k_{eg} \mathbb{\bar{B}}_e$, is defined from the octonion field potential $\mathbb{\bar{A}} = \mathbb{\bar{A}}_g + k_{eg} \mathbb{\bar{A}}_e$ still. Herein the gravitational strength is $\mathbb{\bar{B}}_g = \Sigma ( \bar{h}_i \emph{\textbf{i}}_i) = \lozenge \circ \mathbb{\bar{A}}_g + k_a ( \mathbb{\bar{A}}_g \circ \mathbb{\bar{A}}_g + k_{eg}^2 \mathbb{\bar{A}}_e \circ \mathbb{\bar{A}}_e )$, while the electromagnetic strength is $\mathbb{\bar{B}}_e = \Sigma (\bar{H}_i \emph{\textbf{I}}_i) = \lozenge \circ \mathbb{\bar{A}}_e + k_a ( \mathbb{\bar{A}}_e \circ \mathbb{\bar{A}}_g + \mathbb{\bar{A}}_g \circ \mathbb{\bar{A}}_e)$. Meanwhile the gauge equations are $\bar{h}_0 = 0$ and $\bar{H}_0 = 0$ respectively.

The linear momentum density $\mathbb{\bar{S}}_g$ is the source for the gravitational field, and the electric current density $\mathbb{\bar{S}}_e$ is that for the electromagnetic field. The octonion field source $\mathbb{\bar{S}}$ satisfies
\begin{eqnarray}
\mu \mathbb{\bar{S}} = - (\lozenge + k_a \mathbb{\bar{A}} + k_b \mathbb{\bar{B}})^* \circ \mathbb{\bar{B}} = \mu_g \mathbb{\bar{S}}_g + k_{eg} \mu_e \mathbb{\bar{S}}_e - k_a \mathbb{\bar{A}}^* \circ \mathbb{\bar{B}} - k_b \mathbb{\bar{B}}^* \circ \mathbb{\bar{B}} ~,
\end{eqnarray}
where $(k_a \mathbb{\bar{A}}^* \cdot \mathbb{\bar{B}} + k_b \mathbb{\bar{B}}^* \cdot \mathbb{\bar{B}})$ is the field strength helicity, and $(k_a \mathbb{\bar{A}}^* \cdot \mathbb{\bar{B}} + k_b \mathbb{\bar{B}}^* \cdot \mathbb{\bar{B}}) / (2 \mu_g)$ is the field energy density. $\mathbb{\bar{A}}^* \cdot \mathbb{\bar{B}}$ denotes the scalar part of the octonion $\mathbb{\bar{A}}^* \circ \mathbb{\bar{B}}$.

The octonion linear momentum density is $\mathbb{\bar{P}} = \mu \mathbb{\bar{S}} / \mu_g = \Sigma (\bar{p}_i \emph{\textbf{i}}_i + \bar{P}_i \emph{\textbf{I}}_i)$, and the octonion angular momentum density is $\mathbb{\bar{L}} = \mathbb{\bar{R}} \circ \mathbb{\bar{P}} = \Sigma (\bar{l}_i \emph{\textbf{i}}_i + \bar{L}_i \emph{\textbf{I}}_i)$. The octonion torque-energy density $\mathbb{\bar{W}}$ is defined from the octonion angular momentum density $\mathbb{\bar{L}}$, the octonion field potential $\mathbb{\bar{A}}$, and the octonion field strength $\mathbb{\bar{B}}$,
\begin{eqnarray}
\mathbb{\bar{W}} = v_0 (\lozenge + k_a \mathbb{\bar{A}} + k_b \mathbb{\bar{B}}) \circ \mathbb{\bar{L}}~,
\end{eqnarray}
where $\mathbb{\bar{W}} = \Sigma (\bar{w}_i \emph{\textbf{i}}_i ) + \Sigma (\bar{W}_i \emph{\textbf{I}}_i )$; the $-\bar{w}_0/2$ is the energy density, while the $\bar{\textbf{w}}/2 = \Sigma (\bar{w}_j \emph{\textbf{i}}_j )/2$ is the torque density. $\bar{\textbf{p}} = \Sigma (\bar{p}_j \emph{\textbf{i}}_j)$; $\bar{\textbf{P}}_0 = \bar{P}_0 \emph{\textbf{I}}_0 $; $\bar{\textbf{P}} = \Sigma (\bar{P}_j \emph{\textbf{I}}_j )$.

The scalar $\bar{w}_0$ of $\mathbb{\bar{W}}$ is written as,
\begin{eqnarray}
\bar{w}_0 / v_0 = \partial_0 \bar{l}_0 + \nabla \cdot \bar{\textbf{l}} + k_a \bar{\textbf{a}} \cdot \bar{\textbf{l}} + k_a \bar{a}_0 \bar{l}_0 + k_{eg} k_a ( \bar{\textbf{A}} \cdot \bar{\textbf{L}} + \bar{\textbf{A}}_0 \circ \bar{\textbf{L}}_0 ) + k_b ( \bar{\textbf{h}} \cdot \bar{\textbf{l}} + k_{eg} \bar{\textbf{H}} \cdot \bar{\textbf{L}} )~,
\end{eqnarray}
where $-\bar{w}_0/2$ includes the kinetic energy, the gravitational potential energy, the electric potential energy, the magnetic potential energy, the field energy, the work, the interacting energy between the dipole moment with the fields, and some new energy terms. $\bar{\textbf{a}} = \Sigma (\bar{a}_j \emph{\textbf{i}}_j)$, $\bar{\textbf{A}}_0 = \bar{A}_0 \emph{\textbf{I}}_0$, $\bar{\textbf{A}} = \Sigma (\bar{A}_j \emph{\textbf{I}}_j)$. $\bar{\textbf{l}} = \Sigma (\bar{l}_j \emph{\textbf{i}}_j)$, $\bar{\textbf{L}}_0 = \bar{L}_0 \emph{\textbf{I}}_0$, $\bar{\textbf{L}} = \Sigma (\bar{L}_j \emph{\textbf{I}}_j)$.

In a similar way, expressing the torque density $\bar{\textbf{w}}$ of $\mathbb{\bar{W}}$ as
\begin{align}
\bar{\textbf{w}} / v_0 = & ~ \partial_0 \bar{\textbf{l}} + \nabla \bar{l}_0 + \nabla \times \bar{\textbf{l}} + k_b \bar{l}_0 \bar{\textbf{h}} + k_b \bar{\textbf{h}} \times \bar{\textbf{l}} + k_{eg} k_b (\bar{\textbf{H}} \times \bar{\textbf{L}} + \bar{\textbf{H}} \circ \bar{\textbf{L}}_0)
\nonumber
\\
& + k_a \bar{\textbf{a}} \bar{l}_0 + k_a \bar{\textbf{a}} \times \bar{\textbf{l}} + k_a \bar{a}_0 \bar{\textbf{l}} + k_{eg} k_a ( \bar{\textbf{A}} \times \bar{\textbf{L}} + \bar{\textbf{A}}_0 \circ \bar{\textbf{L}} + \bar{\textbf{A}} \circ \bar{\textbf{L}}_0)~,
\end{align}
where the above encompasses some new terms of the torque density.

\subsection{Force}

In the octonion compounding space with the octonion operator $(\lozenge + k_a \mathbb{\bar{A}} + k_b \mathbb{\bar{B}})$, the octonion force-power density, $\mathbb{\bar{N}} = \Sigma (\bar{n}_i \emph{\textbf{i}}_i) + \Sigma (\bar{N}_i \emph{\textbf{I}}_i)$, is defined from the torque-energy density $\mathbb{\bar{W}}$,
\begin{eqnarray}
\mathbb{\bar{N}} = v_0 (\lozenge + k_a \mathbb{\bar{A}} + k_b \mathbb{\bar{B}})^* \circ \mathbb{\bar{W}}~,
\end{eqnarray}
where the power density is $\bar{f}_0 = - \bar{n}_0/(2 v_0)$, and the force density is $\bar{\textbf{f}} = - \bar{\textbf{n}} / (2 v_0)$. The vectorial parts are $\bar{\textbf{n}} = \Sigma (\bar{n}_j \emph{\textbf{i}}_j )$, $\bar{\textbf{N}}_0 = \bar{N}_0 \emph{\textbf{I}}_0$, and $\bar{\textbf{N}} = \Sigma (\bar{N}_j \emph{\textbf{I}}_j)$.

Further expressing the scalar $\bar{n}_0$ of the octonion force-power density $\mathbb{\bar{N}}$ as
\begin{eqnarray}
\bar{n}_0 / v_0 =  \partial_0 \bar{w}_0 + \nabla^* \cdot \bar{\textbf{w}} + k_a ( \bar{\textbf{a}}^* \cdot \bar{\textbf{w}} + \bar{a}_0 \bar{w}_0 )
+ k_{eg} k_a ( \bar{\textbf{A}}^* \cdot \bar{\textbf{W}} + \bar{\textbf{A}}_0^* \circ \bar{\textbf{W}}_0 ) + k_b ( \bar{\textbf{h}}^* \cdot \bar{\textbf{w}} + k_{eg} \bar{\textbf{H}}^* \cdot \bar{\textbf{W}} )~.
\end{eqnarray}

In the gravitational field and the electromagnetic field with the operator $(\lozenge + k_a \mathbb{\bar{A}} + k_b \mathbb{\bar{B}})$, the force density $\bar{\textbf{f}}$ can be defined from the vectorial part $\bar{\textbf{n}}$ of the octonion force-power density $\mathbb{\bar{N}}$ ,
\begin{align}
- 2 \bar{\textbf{f}} = & ~ \partial_0 \bar{\textbf{w}} + \nabla^* \bar{w}_0 + \nabla^* \times \bar{\textbf{w}} + k_b \bar{w}_0 \bar{\textbf{h}}^* + k_b \bar{\textbf{h}}^* \times \bar{\textbf{w}} + k_{eg} k_b (\bar{\textbf{H}}^* \times \bar{\textbf{W}} + \bar{\textbf{H}}^* \circ \bar{\textbf{W}}_0)
\nonumber\\
& + k_a ( \bar{\textbf{a}}^* \bar{w}_0 + \bar{\textbf{a}}^* \times \bar{\textbf{w}} + \bar{a}_0 \bar{\textbf{w}})
+ k_{eg} k_a ( \bar{\textbf{A}}^* \times \bar{\textbf{W}} + \bar{\textbf{A}}_0^* \circ \bar{\textbf{W}} + \bar{\textbf{A}}^* \circ \bar{\textbf{W}}_0)~,
\end{align}
where the force density $\bar{\textbf{f}}$ includes that of the inertial force, the gravitational force, the gradient of energy, Lorentz force, and the interacting force between the dipole moment with the fields etc. This force definition is much more complex than that in the classical field theory, and includes more new force terms related to the gradient of energy, the field potential, and the angular velocity etc. In case the vorticity $\mathbb{U}$ is quick enough, the related force terms about the vorticity will be detected.

\subsection{Helicity}

In the electromagnetic field and the gravitational field, as one necessary part of the octonion compounding space with the octonion operator $(\lozenge + k_a \mathbb{\bar{A}} + k_b \mathbb{\bar{B}})$, the helicity \cite{irvine} will impact the gravitational mass density, the charge continuity equation, and the mass continuity equation etc.

\subsubsection{Field strength helicity}

The octonion linear momentum density $\mathbb{\bar{P}} = \mu \mathbb{\bar{S}} / \mu_g$ can be defined from the octonion field source $\mathbb{\bar{S}}$ in the octonion compounding space with the operator $(\lozenge + k_a \mathbb{\bar{A}} + k_b \mathbb{\bar{B}})$,
\begin{eqnarray}
\mathbb{\bar{P}} = \widehat{m} \bar{v}_0 + \Sigma (m \bar{v}_j \emph{\textbf{i}}_j) + \Sigma (M \bar{V}_i \emph{\textbf{I}}_i)~,
\end{eqnarray}
where $\widehat{m} = m (\bar{v}_0^\delta / \bar{v}_0) + \bigtriangleup m$, $\bigtriangleup m = - (k_a \mathbb{\bar{A}}^* \cdot \mathbb{\bar{B}} + k_b \mathbb{\bar{B}}^* \cdot \mathbb{\bar{B}}) / ( \bar{v}_0 \mu_g )$ .

According to the octonion features, the gravitational mass $\widehat{m}$ is one reserved scalar in the above, and is changed with the field strength $\mathbb{\bar{B}}_g$ and $\mathbb{\bar{B}}_e$, the field potential $\mathbb{\bar{A}}_g$ and $\mathbb{\bar{A}}_e$, the enstrophy $\textbf{u}^* \cdot \textbf{u} / 8$, and the helicity $(k_a \mathbb{\bar{A}}^* \cdot \mathbb{\bar{B}} + k_b \mathbb{\bar{B}}^* \cdot \mathbb{\bar{B}})$ etc. The helicity covers the magnetic helicity $ \bar{\textbf{A}} \cdot \bar{\textbf{B}} $, the kinetic helicity $ \bar{\textbf{v}} \cdot \bar{\textbf{u}} $, the cross helicity $ \bar{\textbf{V}} \cdot \bar{\textbf{B}} $, and one new helicity term $ \bar{\textbf{A}} \cdot \bar{\textbf{U}} $ etc. Herein $ \bar{\textbf{A}} = \Sigma ( \bar{A}_j \emph{\textbf{I}}_j )$, $ \bar{\textbf{B}} = \Sigma ( \bar{B}_j \emph{\textbf{I}}_j )$, $ \bar{\textbf{V}} = \Sigma ( \bar{V}_j \emph{\textbf{I}}_j )$, $ \bar{\textbf{U}} = \Sigma ( \bar{U}_j \emph{\textbf{I}}_j )$.

\subsubsection{Field source helicity}

The part force density $\mathbb{\bar{F}}$ is one part of the octonion force-power density $\mathbb{\bar{N}}$, and is defined from the octonion linear momentum density $\mathbb{\bar{P}}$ ,
\begin{eqnarray}
\mathbb{\bar{F}} = v_0 (\lozenge + k_a \mathbb{\bar{A}} + k_b \mathbb{\bar{B}})^* \circ \mathbb{\bar{P}}~,
\end{eqnarray}
where the above part force density includes that of the inertial force, the gravitational force, Lorentz force, and the interacting force between the fields with the dipoles etc.

The scalar $\bar{f}_0$ of $\mathbb{\bar{F}}$ is written as,
\begin{eqnarray}
\bar{f}_0 / v_0 = \partial_0 \bar{p}_0 + \nabla^* \cdot \bar{\textbf{p}} + k_a ( \bar{\textbf{a}}^* \cdot \bar{\textbf{p}} + \bar{a}_0 \bar{p}_0 )  + k_{eg} k_a ( \bar{\textbf{A}}^* \cdot \bar{\textbf{P}} + \bar{\textbf{A}}_0^* \circ \bar{\textbf{P}}_0 )
+ k_b ( \bar{\textbf{h}}^* \cdot \bar{\textbf{p}} + k_{eg} \bar{\textbf{H}}^* \cdot \bar{\textbf{P}} )~,
\end{eqnarray}
where the terms $( \bar{\textbf{a}}^* \cdot \bar{\textbf{p}} + \bar{a}_0 \bar{p}_0 )$, $( \bar{\textbf{A}}^* \cdot \bar{\textbf{P}} + \bar{\textbf{A}}_0^* \circ \bar{\textbf{P}}_0 )$, and $( \bar{\textbf{h}}^* \cdot \bar{\textbf{p}} + k_{eg} \bar{\textbf{H}}^* \cdot \bar{\textbf{P}} )$ are the helicity terms in the two fields with the operator $(\lozenge + k_a \mathbb{\bar{A}} + k_b \mathbb{\bar{B}})$. The helicity covers the magnetic helicity $ \textbf{A} \cdot \textbf{B} $, the kinetic helicity $ \textbf{v} \cdot \textbf{u} $, the cross helicity $ \textbf{V} \cdot \textbf{B} $, and the current helicity $ \textbf{B}^* \cdot \textbf{P} $ etc.

The above is the mass continuity equation in the electromagnetic field and the gravitational field with the operator $(\lozenge + k_a \mathbb{\bar{A}} + k_b \mathbb{\bar{B}})$, and it is influenced by the helicity of the electromagnetic field and of the gravitational field. The impact of the helicity may be significant especially in the strong fields.

A new physical quantity $\mathbb{\bar{F}}_q$ can be defined from the part force density $\mathbb{\bar{F}}$,
\begin{eqnarray}
\mathbb{\bar{F}}_q = \mathbb{\bar{F}} \circ \emph{\textbf{I}}_0^*~.
\end{eqnarray}

The scalar part $\bar{F}_0$ of the $\mathbb{\bar{F}}_q$ is written as,
\begin{align}
\bar{F}_0 / v_0 = & ~ (\partial_0 \bar{\textbf{P}}_0 + \nabla^* \cdot \bar{\textbf{P}}) \circ \emph{\textbf{I}}_0^* + k_a ( \bar{\textbf{a}}^* \cdot \bar{\textbf{P}} + a_0 \bar{\textbf{P}}_0 ) \circ \emph{\textbf{I}}_0^*
\nonumber
\\
& + k_{eg} k_a ( \bar{\textbf{A}}^* \cdot \bar{\textbf{p}} + p_0 \bar{\textbf{A}}_0^* ) \circ \emph{\textbf{I}}_0^*  + k_b ( \bar{\textbf{h}}^* \cdot \bar{\textbf{P}} + k_{eg} \bar{\textbf{H}}^* \cdot \bar{\textbf{p}} ) \circ \emph{\textbf{I}}_0^*
~,
\end{align}
where the last three terms are the helicity terms in the gravitational field and the electromagnetic field.

The above is the charge continuity equation in the case for coexistence of the gravitational field and electromagnetic field with the octonion operator $(\lozenge + k_a \mathbb{\bar{A}} + k_b \mathbb{\bar{B}})$ when $F_0 = 0$. And this charge continuity equation is the invariant under the octonion coordinate transformation. It states that the potential and the strength of the gravitational and electromagnetic fields have the influence on the charge continuity equation, although the terms $( \bar{\textbf{a}}^* \cdot \bar{\textbf{P}} + a_0 \bar{\textbf{P}}_0 ) \circ \emph{\textbf{I}}_0^*$, $( \bar{\textbf{A}}^* \cdot \bar{\textbf{p}} + p_0 \bar{\textbf{A}}_0^* ) \circ \emph{\textbf{I}}_0^*$, $( \bar{\textbf{h}}^* \cdot \bar{\textbf{P}} + k_{eg} \bar{\textbf{H}}^* \cdot \bar{\textbf{p}} ) \circ \emph{\textbf{I}}_0^*$, and $\triangle m$ are usually trifling when the fields are weak enough.

\begin{table}[h]
\caption{Some physical quantities in the octonion compounding spaces with the operator ($\lozenge + k_a \mathbb{\bar{A}} + k_b \mathbb{\bar{B}}$).}
\label{tab:table3}
\centering
\begin{tabular}{ll}
\hline\hline
$ definitions $                                                                                                      & $ meanings $ \\
\hline
$\mathbb{\bar{X}}$                                                                                                   & field quantity \\
$\mathbb{\bar{A}} = \lozenge \circ \mathbb{\bar{X}}$                                                                 & field potential \\
$\mathbb{\bar{B}} = (\lozenge + k_a \mathbb{\bar{A}}) \circ \mathbb{\bar{A}}$                                        & field strength \\
$\mathbb{\bar{R}}$                                                                                                   & radius vector \\
$\mathbb{\bar{V}} = v_0 \lozenge \circ \mathbb{\bar{R}}$                                                             & velocity \\
$\mathbb{\bar{U}} = \lozenge \circ \mathbb{\bar{V}}$                                                                 & velocity curl \\
$\mu \mathbb{\bar{S}} = - ( \lozenge + k_a \mathbb{\bar{A}} + k_b \mathbb{\bar{B}} )^* \circ \mathbb{\bar{B}}$       & field source \\
$\mathbb{\bar{H}}_b = (k_a \mathbb{\bar{A}} + k_b \mathbb{\bar{B}})^* \cdot \mathbb{\bar{B}}$                        & field strength helicity\\
$\mathbb{\bar{P}} = \mu \mathbb{\bar{S}} / \mu_g$                                                                    & linear momentum density \\
$\mathbb{\bar{R}} = \mathbb{R} + k_{rx} \mathbb{X}$                                                                  & compounding radius vector \\
$\mathbb{\bar{L}} = \mathbb{\bar{R}} \circ \mathbb{\bar{P}}$                                                         & angular momentum density \\
$\mathbb{\bar{W}} = v_0 ( \lozenge + k_a \mathbb{\bar{A}} + k_b \mathbb{\bar{B}} ) \circ \mathbb{\bar{L}}$           & torque-energy densities \\
$\mathbb{\bar{N}} = v_0 ( \lozenge + k_a \mathbb{\bar{A}} + k_b \mathbb{\bar{B}} )^* \circ \mathbb{\bar{W}}$         & force-power density \\
$\mathbb{\bar{F}} = - \mathbb{\bar{N}} / (2v_0)$                                                                     & force density \\
$\mathbb{\bar{H}}_s = ( k_a \mathbb{\bar{A}} + k_b \mathbb{\bar{B}} )^*\cdot \mathbb{\bar{P}}$                      & field source helicity\\
\hline\hline
\end{tabular}
\end{table}

%--7--%

\section{The fields regarding the operator ($\lozenge, \mathbb{\bar{X}}, \mathbb{\bar{A}}, \mathbb{\bar{B}}$) }

In the electromagnetic field and the gravitational field, the operator ($\lozenge + k_a \mathbb{\bar{A}} + k_b \mathbb{\bar{B}}$) can conclude the most of the two fields' properties, including the angular momentum, the current helicity, the magnetic helicity, the cross helicity, and the kinetic helicity etc of the rotational objects and spinning charged objects. The results arouse us to speculate that there may be other physical quantities affecting the helicities. In this section, the operator ($\lozenge + k_a \mathbb{\bar{A}} + k_b \mathbb{\bar{B}}$) should be replaced by one new combined operator ($\lozenge + k_x \mathbb{\bar{X}} + k_a \mathbb{\bar{A}} + k_b \mathbb{\bar{B}}$) to contain more characteristics of the electromagnetic field and gravitational field simultaneously, with the $k_x$ being the coefficient for the dimensional homogeneity.

By means of the operator ($\lozenge + k_x \mathbb{\bar{X}} + k_a \mathbb{\bar{A}} + k_b \mathbb{\bar{B}}$) in the electromagnetic and gravitational fields, we can depict the field strength, the field source, the linear momentum, the energy, the torque, the force, and the helicity, including the influence of $\mathbb{X}$ on the gravitational mass, the charge continuity equation, and the mass continuity equation.

\subsection{Field source and torque}

In the electromagnetic field and the gravitational field with the operator ($\lozenge + k_x \mathbb{\bar{X}} + k_a \mathbb{\bar{A}} + k_b \mathbb{\bar{B}}$), some quantities will keep unchanged, except for the gravitational mass density and the helicity etc.

In the octonion compounding space with the operator ($\lozenge + k_x \mathbb{\bar{X}} + k_a \mathbb{\bar{A}} + k_b \mathbb{\bar{B}}$), the octonion basis vector is $ \mathbb{E} = (1, \emph{\textbf{i}}_1, \emph{\textbf{i}}_2, \emph{\textbf{i}}_3, \emph{\textbf{I}}_0, \emph{\textbf{I}}_1, \emph{\textbf{I}}_2, \emph{\textbf{I}}_3) $, the radius vector is $\mathbb{\bar{R}} = \Sigma ( \emph{\textbf{i}}_i \bar{r}_i + k_{eg} \emph{\textbf{I}}_i \bar{R}_i)$, and the velocity is $\mathbb{\bar{V}} = \Sigma ( \emph{\textbf{i}}_i \bar{v}_i + k_{eg} \emph{\textbf{I}}_i \bar{V}_i)$. The octonion field potential is defined as $\mathbb{\bar{A}} = (\lozenge + k_x \mathbb{\bar{X}}) \circ \mathbb{\bar{X}} = \mathbb{\bar{A}}_g + k_{eg} \mathbb{\bar{A}}_e$. The gravitational potential is $\mathbb{\bar{A}}_g = \Sigma ( \bar{x}_i \emph{\textbf{i}}_i) = \lozenge \circ \mathbb{\bar{X}}_g + k_x ( \mathbb{\bar{X}}_g \circ \mathbb{\bar{X}}_g + k_{eg}^2 \mathbb{\bar{X}}_e \circ \mathbb{\bar{X}}_e )$, while the electromagnetic potential is $\mathbb{\bar{A}}_e = \Sigma (\bar{X}_i \emph{\textbf{I}}_i) = \lozenge \circ \mathbb{\bar{X}}_e + k_x ( \mathbb{\bar{X}}_e \circ \mathbb{\bar{X}}_g + \mathbb{\bar{X}}_g \circ \mathbb{\bar{X}}_e)$. Meanwhile the octonion field strength is defined as $\mathbb{\bar{B}} = (\lozenge + k_x \mathbb{\bar{X}} + k_a \mathbb{\bar{A}}) \circ \mathbb{\bar{A}} = \mathbb{\bar{B}}_g + k_{eg} \mathbb{\bar{B}}_e$. The gravitational strength is $\mathbb{\bar{B}}_g = \Sigma ( \bar{h}_i \emph{\textbf{i}}_i) = \lozenge \circ \mathbb{\bar{A}}_g + k_x ( \mathbb{\bar{X}}_g \circ \mathbb{\bar{A}}_g + k_{eg}^2 \mathbb{\bar{X}}_e \circ \mathbb{\bar{A}}_e ) + k_a ( \mathbb{\bar{A}}_g \circ \mathbb{\bar{A}}_g + k_{eg}^2 \mathbb{\bar{A}}_e \circ \mathbb{\bar{A}}_e )$, while the electromagnetic strength is $\mathbb{\bar{B}}_e = \Sigma (\bar{H}_i \emph{\textbf{I}}_i) = \lozenge \circ \mathbb{\bar{A}}_e + k_x ( \mathbb{\bar{X}}_e \circ \mathbb{\bar{A}}_g + \mathbb{\bar{X}}_g \circ \mathbb{\bar{A}}_e) + k_a ( \mathbb{\bar{A}}_e \circ \mathbb{\bar{A}}_g + \mathbb{\bar{A}}_g \circ \mathbb{\bar{A}}_e)$. The gauge equations are $\bar{h}_0 = 0$ and $\bar{H}_0 = 0$.

The linear momentum density $\mathbb{\bar{S}}_g$ is the source for the gravitational field, and the electric current density $\mathbb{\bar{S}}_e$ is that for the electromagnetic field. The octonion field source $\mathbb{\bar{S}}$ satisfies
\begin{eqnarray}
\mu \mathbb{\bar{S}} =  - (\lozenge + k_x \mathbb{\bar{X}} + k_a \mathbb{\bar{A}} + k_b \mathbb{\bar{B}})^* \circ \mathbb{\bar{B}}
=  \mu_g \mathbb{\bar{S}}_g + k_{eg} \mu_e \mathbb{\bar{S}}_e - k_x \mathbb{\bar{X}}^* \circ \mathbb{\bar{B}} - k_a \mathbb{\bar{A}}^* \circ \mathbb{\bar{B}} - k_b \mathbb{\bar{B}}^* \circ \mathbb{\bar{B}} ~,
\end{eqnarray}
where $(k_x \mathbb{\bar{X}}^* \cdot \mathbb{\bar{B}} + k_a \mathbb{\bar{A}}^* \cdot \mathbb{\bar{B}} + k_b \mathbb{\bar{B}}^* \cdot \mathbb{\bar{B}})$ is the field strength helicity.

The octonion linear momentum density is $\mathbb{\bar{P}} = \mu \mathbb{\bar{S}} / \mu_g = \Sigma (\bar{p}_i \emph{\textbf{i}}_i + \bar{P}_i \emph{\textbf{I}}_i)$, and the octonion angular momentum density is $\mathbb{\bar{L}} = \mathbb{\bar{R}} \circ \mathbb{\bar{P}} = \Sigma (\bar{l}_i \emph{\textbf{i}}_i + \bar{L}_i \emph{\textbf{I}}_i)$. The octonion torque-energy density $\mathbb{\bar{W}}$ is defined from the octonion angular momentum density $\mathbb{\bar{L}}$, the octonion field potential $\mathbb{\bar{A}}$, and the octonion field strength $\mathbb{\bar{B}}$ etc,
\begin{eqnarray}
\mathbb{\bar{W}} = v_0 (\lozenge + k_x \mathbb{\bar{X}} + k_a \mathbb{\bar{A}} + k_b \mathbb{\bar{B}}) \circ \mathbb{\bar{L}}~,
\end{eqnarray}
where the octonion $\mathbb{\bar{W}} = \Sigma (\bar{w}_i \emph{\textbf{i}}_i ) + \Sigma (\bar{W}_i \emph{\textbf{I}}_i )$; the $-\bar{w}_0/2$ is the energy density, and the $\bar{\textbf{w}}/2 = \Sigma (\bar{w}_j \emph{\textbf{i}}_j )/2$ is the torque density. $\bar{\textbf{p}} = \Sigma (\bar{p}_j \emph{\textbf{i}}_j)$; $\bar{\textbf{P}}_0 = \bar{P}_0 \emph{\textbf{I}}_0 $; $\bar{\textbf{P}} = \Sigma (\bar{P}_j \emph{\textbf{I}}_j )$.

The scalar $\bar{w}_0$ of $\mathbb{\bar{W}}$ is written as,
\begin{align}
\bar{w}_0 / v_0 = & ~ \partial_0 \bar{l}_0 + \nabla \cdot \bar{\textbf{l}} + k_a \bar{\textbf{a}} \cdot \bar{\textbf{l}} + k_a \bar{a}_0 \bar{l}_0 + k_{eg} k_a ( \bar{\textbf{A}} \cdot \bar{\textbf{L}} + \bar{\textbf{A}}_0 \circ \bar{\textbf{L}}_0 ) + k_b ( \bar{\textbf{h}} \cdot \bar{\textbf{l}} + k_{eg} \bar{\textbf{H}} \cdot \bar{\textbf{L}} )
\nonumber\\
&
+ k_x \bar{\textbf{x}} \cdot \bar{\textbf{l}} + k_x \bar{x}_0 \bar{l}_0 + k_{eg} k_x ( \bar{\textbf{X}} \cdot \bar{\textbf{L}} + \bar{\textbf{X}}_0 \circ \bar{\textbf{L}}_0 )~,
\end{align}
where $-\bar{w}_0/2$ includes the kinetic energy, gravitational potential energy, electric potential energy, magnetic potential energy, field energy, work, the interacting energy between the dipole moment with the fields, and some new energy terms. $\bar{\textbf{x}} = \Sigma (\bar{x}_j \emph{\textbf{i}}_j)$, $\bar{\textbf{X}}_0 = \bar{X}_0 \emph{\textbf{I}}_0$, $\bar{\textbf{X}} = \Sigma (\bar{X}_j \emph{\textbf{I}}_j)$.

In a similar way, expressing the torque density $\bar{\textbf{w}}$ of $\mathbb{\bar{W}}$ as
\begin{align}
\bar{\textbf{w}} / v_0 = & ~ \partial_0 \bar{\textbf{l}} + \nabla \bar{l}_0 + \nabla \times \bar{\textbf{l}} + k_b \bar{l}_0 \bar{\textbf{h}} + k_b \bar{\textbf{h}} \times \bar{\textbf{l}} + k_{eg} k_b (\bar{\textbf{H}} \times \bar{\textbf{L}} + \bar{\textbf{H}} \circ \bar{\textbf{L}}_0)
\nonumber
\\
& + k_a \bar{\textbf{a}} \bar{l}_0 + k_a \bar{\textbf{a}} \times \bar{\textbf{l}} + k_a \bar{a}_0 \bar{\textbf{l}} + k_{eg} k_a ( \bar{\textbf{A}} \times \bar{\textbf{L}} + \bar{\textbf{A}}_0 \circ \bar{\textbf{L}} + \bar{\textbf{A}} \circ \bar{\textbf{L}}_0)\nonumber
\\
& + k_x \bar{\textbf{x}} \bar{l}_0 + k_x \bar{\textbf{x}} \times \bar{\textbf{l}} + k_x \bar{x}_0 \bar{\textbf{l}} + k_{eg} k_x ( \bar{\textbf{X}} \times \bar{\textbf{L}} + \bar{\textbf{X}}_0 \circ \bar{\textbf{L}} + \bar{\textbf{X}} \circ \bar{\textbf{L}}_0)~,
\end{align}
where the above encompasses some new terms of the torque density.

\subsection{Force}

In the octonion compounding space with the operator $(\lozenge + k_x \mathbb{\bar{X}} + k_a \mathbb{\bar{A}} + k_b \mathbb{\bar{B}})$, the octonion force-power density, $\mathbb{\bar{N}} = \Sigma (\bar{n}_i \emph{\textbf{i}}_i) + \Sigma (\bar{N}_i \emph{\textbf{I}}_i)$, is defined from the torque-energy density $\mathbb{\bar{W}}$,
\begin{eqnarray}
\mathbb{\bar{N}} = v_0 (\lozenge + k_x \mathbb{\bar{X}} + k_a \mathbb{\bar{A}} + k_b \mathbb{\bar{B}})^* \circ \mathbb{\bar{W}}~,
\end{eqnarray}
where the power density is $\bar{f}_0 = - \bar{n}_0/(2 v_0)$, and the force density is $\bar{\textbf{f}} = - \bar{\textbf{n}} / (2 v_0)$. The vectorial parts are $\bar{\textbf{n}} = \Sigma (\bar{n}_j \emph{\textbf{i}}_j )$, $\bar{\textbf{N}}_0 = \bar{N}_0 \emph{\textbf{I}}_0$, and $\bar{\textbf{N}} = \Sigma (\bar{N}_j \emph{\textbf{I}}_j)$.

Further expressing the scalar $\bar{n}_0$ of the octonion force-power density $\mathbb{\bar{N}}$ as
\begin{align}
\bar{n}_0 / v_0 = & ~ \partial_0 \bar{w}_0 + \nabla^* \cdot \bar{\textbf{w}} + k_b ( \bar{\textbf{h}}^* \cdot \bar{\textbf{w}} + k_{eg} \bar{\textbf{H}}^* \cdot \bar{\textbf{W}} )
+ k_a ( \bar{\textbf{a}}^* \cdot \bar{\textbf{w}} + \bar{a}_0 \bar{w}_0 )
+ k_{eg} k_a ( \bar{\textbf{A}}^* \cdot \bar{\textbf{W}} + \bar{\textbf{A}}_0^* \circ \bar{\textbf{W}}_0 )
\nonumber\\
&
+ k_x ( \bar{\textbf{x}}^* \cdot \bar{\textbf{w}} + \bar{x}_0 \bar{w}_0 )
+ k_{eg} k_x ( \bar{\textbf{X}}^* \cdot \bar{\textbf{W}} + \bar{\textbf{X}}_0^* \circ \bar{\textbf{W}}_0 )
~.
\end{align}

In the gravitational field and the electromagnetic field with the operator $(\lozenge + k_x \mathbb{\bar{X}} + k_a \mathbb{\bar{A}} + k_b \mathbb{\bar{B}})$, the force density $\bar{\textbf{f}}$ can be defined from the vectorial part $\bar{\textbf{n}}$ of the octonion force-power density $\mathbb{\bar{N}}$ ,
\begin{align}
- 2 \bar{\textbf{f}} = & ~ \partial_0 \bar{\textbf{w}} + \nabla^* \bar{w}_0 + \nabla^* \times \bar{\textbf{w}} + k_b \bar{w}_0 \bar{\textbf{h}}^* + k_b \bar{\textbf{h}}^* \times \bar{\textbf{w}} + k_{eg} k_b (\bar{\textbf{H}}^* \times \bar{\textbf{W}} + \bar{\textbf{H}}^* \circ \bar{\textbf{W}}_0)
\nonumber\\
& + k_a ( \bar{\textbf{a}}^* \bar{w}_0 + \bar{\textbf{a}}^* \times \bar{\textbf{w}} + \bar{a}_0 \bar{\textbf{w}})
+ k_{eg} k_a ( \bar{\textbf{A}}^* \times \bar{\textbf{W}} + \bar{\textbf{A}}_0^* \circ \bar{\textbf{W}} + \bar{\textbf{A}}^* \circ \bar{\textbf{W}}_0)
\nonumber\\
& + k_x ( \bar{\textbf{x}}^* \bar{w}_0 + \bar{\textbf{x}}^* \times \bar{\textbf{w}} + \bar{x}_0 \bar{\textbf{w}})
+ k_{eg} k_x ( \bar{\textbf{X}}^* \times \bar{\textbf{W}} + \bar{\textbf{X}}_0^* \circ \bar{\textbf{W}} + \bar{\textbf{X}}^* \circ \bar{\textbf{W}}_0)~,
\end{align}
where the force density $\bar{\textbf{f}}$ includes that of the inertial force, gravitational force, gradient of energy, Lorentz force, and the interacting force between the dipole moment with the fields etc. This force definition includes more new force terms regarding the gradient of energy, the field potential, and the angular velocity etc. In case the octonion $\mathbb{\bar{X}}$ is distinct enough, the related force terms could be detected.

\subsection{Helicity}

In the octonion compounding space with the operator $(\lozenge + k_x \mathbb{\bar{X}} + k_a \mathbb{\bar{A}} + k_b \mathbb{\bar{B}})$, some helicity terms will impact the gravitational mass density, the charge continuity equation, and the mass continuity equation etc.

\subsubsection{Field strength helicity}

The octonion linear momentum density $\mathbb{\bar{P}} = \mu \mathbb{\bar{S}} / \mu_g$ can be defined from the octonion field source $\mathbb{\bar{S}}$ in the octonion compounding space with the operator $(\lozenge + k_x \mathbb{\bar{X}} + k_a \mathbb{\bar{A}} + k_b \mathbb{\bar{B}})$,
\begin{eqnarray}
\mathbb{\bar{P}} = \widehat{m} \bar{v}_0 + \Sigma (m \bar{v}_j \emph{\textbf{i}}_j) + \Sigma (M \bar{V}_i \emph{\textbf{I}}_i)~,
\end{eqnarray}
where $\widehat{m} = m (\bar{v}_0^\delta / \bar{v}_0) + \bigtriangleup m$, $\bigtriangleup m = - (k_x \mathbb{\bar{X}}^* \cdot \mathbb{\bar{B}} + k_a \mathbb{\bar{A}}^* \cdot \mathbb{\bar{B}} + k_b \mathbb{\bar{B}}^* \cdot \mathbb{\bar{B}}) / ( \bar{v}_0 \mu_g )$ .

According to the octonion features, the gravitational mass $\widehat{m}$ is one reserved scalar in the above, and is changed with the field strength $\mathbb{\bar{B}}_g$ and $\mathbb{\bar{B}}_e$, the field potential $\mathbb{\bar{A}}_g$ and $\mathbb{\bar{A}}_e$, the helicity $(k_x \mathbb{\bar{X}}^* \cdot \mathbb{\bar{B}} + k_a \mathbb{\bar{A}}^* \cdot \mathbb{\bar{B}} + k_b \mathbb{\bar{B}}^* \cdot \mathbb{\bar{B}})$, the enstrophy $\textbf{u}^* \cdot \textbf{u} / 8$, and the octonion $\mathbb{\bar{X}}_g$ and $\mathbb{\bar{X}}_e$ etc in the octonion compounding space. The helicity covers the magnetic helicity $ \bar{\textbf{A}} \cdot \bar{\textbf{B}} $, the kinetic helicity $ \bar{\textbf{v}} \cdot \bar{\textbf{u}} $, the cross helicity $ \bar{\textbf{V}} \cdot \bar{\textbf{B}} $, and one new helicity term $ \bar{\textbf{A}} \cdot \bar{\textbf{U}} $ etc.

\begin{table}[b]
\caption{Some quantities in the octonion compounding spaces with the operator ($\lozenge + k_x \mathbb{\bar{X}} + k_a \mathbb{\bar{A}} + k_b \mathbb{\bar{B}}$).}
\label{tab:table3}
\centering
\begin{tabular}{ll}
\hline\hline
$ definitions $                                                                                                                          & $ meanings $ \\
\hline
$\mathbb{\bar{X}}$                                                                                                                       & field quantity \\
$\mathbb{\bar{A}} = (\lozenge + k_x \mathbb{\bar{X}}) \circ \mathbb{\bar{X}}$                                                            & field potential \\
$\mathbb{\bar{B}} = (\lozenge + k_x \mathbb{\bar{X}} + k_a \mathbb{\bar{A}}) \circ \mathbb{\bar{A}}$                                     & field strength \\
$\mathbb{\bar{R}}$                                                                                                                       & radius vector \\
$\mathbb{\bar{V}} = v_0 \lozenge \circ \mathbb{\bar{R}}$                                                                                 & velocity \\
$\mathbb{\bar{U}} = \lozenge \circ \mathbb{\bar{V}}$                                                                                     & velocity curl \\
$\mu \mathbb{\bar{S}} = - ( \lozenge + k_x \mathbb{\bar{X}} + k_a \mathbb{\bar{A}} + k_b \mathbb{\bar{B}} )^* \circ \mathbb{\bar{B}}$    & field source \\
$\mathbb{\bar{H}}_b = (k_x \mathbb{\bar{X}} + k_a \mathbb{\bar{A}} + k_b \mathbb{\bar{B}})^* \cdot \mathbb{\bar{B}}$                     & field strength helicity\\
$\mathbb{\bar{P}} = \mu \mathbb{\bar{S}} / \mu_g$                                                                                        & linear momentum density \\
$\mathbb{\bar{R}} = \mathbb{R} + k_{rx} \mathbb{X}$                                                                                      & compounding radius vector \\
$\mathbb{\bar{L}} = \mathbb{\bar{R}} \circ \mathbb{\bar{P}}$                                                                             & angular momentum density \\
$\mathbb{\bar{W}} = v_0 ( \lozenge + k_x \mathbb{\bar{X}} + k_a \mathbb{\bar{A}} + k_b \mathbb{\bar{B}} ) \circ \mathbb{\bar{L}}$        & torque-energy densities \\
$\mathbb{\bar{N}} = v_0 ( \lozenge + k_x \mathbb{\bar{X}} + k_a \mathbb{\bar{A}} + k_b \mathbb{\bar{B}} )^* \circ \mathbb{\bar{W}}$      & force-power density \\
$\mathbb{\bar{F}} = - \mathbb{\bar{N}} / (2v_0)$                                                                                         & force density \\
$\mathbb{\bar{H}}_s = ( k_x \mathbb{\bar{X}} + k_a \mathbb{\bar{A}} + k_b \mathbb{\bar{B}} )^* \cdot \mathbb{\bar{P}}$                   & field source helicity\\
\hline\hline
\end{tabular}
\end{table}

\subsubsection{Field source helicity}

The part force density $\mathbb{\bar{F}}$ is one part of the octonion force-power density $\mathbb{\bar{N}}$, and is defined from the octonion linear momentum density $\mathbb{\bar{P}}$ ,
\begin{eqnarray}
\mathbb{\bar{F}} = v_0 (\lozenge + k_x \mathbb{\bar{X}} + k_a \mathbb{\bar{A}} + k_b \mathbb{\bar{B}})^* \circ \mathbb{\bar{P}}~,
\end{eqnarray}
where the part force density includes that of the inertial force, gravitational force, Lorentz force, and the interacting force between the fields with the dipoles etc.

The scalar $\bar{f}_0$ of $\mathbb{\bar{F}}$ is written as,
\begin{align}
\bar{f}_0 / v_0 = & ~ \partial_0 \bar{p}_0 + \nabla^* \cdot \bar{\textbf{p}} + k_b ( \bar{\textbf{h}}^* \cdot \bar{\textbf{p}} + k_{eg} \bar{\textbf{H}}^* \cdot \bar{\textbf{P}} )
\nonumber\\
&
+ k_a ( \bar{\textbf{a}}^* \cdot \bar{\textbf{p}} + \bar{a}_0 \bar{p}_0 ) + k_{eg} k_a ( \bar{\textbf{A}}^* \cdot \bar{\textbf{P}} + \bar{\textbf{A}}_0^* \circ \bar{\textbf{P}}_0 )
\nonumber\\
&
+ k_x ( \bar{\textbf{x}}^* \cdot \bar{\textbf{p}} + \bar{x}_0 \bar{p}_0 ) + k_{eg} k_x ( \bar{\textbf{X}}^* \cdot \bar{\textbf{P}} + \bar{\textbf{X}}_0^* \circ \bar{\textbf{P}}_0 )~,
\end{align}
where the field source helicity in the fields with the operator $(\lozenge + k_x \mathbb{\bar{X}} + k_a \mathbb{\bar{A}} + k_b \mathbb{\bar{B}})$ covers the terms $( \bar{\textbf{a}}^* \cdot \bar{\textbf{p}} + \bar{a}_0 \bar{p}_0 )$, $( \bar{\textbf{A}}^* \cdot \bar{\textbf{P}} + \bar{\textbf{A}}_0^* \circ \bar{\textbf{P}}_0 )$, $( \bar{\textbf{h}}^* \cdot \bar{\textbf{p}} + k_{eg} \bar{\textbf{H}}^* \cdot \bar{\textbf{P}} )$, $(\bar{\textbf{x}}^* \cdot \bar{\textbf{p}} + \bar{x}_0 \bar{p}_0)$, and $(\bar{\textbf{X}}^* \cdot \bar{\textbf{P}} + \bar{\textbf{X}}_0^* \circ \bar{\textbf{P}}_0)$ etc. And they include the magnetic helicity $ \textbf{A} \cdot \textbf{B} $, the kinetic helicity $ \textbf{v} \cdot \textbf{u} $, the cross helicity $ \textbf{V} \cdot \textbf{B} $, and the current helicity $ \textbf{B}^* \cdot \textbf{P} $ etc.

The above is the mass continuity equation in the electromagnetic field and the gravitational field with the operator $(\lozenge + k_x \mathbb{\bar{X}} + k_a \mathbb{\bar{A}} + k_b \mathbb{\bar{B}})$, and it is influenced by the helicity \cite{aluie} of electromagnetic field and of gravitational field. The impact of the helicity may be detected in the strong fields.

A new physical quantity $\mathbb{\bar{F}}_q$ can be defined from the part force density $\mathbb{\bar{F}}$,
\begin{eqnarray}
\mathbb{\bar{F}}_q = \mathbb{\bar{F}} \circ \emph{\textbf{I}}_0^*~.
\end{eqnarray}

The scalar part $\bar{F}_0$ of the $\mathbb{\bar{F}}_q$ is written as,
\begin{align}
\bar{F}_0 / v_0 = & ~ (\partial_0 \bar{\textbf{P}}_0 + \nabla^* \cdot \bar{\textbf{P}}) \circ \emph{\textbf{I}}_0^* + k_a ( \bar{\textbf{a}}^* \cdot \bar{\textbf{P}} + a_0 \bar{\textbf{P}}_0 ) \circ \emph{\textbf{I}}_0^* + k_{eg} k_a ( \bar{\textbf{A}}^* \cdot \bar{\textbf{p}} + p_0 \bar{\textbf{A}}_0^* ) \circ \emph{\textbf{I}}_0^*
\nonumber
\\
&
+ k_x ( \bar{\textbf{x}}^* \cdot \bar{\textbf{P}} + x_0 \bar{\textbf{P}}_0 ) \circ \emph{\textbf{I}}_0^* + k_{eg} k_x ( \bar{\textbf{X}}^* \cdot \bar{\textbf{p}} + p_0 \bar{\textbf{X}}_0^* ) \circ \emph{\textbf{I}}_0^*
+ k_b ( \bar{\textbf{h}}^* \cdot \bar{\textbf{P}} + k_{eg} \bar{\textbf{H}}^* \cdot \bar{\textbf{p}} ) \circ \emph{\textbf{I}}_0^*
~,
\end{align}
where the last five terms are the helicity in the gravitational field and electromagnetic field.

The above is the charge continuity equation in the case for coexistence of the gravitational field and electromagnetic field with the octonion operator $(\lozenge + k_x \mathbb{\bar{X}} + k_a \mathbb{\bar{A}} + k_b \mathbb{\bar{B}})$ when $F_0 = 0$. And this charge continuity equation is the invariant under the octonion coordinate transformation. It states that the potential $\mathbb{\bar{A}}$, the strength $\mathbb{\bar{B}}$ and the octonion $\mathbb{\bar{X}}$ of the gravitational and electromagnetic fields have the influence on the charge continuity equation, although the terms $( \bar{\textbf{a}}^* \cdot \bar{\textbf{P}} + a_0 \bar{\textbf{P}}_0 ) \circ \emph{\textbf{I}}_0^*$, $( \bar{\textbf{A}}^* \cdot \bar{\textbf{p}} + p_0 \bar{\textbf{A}}_0^* ) \circ \emph{\textbf{I}}_0^*$, $( \bar{\textbf{h}}^* \cdot \bar{\textbf{P}} + k_{eg} \bar{\textbf{H}}^* \cdot \bar{\textbf{p}} ) \circ \emph{\textbf{I}}_0^*$, $( \bar{\textbf{x}}^* \cdot \bar{\textbf{P}} + x_0 \bar{\textbf{P}}_0 ) \circ \emph{\textbf{I}}_0^*$, $( \bar{\textbf{X}}^* \cdot \bar{\textbf{p}} + p_0 \bar{\textbf{X}}_0^* ) \circ \emph{\textbf{I}}_0^*$, and $\triangle m$ are usually trifling when the fields are faint enough.

%--8--%

\section{The fields regarding the operator ($\lozenge, \mathbb{\bar{X}}, \mathbb{\bar{A}}, \mathbb{\bar{B}}, \mathbb{\bar{S}}$) }

In the electromagnetic and gravitational fields, the octonion operator ($\lozenge + k_x \mathbb{\bar{X}} + k_a \mathbb{\bar{A}} + k_b \mathbb{\bar{B}}$) can deduce some physical properties of two fields, including the linear momentum, the angular momentum, and some helicities of the rotational objects as well as the spinning charged objects. Those researches inspirit us further to presume that the $\mathbb{\bar{S}}$ may impact the helicities. In this section, the octonion operator ($\lozenge + k_x \mathbb{\bar{X}} + k_a \mathbb{\bar{A}} + k_b \mathbb{\bar{B}}$) should be replaced by one combined operator ($\lozenge + k_x \mathbb{\bar{X}} + k_a \mathbb{\bar{A}} + k_b \mathbb{\bar{B}} + k_s \mathbb{\bar{S}}$) to comprise more features of the electromagnetic field and gravitational field simultaneously, with the $k_s$ being the coefficient for the dimensional homogeneity.

By means of the octonion operator ($\lozenge + k_x \mathbb{\bar{X}} + k_a \mathbb{\bar{A}} + k_b \mathbb{\bar{B}} + k_s \mathbb{\bar{S}}$) in the electromagnetic and gravitational fields, we can bewrite the field strength, the field source, the linear momentum, the energy, the torque, the force, and the helicity, including the influence of $\mathbb{S}$ on the torque, the force, and the helicity etc.

\subsection{Field source and torque}

In the electromagnetic field and the gravitational field with the octonion operator ($\lozenge + k_x \mathbb{\bar{X}} + k_a \mathbb{\bar{A}} + k_b \mathbb{\bar{B}} + k_s \mathbb{\bar{S}}$), the most of physical quantities will keep unchanged, except for the gravitational mass density and the helicity terms etc. Some octonion physical quantities in the octonion compounding space with the operator ($\lozenge + k_x \mathbb{\bar{X}} + k_a \mathbb{\bar{A}} + k_b \mathbb{\bar{B}} + k_s \mathbb{\bar{S}}$) are the same as that with the octonion operator ($\lozenge + k_x \mathbb{\bar{X}} + k_a \mathbb{\bar{A}} + k_b \mathbb{\bar{B}}$), including the octonion basis vector, the octonion radius vector, the octonion velocity, the octonion velocity curl, the field potential, the gravitational potential, the electromagnetic potential, the octonion field strength, the gravitational strength, the electromagnetic strength, the field source, the linear momentum, the electric current, the gauge equations, and the field strength helicity etc.

The octonion linear momentum density is $\mathbb{\bar{P}} = \mu \mathbb{\bar{S}} / \mu_g = \Sigma (\bar{p}_i \emph{\textbf{i}}_i + \bar{P}_i \emph{\textbf{I}}_i)$, and the octonion angular momentum density is $\mathbb{\bar{L}} = \mathbb{\bar{R}} \circ \mathbb{\bar{P}} = \Sigma (\bar{l}_i \emph{\textbf{i}}_i + \bar{L}_i \emph{\textbf{I}}_i)$. Meanwhile the octonion torque-energy density $\mathbb{\bar{W}}$ is defined from the angular momentum density $\mathbb{\bar{L}}$, the field strength $\mathbb{\bar{B}}$, and the field source $\mathbb{\bar{S}}$ etc,
\begin{eqnarray}
\mathbb{\bar{W}} = v_0 (\lozenge + k_x \mathbb{\bar{X}} + k_a \mathbb{\bar{A}} + k_b \mathbb{\bar{B}} + k_s \mathbb{\bar{S}}) \circ \mathbb{\bar{L}}~,
\end{eqnarray}
where $\mathbb{\bar{S}} = \Sigma (\bar{s}_i \emph{\textbf{i}}_i ) + k_{eg} \Sigma (\bar{S}_i \emph{\textbf{I}}_i )$; $-\bar{w}_0/2$ is the energy density, $\bar{\textbf{w}}/2 = \Sigma (\bar{w}_j \emph{\textbf{i}}_j )/2$ is the torque density.

The scalar $\bar{w}_0$ of the octonion $\mathbb{\bar{W}} = \Sigma (\bar{w}_i \emph{\textbf{i}}_i ) + \Sigma (\bar{W}_i \emph{\textbf{I}}_i )$ is written as,
\begin{align}
\bar{w}_0 / v_0 = & ~ \partial_0 \bar{l}_0 + \nabla \cdot \bar{\textbf{l}} + k_b ( \bar{\textbf{h}} \cdot \bar{\textbf{l}} + k_{eg} \bar{\textbf{H}} \cdot \bar{\textbf{L}} )
\nonumber\\
&
+ k_a \bar{\textbf{a}} \cdot \bar{\textbf{l}} + k_a \bar{a}_0 \bar{l}_0 + k_{eg} k_a ( \bar{\textbf{A}} \cdot \bar{\textbf{L}} + \bar{\textbf{A}}_0 \circ \bar{\textbf{L}}_0 )
\nonumber\\
&
+ k_x \bar{\textbf{x}} \cdot \bar{\textbf{l}} + k_x \bar{x}_0 \bar{l}_0 + k_{eg} k_x ( \bar{\textbf{X}} \cdot \bar{\textbf{L}} + \bar{\textbf{X}}_0 \circ \bar{\textbf{L}}_0 )
\nonumber\\
&
+ k_s \bar{\textbf{s}} \cdot \bar{\textbf{l}} + k_s \bar{s}_0 \bar{l}_0 + k_{eg} k_s ( \bar{\textbf{S}} \cdot \bar{\textbf{L}} + \bar{\textbf{S}}_0 \circ \bar{\textbf{L}}_0 )
~,
\end{align}
where $-\bar{w}_0/2$ includes the kinetic energy, gravitational potential energy, electric potential energy, magnetic potential energy, field energy, work, the interacting energy between the dipole moment with the fields, and some new energy terms. $\bar{\textbf{s}} = \Sigma (\bar{s}_j \emph{\textbf{i}}_j)$, $\bar{\textbf{S}}_0 = \bar{S}_0 \emph{\textbf{I}}_0$, $\bar{\textbf{S}} = \Sigma (\bar{S}_j \emph{\textbf{I}}_j)$.

In a similar way, expressing the torque density $\bar{\textbf{w}}$ of $\mathbb{\bar{W}}$ as
\begin{align}
\bar{\textbf{w}} / v_0 = & ~ \partial_0 \bar{\textbf{l}} + \nabla \bar{l}_0 + \nabla \times \bar{\textbf{l}} + k_b \bar{l}_0 \bar{\textbf{h}} + k_b \bar{\textbf{h}} \times \bar{\textbf{l}} + k_{eg} k_b (\bar{\textbf{H}} \times \bar{\textbf{L}} + \bar{\textbf{H}} \circ \bar{\textbf{L}}_0)
\nonumber
\\
& + k_a \bar{\textbf{a}} \bar{l}_0 + k_a \bar{\textbf{a}} \times \bar{\textbf{l}} + k_a \bar{a}_0 \bar{\textbf{l}} + k_{eg} k_a ( \bar{\textbf{A}} \times \bar{\textbf{L}} + \bar{\textbf{A}}_0 \circ \bar{\textbf{L}} + \bar{\textbf{A}} \circ \bar{\textbf{L}}_0)
\nonumber
\\
& + k_x \bar{\textbf{x}} \bar{l}_0 + k_x \bar{\textbf{x}} \times \bar{\textbf{l}} + k_x \bar{x}_0 \bar{\textbf{l}} + k_{eg} k_x ( \bar{\textbf{X}} \times \bar{\textbf{L}} + \bar{\textbf{X}}_0 \circ \bar{\textbf{L}} + \bar{\textbf{X}} \circ \bar{\textbf{L}}_0)
\nonumber
\\
& + k_s \bar{\textbf{s}} \bar{l}_0 + k_s \bar{\textbf{s}} \times \bar{\textbf{l}} + k_s \bar{s}_0 \bar{\textbf{l}} + k_{eg} k_s ( \bar{\textbf{S}} \times \bar{\textbf{L}} + \bar{\textbf{S}}_0 \circ \bar{\textbf{L}} + \bar{\textbf{S}} \circ \bar{\textbf{L}}_0)~,
\end{align}
where the above encompasses some new terms of the torque density.

\subsection{Force}

In the octonion compounding space with the operator $(\lozenge + k_x \mathbb{\bar{X}} + k_a \mathbb{\bar{A}} + k_b \mathbb{\bar{B}} + k_s \mathbb{\bar{S}})$, the octonion force-power density, $\mathbb{\bar{N}} = \Sigma (\bar{n}_i \emph{\textbf{i}}_i) + \Sigma (\bar{N}_i \emph{\textbf{I}}_i)$, is defined from the torque-energy density $\mathbb{\bar{W}}$,
\begin{eqnarray}
\mathbb{\bar{N}} = v_0 (\lozenge + k_x \mathbb{\bar{X}} + k_a \mathbb{\bar{A}} + k_b \mathbb{\bar{B}} + k_s \mathbb{\bar{S}})^* \circ \mathbb{\bar{W}}~,
\end{eqnarray}
where the power density is $\bar{f}_0 = - \bar{n}_0/(2 v_0)$, and the force density is $\bar{\textbf{f}} = - \bar{\textbf{n}} / (2 v_0)$. The vectorial parts are $\bar{\textbf{n}} = \Sigma (\bar{n}_j \emph{\textbf{i}}_j )$, $\bar{\textbf{N}}_0 = \bar{N}_0 \emph{\textbf{I}}_0$, and $\bar{\textbf{N}} = \Sigma (\bar{N}_j \emph{\textbf{I}}_j)$.

Further expressing the scalar $\bar{n}_0$ of the octonion force-power density $\mathbb{\bar{N}}$ as
\begin{align}
\bar{n}_0 / v_0 = & ~ \partial_0 \bar{w}_0 + \nabla^* \cdot \bar{\textbf{w}} + k_b ( \bar{\textbf{h}}^* \cdot \bar{\textbf{w}} + k_{eg} \bar{\textbf{H}}^* \cdot \bar{\textbf{W}} )
\nonumber\\
&
+ k_a ( \bar{\textbf{a}}^* \cdot \bar{\textbf{w}} + \bar{a}_0 \bar{w}_0 )
+ k_{eg} k_a ( \bar{\textbf{A}}^* \cdot \bar{\textbf{W}} + \bar{\textbf{A}}_0^* \circ \bar{\textbf{W}}_0 )
\nonumber\\
&
+ k_x ( \bar{\textbf{x}}^* \cdot \bar{\textbf{w}} + \bar{x}_0 \bar{w}_0 )
+ k_{eg} k_x ( \bar{\textbf{X}}^* \cdot \bar{\textbf{W}} + \bar{\textbf{X}}_0^* \circ \bar{\textbf{W}}_0 )
\nonumber\\
&
+ k_s ( \bar{\textbf{s}}^* \cdot \bar{\textbf{w}} + \bar{s}_0 \bar{w}_0 )
+ k_{eg} k_s ( \bar{\textbf{S}}^* \cdot \bar{\textbf{W}} + \bar{\textbf{S}}_0^* \circ \bar{\textbf{W}}_0 )
~.
\end{align}

In the gravitational field and electromagnetic field with the operator $(\lozenge + k_x \mathbb{\bar{X}} + k_a \mathbb{\bar{A}} + k_b \mathbb{\bar{B}} + k_s \mathbb{\bar{S}})$, the force density $\bar{\textbf{f}}$ can be defined from the vectorial part $\bar{\textbf{n}}$ of the octonion force-power density $\mathbb{\bar{N}}$ ,
\begin{align}
- 2 \bar{\textbf{f}} = & ~ \partial_0 \bar{\textbf{w}} + \nabla^* \bar{w}_0 + \nabla^* \times \bar{\textbf{w}} + k_b \bar{w}_0 \bar{\textbf{h}}^* + k_b \bar{\textbf{h}}^* \times \bar{\textbf{w}} + k_{eg} k_b (\bar{\textbf{H}}^* \times \bar{\textbf{W}} + \bar{\textbf{H}}^* \circ \bar{\textbf{W}}_0)
\nonumber\\
& + k_a ( \bar{\textbf{a}}^* \bar{w}_0 + \bar{\textbf{a}}^* \times \bar{\textbf{w}} + \bar{a}_0 \bar{\textbf{w}})
+ k_{eg} k_a ( \bar{\textbf{A}}^* \times \bar{\textbf{W}} + \bar{\textbf{A}}_0^* \circ \bar{\textbf{W}} + \bar{\textbf{A}}^* \circ \bar{\textbf{W}}_0)
\nonumber\\
& + k_x ( \bar{\textbf{x}}^* \bar{w}_0 + \bar{\textbf{x}}^* \times \bar{\textbf{w}} + \bar{x}_0 \bar{\textbf{w}})
+ k_{eg} k_x ( \bar{\textbf{X}}^* \times \bar{\textbf{W}} + \bar{\textbf{X}}_0^* \circ \bar{\textbf{W}} + \bar{\textbf{X}}^* \circ \bar{\textbf{W}}_0)
\nonumber\\
& + k_s ( \bar{\textbf{s}}^* \bar{w}_0 + \bar{\textbf{s}}^* \times \bar{\textbf{w}} + \bar{s}_0 \bar{\textbf{w}})
+ k_{eg} k_s ( \bar{\textbf{S}}^* \times \bar{\textbf{W}} + \bar{\textbf{S}}_0^* \circ \bar{\textbf{W}} + \bar{\textbf{S}}^* \circ \bar{\textbf{W}}_0)
~,
\end{align}
where the force density $\bar{\textbf{f}}$ includes that of the inertial force, the gravitational force, the gradient of energy, Lorentz force, and the interacting force between the dipole moment with the fields etc. This force definition is more complex than that in the classical field theory, and includes more new force terms regarding the gradient of energy, the field potential, and the angular velocity etc.

\subsection{Helicity}

As the inference of the compounding space with the operator $(\lozenge + k_x \mathbb{\bar{X}} + k_a \mathbb{\bar{A}} + k_b \mathbb{\bar{B}} + k_s \mathbb{\bar{S}})$, the helicity \cite{sur} will impact the gravitational mass, the charge continuity equation, and the mass continuity equation etc.

\subsubsection{Field strength helicity}

The octonion linear momentum density $\mathbb{\bar{P}} = \mu \mathbb{\bar{S}} / \mu_g$ can be defined from the octonion field source $\mathbb{\bar{S}}$ in the octonion compounding space with the operator $(\lozenge + k_x \mathbb{\bar{X}} + k_a \mathbb{\bar{A}} + k_b \mathbb{\bar{B}} + k_s \mathbb{\bar{S}})$,
\begin{eqnarray}
\mathbb{\bar{P}} = \widehat{m} \bar{v}_0 + \Sigma (m \bar{v}_j \emph{\textbf{i}}_j) + \Sigma (M \bar{V}_i \emph{\textbf{I}}_i)~,
\end{eqnarray}
where $\widehat{m} = m (\bar{v}_0^\delta / \bar{v}_0) + \bigtriangleup m$, $\bigtriangleup m = - (k_x \mathbb{\bar{X}}^* \cdot \mathbb{\bar{B}} + k_a \mathbb{\bar{A}}^* \cdot \mathbb{\bar{B}} + k_b \mathbb{\bar{B}}^* \cdot \mathbb{\bar{B}}) / ( \bar{v}_0 \mu_g )$ .

According to the octonion features, the gravitational mass $\widehat{m}$ is one reserved scalar in the above, and is changed with the field strength $\mathbb{\bar{B}}_g$ and $\mathbb{\bar{B}}_e$, the field potential $\mathbb{\bar{A}}_g$ and $\mathbb{\bar{A}}_e$, the helicity $(k_x \mathbb{\bar{X}}^* \cdot \mathbb{\bar{B}} + k_a \mathbb{\bar{A}}^* \cdot \mathbb{\bar{B}} + k_b \mathbb{\bar{B}}^* \cdot \mathbb{\bar{B}})$, the enstrophy $\textbf{u}^* \cdot \textbf{u} / 8$, and the octonion $\mathbb{\bar{X}}_g$ and $\mathbb{\bar{X}}_e$ etc in the octonion compounding space. The helicity covers the magnetic helicity $ \bar{\textbf{A}} \cdot \bar{\textbf{B}} $, the kinetic helicity $ \bar{\textbf{v}} \cdot \bar{\textbf{u}} $, the cross helicity $ \bar{\textbf{V}} \cdot \bar{\textbf{B}} $, and one new helicity term $ \bar{\textbf{A}} \cdot \bar{\textbf{U}} $ etc. But the field source $\mathbb{\bar{S}}$ does not impact the field strength helicity in the octonion compounding space with the octonion operator $(\lozenge + k_x \mathbb{\bar{X}} + k_a \mathbb{\bar{A}} + k_b \mathbb{\bar{B}} + k_s \mathbb{\bar{S}})$. And that this field can not be distinguished from the compounding space with the octonion operator $(\lozenge + k_x \mathbb{\bar{X}} + k_a \mathbb{\bar{A}} + k_b \mathbb{\bar{B}})$, according to the viewpoint of the field strength helicity.

\begin{table}[h]
\caption{Some quantities with the octonion operator ($\lozenge + k_x \mathbb{\bar{X}} + k_a \mathbb{\bar{A}} + k_b \mathbb{\bar{B}} + k_s \mathbb{\bar{S}}$).}
\label{tab:table3}
\centering
\begin{tabular}{ll}
\hline\hline
$ definitions $                                                                                                                          & $ meanings $ \\
\hline
$\mathbb{\bar{X}}$                                                                                                                       & field quantity \\
$\mathbb{\bar{A}} = (\lozenge + k_x \mathbb{\bar{X}}) \circ \mathbb{\bar{X}}$                                                            & field potential \\
$\mathbb{\bar{B}} = (\lozenge + k_x \mathbb{\bar{X}} + k_a \mathbb{\bar{A}}) \circ \mathbb{\bar{A}}$                                     & field strength \\
$\mathbb{\bar{R}}$                                                                                                                       & radius vector \\
$\mathbb{\bar{V}} = v_0 \lozenge \circ \mathbb{\bar{R}}$                                                                                 & velocity \\
$\mathbb{\bar{U}} = \lozenge \circ \mathbb{\bar{V}}$                                                                                     & velocity curl \\
$\mu \mathbb{\bar{S}} = - ( \lozenge + k_x \mathbb{\bar{X}} + k_a \mathbb{\bar{A}} + k_b \mathbb{\bar{B}} )^* \circ \mathbb{\bar{B}}$    & field source \\
$\mathbb{\bar{H}}_b = (k_x \mathbb{\bar{X}} + k_a \mathbb{\bar{A}} + k_b \mathbb{\bar{B}})^* \cdot \mathbb{\bar{B}}$                     & field strength helicity\\
$\mathbb{\bar{P}} = \mu \mathbb{\bar{S}} / \mu_g$                                                                                        & linear momentum density \\
$\mathbb{\bar{R}} = \mathbb{R} + k_{rx} \mathbb{X}$                                                                                      & compounding radius vector \\
$\mathbb{\bar{L}} = \mathbb{\bar{R}} \circ \mathbb{\bar{P}}$                                                                             & angular momentum density \\
$\mathbb{\bar{W}} = v_0 ( \lozenge + k_x \mathbb{\bar{X}} + k_a \mathbb{\bar{A}} + k_b \mathbb{\bar{B}} + k_s \mathbb{\bar{S}} ) \circ \mathbb{\bar{L}}$
                                                                                                                                         & torque-energy densities \\
$\mathbb{\bar{N}} = v_0 ( \lozenge + k_x \mathbb{\bar{X}} + k_a \mathbb{\bar{A}} + k_b \mathbb{\bar{B}} + k_s \mathbb{\bar{S}} )^* \circ \mathbb{\bar{W}}$
                                                                                                                                         & force-power density \\
$\mathbb{\bar{F}} = - \mathbb{\bar{N}} / (2v_0)$                                                                                         & force density \\
$\mathbb{\bar{H}}_s = ( k_x \mathbb{\bar{X}} + k_a \mathbb{\bar{A}} + k_b \mathbb{\bar{B}} + k_s \mathbb{\bar{S}} )^* \cdot \mathbb{\bar{P}}$
                                                                                                                                         & field source helicity\\
\hline\hline
\end{tabular}
\end{table}

\subsubsection{Field source helicity}

The part force density $\mathbb{\bar{F}}$ is one part of the octonion force-power density $\mathbb{\bar{N}}$, and is defined from the octonion linear momentum density $\mathbb{\bar{P}}$ ,
\begin{eqnarray}
\mathbb{\bar{F}} = v_0 (\lozenge + k_x \mathbb{\bar{X}} + k_a \mathbb{\bar{A}} + k_b \mathbb{\bar{B}} + k_s \mathbb{\bar{S}})^* \circ \mathbb{\bar{P}}~,
\end{eqnarray}
where the part force density $\mathbb{\bar{F}}$ includes the density of the inertial force, the gravitational force, Lorentz force, and the interacting force between the fields with the dipoles etc.

The scalar $\bar{f}_0$ of $\mathbb{\bar{F}}$ is written as,
\begin{align}
\bar{f}_0 / v_0 = & ~ \partial_0 \bar{p}_0 + \nabla^* \cdot \bar{\textbf{p}} + k_b ( \bar{\textbf{h}}^* \cdot \bar{\textbf{p}} + k_{eg} \bar{\textbf{H}}^* \cdot \bar{\textbf{P}} )
\nonumber\\
&
+ k_a ( \bar{\textbf{a}}^* \cdot \bar{\textbf{p}} + \bar{a}_0 \bar{p}_0 ) + k_{eg} k_a ( \bar{\textbf{A}}^* \cdot \bar{\textbf{P}} + \bar{\textbf{A}}_0^* \circ \bar{\textbf{P}}_0 )
\nonumber\\
& + k_x ( \bar{\textbf{x}}^* \cdot \bar{\textbf{p}} + \bar{x}_0 \bar{p}_0 ) + k_{eg} k_x ( \bar{\textbf{X}}^* \cdot \bar{\textbf{P}} + \bar{\textbf{X}}_0^* \circ \bar{\textbf{P}}_0 )
\nonumber\\
&
+ k_s ( \bar{\textbf{s}}^* \cdot \bar{\textbf{p}} + \bar{s}_0 \bar{p}_0 ) + k_{eg} k_s ( \bar{\textbf{S}}^* \cdot \bar{\textbf{P}} + \bar{\textbf{S}}_0^* \circ \bar{\textbf{P}}_0 )
~,
\end{align}
where the field source helicity in the fields with the octonion operator $(\lozenge + k_x \mathbb{\bar{X}} + k_a \mathbb{\bar{A}} + k_b \mathbb{\bar{B}} + k_s \mathbb{\bar{S}})$ covers the helicity terms $( \bar{\textbf{a}}^* \cdot \bar{\textbf{p}} + \bar{a}_0 \bar{p}_0 )$, $( \bar{\textbf{A}}^* \cdot \bar{\textbf{P}} + \bar{\textbf{A}}_0^* \circ \bar{\textbf{P}}_0 )$, $( \bar{\textbf{h}}^* \cdot \bar{\textbf{p}} + k_{eg} \bar{\textbf{H}}^* \cdot \bar{\textbf{P}} )$, $(\bar{\textbf{x}}^* \cdot \bar{\textbf{p}} + \bar{x}_0 \bar{p}_0)$, $(\bar{\textbf{X}}^* \cdot \bar{\textbf{P}} + \bar{\textbf{X}}_0^* \circ \bar{\textbf{P}}_0)$, $( \bar{\textbf{s}}^* \cdot \bar{\textbf{p}} + \bar{s}_0 \bar{p}_0 )$, and $( \bar{\textbf{S}}^* \cdot \bar{\textbf{P}} + \bar{\textbf{S}}_0^* \circ \bar{\textbf{P}}_0 )$ etc. And they include the magnetic helicity $ \textbf{A} \cdot \textbf{B} $, the kinetic helicity $ \textbf{v} \cdot \textbf{u} $, the cross helicity $ \textbf{V} \cdot \textbf{B} $, and the current helicity $ \textbf{B}^* \cdot \textbf{P} $ etc.

The above is the mass continuity equation in the electromagnetic field and the gravitational field with the operator $(\lozenge + k_x \mathbb{\bar{X}} + k_a \mathbb{\bar{A}} + k_b \mathbb{\bar{B}} + k_s \mathbb{\bar{S}})$, and is influenced by the helicity of electromagnetic field and the gravitational field. The impact of the helicity related to the $\mathbb{\bar{S}}$ may be detected sometimes.

A new physical quantity $\mathbb{\bar{F}}_q$ can be defined from the part force density $\mathbb{\bar{F}}$,
\begin{eqnarray}
\mathbb{\bar{F}}_q = \mathbb{\bar{F}} \circ \emph{\textbf{I}}_0^*~.
\end{eqnarray}

The scalar part $\bar{F}_0$ of the $\mathbb{\bar{F}}_q$ is written as,
\begin{align}
\bar{F}_0 / v_0 = & ~ (\partial_0 \bar{\textbf{P}}_0 + \nabla^* \cdot \bar{\textbf{P}}) \circ \emph{\textbf{I}}_0^* + k_b ( \bar{\textbf{h}}^* \cdot \bar{\textbf{P}} + k_{eg} \bar{\textbf{H}}^* \cdot \bar{\textbf{p}} ) \circ \emph{\textbf{I}}_0^*
\nonumber
\\
&
+ k_a ( \bar{\textbf{a}}^* \cdot \bar{\textbf{P}} + a_0 \bar{\textbf{P}}_0 ) \circ \emph{\textbf{I}}_0^* + k_{eg} k_a ( \bar{\textbf{A}}^* \cdot \bar{\textbf{p}} + p_0 \bar{\textbf{A}}_0^* ) \circ \emph{\textbf{I}}_0^*
\nonumber
\\
&
+ k_x ( \bar{\textbf{x}}^* \cdot \bar{\textbf{P}} + x_0 \bar{\textbf{P}}_0 ) \circ \emph{\textbf{I}}_0^* + k_{eg} k_x ( \bar{\textbf{X}}^* \cdot \bar{\textbf{p}} + p_0 \bar{\textbf{X}}_0^* ) \circ \emph{\textbf{I}}_0^*
\nonumber
\\
&
+ k_s ( \bar{\textbf{s}}^* \cdot \bar{\textbf{P}} + s_0 \bar{\textbf{P}}_0 ) \circ \emph{\textbf{I}}_0^* + k_{eg} k_s ( \bar{\textbf{S}}^* \cdot \bar{\textbf{p}} + p_0 \bar{\textbf{S}}_0^* ) \circ \emph{\textbf{I}}_0^*
~,
\end{align}
where the last seven terms are the helicity in the gravitational field and electromagnetic field.

The above is the charge continuity equation in the case for coexistence of the gravitational field and electromagnetic field with the octonion operator $(\lozenge + k_x \mathbb{\bar{X}} + k_a \mathbb{\bar{A}} + k_b \mathbb{\bar{B}} + k_s \mathbb{\bar{S}} )$ when $F_0 = 0$. And this charge continuity equation is the invariant under the octonion coordinate transformation. It states that the physical quantity $\mathbb{\bar{X}}$, $\mathbb{\bar{A}}$, $\mathbb{\bar{B}}$, and $\mathbb{\bar{S}}$ of the gravitational field and the electromagnetic field have the influence on the charge continuity equation, although the terms $( \bar{\textbf{a}}^* \cdot \bar{\textbf{P}} + a_0 \bar{\textbf{P}}_0 ) \circ \emph{\textbf{I}}_0^*$, $( \bar{\textbf{A}}^* \cdot \bar{\textbf{p}} + p_0 \bar{\textbf{A}}_0^* ) \circ \emph{\textbf{I}}_0^*$, $( \bar{\textbf{h}}^* \cdot \bar{\textbf{P}} + k_{eg} \bar{\textbf{H}}^* \cdot \bar{\textbf{p}} ) \circ \emph{\textbf{I}}_0^*$, $( \bar{\textbf{x}}^* \cdot \bar{\textbf{P}} + x_0 \bar{\textbf{P}}_0 ) \circ \emph{\textbf{I}}_0^*$, $( \bar{\textbf{X}}^* \cdot \bar{\textbf{p}} + p_0 \bar{\textbf{X}}_0^* ) \circ \emph{\textbf{I}}_0^*$, $( \bar{\textbf{s}}^* \cdot \bar{\textbf{P}} + s_0 \bar{\textbf{P}}_0 ) \circ \emph{\textbf{I}}_0^*$, $( \bar{\textbf{S}}^* \cdot \bar{\textbf{p}} + p_0 \bar{\textbf{S}}_0^* ) \circ \emph{\textbf{I}}_0^*$, and $\triangle m$ are usually trifling when the fields are feeble enough.

%--9--%

\section{The fields regarding the operator ($\lozenge, \mathbb{\bar{X}}, \mathbb{\bar{A}}, \mathbb{\bar{B}}, \mathbb{\bar{S}}, \mathbb{\bar{L}}$) }

In the electromagnetic field and gravitational field, the octonion operator ($\lozenge + k_x \mathbb{\bar{X}} + k_a \mathbb{\bar{A}} + k_b \mathbb{\bar{B}} + k_s \mathbb{\bar{S}}$) can derive the physical properties of two fields, including the octonion linear momentum, the octonion angular momentum, the energy, the torque, the power, the force, and some helicities of the rotational objects as well as the spinning charged objects. Further those results lead us to assume that the angular momentum $\mathbb{\bar{L}}$ may exert an influence on the helicities. In this section, one new octonion operator ($\lozenge + k_x \mathbb{\bar{X}} + k_a \mathbb{\bar{A}} + k_b \mathbb{\bar{B}} + k_s \mathbb{\bar{S}} + k_l \mathbb{\bar{L}}$) will instead of the octonion operator ($\lozenge + k_x \mathbb{\bar{X}} + k_a \mathbb{\bar{A}} + k_b \mathbb{\bar{B}} + k_s \mathbb{\bar{S}}$) to include more properties of the electromagnetic and gravitational fields simultaneously, with the $k_l$ being the coefficient for the dimensional homogeneity.

By means of the octonion operator ($\lozenge + k_x \mathbb{\bar{X}} + k_a \mathbb{\bar{A}} + k_b \mathbb{\bar{B}} + k_s \mathbb{\bar{S}} + k_l \mathbb{\bar{L}}$) in the electromagnetic field and gravitational field, we can represent the octonion field potential, the octonion field strength, the octonion field source, the octonion linear momentum, the energy, the torque, the force, and some helicities, including the influence of the $\mathbb{L}$ on the torque, the force, and the helicity etc.

\subsection{Field source and torque}

In the electromagnetic and gravitational fields with the operator ($\lozenge + k_x \mathbb{\bar{X}} + k_a \mathbb{\bar{A}} + k_b \mathbb{\bar{B}} + k_s \mathbb{\bar{S}} + k_l \mathbb{\bar{L}}$), the majority physical quantities will remain the same, except for the gravitational mass density, the charge continuity equation, the mass continuity equation, and the helicity terms etc. Some octonion physical quantities in the octonion compounding space with the octonion operator ($\lozenge + k_x \mathbb{\bar{X}} + k_a \mathbb{\bar{A}} + k_b \mathbb{\bar{B}} + k_s \mathbb{\bar{S}} + k_l \mathbb{\bar{L}}$) are the same as that in the space with the operator ($\lozenge + k_x \mathbb{\bar{X}} + k_a \mathbb{\bar{A}} + k_b \mathbb{\bar{B}} + k_s \mathbb{\bar{S}}$) respectively, including the octonion basis vector, the octonion radius vector, the octonion velocity, the octonion vorticity, the octonion field potential, the gravitational potential, the electromagnetic potential, the octonion field strength, the gravitational strength, the electromagnetic strength, the octonion field source, the octonion linear momentum, the electric current, and the field strength helicity etc.

The octonion linear momentum density is $\mathbb{\bar{P}} = \mu \mathbb{\bar{S}} / \mu_g = \Sigma (\bar{p}_i \emph{\textbf{i}}_i + \bar{P}_i \emph{\textbf{I}}_i)$, and the octonion angular momentum density is $\mathbb{\bar{L}} = \mathbb{\bar{R}} \circ \mathbb{\bar{P}} = \Sigma (\bar{l}_i \emph{\textbf{i}}_i + \bar{L}_i \emph{\textbf{I}}_i)$.  Meanwhile the octonion torque-energy density $\mathbb{\bar{W}}$ is defined from the angular momentum density $\mathbb{\bar{L}}$, the field strength $\mathbb{\bar{B}}$, and the field source $\mathbb{\bar{S}}$ etc,
\begin{eqnarray}
\mathbb{\bar{W}} = v_0 (\lozenge + k_x \mathbb{\bar{X}} + k_a \mathbb{\bar{A}} + k_b \mathbb{\bar{B}} + k_s \mathbb{\bar{S}} + k_l \mathbb{\bar{L}}) \circ \mathbb{\bar{L}}~,
\end{eqnarray}
where $\mathbb{\bar{S}} = \Sigma (\bar{s}_i \emph{\textbf{i}}_i ) + k_{eg} \Sigma (\bar{S}_i \emph{\textbf{I}}_i )$; $-\bar{w}_0/2$ is the energy density, $\bar{\textbf{w}}/2 = \Sigma (\bar{w}_j \emph{\textbf{i}}_j )/2$ is the torque density.

The scalar $\bar{w}_0$ of the angular momentum density $\mathbb{\bar{W}} = \Sigma (\bar{w}_i \emph{\textbf{i}}_i ) + \Sigma (\bar{W}_i \emph{\textbf{I}}_i )$ is written as,
\begin{align}
\bar{w}_0 / v_0 = & ~ \partial_0 \bar{l}_0 + \nabla \cdot \bar{\textbf{l}}
+ k_b ( \bar{\textbf{h}} \cdot \bar{\textbf{l}} + k_{eg} \bar{\textbf{H}} \cdot \bar{\textbf{L}} )
\nonumber\\
&
+ k_a \bar{\textbf{a}} \cdot \bar{\textbf{l}} + k_a \bar{a}_0 \bar{l}_0 + k_{eg} k_a ( \bar{\textbf{A}} \cdot \bar{\textbf{L}} + \bar{\textbf{A}}_0 \circ \bar{\textbf{L}}_0 )
\nonumber\\
&
+ k_x \bar{\textbf{x}} \cdot \bar{\textbf{l}} + k_x \bar{x}_0 \bar{l}_0 + k_{eg} k_x ( \bar{\textbf{X}} \cdot \bar{\textbf{L}} + \bar{\textbf{X}}_0 \circ \bar{\textbf{L}}_0 )
\nonumber\\
&
+ k_s \bar{\textbf{s}} \cdot \bar{\textbf{l}} + k_s \bar{s}_0 \bar{l}_0 + k_{eg} k_s ( \bar{\textbf{S}} \cdot \bar{\textbf{L}} + \bar{\textbf{S}}_0 \circ \bar{\textbf{L}}_0 )
\nonumber\\
&
+ k_l \bar{\textbf{l}} \cdot \bar{\textbf{l}} + k_l \bar{l}_0^2 + k_{eg} k_l ( \bar{\textbf{L}} \cdot \bar{\textbf{L}} + \bar{\textbf{L}}_0 \circ \bar{\textbf{L}}_0 )
~,
\end{align}
where $-\bar{w}_0/2$ includes the kinetic energy, the gravitational potential energy, the electric potential energy, the magnetic potential energy, the field energy, the work, the interacting energy between the dipole moment with the fields, and some new energy terms. $\bar{\textbf{l}} = \Sigma (\bar{l}_j \emph{\textbf{i}}_j)$, $\bar{\textbf{L}}_0 = \bar{L}_0 \emph{\textbf{I}}_0$, $\bar{\textbf{L}} = \Sigma (\bar{L}_j \emph{\textbf{I}}_j)$. $\lozenge = \Sigma (\emph{\textbf{i}}_i \partial_i)$.

In a similar way, expressing the torque density $\bar{\textbf{w}}$ of the angular momentum density $\mathbb{\bar{W}}$ as
\begin{align}
\bar{\textbf{w}} / v_0 = & ~ \partial_0 \bar{\textbf{l}} + \nabla \bar{l}_0 + \nabla \times \bar{\textbf{l}} + k_b \bar{l}_0 \bar{\textbf{h}} + k_b \bar{\textbf{h}} \times \bar{\textbf{l}} + k_{eg} k_b (\bar{\textbf{H}} \times \bar{\textbf{L}} + \bar{\textbf{H}} \circ \bar{\textbf{L}}_0)
\nonumber
\\
& + k_a \bar{\textbf{a}} \bar{l}_0 + k_a \bar{\textbf{a}} \times \bar{\textbf{l}} + k_a \bar{a}_0 \bar{\textbf{l}} + k_{eg} k_a ( \bar{\textbf{A}} \times \bar{\textbf{L}} + \bar{\textbf{A}}_0 \circ \bar{\textbf{L}} + \bar{\textbf{A}} \circ \bar{\textbf{L}}_0)
\nonumber
\\
& + k_x \bar{\textbf{x}} \bar{l}_0 + k_x \bar{\textbf{x}} \times \bar{\textbf{l}} + k_x \bar{x}_0 \bar{\textbf{l}} + k_{eg} k_x ( \bar{\textbf{X}} \times \bar{\textbf{L}} + \bar{\textbf{X}}_0 \circ \bar{\textbf{L}} + \bar{\textbf{X}} \circ \bar{\textbf{L}}_0)
\nonumber
\\
& + k_s \bar{\textbf{s}} \bar{l}_0 + k_s \bar{\textbf{s}} \times \bar{\textbf{l}} + k_s \bar{s}_0 \bar{\textbf{l}} + k_{eg} k_s ( \bar{\textbf{S}} \times \bar{\textbf{L}} + \bar{\textbf{S}}_0 \circ \bar{\textbf{L}} + \bar{\textbf{S}} \circ \bar{\textbf{L}}_0)
+ 2 k_l \bar{l}_0 \bar{\textbf{l}}
~,
\end{align}
where the above includes some new terms of the torque density.

\subsection{Force}

In the octonion compounding space with the operator $(\lozenge + k_x \mathbb{\bar{X}} + k_a \mathbb{\bar{A}} + k_b \mathbb{\bar{B}} + k_s \mathbb{\bar{S}} + k_l \mathbb{\bar{L}})$, the octonion force-power density, $\mathbb{\bar{N}} = \Sigma (\bar{n}_i \emph{\textbf{i}}_i) + \Sigma (\bar{N}_i \emph{\textbf{I}}_i)$, is defined as follows,
\begin{eqnarray}
\mathbb{\bar{N}} = v_0 (\lozenge + k_x \mathbb{\bar{X}} + k_a \mathbb{\bar{A}} + k_b \mathbb{\bar{B}} + k_s \mathbb{\bar{S}} + k_l \mathbb{\bar{L}})^* \circ \mathbb{\bar{W}}~,
\end{eqnarray}
where the power density is $\bar{f}_0 = - \bar{n}_0/(2 v_0)$, and the force density is $\bar{\textbf{f}} = - \bar{\textbf{n}} / (2 v_0)$. The vectorial parts are $\bar{\textbf{n}} = \Sigma (\bar{n}_j \emph{\textbf{i}}_j )$, $\bar{\textbf{N}}_0 = \bar{N}_0 \emph{\textbf{I}}_0$, and $\bar{\textbf{N}} = \Sigma (\bar{N}_j \emph{\textbf{I}}_j)$.

Further expressing the scalar $\bar{n}_0$ of the octonion $\mathbb{\bar{N}}$ as
\begin{align}
\bar{n}_0 / v_0 = & ~ \partial_0 \bar{w}_0 + \nabla^* \cdot \bar{\textbf{w}} + k_b ( \bar{\textbf{h}}^* \cdot \bar{\textbf{w}} + k_{eg} \bar{\textbf{H}}^* \cdot \bar{\textbf{W}} )
\nonumber\\
&
+ k_a ( \bar{\textbf{a}}^* \cdot \bar{\textbf{w}} + \bar{a}_0 \bar{w}_0 )
+ k_{eg} k_a ( \bar{\textbf{A}}^* \cdot \bar{\textbf{W}} + \bar{\textbf{A}}_0^* \circ \bar{\textbf{W}}_0 )
\nonumber\\
&
+ k_x ( \bar{\textbf{x}}^* \cdot \bar{\textbf{w}} + \bar{x}_0 \bar{w}_0 )
+ k_{eg} k_x ( \bar{\textbf{X}}^* \cdot \bar{\textbf{W}} + \bar{\textbf{X}}_0^* \circ \bar{\textbf{W}}_0 )
\nonumber\\
&
+ k_s ( \bar{\textbf{s}}^* \cdot \bar{\textbf{w}} + \bar{s}_0 \bar{w}_0 )
+ k_{eg} k_s ( \bar{\textbf{S}}^* \cdot \bar{\textbf{W}} + \bar{\textbf{S}}_0^* \circ \bar{\textbf{W}}_0 )
\nonumber\\
&
+ k_l ( \bar{\textbf{l}}^* \cdot \bar{\textbf{w}} + \bar{l}_0 \bar{w}_0 )
+ k_{eg} k_l ( \bar{\textbf{L}}^* \cdot \bar{\textbf{W}} + \bar{\textbf{L}}_0^* \circ \bar{\textbf{W}}_0 )
~.
\end{align}

In the gravitational field and electromagnetic field with the octonion operator $(\lozenge + k_x \mathbb{\bar{X}} + k_a \mathbb{\bar{A}} + k_b \mathbb{\bar{B}} + k_s \mathbb{\bar{S}} + k_l \mathbb{\bar{L}})$, the force density $\bar{\textbf{f}}$ can be defined from the vectorial part $\bar{\textbf{n}}$ of $\mathbb{\bar{N}}$ ,
\begin{align}
- 2 \bar{\textbf{f}} = & ~ \partial_0 \bar{\textbf{w}} + \nabla^* \bar{w}_0 + \nabla^* \times \bar{\textbf{w}} + k_b \bar{w}_0 \bar{\textbf{h}}^* + k_b \bar{\textbf{h}}^* \times \bar{\textbf{w}} + k_{eg} k_b (\bar{\textbf{H}}^* \times \bar{\textbf{W}} + \bar{\textbf{H}}^* \circ \bar{\textbf{W}}_0)
\nonumber\\
& + k_a ( \bar{\textbf{a}}^* \bar{w}_0 + \bar{\textbf{a}}^* \times \bar{\textbf{w}} + \bar{a}_0 \bar{\textbf{w}})
+ k_{eg} k_a ( \bar{\textbf{A}}^* \times \bar{\textbf{W}} + \bar{\textbf{A}}_0^* \circ \bar{\textbf{W}} + \bar{\textbf{A}}^* \circ \bar{\textbf{W}}_0)
\nonumber\\
& + k_x ( \bar{\textbf{x}}^* \bar{w}_0 + \bar{\textbf{x}}^* \times \bar{\textbf{w}} + \bar{x}_0 \bar{\textbf{w}})
+ k_{eg} k_x ( \bar{\textbf{X}}^* \times \bar{\textbf{W}} + \bar{\textbf{X}}_0^* \circ \bar{\textbf{W}} + \bar{\textbf{X}}^* \circ \bar{\textbf{W}}_0)
\nonumber\\
& + k_s ( \bar{\textbf{s}}^* \bar{w}_0 + \bar{\textbf{s}}^* \times \bar{\textbf{w}} + \bar{s}_0 \bar{\textbf{w}})
+ k_{eg} k_s ( \bar{\textbf{S}}^* \times \bar{\textbf{W}} + \bar{\textbf{S}}_0^* \circ \bar{\textbf{W}} + \bar{\textbf{S}}^* \circ \bar{\textbf{W}}_0)
\nonumber\\
& + k_l ( \bar{\textbf{l}}^* \bar{w}_0 + \bar{\textbf{l}}^* \times \bar{\textbf{w}} + \bar{l}_0 \bar{\textbf{w}})
+ k_{eg} k_l ( \bar{\textbf{L}}^* \times \bar{\textbf{W}} + \bar{\textbf{L}}_0^* \circ \bar{\textbf{W}} + \bar{\textbf{L}}^* \circ \bar{\textbf{W}}_0)
~,
\end{align}
where the force density $\bar{\textbf{f}}$ includes that of the inertial force, the gravitational force, the gradient of energy, Lorentz force, and the interacting force between the dipole moment with the two fields etc. The above force is much longer than that in the classical field theory, and includes more new force terms related to the gradient of energy, the field potential, and the angular velocity etc.

\subsection{Helicity}

In the compounding space with the octonion operator $(\lozenge + k_x \mathbb{\bar{X}} + k_a \mathbb{\bar{A}} + k_b \mathbb{\bar{B}} + k_s \mathbb{\bar{S}} + k_l \mathbb{\bar{L}})$, some helicities \cite{pogosian} and the speed of light will impact the mass continuity equation and the charge continuity equation etc.

\subsubsection{Field strength helicity}

The octonion linear momentum density $\mathbb{\bar{P}} = \mu \mathbb{\bar{S}} / \mu_g$ is defined from octonion the field source $\mathbb{\bar{S}}$ in the compounding field with the operator $(\lozenge + k_x \mathbb{\bar{X}} + k_a \mathbb{\bar{A}} + k_b \mathbb{\bar{B}} + k_s \mathbb{\bar{S}} + k_l \mathbb{\bar{L}})$,
\begin{eqnarray}
\mathbb{\bar{P}} = \widehat{m} \bar{v}_0 + \Sigma (m \bar{v}_j \emph{\textbf{i}}_j) + \Sigma (M \bar{V}_i \emph{\textbf{I}}_i)~,
\end{eqnarray}
where $\widehat{m} = m (\bar{v}_0^\delta / \bar{v}_0) + \bigtriangleup m$, $\bigtriangleup m = - (k_x \mathbb{\bar{X}}^* \cdot \mathbb{\bar{B}} + k_a \mathbb{\bar{A}}^* \cdot \mathbb{\bar{B}} + k_b \mathbb{\bar{B}}^* \cdot \mathbb{\bar{B}}) / ( \bar{v}_0 \mu_g )$.

According to the octonion features, the gravitational mass $\widehat{m}$ is one reserved scalar in the above, and is changed with the field strength $\mathbb{\bar{B}}_g$ and $\mathbb{\bar{B}}_e$, the field potential $\mathbb{\bar{A}}_g$ and $\mathbb{\bar{A}}_e$, the helicity $(k_x \mathbb{\bar{X}}^* \cdot \mathbb{\bar{B}} + k_a \mathbb{\bar{A}}^* \cdot \mathbb{\bar{B}} + k_b \mathbb{\bar{B}}^* \cdot \mathbb{\bar{B}})$, the enstrophy $\textbf{u}^* \cdot \textbf{u} / 8$, and the octonion $\mathbb{\bar{X}}_g$ and $\mathbb{\bar{X}}_e$ etc in the octonion compounding space. The helicity covers the magnetic helicity $ \bar{\textbf{A}} \cdot \bar{\textbf{B}} $, the kinetic helicity $ \bar{\textbf{v}} \cdot \bar{\textbf{u}} $, the cross helicity $ \bar{\textbf{V}} \cdot \bar{\textbf{B}} $, and one new helicity term $ \bar{\textbf{A}} \cdot \bar{\textbf{U}} $ etc.

However the octonion angular momentum $\mathbb{\bar{L}}$ does not influence the field strength helicity in the compounding space with the octonion operator $(\lozenge + k_x \mathbb{\bar{X}} + k_a \mathbb{\bar{A}} + k_b \mathbb{\bar{B}} + k_s \mathbb{\bar{S}} + k_l \mathbb{\bar{L}})$. Consequently we can not distinguish this field from some fields with other octonion operators, according to the viewpoint of the field strength helicity.

\begin{table}[t]
\caption{Some quantities with the octonion operator ($\lozenge + k_x \mathbb{\bar{X}} + k_a \mathbb{\bar{A}} + k_b \mathbb{\bar{B}} + k_s \mathbb{\bar{S}} + k_l \mathbb{\bar{L}}$).}
\label{tab:table3}
\centering
\begin{tabular}{ll}
\hline\hline
$ definitions $                                                                                                                          & $ meanings $ \\
\hline
$\mathbb{\bar{X}}$                                                                                                                       & field quantity \\
$\mathbb{\bar{A}} = (\lozenge + k_x \mathbb{\bar{X}}) \circ \mathbb{\bar{X}}$                                                            & field potential \\
$\mathbb{\bar{B}} = (\lozenge + k_x \mathbb{\bar{X}} + k_a \mathbb{\bar{A}}) \circ \mathbb{\bar{A}}$                                     & field strength \\
$\mathbb{\bar{R}}$                                                                                                                       & radius vector \\
$\mathbb{\bar{V}} = v_0 \lozenge \circ \mathbb{\bar{R}}$                                                                                 & velocity \\
$\mathbb{\bar{U}} = \lozenge \circ \mathbb{\bar{V}}$                                                                                     & velocity curl \\
$\mu \mathbb{\bar{S}} = - ( \lozenge + k_x \mathbb{\bar{X}} + k_a \mathbb{\bar{A}} + k_b \mathbb{\bar{B}} )^* \circ \mathbb{\bar{B}}$    & field source \\
$\mathbb{\bar{H}}_b = (k_x \mathbb{\bar{X}} + k_a \mathbb{\bar{A}} + k_b \mathbb{\bar{B}})^* \cdot \mathbb{\bar{B}}$                     & field strength helicity\\
$\mathbb{\bar{P}} = \mu \mathbb{\bar{S}} / \mu_g$                                                                                        & linear momentum density \\
$\mathbb{\bar{R}} = \mathbb{R} + k_{rx} \mathbb{X}$                                                                                      & compounding radius vector \\
$\mathbb{\bar{L}} = \mathbb{\bar{R}} \circ \mathbb{\bar{P}}$                                                                             & angular momentum density \\
$\mathbb{\bar{W}} = v_0 ( \lozenge + k_x \mathbb{\bar{X}} + k_a \mathbb{\bar{A}} + k_b \mathbb{\bar{B}} + k_s \mathbb{\bar{S}} + k_l \mathbb{\bar{L}} ) \circ \mathbb{\bar{L}}$                                                                                                                        & torque-energy densities \\
$\mathbb{\bar{N}} = v_0 ( \lozenge + k_x \mathbb{\bar{X}} + k_a \mathbb{\bar{A}} + k_b \mathbb{\bar{B}} + k_s \mathbb{\bar{S}} + k_l \mathbb{\bar{L}} )^* \circ \mathbb{\bar{W}}$                                                                                                                        & force-power density \\
$\mathbb{\bar{F}} = - \mathbb{\bar{N}} / (2v_0)$                                                                                         & force density \\
$\mathbb{\bar{H}}_s = ( k_x \mathbb{\bar{X}} + k_a \mathbb{\bar{A}} + k_b \mathbb{\bar{B}} + k_s \mathbb{\bar{S}} + k_l \mathbb{\bar{L}})^* \cdot \mathbb{\bar{P}}$
                                                                                                                                         & field source helicity\\
\hline\hline
\end{tabular}
\end{table}

\subsubsection{Field source helicity}

The part force density $\mathbb{\bar{F}}$ is one part of the octonion force-power density $\mathbb{\bar{N}}$, and is defined from the octonion linear momentum density $\mathbb{\bar{P}}$ ,
\begin{eqnarray}
\mathbb{\bar{F}} = v_0 (\lozenge + k_x \mathbb{\bar{X}} + k_a \mathbb{\bar{A}} + k_b \mathbb{\bar{B}} + k_s \mathbb{\bar{S}} + k_l \mathbb{\bar{L}})^* \circ \mathbb{\bar{P}}~,
\end{eqnarray}
where the part force density includes that of the inertial force, gravitational force, Lorentz force, and the interacting force between the fields with the dipoles etc.

The scalar $\bar{f}_0$ of the octonion $\mathbb{\bar{F}}$ is written as,
\begin{align}
\bar{f}_0 / v_0 = & ~ \partial_0 \bar{p}_0 + \nabla^* \cdot \bar{\textbf{p}} + k_b ( \bar{\textbf{h}}^* \cdot \bar{\textbf{p}} + k_{eg} \bar{\textbf{H}}^* \cdot \bar{\textbf{P}} )
\nonumber\\
&
+ k_a ( \bar{\textbf{a}}^* \cdot \bar{\textbf{p}} + \bar{a}_0 \bar{p}_0 ) + k_{eg} k_a ( \bar{\textbf{A}}^* \cdot \bar{\textbf{P}} + \bar{\textbf{A}}_0^* \circ \bar{\textbf{P}}_0 )
\nonumber\\
& + k_x ( \bar{\textbf{x}}^* \cdot \bar{\textbf{p}} + \bar{x}_0 \bar{p}_0 ) + k_{eg} k_x ( \bar{\textbf{X}}^* \cdot \bar{\textbf{P}} + \bar{\textbf{X}}_0^* \circ \bar{\textbf{P}}_0 )
\nonumber\\
&
+ k_s ( \bar{\textbf{s}}^* \cdot \bar{\textbf{p}} + \bar{s}_0 \bar{p}_0 ) + k_{eg} k_s ( \bar{\textbf{S}}^* \cdot \bar{\textbf{P}} + \bar{\textbf{S}}_0^* \circ \bar{\textbf{P}}_0 )
\nonumber\\
& + k_l ( \bar{\textbf{l}}^* \cdot \bar{\textbf{p}} + \bar{l}_0 \bar{p}_0 ) + k_{eg} k_l ( \bar{\textbf{L}}^* \cdot \bar{\textbf{P}} + \bar{\textbf{L}}_0^* \circ \bar{\textbf{P}}_0 )
~,
\end{align}
where the field source helicity in the two fields with the octonion operator $(\lozenge + k_x \mathbb{\bar{X}} + k_a \mathbb{\bar{A}} + k_b \mathbb{\bar{B}} + k_s \mathbb{\bar{S}} + k_l \mathbb{\bar{L}})$ covers the terms $( \bar{\textbf{a}}^* \cdot \bar{\textbf{p}} + \bar{a}_0 \bar{p}_0 )$, $( \bar{\textbf{A}}^* \cdot \bar{\textbf{P}} + \bar{\textbf{A}}_0^* \circ \bar{\textbf{P}}_0 )$, $( \bar{\textbf{h}}^* \cdot \bar{\textbf{p}} + k_{eg} \bar{\textbf{H}}^* \cdot \bar{\textbf{P}} )$, $(\bar{\textbf{x}}^* \cdot \bar{\textbf{p}} + \bar{x}_0 \bar{p}_0)$, $(\bar{\textbf{X}}^* \cdot \bar{\textbf{P}} + \bar{\textbf{X}}_0^* \circ \bar{\textbf{P}}_0)$, $( \bar{\textbf{s}}^* \cdot \bar{\textbf{p}} + \bar{s}_0 \bar{p}_0 )$, $( \bar{\textbf{S}}^* \cdot \bar{\textbf{P}} + \bar{\textbf{S}}_0^* \circ \bar{\textbf{P}}_0 )$, $( \bar{\textbf{l}}^* \cdot \bar{\textbf{p}} + \bar{l}_0 \bar{p}_0 )$, and $( \bar{\textbf{L}}^* \cdot \bar{\textbf{P}} + \bar{\textbf{L}}_0^* \circ \bar{\textbf{P}}_0 )$ etc. And that they include the magnetic helicity $ \textbf{A} \cdot \textbf{B} $, the kinetic helicity $ \textbf{v} \cdot \textbf{u} $, the cross helicity $ \textbf{V} \cdot \textbf{B} $, the current helicity $ \textbf{B}^* \cdot \textbf{P} $, and some other new helicity terms etc.

The above is the mass continuity equation in the electromagnetic field and the gravitational field with the octonion operator $(\lozenge + k_x \mathbb{\bar{X}} + k_a \mathbb{\bar{A}} + k_b \mathbb{\bar{B}} + k_s \mathbb{\bar{S}} + k_l \mathbb{\bar{L}})$, and is effected by the speed of light $\bar{v}_0$ and some helicity terms of the rotational objects as well as of the spinning charged objects. The impact of the field source helicity about the $\mathbb{\bar{L}}$ may be detected sometimes.

A new physical quantity $\mathbb{\bar{F}}_q$ can be defined from the part force density $\mathbb{\bar{F}}$,
\begin{eqnarray}
\mathbb{\bar{F}}_q = \mathbb{\bar{F}} \circ \emph{\textbf{I}}_0^*~.
\end{eqnarray}

The scalar part $\bar{F}_0$ of the $\mathbb{\bar{F}}_q$ is written as,
\begin{align}
\bar{F}_0 / v_0 = & ~ (\partial_0 \bar{\textbf{P}}_0 + \nabla^* \cdot \bar{\textbf{P}}) \circ \emph{\textbf{I}}_0^* + k_b ( \bar{\textbf{h}}^* \cdot \bar{\textbf{P}} + k_{eg} \bar{\textbf{H}}^* \cdot \bar{\textbf{p}} ) \circ \emph{\textbf{I}}_0^*
\nonumber
\\
&
+ k_a ( \bar{\textbf{a}}^* \cdot \bar{\textbf{P}} + a_0 \bar{\textbf{P}}_0 ) \circ \emph{\textbf{I}}_0^* + k_{eg} k_a ( \bar{\textbf{A}}^* \cdot \bar{\textbf{p}} + p_0 \bar{\textbf{A}}_0^* ) \circ \emph{\textbf{I}}_0^*
\nonumber
\\
&
+ k_x ( \bar{\textbf{x}}^* \cdot \bar{\textbf{P}} + x_0 \bar{\textbf{P}}_0 ) \circ \emph{\textbf{I}}_0^* + k_{eg} k_x ( \bar{\textbf{X}}^* \cdot \bar{\textbf{p}} + p_0 \bar{\textbf{X}}_0^* ) \circ \emph{\textbf{I}}_0^*
\nonumber
\\
&
+ k_s ( \bar{\textbf{s}}^* \cdot \bar{\textbf{P}} + s_0 \bar{\textbf{P}}_0 ) \circ \emph{\textbf{I}}_0^* + k_{eg} k_s ( \bar{\textbf{S}}^* \cdot \bar{\textbf{p}} + p_0 \bar{\textbf{S}}_0^* ) \circ \emph{\textbf{I}}_0^*
\nonumber
\\
&
+ k_l ( \bar{\textbf{l}}^* \cdot \bar{\textbf{P}} + l_0 \bar{\textbf{P}}_0 ) \circ \emph{\textbf{I}}_0^* + k_{eg} k_l ( \bar{\textbf{L}}^* \cdot \bar{\textbf{p}} + p_0 \bar{\textbf{L}}_0^* ) \circ \emph{\textbf{I}}_0^*
~,
\end{align}
where the last nine terms are the helicity in the gravitational field and electromagnetic field.

The above is the charge continuity equation in the case for coexistence of the gravitational field and electromagnetic field with the octonion operator $(\lozenge + k_x \mathbb{\bar{X}} + k_a \mathbb{\bar{A}} + k_b \mathbb{\bar{B}} + k_s \mathbb{\bar{S}} + k_l \mathbb{\bar{L}} )$ when the scalar $F_0 = 0$. And this charge continuity equation is the invariant under the octonion coordinate transformation. It states that the $\mathbb{\bar{X}}$, $\mathbb{\bar{A}}$, $\mathbb{\bar{B}}$, $\mathbb{\bar{S}}$, and $\mathbb{\bar{L}}$ of the gravitational field and electromagnetic field have the influence on the charge continuity equation, although the terms $( \bar{\textbf{a}}^* \cdot \bar{\textbf{P}} + a_0 \bar{\textbf{P}}_0 ) \circ \emph{\textbf{I}}_0^*$, $( \bar{\textbf{A}}^* \cdot \bar{\textbf{p}} + p_0 \bar{\textbf{A}}_0^* ) \circ \emph{\textbf{I}}_0^*$, $( \bar{\textbf{h}}^* \cdot \bar{\textbf{P}} + k_{eg} \bar{\textbf{H}}^* \cdot \bar{\textbf{p}} ) \circ \emph{\textbf{I}}_0^*$, $( \bar{\textbf{x}}^* \cdot \bar{\textbf{P}} + x_0 \bar{\textbf{P}}_0 ) \circ \emph{\textbf{I}}_0^*$, $( \bar{\textbf{X}}^* \cdot \bar{\textbf{p}} + p_0 \bar{\textbf{X}}_0^* ) \circ \emph{\textbf{I}}_0^*$, $( \bar{\textbf{s}}^* \cdot \bar{\textbf{P}} + s_0 \bar{\textbf{P}}_0 ) \circ \emph{\textbf{I}}_0^*$, $( \bar{\textbf{S}}^* \cdot \bar{\textbf{p}} + p_0 \bar{\textbf{S}}_0^* ) \circ \emph{\textbf{I}}_0^*$, $( \bar{\textbf{l}}^* \cdot \bar{\textbf{P}} + l_0 \bar{\textbf{P}}_0 ) \circ \emph{\textbf{I}}_0^*$, $( \bar{\textbf{L}}^* \cdot \bar{\textbf{p}} + p_0 \bar{\textbf{L}}_0^* ) \circ \emph{\textbf{I}}_0^*$, and $\triangle m$ are usually petty when the electromagnetic field and the gravitational field are ignorable.

%--10--%

\section{The fields regarding the operator ($\lozenge, \mathbb{\bar{X}}, \mathbb{\bar{A}}, \mathbb{\bar{B}}, \mathbb{\bar{S}}, \mathbb{\bar{L}}, \mathbb{\bar{W}}$) }

In the electromagnetic and gravitational fields, the octonion operator ($\lozenge + k_x \mathbb{\bar{X}} + k_a \mathbb{\bar{A}} + k_b \mathbb{\bar{B}} + k_s \mathbb{\bar{S}} + k_l \mathbb{\bar{L}}$) concludes the physical properties of two fields, including the linear momentum, the angular momentum, the energy, the torque, the power, the force, and some helicities of the rotational objects and of the spinning charged objects. Further those inferences lead us to suppose that the angular momentum $\mathbb{\bar{W}}$ may have an influence on the helicities. In this section, one new operator ($\lozenge + k_x \mathbb{\bar{X}} + k_a \mathbb{\bar{A}} + k_b \mathbb{\bar{B}} + k_s \mathbb{\bar{S}} + k_l \mathbb{\bar{L}} + k_w \mathbb{\bar{W}}$) will substitute the operator ($\lozenge + k_x \mathbb{\bar{X}} + k_a \mathbb{\bar{A}} + k_b \mathbb{\bar{B}} + k_s \mathbb{\bar{S}} + k_l \mathbb{\bar{L}}$) to include more properties of the electromagnetic and gravitational fields simultaneously, herein the $k_w = 2 \pi / (v_0 h)$ is the coefficient, with the $h$ being the Planck constant by comparing with the quantum mechanics.

By means of the new octonion operator ($\lozenge + k_x \mathbb{\bar{X}} + k_a \mathbb{\bar{A}} + k_b \mathbb{\bar{B}} + k_s \mathbb{\bar{S}} + k_l \mathbb{\bar{L}} + k_w \mathbb{\bar{W}}$) in the electromagnetic field and the gravitational field, we can represent the octonion field potential, the octonion field strength, the field source, the linear momentum, the angular momentum, the energy, the torque, the force, and some helicities, including the influence of the $\mathbb{W}$ on the torque, the force, and the helicity etc.

\subsection{Field source and torque}

In the electromagnetic and gravitational fields with the octonion operator ($\lozenge + k_x \mathbb{\bar{X}} + k_a \mathbb{\bar{A}} + k_b \mathbb{\bar{B}} + k_s \mathbb{\bar{S}} + k_l \mathbb{\bar{L}} + k_w \mathbb{\bar{W}}$), the majority physical quantities will remain the same, except for the gravitational mass density, the mass continuity equation, the charge continuity equation, and the helicity terms etc. Some octonion physical quantities in the octonion compounding space with the octonion operator ($\lozenge + k_x \mathbb{\bar{X}} + k_a \mathbb{\bar{A}} + k_b \mathbb{\bar{B}} + k_s \mathbb{\bar{S}} + k_l \mathbb{\bar{L}} + k_w \mathbb{\bar{W}}$) are the same as that in the space with the octonion operator ($\lozenge + k_x \mathbb{\bar{X}} + k_a \mathbb{\bar{A}} + k_b \mathbb{\bar{B}} + k_s \mathbb{\bar{S}} + k_l \mathbb{\bar{L}}$) respectively, including the octonion basis vector, the octonion radius vector, the octonion velocity, the octonion vorticity, the octonion field potential, the compounding gravitational potential, the compounding electromagnetic potential, the gravitational strength, the electromagnetic strength, the octonion field strength, the octonion field source, the octonion linear momentum, the octonion angular momentum, the gauge equations, and the field strength helicity etc.

\subsection{Force}

In the compounding space with the octonion operator $(\lozenge + k_x \mathbb{\bar{X}} + k_a \mathbb{\bar{A}} + k_b \mathbb{\bar{B}} + k_s \mathbb{\bar{S}} + k_l \mathbb{\bar{L}} + k_w \mathbb{\bar{W}})$, the octonion force-power density, $\mathbb{\bar{N}} = \Sigma (\bar{n}_i \emph{\textbf{i}}_i) + \Sigma (\bar{N}_i \emph{\textbf{I}}_i)$, is defined as follows,
\begin{eqnarray}
\mathbb{\bar{N}} = v_0 (\lozenge + k_x \mathbb{\bar{X}} + k_a \mathbb{\bar{A}} + k_b \mathbb{\bar{B}} + k_s \mathbb{\bar{S}} + k_l \mathbb{\bar{L}} + k_w \mathbb{\bar{W}})^* \circ \mathbb{\bar{W}}~,
\end{eqnarray}
where the power density is $\bar{f}_0 = - \bar{n}_0/(2 v_0)$, and the force density is $\bar{\textbf{f}} = - \bar{\textbf{n}} / (2 v_0)$. The vectorial parts are $\bar{\textbf{n}} = \Sigma (\bar{n}_j \emph{\textbf{i}}_j )$, $\bar{\textbf{N}}_0 = \bar{N}_0 \emph{\textbf{I}}_0$, and $\bar{\textbf{N}} = \Sigma (\bar{N}_j \emph{\textbf{I}}_j)$.

Further expressing the scalar $\bar{n}_0$ of the octonion $\mathbb{\bar{N}}$ as
\begin{align}
\bar{n}_0 / v_0 = & ~ \partial_0 \bar{w}_0 + \nabla^* \cdot \bar{\textbf{w}} + k_b ( \bar{\textbf{h}}^* \cdot \bar{\textbf{w}} + k_{eg} \bar{\textbf{H}}^* \cdot \bar{\textbf{W}} )
\nonumber\\
&
+ k_a ( \bar{\textbf{a}}^* \cdot \bar{\textbf{w}} + \bar{a}_0 \bar{w}_0 )
+ k_{eg} k_a ( \bar{\textbf{A}}^* \cdot \bar{\textbf{W}} + \bar{\textbf{A}}_0^* \circ \bar{\textbf{W}}_0 )
\nonumber\\
&
+ k_x ( \bar{\textbf{x}}^* \cdot \bar{\textbf{w}} + \bar{x}_0 \bar{w}_0 )
+ k_{eg} k_x ( \bar{\textbf{X}}^* \cdot \bar{\textbf{W}} + \bar{\textbf{X}}_0^* \circ \bar{\textbf{W}}_0 )
\nonumber\\
&
+ k_s ( \bar{\textbf{s}}^* \cdot \bar{\textbf{w}} + \bar{s}_0 \bar{w}_0 )
+ k_{eg} k_s ( \bar{\textbf{S}}^* \cdot \bar{\textbf{W}} + \bar{\textbf{S}}_0^* \circ \bar{\textbf{W}}_0 )
\nonumber\\
&
+ k_l ( \bar{\textbf{l}}^* \cdot \bar{\textbf{w}} + \bar{l}_0 \bar{w}_0 )
+ k_{eg} k_l ( \bar{\textbf{L}}^* \cdot \bar{\textbf{W}} + \bar{\textbf{L}}_0^* \circ \bar{\textbf{W}}_0 )
\nonumber\\
&
+ k_w ( \bar{\textbf{w}}^* \cdot \bar{\textbf{w}} + \bar{w}_0^2 )
+ k_{eg} k_w ( \bar{\textbf{W}}^* \cdot \bar{\textbf{W}} + \bar{\textbf{W}}_0^* \circ \bar{\textbf{W}}_0 )
~.
\end{align}

In the gravitational and electromagnetic fields with the octonion operator $(\lozenge + k_x \mathbb{\bar{X}} + k_a \mathbb{\bar{A}} + k_b \mathbb{\bar{B}} + k_s \mathbb{\bar{S}} + k_l \mathbb{\bar{L}} + k_w \mathbb{\bar{W}})$, the force density $\bar{\textbf{f}}$ can be defined from the vectorial part $\bar{\textbf{n}}$ of the octonion $\mathbb{\bar{N}}$ ,
\begin{align}
- 2 \bar{\textbf{f}} = & ~ \partial_0 \bar{\textbf{w}} + \nabla^* \bar{w}_0 + \nabla^* \times \bar{\textbf{w}} + k_b \bar{w}_0 \bar{\textbf{h}}^* + k_b \bar{\textbf{h}}^* \times \bar{\textbf{w}} + k_{eg} k_b (\bar{\textbf{H}}^* \times \bar{\textbf{W}} + \bar{\textbf{H}}^* \circ \bar{\textbf{W}}_0)
\nonumber\\
& + k_a ( \bar{\textbf{a}}^* \bar{w}_0 + \bar{\textbf{a}}^* \times \bar{\textbf{w}} + \bar{a}_0 \bar{\textbf{w}})
+ k_{eg} k_a ( \bar{\textbf{A}}^* \times \bar{\textbf{W}} + \bar{\textbf{A}}_0^* \circ \bar{\textbf{W}} + \bar{\textbf{A}}^* \circ \bar{\textbf{W}}_0)
\nonumber\\
& + k_x ( \bar{\textbf{x}}^* \bar{w}_0 + \bar{\textbf{x}}^* \times \bar{\textbf{w}} + \bar{x}_0 \bar{\textbf{w}})
+ k_{eg} k_x ( \bar{\textbf{X}}^* \times \bar{\textbf{W}} + \bar{\textbf{X}}_0^* \circ \bar{\textbf{W}} + \bar{\textbf{X}}^* \circ \bar{\textbf{W}}_0)
\nonumber\\
& + k_s ( \bar{\textbf{s}}^* \bar{w}_0 + \bar{\textbf{s}}^* \times \bar{\textbf{w}} + \bar{s}_0 \bar{\textbf{w}})
+ k_{eg} k_s ( \bar{\textbf{S}}^* \times \bar{\textbf{W}} + \bar{\textbf{S}}_0^* \circ \bar{\textbf{W}} + \bar{\textbf{S}}^* \circ \bar{\textbf{W}}_0)
\nonumber\\
& + k_l ( \bar{\textbf{l}}^* \bar{w}_0 + \bar{\textbf{l}}^* \times \bar{\textbf{w}} + \bar{l}_0 \bar{\textbf{w}})
+ k_{eg} k_l ( \bar{\textbf{L}}^* \times \bar{\textbf{W}} + \bar{\textbf{L}}_0^* \circ \bar{\textbf{W}} + \bar{\textbf{L}}^* \circ \bar{\textbf{W}}_0)
~,
\end{align}
where the force density $\bar{\textbf{f}}$ includes that of the inertial force, the gravitational force, the gradient of energy, Lorentz force, and the interacting force between the dipole moment with the two fields etc. The above force definition is much longer than that in the classical field theory, and includes more new force terms related to the gradient of energy, the field potential, and the angular velocity etc.

\subsection{Helicity}

As the derivation of the compounding space with the operator $(\lozenge + k_x \mathbb{\bar{X}} + k_a \mathbb{\bar{A}} + k_b \mathbb{\bar{B}} + k_s \mathbb{\bar{S}} + k_l \mathbb{\bar{L}} + k_w \mathbb{\bar{W}})$, some helicity terms \cite{liu1, liu2} will impact the mass continuity equation and the charge continuity equation etc.

\subsubsection{Field strength helicity}

The octonion linear momentum density, $\mathbb{\bar{P}} = \mu \mathbb{\bar{S}} / \mu_g$, can be defined from the octonion field source $\mathbb{\bar{S}}$ in the octonion compounding space with the operator $(\lozenge + k_x \mathbb{\bar{X}} + k_a \mathbb{\bar{A}} + k_b \mathbb{\bar{B}} + k_s \mathbb{\bar{S}} + k_l \mathbb{\bar{L}} + k_w \mathbb{\bar{W}})$,
\begin{eqnarray}
\mathbb{\bar{P}} = \widehat{m} \bar{v}_0 + \Sigma (m \bar{v}_j \emph{\textbf{i}}_j) + \Sigma (M \bar{V}_i \emph{\textbf{I}}_i)~,
\end{eqnarray}
where $\widehat{m} = m (\bar{v}_0^\delta / \bar{v}_0) + \bigtriangleup m$, $\bigtriangleup m = - (k_x \mathbb{\bar{X}}^* \cdot \mathbb{\bar{B}} + k_a \mathbb{\bar{A}}^* \cdot \mathbb{\bar{B}} + k_b \mathbb{\bar{B}}^* \cdot \mathbb{\bar{B}}) / ( \bar{v}_0 \mu_g )$ .

According to the octonion features, the gravitational mass $\widehat{m}$ is one reserved scalar, and is changed with the field strength $\mathbb{\bar{B}}_g$ and $\mathbb{\bar{B}}_e$, the field potential $\mathbb{\bar{A}}_g$ and $\mathbb{\bar{A}}_e$, the helicity $(k_x \mathbb{\bar{X}}^* \cdot \mathbb{\bar{B}} + k_a \mathbb{\bar{A}}^* \cdot \mathbb{\bar{B}} + k_b \mathbb{\bar{B}}^* \cdot \mathbb{\bar{B}})$, and the octonion $\mathbb{\bar{X}}_g$ and $\mathbb{\bar{X}}_e$ etc in the octonion compounding space. The helicity includes the magnetic helicity $ \bar{\textbf{A}} \cdot \bar{\textbf{B}} $, the kinetic helicity $ \bar{\textbf{v}} \cdot \bar{\textbf{u}} $, the cross helicity $ \bar{\textbf{V}} \cdot \bar{\textbf{B}} $, and one new helicity term $ \bar{\textbf{A}} \cdot \bar{\textbf{U}} $ etc.

Similarly to the $\mathbb{\bar{S}}$ and the $\mathbb{\bar{L}}$, the octonion angular momentum $\mathbb{\bar{W}}$ does not exert an influence on the field strength helicity in the octonion compounding space with the operator $(\lozenge + k_x \mathbb{\bar{X}} + k_a \mathbb{\bar{A}} + k_b \mathbb{\bar{B}} + k_s \mathbb{\bar{S}} + k_l \mathbb{\bar{L}} + k_w \mathbb{\bar{W}})$. Consequently we can not distinguish this field from some fields with other octonion operators, according to the viewpoint of the field strength helicity.

\subsubsection{Field source helicity}

The part force density $\mathbb{\bar{F}}$ is one part of the octonion force-power density $\mathbb{\bar{N}}$, and is defined from the octonion linear momentum density $\mathbb{\bar{P}}$ ,
\begin{eqnarray}
\mathbb{\bar{F}} = v_0 (\lozenge + k_x \mathbb{\bar{X}} + k_a \mathbb{\bar{A}} + k_b \mathbb{\bar{B}} + k_s \mathbb{\bar{S}} + k_l \mathbb{\bar{L}} + k_w \mathbb{\bar{W}})^* \circ \mathbb{\bar{P}}~,
\end{eqnarray}
where the part force density $\mathbb{\bar{F}}$ includes the density of the inertial force, the gravitational force, Lorentz force, and the interacting force between the two fields with the dipoles etc.

The scalar $\bar{f}_0$ of the part force density $\mathbb{\bar{F}}$ is written as,
\begin{align}
\bar{f}_0 / v_0 = & ~ \partial_0 \bar{p}_0 + \nabla^* \cdot \bar{\textbf{p}} + k_b ( \bar{\textbf{h}}^* \cdot \bar{\textbf{p}} + k_{eg} \bar{\textbf{H}}^* \cdot \bar{\textbf{P}} )
\nonumber\\
&
+ k_a ( \bar{\textbf{a}}^* \cdot \bar{\textbf{p}} + \bar{a}_0 \bar{p}_0 ) + k_{eg} k_a ( \bar{\textbf{A}}^* \cdot \bar{\textbf{P}} + \bar{\textbf{A}}_0^* \circ \bar{\textbf{P}}_0 )
\nonumber\\
& + k_x ( \bar{\textbf{x}}^* \cdot \bar{\textbf{p}} + \bar{x}_0 \bar{p}_0 ) + k_{eg} k_x ( \bar{\textbf{X}}^* \cdot \bar{\textbf{P}} + \bar{\textbf{X}}_0^* \circ \bar{\textbf{P}}_0 )
\nonumber\\
&
+ k_s ( \bar{\textbf{s}}^* \cdot \bar{\textbf{p}} + \bar{s}_0 \bar{p}_0 ) + k_{eg} k_s ( \bar{\textbf{S}}^* \cdot \bar{\textbf{P}} + \bar{\textbf{S}}_0^* \circ \bar{\textbf{P}}_0 )
\nonumber\\
& + k_l ( \bar{\textbf{l}}^* \cdot \bar{\textbf{p}} + \bar{l}_0 \bar{p}_0 ) + k_{eg} k_l ( \bar{\textbf{L}}^* \cdot \bar{\textbf{P}} + \bar{\textbf{L}}_0^* \circ \bar{\textbf{P}}_0 )
\nonumber\\
&
+ k_w ( \bar{\textbf{w}}^* \cdot \bar{\textbf{p}} + \bar{w}_0 \bar{p}_0 ) + k_{eg} k_w ( \bar{\textbf{W}}^* \cdot \bar{\textbf{P}} + \bar{\textbf{W}}_0^* \circ \bar{\textbf{P}}_0 )
~,
\end{align}
where the field source helicity in the two fields with the operator $(\lozenge + k_x \mathbb{\bar{X}} + k_a \mathbb{\bar{A}} + k_b \mathbb{\bar{B}} + k_s \mathbb{\bar{S}} + k_l \mathbb{\bar{L}} + k_w \mathbb{\bar{W}})$ includes the helicity terms $( \bar{\textbf{a}}^* \cdot \bar{\textbf{p}} + \bar{a}_0 \bar{p}_0 )$, $( \bar{\textbf{A}}^* \cdot \bar{\textbf{P}} + \bar{\textbf{A}}_0^* \circ \bar{\textbf{P}}_0 )$, $( \bar{\textbf{h}}^* \cdot \bar{\textbf{p}} + k_{eg} \bar{\textbf{H}}^* \cdot \bar{\textbf{P}} )$, $(\bar{\textbf{x}}^* \cdot \bar{\textbf{p}} + \bar{x}_0 \bar{p}_0)$, $(\bar{\textbf{X}}^* \cdot \bar{\textbf{P}} + \bar{\textbf{X}}_0^* \circ \bar{\textbf{P}}_0)$, $( \bar{\textbf{s}}^* \cdot \bar{\textbf{p}} + \bar{s}_0 \bar{p}_0 )$, $( \bar{\textbf{S}}^* \cdot \bar{\textbf{P}} + \bar{\textbf{S}}_0^* \circ \bar{\textbf{P}}_0 )$, $( \bar{\textbf{l}}^* \cdot \bar{\textbf{p}} + \bar{l}_0 \bar{p}_0 )$, $( \bar{\textbf{L}}^* \cdot \bar{\textbf{P}} + \bar{\textbf{L}}_0^* \circ \bar{\textbf{P}}_0 )$, $( \bar{\textbf{w}}^* \cdot \bar{\textbf{p}} + \bar{w}_0 \bar{p}_0 )$, and $( \bar{\textbf{W}}^* \cdot \bar{\textbf{P}} + \bar{\textbf{W}}_0^* \circ \bar{\textbf{P}}_0 )$ etc. And that they include the magnetic helicity $ \textbf{A} \cdot \textbf{B} $, the kinetic helicity $ \textbf{v} \cdot \textbf{u} $, the cross helicity $ \textbf{V} \cdot \textbf{B} $, the current helicity $ \textbf{B}^* \cdot \textbf{P} $, and some other new helicity terms in the octonion space.

The above is the mass continuity equation in the electromagnetic field and the gravitational field with the octonion operator $(\lozenge + k_x \mathbb{\bar{X}} + k_a \mathbb{\bar{A}} + k_b \mathbb{\bar{B}} + k_s \mathbb{\bar{S}} + k_l \mathbb{\bar{L}} + k_w \mathbb{\bar{W}})$, and is effected by the speed of light $\bar{v}_0$ and some helicity terms of the rotational objects and of the spinning charged objects. The impact of the field source helicity about the octonion torque-energy $\mathbb{\bar{W}}$ may be detected sometimes.

A new physical quantity $\mathbb{\bar{F}}_q$ can be defined from the part force density $\mathbb{\bar{F}}$,
\begin{eqnarray}
\mathbb{\bar{F}}_q = \mathbb{\bar{F}} \circ \emph{\textbf{I}}_0^*~.
\end{eqnarray}

The scalar part $\bar{F}_0$ of the $\mathbb{\bar{F}}_q$ is written as,
\begin{align}
\bar{F}_0 / v_0 = & ~ (\partial_0 \bar{\textbf{P}}_0 + \nabla^* \cdot \bar{\textbf{P}}) \circ \emph{\textbf{I}}_0^* + k_b ( \bar{\textbf{h}}^* \cdot \bar{\textbf{P}} + k_{eg} \bar{\textbf{H}}^* \cdot \bar{\textbf{p}} ) \circ \emph{\textbf{I}}_0^*
\nonumber
\\
&
+ k_a ( \bar{\textbf{a}}^* \cdot \bar{\textbf{P}} + a_0 \bar{\textbf{P}}_0 ) \circ \emph{\textbf{I}}_0^* + k_{eg} k_a ( \bar{\textbf{A}}^* \cdot \bar{\textbf{p}} + p_0 \bar{\textbf{A}}_0^* ) \circ \emph{\textbf{I}}_0^*
\nonumber
\\
&
+ k_x ( \bar{\textbf{x}}^* \cdot \bar{\textbf{P}} + x_0 \bar{\textbf{P}}_0 ) \circ \emph{\textbf{I}}_0^* + k_{eg} k_x ( \bar{\textbf{X}}^* \cdot \bar{\textbf{p}} + p_0 \bar{\textbf{X}}_0^* ) \circ \emph{\textbf{I}}_0^*
\nonumber
\\
&
+ k_s ( \bar{\textbf{s}}^* \cdot \bar{\textbf{P}} + s_0 \bar{\textbf{P}}_0 ) \circ \emph{\textbf{I}}_0^* + k_{eg} k_s ( \bar{\textbf{S}}^* \cdot \bar{\textbf{p}} + p_0 \bar{\textbf{S}}_0^* ) \circ \emph{\textbf{I}}_0^*
\nonumber
\\
&
+ k_l ( \bar{\textbf{l}}^* \cdot \bar{\textbf{P}} + l_0 \bar{\textbf{P}}_0 ) \circ \emph{\textbf{I}}_0^* + k_{eg} k_l ( \bar{\textbf{L}}^* \cdot \bar{\textbf{p}} + p_0 \bar{\textbf{L}}_0^* ) \circ \emph{\textbf{I}}_0^*
\nonumber
\\
&
+ k_w ( \bar{\textbf{w}}^* \cdot \bar{\textbf{P}} + w_0 \bar{\textbf{P}}_0 ) \circ \emph{\textbf{I}}_0^* + k_{eg} k_w ( \bar{\textbf{W}}^* \cdot \bar{\textbf{p}} + p_0 \bar{\textbf{W}}_0^* ) \circ \emph{\textbf{I}}_0^*
~,
\end{align}
where the last two terms are the new kinds of helicity terms in the gravitational field and electromagnetic field.

The above is the charge continuity equation in the presence of the gravitational field and electromagnetic field with the octonion operator $(\lozenge + k_x \mathbb{\bar{X}} + k_a \mathbb{\bar{A}} + k_b \mathbb{\bar{B}} + k_s \mathbb{\bar{S}} + k_l \mathbb{\bar{L}} + k_w \mathbb{\bar{W}} )$ when the scalar $F_0 = 0$. And this charge continuity equation is the invariant under the octonion coordinate transformation. It states that the $\mathbb{\bar{X}}$, $\mathbb{\bar{A}}$, $\mathbb{\bar{B}}$, $\mathbb{\bar{S}}$, $\mathbb{\bar{L}}$, and $\mathbb{\bar{W}}$ of the gravitational field and the electromagnetic field have the influence on the charge continuity equation, although the terms $( \bar{\textbf{a}}^* \cdot \bar{\textbf{P}} + a_0 \bar{\textbf{P}}_0 ) \circ \emph{\textbf{I}}_0^*$, $( \bar{\textbf{A}}^* \cdot \bar{\textbf{p}} + p_0 \bar{\textbf{A}}_0^* ) \circ \emph{\textbf{I}}_0^*$, $( \bar{\textbf{h}}^* \cdot \bar{\textbf{P}} + k_{eg} \bar{\textbf{H}}^* \cdot \bar{\textbf{p}} ) \circ \emph{\textbf{I}}_0^*$, $( \bar{\textbf{x}}^* \cdot \bar{\textbf{P}} + x_0 \bar{\textbf{P}}_0 ) \circ \emph{\textbf{I}}_0^*$, $( \bar{\textbf{X}}^* \cdot \bar{\textbf{p}} + p_0 \bar{\textbf{X}}_0^* ) \circ \emph{\textbf{I}}_0^*$, $( \bar{\textbf{s}}^* \cdot \bar{\textbf{P}} + s_0 \bar{\textbf{P}}_0 ) \circ \emph{\textbf{I}}_0^*$, $( \bar{\textbf{S}}^* \cdot \bar{\textbf{p}} + p_0 \bar{\textbf{S}}_0^* ) \circ \emph{\textbf{I}}_0^*$, $( \bar{\textbf{l}}^* \cdot \bar{\textbf{P}} + l_0 \bar{\textbf{P}}_0 ) \circ \emph{\textbf{I}}_0^*$, $( \bar{\textbf{L}}^* \cdot \bar{\textbf{p}} + p_0 \bar{\textbf{L}}_0^* ) \circ \emph{\textbf{I}}_0^*$, $( \bar{\textbf{w}}^* \cdot \bar{\textbf{P}} + w_0 \bar{\textbf{P}}_0 ) \circ \emph{\textbf{I}}_0^*$, $( \bar{\textbf{W}}^* \cdot \bar{\textbf{p}} + p_0 \bar{\textbf{W}}_0^* ) \circ \emph{\textbf{I}}_0^*$, and $\triangle m$ are usually paltry when the electromagnetic field and the gravitational field are slender.

\begin{table}[h]
\caption{Some quantities with the octonion operator ($\lozenge + k_x \mathbb{\bar{X}} + k_a \mathbb{\bar{A}} + k_b \mathbb{\bar{B}} + k_s \mathbb{\bar{S}} + k_l \mathbb{\bar{L}} + k_w \mathbb{\bar{W}}$).}
\label{tab:table3}
\centering
\begin{tabular}{ll}
\hline\hline
$ definitions $                                                                                                                          & $ meanings $ \\
\hline
$\mathbb{\bar{X}}$                                                                                                                       & field quantity \\
$\mathbb{\bar{A}} = (\lozenge + k_x \mathbb{\bar{X}}) \circ \mathbb{\bar{X}}$                                                            & field potential \\
$\mathbb{\bar{B}} = (\lozenge + k_x \mathbb{\bar{X}} + k_a \mathbb{\bar{A}}) \circ \mathbb{\bar{A}}$                                     & field strength \\
$\mathbb{\bar{R}}$                                                                                                                       & radius vector \\
$\mathbb{\bar{V}} = v_0 \lozenge \circ \mathbb{\bar{R}}$                                                                                 & velocity \\
$\mathbb{\bar{U}} = \lozenge \circ \mathbb{\bar{V}}$                                                                                     & velocity curl \\
$\mu \mathbb{\bar{S}} = - ( \lozenge + k_x \mathbb{\bar{X}} + k_a \mathbb{\bar{A}} + k_b \mathbb{\bar{B}} )^* \circ \mathbb{\bar{B}}$    & field source \\
$\mathbb{\bar{H}}_b = (k_x \mathbb{\bar{X}} + k_a \mathbb{\bar{A}} + k_b \mathbb{\bar{B}})^* \cdot \mathbb{\bar{B}}$                     & field strength helicity\\
$\mathbb{\bar{P}} = \mu \mathbb{\bar{S}} / \mu_g$                                                                                        & linear momentum density \\
$\mathbb{\bar{R}} = \mathbb{R} + k_{rx} \mathbb{X}$                                                                                      & compounding radius vector \\
$\mathbb{\bar{L}} = \mathbb{\bar{R}} \circ \mathbb{\bar{P}}$                                                                             & angular momentum density \\
$\mathbb{\bar{W}} = v_0 ( \lozenge + k_x \mathbb{\bar{X}} + k_a \mathbb{\bar{A}} + k_b \mathbb{\bar{B}} + k_s \mathbb{\bar{S}} + k_l \mathbb{\bar{L}} ) \circ \mathbb{\bar{L}}$                                                                                                                        & torque-energy densities \\
$\mathbb{\bar{N}} = v_0 ( \lozenge + k_x \mathbb{\bar{X}} + k_a \mathbb{\bar{A}} + k_b \mathbb{\bar{B}} + k_s \mathbb{\bar{S}} + k_l \mathbb{\bar{L}} + k_w \mathbb{\bar{W}} )^* \circ \mathbb{\bar{W}}$                                                                                             & force-power density \\
$\mathbb{\bar{F}} = - \mathbb{\bar{N}} / (2v_0)$                                                                                         & force density \\
$\mathbb{\bar{H}}_s = ( k_x \mathbb{\bar{X}} + k_a \mathbb{\bar{A}} + k_b \mathbb{\bar{B}} + k_s \mathbb{\bar{S}} + k_l \mathbb{\bar{L}} + k_w \mathbb{\bar{W}})^* \cdot \mathbb{\bar{P}}$                                                                                                                        & field source helicity\\
\hline\hline
\end{tabular}
\end{table}

%--11--%

\section{The fields regarding the octonion operator $\lozenge_8$}

In the electromagnetic and gravitational fields, the octonion operator ($\lozenge_8 + k_x \mathbb{\bar{X}} + k_a \mathbb{\bar{A}} + k_b \mathbb{\bar{B}} + k_s \mathbb{\bar{S}} + k_l \mathbb{\bar{L}} + k_w \mathbb{\bar{W}}$) can derive the octonion physical properties of the two fields, including the octonion linear momentum, the octonion angular momentum, the energy, the torque, the power, the force, and some helicity terms of the rotational objects and the spinning charged objects. However these inferences can not involve the adjoint fields and their related helicities of the electromagnetic field and the gravitational field etc. In this section, the octonion operator will substitute the $\lozenge_8 = \lozenge_g + d_e \lozenge_e$ for the $\lozenge_g$, to include more properties of the electromagnetic and gravitational fields simultaneously. Herein $\lozenge_g = \Sigma ( \emph{\textbf{i}}_i \partial_{gi})$, $\lozenge_e = \Sigma ( \emph{\textbf{I}}_i \partial_{ei})$; $\partial_{gi} = \partial/\partial r_i$, $\partial_{ei} = \partial/\partial R_i$. $d_e$ is the coefficient.

With the octonion operator $\lozenge_8$ in the electromagnetic and gravitational fields, we can represent the field potential, the field strength, the field source, the linear momentum, the angular momentum, the energy, the torque, the force, and some helicity terms, including the influence of the adjoint fields on the torque, the force, and the helicity etc.

\subsection{Field source and torque}

In the electromagnetic field and the gravitational field with the operator ($\lozenge_8 + k_x \mathbb{\bar{X}} + k_a \mathbb{\bar{A}} + k_b \mathbb{\bar{B}} + k_s \mathbb{\bar{S}} + k_l \mathbb{\bar{L}} + k_w \mathbb{\bar{W}}$), the most of physical quantities will keep unchanged, except for the gravitational mass density and the helicity etc. In the compounding space with the octonion operator ($\lozenge_8 + k_x \mathbb{\bar{X}} + k_a \mathbb{\bar{A}} + k_b \mathbb{\bar{B}} + k_s \mathbb{\bar{S}} + k_l \mathbb{\bar{L}} + k_w \mathbb{\bar{W}}$), the octonion basis vector is $ \mathbb{E} = (1, \emph{\textbf{i}}_1, \emph{\textbf{i}}_2, \emph{\textbf{i}}_3, \emph{\textbf{I}}_0, \emph{\textbf{I}}_1, \emph{\textbf{I}}_2, \emph{\textbf{I}}_3) $, the octonion radius vector is $\mathbb{\bar{R}} = \Sigma ( \emph{\textbf{i}}_i \bar{r}_i + k_{eg} \emph{\textbf{I}}_i \bar{R}_i)$, and the octonion velocity is $\mathbb{\bar{V}} = \Sigma ( \emph{\textbf{i}}_i \bar{v}_i + k_{eg} \emph{\textbf{I}}_i \bar{V}_i)$. The octonion field potential is $\mathbb{\bar{A}} = (\lozenge_8 + k_x \mathbb{\bar{X}}) \circ \mathbb{\bar{X}} = \mathbb{\bar{A}}_g + k_{eg} \mathbb{\bar{A}}_e$. The octonion compounding gravitational potential is
\begin{eqnarray}
\mathbb{\bar{A}}_g = ( \lozenge_g \circ \mathbb{\bar{X}}_g + d_e k_{eg} \lozenge_e \circ \mathbb{\bar{X}}_e ) + k_x ( \mathbb{\bar{X}}_g \circ \mathbb{\bar{X}}_g + k_{eg}^2 \mathbb{\bar{X}}_e \circ \mathbb{\bar{X}}_e )~,
\end{eqnarray}
and the compounding electromagnetic potential is
\begin{eqnarray}
\mathbb{\bar{A}}_e = ( \lozenge_g \circ \mathbb{\bar{X}}_e + d_e k_{eg}^{-1} \lozenge_e \circ \mathbb{\bar{X}}_g ) + k_x ( \mathbb{\bar{X}}_e \circ \mathbb{\bar{X}}_g + \mathbb{\bar{X}}_g \circ \mathbb{\bar{X}}_e)~,
\end{eqnarray}
where $\mathbb{\bar{X}}_g = \Sigma ( \bar{x}_i \emph{\textbf{i}}_i) $, and $\mathbb{\bar{X}}_e = \Sigma (\bar{X}_i \emph{\textbf{I}}_i)$.

The octonion field strength is defined as $\mathbb{\bar{B}} = (\lozenge_8 + k_x \mathbb{\bar{X}} + k_a \mathbb{\bar{A}}) \circ \mathbb{\bar{A}} = \mathbb{\bar{B}}_g + k_{eg} \mathbb{\bar{B}}_e$ . And the octonion compounding gravitational strength is
\begin{eqnarray}
\mathbb{\bar{B}}_g = ( \lozenge_g \circ \mathbb{\bar{A}}_g + d_e k_{eg} \lozenge_e \circ \mathbb{\bar{A}}_e ) + k_x ( \mathbb{\bar{X}}_g \circ \mathbb{\bar{A}}_g + k_{eg}^2 \mathbb{\bar{X}}_e \circ \mathbb{\bar{A}}_e )+ k_a ( \mathbb{\bar{A}}_g \circ \mathbb{\bar{A}}_g + k_{eg}^2 \mathbb{\bar{A}}_e \circ \mathbb{\bar{A}}_e )~,
\end{eqnarray}
while the compounding electromagnetic strength is
\begin{eqnarray}
\mathbb{\bar{B}}_e = ( \lozenge_g \circ \mathbb{\bar{A}}_e  + d_e k_{eg}^{-1} \lozenge_e \circ \mathbb{\bar{A}}_g ) + k_x ( \mathbb{\bar{X}}_e \circ \mathbb{\bar{A}}_g + \mathbb{\bar{X}}_g \circ \mathbb{\bar{A}}_e) + k_a ( \mathbb{\bar{A}}_e \circ \mathbb{\bar{A}}_g + \mathbb{\bar{A}}_g \circ \mathbb{\bar{A}}_e)~,
\end{eqnarray}
where the gauge equations are $\bar{h}_0 = 0$ and $\bar{H}_0 = 0$. $\mathbb{\bar{B}}_g = \Sigma ( \bar{h}_i \emph{\textbf{i}}_i)$, and $\mathbb{\bar{B}}_e = \Sigma (\bar{H}_i \emph{\textbf{I}}_i)$.

In the octonion compounding space, the linear momentum density $\mathbb{\bar{S}}_{gg}$ is the source of the gravitational field, and the $\mathbb{\bar{S}}_{ge}$ is the source of the gravitational adjoint field,
\begin{eqnarray}
-\mu_{gg} \mathbb{\bar{S}}_{gg} = \lozenge_g^* \circ \mathbb{\bar{B}}_g~,~-\mu_{ge} \mathbb{\bar{S}}_{ge} = \lozenge_e^* \circ \mathbb{\bar{B}}_g~,
\end{eqnarray}
meanwhile the electric current density $\mathbb{\bar{S}}_{ge}$ is the source of the electromagnetic field, and the $\mathbb{\bar{S}}_{ee}$ is the source of the electromagnetic adjoint field,
\begin{eqnarray}
-\mu_{eg} \mathbb{\bar{S}}_{eg} = \lozenge_g^* \circ \mathbb{\bar{B}}_e~,~-\mu_{ee} \mathbb{\bar{S}}_{ee} = \lozenge_e^* \circ \mathbb{\bar{B}}_e~.
\end{eqnarray}

The $\mathbb{\bar{S}}_{eg}$ possesses the properties of the electromagnetic field, and the $\mathbb{\bar{S}}_{ee}$ seizes on the features of the gravitational field. It means that the latter can be considered as the candidate of the dark matters.

And then the octonion field source $\mathbb{\bar{S}}$ can be defined from the above
\begin{align}
\mu \mathbb{\bar{S}} = & ~ - (\lozenge_8 + k_x \mathbb{\bar{X}} + k_a \mathbb{\bar{A}} + k_b \mathbb{\bar{B}})^* \circ \mathbb{\bar{B}}
\nonumber\\
= & ~ \mu_{gg} \mathbb{\bar{S}}_{gg} + k_{eg} \mu_{ge} \mathbb{\bar{S}}_{ge} + d_e (\mu_{eg} \mathbb{\bar{S}}_{eg} + k_{eg} \mu_{ee} \mathbb{\bar{S}}_{ee})
- k_x \mathbb{\bar{X}}^* \circ \mathbb{\bar{B}} - k_a \mathbb{\bar{A}}^* \circ \mathbb{\bar{B}} - k_b \mathbb{\bar{B}}^* \circ \mathbb{\bar{B}} ~,
\end{align}
where $(k_x \mathbb{\bar{X}}^* \cdot \mathbb{\bar{B}} + k_a \mathbb{\bar{A}}^* \cdot \mathbb{\bar{B}} + k_b \mathbb{\bar{B}}^* \cdot \mathbb{\bar{B}})$ is the field strength helicity. $\mu$, $\mu_{gg}$, $\mu_{ge}$, $\mu_{eg}$, and $\mu_{ee}$ are the coefficients. The $\mu_{gg}$ and $\mu_{eg}$ are the gravitational and electromagnetic constants respectively. $\mathbb{\bar{X}}^* \cdot \mathbb{\bar{B}}$ is the scalar part of $\mathbb{\bar{X}}^* \circ \mathbb{\bar{B}}$.

The octonion linear momentum density is $\mathbb{\bar{P}} = \mu \mathbb{\bar{S}} / \mu_{gg} = \Sigma (\bar{p}_i \emph{\textbf{i}}_i + \bar{P}_i \emph{\textbf{I}}_i)$, while the octonion angular momentum density is $\mathbb{\bar{L}} = \mathbb{\bar{R}} \circ \mathbb{\bar{P}} = \Sigma (\bar{l}_i \emph{\textbf{i}}_i + \bar{L}_i \emph{\textbf{I}}_i)$. The octonion torque-energy density $\mathbb{\bar{W}}$ is defined from the octonion angular momentum density $\mathbb{\bar{L}}$, the field strength $\mathbb{\bar{B}}$, and the field source $\mathbb{\bar{S}}$ etc,
\begin{eqnarray}
\mathbb{\bar{W}} = v_0 (\lozenge_8 + k_x \mathbb{\bar{X}} + k_a \mathbb{\bar{A}} + k_b \mathbb{\bar{B}} + k_s \mathbb{\bar{S}} + k_l \mathbb{\bar{L}}) \circ \mathbb{\bar{L}}~,
\end{eqnarray}
where $\mathbb{\bar{S}} = \Sigma (\bar{s}_i \emph{\textbf{i}}_i ) + k_{eg} \Sigma (\bar{S}_i \emph{\textbf{I}}_i )$; $-\bar{w}_0/2$ is the energy density, $\bar{\textbf{w}}/2 = \Sigma (\bar{w}_j \emph{\textbf{i}}_j )/2$ is the torque density.

The scalar $\bar{w}_0$ of the octonion torque-energy density $\mathbb{\bar{W}} = \Sigma (\bar{w}_i \emph{\textbf{i}}_i ) + \Sigma (\bar{W}_i \emph{\textbf{I}}_i )$ is written as,
\begin{align}
\bar{w}_0 / v_0 = & ~ \partial_{g0} \bar{l}_0 + \nabla_g \cdot \bar{\textbf{l}} + \emph{\textbf{I}}_0 \circ \partial_{e0} \bar{\textbf{L}}_0 + \nabla_e \cdot \bar{\textbf{L}}
+ k_b ( \bar{\textbf{h}} \cdot \bar{\textbf{l}} + k_{eg} \bar{\textbf{H}} \cdot \bar{\textbf{L}} )
\nonumber\\
&
+ k_a \bar{\textbf{a}} \cdot \bar{\textbf{l}} + k_a \bar{a}_0 \bar{l}_0 + k_{eg} k_a ( \bar{\textbf{A}} \cdot \bar{\textbf{L}} + \bar{\textbf{A}}_0 \circ \bar{\textbf{L}}_0 )
\nonumber\\
&
+ k_x \bar{\textbf{x}} \cdot \bar{\textbf{l}} + k_x \bar{x}_0 \bar{l}_0 + k_{eg} k_x ( \bar{\textbf{X}} \cdot \bar{\textbf{L}} + \bar{\textbf{X}}_0 \circ \bar{\textbf{L}}_0 )
\nonumber\\
&
+ k_s \bar{\textbf{s}} \cdot \bar{\textbf{l}} + k_s \bar{s}_0 \bar{l}_0 + k_{eg} k_s ( \bar{\textbf{S}} \cdot \bar{\textbf{L}} + \bar{\textbf{S}}_0 \circ \bar{\textbf{L}}_0 )
\nonumber\\
&
+ k_l \bar{\textbf{l}} \cdot \bar{\textbf{l}} + k_l \bar{l}_0^2 + k_{eg} k_l ( \bar{\textbf{L}} \cdot \bar{\textbf{L}} + \bar{\textbf{L}}_0 \circ \bar{\textbf{L}}_0 )
~,
\end{align}
where $-\bar{w}_0/2$ includes the kinetic energy, the gravitational potential energy, the electric potential energy, the magnetic potential energy, the field energy, the work, the interacting energy between the dipole moment with the two fields, and some new energy terms. $\nabla_g = \Sigma ( \emph{\textbf{i}}_j \partial_{gj})$, $\nabla_e = \Sigma ( \emph{\textbf{I}}_j \partial_{ej})$.

In a similar way, expressing the torque density $\bar{\textbf{w}}$ of the octonion torque-energy density $\mathbb{\bar{W}}$ as
\begin{align}
\bar{\textbf{w}} / v_0 = & ~ \partial_{g0} \bar{\textbf{l}} + \nabla_g \bar{l}_0 + \nabla_g \times \bar{\textbf{l}}
+ \emph{\textbf{I}}_0 \circ \partial_{e0} \bar{\textbf{L}} + \nabla_e \circ \bar{\textbf{L}}_0 + \nabla_e \times \bar{\textbf{L}}
\nonumber
\\
&
+ k_b \bar{l}_0 \bar{\textbf{h}} + k_b \bar{\textbf{h}} \times \bar{\textbf{l}} + k_{eg} k_b (\bar{\textbf{H}} \times \bar{\textbf{L}} + \bar{\textbf{H}} \circ \bar{\textbf{L}}_0)
\nonumber
\\
& + k_a \bar{\textbf{a}} \bar{l}_0 + k_a \bar{\textbf{a}} \times \bar{\textbf{l}} + k_a \bar{a}_0 \bar{\textbf{l}} + k_{eg} k_a ( \bar{\textbf{A}} \times \bar{\textbf{L}} + \bar{\textbf{A}}_0 \circ \bar{\textbf{L}} + \bar{\textbf{A}} \circ \bar{\textbf{L}}_0)
\nonumber
\\
& + k_x \bar{\textbf{x}} \bar{l}_0 + k_x \bar{\textbf{x}} \times \bar{\textbf{l}} + k_x \bar{x}_0 \bar{\textbf{l}} + k_{eg} k_x ( \bar{\textbf{X}} \times \bar{\textbf{L}} + \bar{\textbf{X}}_0 \circ \bar{\textbf{L}} + \bar{\textbf{X}} \circ \bar{\textbf{L}}_0)
\nonumber
\\
& + k_s \bar{\textbf{s}} \bar{l}_0 + k_s \bar{\textbf{s}} \times \bar{\textbf{l}} + k_s \bar{s}_0 \bar{\textbf{l}} + k_{eg} k_s ( \bar{\textbf{S}} \times \bar{\textbf{L}} + \bar{\textbf{S}}_0 \circ \bar{\textbf{L}} + \bar{\textbf{S}} \circ \bar{\textbf{L}}_0)
+ 2 k_l \bar{l}_0 \bar{\textbf{l}}
~,
\end{align}
where the above includes some new terms of the torque density.

\subsection{Force}

In the octonion compounding space with the operator $(\lozenge_8 + k_x \mathbb{\bar{X}} + k_a \mathbb{\bar{A}} + k_b \mathbb{\bar{B}} + k_s \mathbb{\bar{S}} + k_l \mathbb{\bar{L}} + k_w \mathbb{\bar{W}})$, the octonion force-power density, $\mathbb{\bar{N}} = \Sigma (\bar{n}_i \emph{\textbf{i}}_i) + \Sigma (\bar{N}_i \emph{\textbf{I}}_i)$, is defined as follows,
\begin{eqnarray}
\mathbb{\bar{N}} = v_0 (\lozenge_8 + k_x \mathbb{\bar{X}} + k_a \mathbb{\bar{A}} + k_b \mathbb{\bar{B}} + k_s \mathbb{\bar{S}} + k_l \mathbb{\bar{L}} + k_w \mathbb{\bar{W}})^* \circ \mathbb{\bar{W}}~,
\end{eqnarray}
where the power density is $\bar{f}_0 = - \bar{n}_0/(2 v_0)$, and the force density is $\bar{\textbf{f}} = - \bar{\textbf{n}} / (2 v_0)$. The vectorial parts are $\bar{\textbf{n}} = \Sigma (\bar{n}_j \emph{\textbf{i}}_j )$, $\bar{\textbf{N}}_0 = \bar{N}_0 \emph{\textbf{I}}_0$, and $\bar{\textbf{N}} = \Sigma (\bar{N}_j \emph{\textbf{I}}_j)$.

Further expressing the scalar $\bar{n}_0$ of the octonion force-power density $\mathbb{\bar{N}}$ as
\begin{align}
\bar{n}_0 / v_0 = & ~ \partial_{g0} \bar{w}_0 + \nabla_g^* \cdot \bar{\textbf{w}} + \emph{\textbf{I}}_0^* \circ \partial_{e0} \bar{\textbf{W}}_0 + \nabla_e^* \cdot \bar{\textbf{W}} + k_b ( \bar{\textbf{h}}^* \cdot \bar{\textbf{w}} + k_{eg} \bar{\textbf{H}}^* \cdot \bar{\textbf{W}} )
\nonumber\\
&
+ k_a ( \bar{\textbf{a}}^* \cdot \bar{\textbf{w}} + \bar{a}_0 \bar{w}_0 )
+ k_{eg} k_a ( \bar{\textbf{A}}^* \cdot \bar{\textbf{W}} + \bar{\textbf{A}}_0^* \circ \bar{\textbf{W}}_0 )
\nonumber\\
&
+ k_x ( \bar{\textbf{x}}^* \cdot \bar{\textbf{w}} + \bar{x}_0 \bar{w}_0 )
+ k_{eg} k_x ( \bar{\textbf{X}}^* \cdot \bar{\textbf{W}} + \bar{\textbf{X}}_0^* \circ \bar{\textbf{W}}_0 )
\nonumber\\
&
+ k_s ( \bar{\textbf{s}}^* \cdot \bar{\textbf{w}} + \bar{s}_0 \bar{w}_0 )
+ k_{eg} k_s ( \bar{\textbf{S}}^* \cdot \bar{\textbf{W}} + \bar{\textbf{S}}_0^* \circ \bar{\textbf{W}}_0 )
\nonumber\\
&
+ k_l ( \bar{\textbf{l}}^* \cdot \bar{\textbf{w}} + \bar{l}_0 \bar{w}_0 )
+ k_{eg} k_l ( \bar{\textbf{L}}^* \cdot \bar{\textbf{W}} + \bar{\textbf{L}}_0^* \circ \bar{\textbf{W}}_0 )
\nonumber\\
&
+ k_w ( \bar{\textbf{w}}^* \cdot \bar{\textbf{w}} + \bar{w}_0^2 )
+ k_{eg} k_w ( \bar{\textbf{W}}^* \cdot \bar{\textbf{W}} + \bar{\textbf{W}}_0^* \circ \bar{\textbf{W}}_0 )
~.
\end{align}

In the gravitational and electromagnetic fields with the operator $(\lozenge_8 + k_x \mathbb{\bar{X}} + k_a \mathbb{\bar{A}} + k_b \mathbb{\bar{B}} + k_s \mathbb{\bar{S}} + k_l \mathbb{\bar{L}} + k_w \mathbb{\bar{W}})$, the force density $\bar{\textbf{f}}$ can be defined from the vectorial part $\bar{\textbf{n}}$ of the octonion force-power density $\mathbb{\bar{N}}$ ,
\begin{align}
- 2 \bar{\textbf{f}} = & ~ \partial_{g0} \bar{\textbf{w}} + \nabla_g^* \bar{w}_0 + \nabla_g^* \times \bar{\textbf{w}}
+ \emph{\textbf{I}}_0^* \circ \partial_{e0} \bar{\textbf{W}} + \nabla_e^* \circ \bar{\textbf{W}}_0 + \nabla_e^* \times \bar{\textbf{W}}
\nonumber\\
&
+ k_b \bar{w}_0 \bar{\textbf{h}}^* + k_b \bar{\textbf{h}}^* \times \bar{\textbf{w}}
+ k_{eg} k_b (\bar{\textbf{H}}^* \times \bar{\textbf{W}} + \bar{\textbf{H}}^* \circ \bar{\textbf{W}}_0)
\nonumber\\
&
+ k_a ( \bar{\textbf{a}}^* \bar{w}_0 + \bar{\textbf{a}}^* \times \bar{\textbf{w}} + \bar{a}_0 \bar{\textbf{w}})
+ k_{eg} k_a ( \bar{\textbf{A}}^* \times \bar{\textbf{W}} + \bar{\textbf{A}}_0^* \circ \bar{\textbf{W}} + \bar{\textbf{A}}^* \circ \bar{\textbf{W}}_0)
\nonumber\\
&
+ k_x ( \bar{\textbf{x}}^* \bar{w}_0 + \bar{\textbf{x}}^* \times \bar{\textbf{w}} + \bar{x}_0 \bar{\textbf{w}})
+ k_{eg} k_x ( \bar{\textbf{X}}^* \times \bar{\textbf{W}} + \bar{\textbf{X}}_0^* \circ \bar{\textbf{W}} + \bar{\textbf{X}}^* \circ \bar{\textbf{W}}_0)
\nonumber\\
&
+ k_s ( \bar{\textbf{s}}^* \bar{w}_0 + \bar{\textbf{s}}^* \times \bar{\textbf{w}} + \bar{s}_0 \bar{\textbf{w}})
+ k_{eg} k_s ( \bar{\textbf{S}}^* \times \bar{\textbf{W}} + \bar{\textbf{S}}_0^* \circ \bar{\textbf{W}} + \bar{\textbf{S}}^* \circ \bar{\textbf{W}}_0)
\nonumber\\
&
+ k_l ( \bar{\textbf{l}}^* \bar{w}_0 + \bar{\textbf{l}}^* \times \bar{\textbf{w}} + \bar{l}_0 \bar{\textbf{w}})
+ k_{eg} k_l ( \bar{\textbf{L}}^* \times \bar{\textbf{W}} + \bar{\textbf{L}}_0^* \circ \bar{\textbf{W}} + \bar{\textbf{L}}^* \circ \bar{\textbf{W}}_0)
~,
\end{align}
where the force density $\bar{\textbf{f}}$ includes that of the inertial force, the gravitational force, the gradient of energy, Lorentz force, and the interacting force between the dipole moment with the two fields, and the forces regarding their adjoint fields etc. The above force is more complex than that in the classical field theory, and includes more new force terms related to the gradient of energy, the field potential, and the angular velocity etc.

\begin{table}[h]
\caption{The field sources of the electromagnetic and gravitational fields and their adjoint fields.}
\label{tab:table3}
\centering
\begin{tabular}{llll}
\hline\hline
$ sources $                              & $ fields $                                 & $ descriptions $                  & $ characterictics $    \\
\hline
$\mathbb{\bar{S}}_{gg}$                  & gravitational~field                        & linear~momentum                   & gravitation            \\
$\mathbb{\bar{S}}_{ge}$                  & gravitational~adjoint~field                & adjoint~linear~momentum           & electromagnetism       \\
$\mathbb{\bar{S}}_{eg}$                  & electromagnetic~field                      & electric~current                  & electromagnetism       \\
$\mathbb{\bar{S}}_{ee}$                  & electromagnetic~adjoint~field              & adjoint~electric~current          & gravitation            \\
\hline\hline
\end{tabular}
\end{table}

\subsection{Helicity}

In the octonion compounding space with the octonion operator $(\lozenge_8 + k_x \mathbb{\bar{X}} + k_a \mathbb{\bar{A}} + k_b \mathbb{\bar{B}} + k_s \mathbb{\bar{S}} + k_l \mathbb{\bar{L}} + k_w \mathbb{\bar{W}})$, some helicities \cite{dun1, dun2} and the speed of light will exert an influence on the mass continuity equation, the charge continuity equation, and the gravitational mass density etc.

\subsubsection{Field strength helicity}

The octonion linear momentum density, $\mathbb{\bar{P}} = \mu \mathbb{\bar{S}} / \mu_{gg}$, can be defined from the octonion field source $\mathbb{\bar{S}}$ in the octonion compounding space with the operator $(\lozenge_8 + k_x \mathbb{\bar{X}} + k_a \mathbb{\bar{A}} + k_b \mathbb{\bar{B}} + k_s \mathbb{\bar{S}} + k_l \mathbb{\bar{L}} + k_w \mathbb{\bar{W}})$,
\begin{eqnarray}
\mathbb{\bar{P}} = \widehat{m} \bar{v}_0 + \Sigma (m \bar{v}_j \emph{\textbf{i}}_j) + \Sigma (M \bar{V}_i \emph{\textbf{I}}_i)~,
\end{eqnarray}
where $\bar{p}_0 = \widehat{m} \bar{v}_0$, $\widehat{m} = m (\bar{v}_0^\delta / \bar{v}_0) + \bigtriangleup m$, $\bigtriangleup m = - (k_x \mathbb{\bar{X}}^* \cdot \mathbb{\bar{B}} + k_a \mathbb{\bar{A}}^* \cdot \mathbb{\bar{B}} + k_b \mathbb{\bar{B}}^* \cdot \mathbb{\bar{B}}) / ( \bar{v}_0 \mu_{gg} )$ . $\bar{v}_0^\delta = v_0 \lozenge_8 \cdot \mathbb{\bar{R}}$. And $\lozenge_8 \cdot \mathbb{\bar{R}}$ is the scalar part of the $\lozenge_8 \circ \mathbb{\bar{R}}$.

According to the octonion properties and the above, the gravitational mass $\widehat{m}$ is one reserved scalar in the above, and is changed with the field strength $\mathbb{\bar{B}}_g$ and $\mathbb{\bar{B}}_e$, the field potential $\mathbb{\bar{A}}_g$ and $\mathbb{\bar{A}}_e$, the helicity $(k_x \mathbb{\bar{X}}^* \cdot \mathbb{\bar{B}} + k_a \mathbb{\bar{A}}^* \cdot \mathbb{\bar{B}} + k_b \mathbb{\bar{B}}^* \cdot \mathbb{\bar{B}})$, and the octonion $\mathbb{\bar{X}}_g$ and $\mathbb{\bar{X}}_e$ etc in the octonion compounding space. The helicity includes the magnetic helicity $ \bar{\textbf{A}} \cdot \bar{\textbf{B}} $, the kinetic helicity $ \bar{\textbf{v}} \cdot \bar{\textbf{u}} $, the cross helicity $ \bar{\textbf{V}} \cdot \bar{\textbf{B}} $, and one new helicity term $ \bar{\textbf{A}} \cdot \bar{\textbf{U}} $ etc.

Similarly to the $\mathbb{\bar{S}}$ and the $\mathbb{\bar{L}}$, the octonion angular momentum $\mathbb{\bar{W}}$ does not exert an influence on the field strength helicity in the octonion compounding space with the octonion operator $(\lozenge_8 + k_x \mathbb{\bar{X}} + k_a \mathbb{\bar{A}} + k_b \mathbb{\bar{B}} + k_s \mathbb{\bar{S}} + k_l \mathbb{\bar{L}} + k_w \mathbb{\bar{W}})$. Consequently we can not distinguish this field from some fields with other octonion operators, according to the viewpoint of the field strength helicity also.

\subsubsection{Field source helicity}

The part force density $\mathbb{\bar{F}}$ is one part of the octonion force-power density $\mathbb{\bar{N}}$, and is defined from the octonion linear momentum density $\mathbb{\bar{P}}$ ,
\begin{eqnarray}
\mathbb{\bar{F}} = v_0 (\lozenge_8 + k_x \mathbb{\bar{X}} + k_a \mathbb{\bar{A}} + k_b \mathbb{\bar{B}} + k_s \mathbb{\bar{S}} + k_l \mathbb{\bar{L}} + k_w \mathbb{\bar{W}})^* \circ \mathbb{\bar{P}}~,
\end{eqnarray}
where the part force density includes that of the inertial force, gravitational force, Lorentz force, and the interacting force between the fields with the dipoles etc.

The scalar $\bar{f}_0$ of the part force density $\mathbb{\bar{F}}$ is written as,
\begin{align}
\bar{f}_0 / v_0 = & ~ \partial_{g0} \bar{p}_0 + \nabla_g^* \cdot \bar{\textbf{p}} + \emph{\textbf{I}}_0^* \circ \partial_{e0} \bar{\textbf{P}}_0 + \nabla_e^* \cdot \bar{\textbf{P}} + k_b ( \bar{\textbf{h}}^* \cdot \bar{\textbf{p}} + k_{eg} \bar{\textbf{H}}^* \cdot \bar{\textbf{P}} )
\nonumber\\
&
+ k_a ( \bar{\textbf{a}}^* \cdot \bar{\textbf{p}} + \bar{a}_0 \bar{p}_0 ) + k_{eg} k_a ( \bar{\textbf{A}}^* \cdot \bar{\textbf{P}} + \bar{\textbf{A}}_0^* \circ \bar{\textbf{P}}_0 )
\nonumber\\
&
+ k_x ( \bar{\textbf{x}}^* \cdot \bar{\textbf{p}} + \bar{x}_0 \bar{p}_0 ) + k_{eg} k_x ( \bar{\textbf{X}}^* \cdot \bar{\textbf{P}} + \bar{\textbf{X}}_0^* \circ \bar{\textbf{P}}_0 )
\nonumber\\
&
+ k_s ( \bar{\textbf{s}}^* \cdot \bar{\textbf{p}} + \bar{s}_0 \bar{p}_0 ) + k_{eg} k_s ( \bar{\textbf{S}}^* \cdot \bar{\textbf{P}} + \bar{\textbf{S}}_0^* \circ \bar{\textbf{P}}_0 )
\nonumber\\
&
+ k_l ( \bar{\textbf{l}}^* \cdot \bar{\textbf{p}} + \bar{l}_0 \bar{p}_0 ) + k_{eg} k_l ( \bar{\textbf{L}}^* \cdot \bar{\textbf{P}} + \bar{\textbf{L}}_0^* \circ \bar{\textbf{P}}_0 )
\nonumber\\
&
+ k_w ( \bar{\textbf{w}}^* \cdot \bar{\textbf{p}} + \bar{w}_0 \bar{p}_0 ) + k_{eg} k_w ( \bar{\textbf{W}}^* \cdot \bar{\textbf{P}} + \bar{\textbf{W}}_0^* \circ \bar{\textbf{P}}_0 )
~,
\end{align}
where the field source helicity in the fields with the $(\lozenge_8 + k_x \mathbb{\bar{X}} + k_a \mathbb{\bar{A}} + k_b \mathbb{\bar{B}} + k_s \mathbb{\bar{S}} + k_l \mathbb{\bar{L}} + k_w \mathbb{\bar{W}})$ covers $( \bar{\textbf{a}}^* \cdot \bar{\textbf{p}} + \bar{a}_0 \bar{p}_0 )$, $( \bar{\textbf{A}}^* \cdot \bar{\textbf{P}} + \bar{\textbf{A}}_0^* \circ \bar{\textbf{P}}_0 )$, $( \bar{\textbf{h}}^* \cdot \bar{\textbf{p}} + k_{eg} \bar{\textbf{H}}^* \cdot \bar{\textbf{P}} )$, $(\bar{\textbf{x}}^* \cdot \bar{\textbf{p}} + \bar{x}_0 \bar{p}_0)$, $(\bar{\textbf{X}}^* \cdot \bar{\textbf{P}} + \bar{\textbf{X}}_0^* \circ \bar{\textbf{P}}_0)$, $( \bar{\textbf{s}}^* \cdot \bar{\textbf{p}} + \bar{s}_0 \bar{p}_0 )$, $( \bar{\textbf{S}}^* \cdot \bar{\textbf{P}} + \bar{\textbf{S}}_0^* \circ \bar{\textbf{P}}_0 )$, $( \bar{\textbf{l}}^* \cdot \bar{\textbf{p}} + \bar{l}_0 \bar{p}_0 )$, $( \bar{\textbf{L}}^* \cdot \bar{\textbf{P}} + \bar{\textbf{L}}_0^* \circ \bar{\textbf{P}}_0 )$, $( \bar{\textbf{w}}^* \cdot \bar{\textbf{p}} + \bar{w}_0 \bar{p}_0 )$, and $( \bar{\textbf{W}}^* \cdot \bar{\textbf{P}} + \bar{\textbf{W}}_0^* \circ \bar{\textbf{P}}_0 )$ etc. And they include the magnetic helicity $ \textbf{A} \cdot \textbf{B} $, the kinetic helicity $ \textbf{v} \cdot \textbf{u} $, the cross helicity $ \textbf{V} \cdot \textbf{B} $, the current helicity $ \textbf{B}^* \cdot \textbf{P} $, and other helicity terms etc.

The above is the mass continuity equation in the electromagnetic and gravitational fields with the octonion operator $(\lozenge_8 + k_x \mathbb{\bar{X}} + k_a \mathbb{\bar{A}} + k_b \mathbb{\bar{B}} + k_s \mathbb{\bar{S}} + k_l \mathbb{\bar{L}} + k_w \mathbb{\bar{W}})$, and is effected by the speed of light $\bar{v}_0$ and some helicity terms of the rotational objects and of the spinning charged objects.

A new physical quantity $\mathbb{\bar{F}}_q$ can be defined from the part force density $\mathbb{\bar{F}}$,
\begin{eqnarray}
\mathbb{\bar{F}}_q = \mathbb{\bar{F}} \circ \emph{\textbf{I}}_0^*~.
\end{eqnarray}

The scalar part $\bar{F}_0$ of the $\mathbb{\bar{F}}_q$ is written as,
\begin{align}
\bar{F}_0 / v_0 = & ~ (\partial_{g0} \bar{\textbf{P}}_0 + \nabla_g^* \cdot \bar{\textbf{P}}) \circ \emph{\textbf{I}}_0^*
+ (\partial_{e0} \bar{p}_0 + \nabla_e^* \cdot \bar{\textbf{p}}) \circ \emph{\textbf{I}}_0^*
+ k_b ( \bar{\textbf{h}}^* \cdot \bar{\textbf{P}} + k_{eg} \bar{\textbf{H}}^* \cdot \bar{\textbf{p}} ) \circ \emph{\textbf{I}}_0^*
\nonumber
\\
&
+ k_a ( \bar{\textbf{a}}^* \cdot \bar{\textbf{P}} + a_0 \bar{\textbf{P}}_0 ) \circ \emph{\textbf{I}}_0^* + k_{eg} k_a ( \bar{\textbf{A}}^* \cdot \bar{\textbf{p}} + p_0 \bar{\textbf{A}}_0^* ) \circ \emph{\textbf{I}}_0^*
\nonumber
\\
&
+ k_x ( \bar{\textbf{x}}^* \cdot \bar{\textbf{P}} + x_0 \bar{\textbf{P}}_0 ) \circ \emph{\textbf{I}}_0^* + k_{eg} k_x ( \bar{\textbf{X}}^* \cdot \bar{\textbf{p}} + p_0 \bar{\textbf{X}}_0^* ) \circ \emph{\textbf{I}}_0^*
\nonumber
\\
&
+ k_s ( \bar{\textbf{s}}^* \cdot \bar{\textbf{P}} + s_0 \bar{\textbf{P}}_0 ) \circ \emph{\textbf{I}}_0^* + k_{eg} k_s ( \bar{\textbf{S}}^* \cdot \bar{\textbf{p}} + p_0 \bar{\textbf{S}}_0^* ) \circ \emph{\textbf{I}}_0^*
\nonumber
\\
&
+ k_l ( \bar{\textbf{l}}^* \cdot \bar{\textbf{P}} + l_0 \bar{\textbf{P}}_0 ) \circ \emph{\textbf{I}}_0^* + k_{eg} k_l ( \bar{\textbf{L}}^* \cdot \bar{\textbf{p}} + p_0 \bar{\textbf{L}}_0^* ) \circ \emph{\textbf{I}}_0^*
\nonumber
\\
&
+ k_w ( \bar{\textbf{w}}^* \cdot \bar{\textbf{P}} + w_0 \bar{\textbf{P}}_0 ) \circ \emph{\textbf{I}}_0^* + k_{eg} k_w ( \bar{\textbf{W}}^* \cdot \bar{\textbf{p}} + p_0 \bar{\textbf{W}}_0^* ) \circ \emph{\textbf{I}}_0^*
~,
\end{align}
where the above helicity terms cover that in the gravitational and electromagnetic fields with their adjoint fields.

The above is the charge continuity equation in the presence of the gravitational field and the electromagnetic field with the octonion operator $(\lozenge_8 + k_x \mathbb{\bar{X}} + k_a \mathbb{\bar{A}} + k_b \mathbb{\bar{B}} + k_s \mathbb{\bar{S}} + k_l \mathbb{\bar{L}} + k_w \mathbb{\bar{W}} )$ when the scalar part $F_0 = 0$. Similarly to the case with the octonion operator $(\lozenge + k_x \mathbb{\bar{X}} + k_a \mathbb{\bar{A}} + k_b \mathbb{\bar{B}} + k_s \mathbb{\bar{S}} + k_l \mathbb{\bar{L}} + k_w \mathbb{\bar{W}} )$, the charge continuity equation is the invariant under the octonion coordinate transformation, meanwhile the $\mathbb{\bar{X}}$, $\mathbb{\bar{A}}$, $\mathbb{\bar{B}}$, $\mathbb{\bar{S}}$, $\mathbb{\bar{L}}$, and $\mathbb{\bar{W}}$ of the gravitational field and the electromagnetic field have the influence on the charge continuity equation, although the impacts are usually paltry when the electromagnetic field and the gravitational field are weak.

\begin{table}[h]
\caption{Some physical quantities of the electromagnetic and gravitational fields with their adjoint fields.}
\label{tab:table3}
\centering
\begin{tabular}{ll}
\hline\hline
$ definitions $                                                                                                                            & $ meanings $ \\
\hline
$\mathbb{\bar{X}}$                                                                                                                         & field quantity \\
$\mathbb{\bar{A}} = (\lozenge_8 + k_x \mathbb{\bar{X}}) \circ \mathbb{\bar{X}}$                                                            & field potential \\
$\mathbb{\bar{B}} = (\lozenge_8 + k_x \mathbb{\bar{X}} + k_a \mathbb{\bar{A}}) \circ \mathbb{\bar{A}}$                                     & field strength \\
$\mathbb{\bar{R}}$                                                                                                                         & radius vector \\
$\mathbb{\bar{V}} = v_0 \lozenge_8 \circ \mathbb{\bar{R}}$                                                                                 & velocity \\
$\mathbb{\bar{U}} = \lozenge_8 \circ \mathbb{\bar{V}}$                                                                                     & velocity curl \\
$\mu \mathbb{\bar{S}} = - ( \lozenge_8 + k_x \mathbb{\bar{X}} + k_a \mathbb{\bar{A}} + k_b \mathbb{\bar{B}} )^* \circ \mathbb{\bar{B}}$    & field source \\
$\mathbb{\bar{H}}_b = (k_x \mathbb{\bar{X}} + k_a \mathbb{\bar{A}} + k_b \mathbb{\bar{B}})^* \cdot \mathbb{\bar{B}}$                       & field strength helicity\\
$\mathbb{\bar{P}} = \mu \mathbb{\bar{S}} / \mu_{gg}$                                                                                       & linear momentum density \\
$\mathbb{\bar{R}} = \mathbb{R} + k_{rx} \mathbb{X}$                                                                                        & compounding radius vector \\
$\mathbb{\bar{L}} = \mathbb{\bar{R}} \circ \mathbb{\bar{P}}$                                                                               & angular momentum density \\
$\mathbb{\bar{W}} = v_0 ( \lozenge_8 + k_x \mathbb{\bar{X}} + k_a \mathbb{\bar{A}} + k_b \mathbb{\bar{B}} + k_s \mathbb{\bar{S}} + k_l \mathbb{\bar{L}} ) \circ \mathbb{\bar{L}}$                                                                                                                          & torque-energy densities \\
$\mathbb{\bar{N}} = v_0 ( \lozenge_8 + k_x \mathbb{\bar{X}} + k_a \mathbb{\bar{A}} + k_b \mathbb{\bar{B}} + k_s \mathbb{\bar{S}} + k_l \mathbb{\bar{L}} + k_w \mathbb{\bar{W}})^* \circ \mathbb{\bar{W}}$                                                                                                & force-power density \\
$\mathbb{\bar{F}} = - \mathbb{\bar{N}} / (2v_0)$                                                                                           & force density \\
$\mathbb{\bar{H}}_s = ( k_x \mathbb{\bar{X}} + k_a \mathbb{\bar{A}} + k_b \mathbb{\bar{B}} + k_s \mathbb{\bar{S}} + k_l \mathbb{\bar{L}})^* \cdot \mathbb{\bar{P}}$
                                                                                                                                           & field source helicity\\
\hline\hline
\end{tabular}
\end{table}

%--12--%

\section{The fields regarding the sedenion operator $\lozenge_{16}$}

In the electromagnetic and gravitational fields, the octonion operator ($\lozenge_8 + k_x \mathbb{\bar{X}} + k_a \mathbb{\bar{A}} + k_b \mathbb{\bar{B}} + k_s \mathbb{\bar{S}} + k_l \mathbb{\bar{L}} + k_w \mathbb{\bar{W}}$) can deduce the octonion physical properties of two fields with their related adjoint fields, including the octonion linear momentum, the octonion angular momentum, the energy, the octonion torque, the power, the octonion force, and some helicities of the rotational objects and the spinning charged objects. However these results can not involve the helicities of the strong nuclear field, the weak nuclear field, and their related adjoint fields etc. In this section, the operator will substitute the sedenion operator $\lozenge_{16}$ for the octonion operator $\lozenge_8$ to cover the physical properties of the electromagnetic field, the gravitational field, the strong nuclear field, and the weak nuclear field simultaneously.

\subsection{Sedenion space}

The sedenion space \cite{bresar, imaeda} consists of four kinds of the independent quaternion spaces. In the quaternion space for the gravitational field, the basis vector is $\mathbb{E}_g$ = ($\emph{\textbf{i}}_0$,  $\emph{\textbf{i}}_1$, $\emph{\textbf{i}}_2$, $\emph{\textbf{i}}_3$), the radius vector is $\mathbb{R}_g$ = ($r_0$, $r_1$, $r_2$, $r_3$), the velocity is $\mathbb{V}_g$ = ($v_0$, $v_1$, $v_2$, $v_3$), and the gravitational potential is $\mathbb{A}_g$ = ($a_0$, $a_1$, $a_2$, $a_3$), with the physical quantity $\mathbb{X}_g$ = ($x_0$, $x_1$, $x_2$, $x_3$). In the quaternion space for the electromagnetic field, the basis vector is $\mathbb{E}_e$ = ($\emph{\textbf{i}}_4$, $\emph{\textbf{i}}_5$, $\emph{\textbf{i}}_6$, $\emph{\textbf{i}}_7$), the radius vector is $\mathbb{R}_e$ = ($r_4$, $r_5$, $r_6$, $r_7$), the velocity is $\mathbb{V}_e$ = ($v_4$, $v_5$, $v_6$, $v_7$), and the electromagnetic potential is $\mathbb{A}_e$ = ($a_4$, $a_5$, $a_6$, $a_7$), with the physical quantity $\mathbb{X}_e$ = ($x_4$, $x_5$, $x_6$, $x_7$). In the quaternion space for the weak nuclear field, the basis vector is $\mathbb{E}_w$ = ($\emph{\textbf{i}}_8$,  $\emph{\textbf{i}}_9$, $\emph{\textbf{i}}_{10}$, $\emph{\textbf{i}}_{11}$), the radius vector is $\mathbb{R}_w$ = ($r_8$, $r_9$, $r_{10}$, $r_{11}$), the velocity is $\mathbb{V}_w$ = ($v_8$, $v_9$, $v_{10}$, $v_{11}$), and the weak nuclear potential is $\mathbb{A}_w$ = ($a_8$, $a_9$, $a_{10}$, $a_{11}$), with the physical quantity $\mathbb{X}_w$ = ($x_8$, $x_9$, $x_{10}$, $x_{11}$). In the quaternion space for the strong nuclear field, the basis vector is $\mathbb{E}_s$ = ($\emph{\textbf{i}}_{12}$, $\emph{\textbf{i}}_{13}$, $\emph{\textbf{i}}_{14}$, $\emph{\textbf{i}}_{15}$), the radius vector is $\mathbb{R}_s$ = ($r_{12}$, $r_{13}$, $r_{14}$, $r_{15}$), and the velocity is $\mathbb{V}_s$ = ($v_{12}$, $v_{13}$, $v_{14}$, $v_{15}$), and the strong nuclear potential is $\mathbb{A}_s$ = ($a_{12}$, $a_{13}$, $a_{14}$, $a_{15}$), with the physical quantity $\mathbb{X}_s$ = ($x_{12}$, $x_{13}$, $x_{14}$, $x_{15}$).

The basis vectors $\mathbb{E}_e$, $\mathbb{E}_g$, $\mathbb{E}_w$, and $\mathbb{E}_s$ are independent to each other, and they can combine together to become the basis vector of sedenion space \cite{kivunge, cawagas1}, $\mathbb{E}_{16}$ = ($\emph{\textbf{i}}_0$, $\emph{\textbf{i}}_1$, $\emph{\textbf{i}}_2$, $\emph{\textbf{i}}_3$, $\emph{\textbf{i}}_4$, $\emph{\textbf{i}}_5$, $\emph{\textbf{i}}_6$, $\emph{\textbf{i}}_7$, $\emph{\textbf{i}}_8$, $\emph{\textbf{i}}_9$, $\emph{\textbf{i}}_{10}$, $\emph{\textbf{i}}_{11}$, $\emph{\textbf{i}}_{12}$, $\emph{\textbf{i}}_{13}$, $\emph{\textbf{i}}_{14}$, $\emph{\textbf{i}}_{15}$). The sedenion radius vector is $\mathbb{R}$ = $\Sigma(\emph{\textbf{i}}_i r_i$ + $f_e \emph{\textbf{i}}_{(i+4)} r_{(i+4)}$ + $f_w \emph{\textbf{i}}_{(i+8)} r_{(i+8)}$ + $f_s \emph{\textbf{i}}_{(i+12)} r_{(i+12)})$, the sedenion velocity is $\mathbb{V}$ = $\Sigma(\emph{\textbf{i}}_i v_i$ + $f_e \emph{\textbf{i}}_{(i+4)} v_{(i+4)}$ + $f_w \emph{\textbf{i}}_{(i+8)} v_{(i+8)}$ + $f_s \emph{\textbf{i}}_{(i+12)} v_{(i+12)})$, and the sedenion field potential is $\mathbb{A}$ = $\Sigma(\emph{\textbf{i}}_i a_i$ + $f_e \emph{\textbf{i}}_{(i+4)} a_{(i+4)}$ + $f_w \emph{\textbf{i}}_{(i+8)} a_{(i+8)}$ + $f_s \emph{\textbf{i}}_{(i+12)} a_{(i+12)})$, with the sedenion physical quantity $\mathbb{X}$ = $\Sigma(\emph{\textbf{i}}_i x_i$ + $f_e \emph{\textbf{i}}_{(i+4)} x_{(i+4)}$ + $f_w \emph{\textbf{i}}_{(i+8)} x_{(i+8)}$ + $f_s \emph{\textbf{i}}_{(i+12)} x_{(i+12)})$. Herein $r_0 = v_0 t$, $t$ is the time; $v_0$ is the speed of light. $f_e$, $f_w$, and $f_s$ are the coefficients for the dimensional homogeneity, and $f_e = k_{eg}$ . $\emph{\textbf{i}}_0 = 1$. $i = 0, 1, 2, 3$.

The sedenion operator possesses four kinds of independent and perpendicular quaternion operators in the sedenion space. The gravitational field operator is $\lozenge_g = \Sigma ( \emph{\textbf{i}}_i \partial_i)$, the electromagnetic field operator is $\lozenge_e = \Sigma ( \emph{\textbf{i}}_{(i+4)} \partial_{(i+4)})$, the weak nuclear field operator is $\lozenge_w = \Sigma ( \emph{\textbf{i}}_{(i+8)} \partial_{(i+8)})$, and the strong nuclear field operator is $\lozenge_s = \Sigma ( \emph{\textbf{i}}_{(i+12)} \partial_{(i+12)})$. Those four operators constitute the sedenion operator $\lozenge_{16} = \lozenge_g + d_e \lozenge_e + d_w \lozenge_w + d_s \lozenge_s$ . Herein $\partial_i = \partial / \partial r_i$ . $d_e$, $d_w$, and $d_s$ are the coefficients for the dimensional homogeneity.

The sedenion field strength is $\mathbb{B} = \mathbb{B}_g + f_e \mathbb{B}_e + f_w \mathbb{B}_w + f_s \mathbb{B}_s$, and it is defined from the sedenion field potential $\mathbb{A} = \mathbb{A}_g + f_e \mathbb{A}_e + f_w \mathbb{A}_w + f_s \mathbb{A}_s$. And that the field strength $\mathbb{B}$ includes the gravitational strength $\mathbb{B}_g = \Sigma ( h_i \emph{\textbf{i}}_i)$, the electromagnetic strength $\mathbb{B}_e = \Sigma ( h_{(i+4)} \emph{\textbf{i}}_{(i+4)})$, the weak nuclear strength $\mathbb{B}_w = \Sigma ( h_{(i+8)} \emph{\textbf{i}}_{(i+8)})$, and the strong nuclear strength $\mathbb{B}_s = \Sigma ( h_{(i+12)} \emph{\textbf{i}}_{(i+12)})$. The gauge equations are $h_0 = 0$, $h_4 = 0$, $h_8 = 0$, and $h_{12} = 0$ respectively. The radius vector $\mathbb{R}$ and the sedenion $\mathbb{X}$ can be combined together to become the compounding radius vector $\mathbb{\bar{R}} = \mathbb{R} + k_{rx} \mathbb{X}$, and the compounding quantity $\mathbb{\bar{X}} = \mathbb{X} + K_{rx} \mathbb{R}$. The related space is called as the sedenion compounding space, which is one kind of function space also. In this space, the compounding field potential is $\mathbb{\bar{A}} = (\lozenge_{16} + k_x \mathbb{\bar{X}}) \circ \mathbb{\bar{X}}$, the compounding field strength is $\mathbb{\bar{B}} = (\lozenge_{16} + k_x \mathbb{\bar{X}} + k_a \mathbb{\bar{A}}) \circ \mathbb{\bar{A}}$, the compounding velocity is $\mathbb{\bar{V}} = \mathbb{V} + v_0 k_{rx} \mathbb{A}$, and the compounding velocity curl is $\mathbb{\bar{U}} = \mathbb{U} + v_0 k_{rx} \mathbb{B}$ .

In terms of the sedenion operator $\lozenge_{16}$ in the above four fields, we can represent the field potential, the field strength, the field source, the linear momentum, the angular momentum, the energy, the torque, the force, and some helicities, including the influence of the adjoint fields on the torque, the force, and the helicity etc.

\begin{table}[b]
\caption{\label{tab:table1}The sedenion multiplication table.}
\centering
\begin{tabular}{c|cccc|cccc||cccc|cccc}
\hline\hline
$ $ & $1$ & $\emph{\textbf{i}}_1$  & $\emph{\textbf{i}}_2$ & $\emph{\textbf{i}}_3$  & $\emph{\textbf{i}}_4$ & $\emph{\textbf{i}}_5$ & $\emph{\textbf{i}}_6$  & $\emph{\textbf{i}}_7$
& $\emph{\textbf{i}}_8$ & $\emph{\textbf{i}}_9$ & $\emph{\textbf{i}}_{10}$  & $\emph{\textbf{i}}_{11}$ & $\emph{\textbf{i}}_{12}$  & $\emph{\textbf{i}}_{13}$ & $\emph{\textbf{i}}_{14}$ & $\emph{\textbf{i}}_{15}$
\\
\hline
$1$ & $\emph{\textbf{i}}_1$ & $\emph{\textbf{i}}_1$  & $\emph{\textbf{i}}_2$ & $\emph{\textbf{i}}_3$  & $\emph{\textbf{i}}_4$  & $\emph{\textbf{i}}_5$ & $\emph{\textbf{i}}_6$  & $\emph{\textbf{i}}_7$
& $\emph{\textbf{i}}_8$ & $\emph{\textbf{i}}_9$ & $\emph{\textbf{i}}_{10}$  & $\emph{\textbf{i}}_{11}$ & $\emph{\textbf{i}}_{12}$  & $\emph{\textbf{i}}_{13}$ & $\emph{\textbf{i}}_{14}$ & $\emph{\textbf{i}}_{15}$
\\
$\emph{\textbf{i}}_1$ & $\emph{\textbf{i}}_1$ & $-1$ & $\emph{\textbf{i}}_3$  & $-\emph{\textbf{i}}_2$ & $\emph{\textbf{i}}_5$ & $-\emph{\textbf{i}}_4$ &  $-\emph{\textbf{i}}_7$ & $\emph{\textbf{i}}_6$
& $\emph{\textbf{i}}_9$ & $-\emph{\textbf{i}}_8$ & $-\emph{\textbf{i}}_{11}$  & $\emph{\textbf{i}}_{10}$ & $-\emph{\textbf{i}}_{13}$  & $\emph{\textbf{i}}_{12}$ & $\emph{\textbf{i}}_{15}$ & $-\emph{\textbf{i}}_{14}$
\\
$\emph{\textbf{i}}_2$ & $\emph{\textbf{i}}_2$ & $-\emph{\textbf{i}}_3$ & $-1$ & $\emph{\textbf{i}}_1$  & $\emph{\textbf{i}}_6$  & $\emph{\textbf{i}}_7$ & $-\emph{\textbf{i}}_4$ & $-\emph{\textbf{i}}_5$
& $\emph{\textbf{i}}_{10}$ & $\emph{\textbf{i}}_{11}$ & $-\emph{\textbf{i}}_8$  & $-\emph{\textbf{i}}_9$ & $-\emph{\textbf{i}}_{14}$  & $-\emph{\textbf{i}}_{15}$ & $\emph{\textbf{i}}_{12}$ & $\emph{\textbf{i}}_{13}$
\\
$\emph{\textbf{i}}_3$ & $\emph{\textbf{i}}_3$ & $\emph{\textbf{i}}_2$ & $-\emph{\textbf{i}}_1$ & $-1$ & $\emph{\textbf{i}}_7$  & $-\emph{\textbf{i}}_6$ & $\emph{\textbf{i}}_5$  & $-\emph{\textbf{i}}_4$
& $\emph{\textbf{i}}_{11}$ & $-\emph{\textbf{i}}_{10}$ & $\emph{\textbf{i}}_9$  & $-\emph{\textbf{i}}_8$ & $-\emph{\textbf{i}}_{15}$  & $\emph{\textbf{i}}_{14}$ & $-\emph{\textbf{i}}_{13}$ & $\emph{\textbf{i}}_{12}$
\\
\hline
$\emph{\textbf{i}}_4$ & $\emph{\textbf{i}}_4$ & $-\emph{\textbf{i}}_5$ & $-\emph{\textbf{i}}_6$ & $-\emph{\textbf{i}}_7$ & $-1$ & $\emph{\textbf{i}}_1$ & $\emph{\textbf{i}}_2$  & $\emph{\textbf{i}}_3$
& $\emph{\textbf{i}}_{12}$ & $\emph{\textbf{i}}_{13}$ & $\emph{\textbf{i}}_{14}$  & $\emph{\textbf{i}}_{15}$ & $-\emph{\textbf{i}}_8$  & $-\emph{\textbf{i}}_9$ & $-\emph{\textbf{i}}_{10}$ & $-\emph{\textbf{i}}_{11}$
\\
$\emph{\textbf{i}}_5$ & $\emph{\textbf{i}}_5$ & $\emph{\textbf{i}}_4$ & $-\emph{\textbf{i}}_7$ & $\emph{\textbf{i}}_6$  & $-\emph{\textbf{i}}_1$ & $-1$ & $-\emph{\textbf{i}}_3$ & $\emph{\textbf{i}}_2$
& $\emph{\textbf{i}}_{13}$ & $-\emph{\textbf{i}}_{12}$ & $\emph{\textbf{i}}_{15}$  & $-\emph{\textbf{i}}_{14}$ & $\emph{\textbf{i}}_9$  & $-\emph{\textbf{i}}_8$ & $\emph{\textbf{i}}_{11}$ & $-\emph{\textbf{i}}_{10}$
\\
$\emph{\textbf{i}}_6$ & $\emph{\textbf{i}}_6$ & $\emph{\textbf{i}}_7$ & $\emph{\textbf{i}}_4$  & $-\emph{\textbf{i}}_5$ & $-\emph{\textbf{i}}_2$ & $\emph{\textbf{i}}_3$  & $-1$ & $-\emph{\textbf{i}}_1$
& $\emph{\textbf{i}}_{14}$ & $-\emph{\textbf{i}}_{15}$ & $-\emph{\textbf{i}}_{12}$  & $\emph{\textbf{i}}_{13}$ & $\emph{\textbf{i}}_{10}$  & $-\emph{\textbf{i}}_{11}$ & $-\emph{\textbf{i}}_8$ & $\emph{\textbf{i}}_9$
\\
$\emph{\textbf{i}}_7$ & $\emph{\textbf{i}}_7$ & $-\emph{\textbf{i}}_6$ & $\emph{\textbf{i}}_5$  & $\emph{\textbf{i}}_4$  & $-\emph{\textbf{i}}_3$ & $-\emph{\textbf{i}}_2$ & $\emph{\textbf{i}}_1$  & $-1$
& $\emph{\textbf{i}}_{15}$ & $\emph{\textbf{i}}_{14}$ & $-\emph{\textbf{i}}_{13}$  & $-\emph{\textbf{i}}_{12}$ & $\emph{\textbf{i}}_{11}$  & $\emph{\textbf{i}}_{10}$ & $-\emph{\textbf{i}}_9$ & $-\emph{\textbf{i}}_8$
%---%
\\
\hline\hline
$\emph{\textbf{i}}_8$ & $\emph{\textbf{i}}_8$ & $-\emph{\textbf{i}}_9$ & $-\emph{\textbf{i}}_{10}$  & $-\emph{\textbf{i}}_{11}$  & $-\emph{\textbf{i}}_{12}$  & $-\emph{\textbf{i}}_{13}$ & $-\emph{\textbf{i}}_{14}$ & $-\emph{\textbf{i}}_{15}$
& $-1$ & $\emph{\textbf{i}}_1$  & $\emph{\textbf{i}}_2$ & $\emph{\textbf{i}}_3$ & $\emph{\textbf{i}}_4$  & $\emph{\textbf{i}}_5$ & $\emph{\textbf{i}}_6$  & $\emph{\textbf{i}}_7$
\\
$\emph{\textbf{i}}_9$ & $\emph{\textbf{i}}_9$ & $\emph{\textbf{i}}_8$ & $-\emph{\textbf{i}}_{11}$  & $\emph{\textbf{i}}_{10}$  & $-\emph{\textbf{i}}_{13}$  & $\emph{\textbf{i}}_{12}$ & $\emph{\textbf{i}}_{15}$ & $-\emph{\textbf{i}}_{14}$
& $-\emph{\textbf{i}}_1$ & $-1$ & $-\emph{\textbf{i}}_3$ & $\emph{\textbf{i}}_2$ & $-\emph{\textbf{i}}_5$  & $\emph{\textbf{i}}_4$ & $\emph{\textbf{i}}_7$  & $-\emph{\textbf{i}}_6$
\\
$\emph{\textbf{i}}_{10}$ & $\emph{\textbf{i}}_{10}$ & $\emph{\textbf{i}}_{11}$ & $\emph{\textbf{i}}_8$  & $-\emph{\textbf{i}}_9$  & $-\emph{\textbf{i}}_{14}$  & $-\emph{\textbf{i}}_{15}$ & $\emph{\textbf{i}}_{12}$ & $\emph{\textbf{i}}_{13}$
& $-\emph{\textbf{i}}_2$  & $\emph{\textbf{i}}_3$ & $-1$ & $-\emph{\textbf{i}}_1$ & $-\emph{\textbf{i}}_6$  & $-\emph{\textbf{i}}_7$ & $\emph{\textbf{i}}_4$  & $\emph{\textbf{i}}_5$
\\
$\emph{\textbf{i}}_{11}$ & $\emph{\textbf{i}}_{11}$ & $-\emph{\textbf{i}}_{10}$ & $\emph{\textbf{i}}_9$  & $\emph{\textbf{i}}_8$  & $-\emph{\textbf{i}}_{15}$  & $\emph{\textbf{i}}_{14}$ & $-\emph{\textbf{i}}_{13}$ & $\emph{\textbf{i}}_{12}$

& $-\emph{\textbf{i}}_3$  & $-\emph{\textbf{i}}_2$ & $\emph{\textbf{i}}_1$ & $-1$ & $-\emph{\textbf{i}}_7$  & $\emph{\textbf{i}}_6$ & $-\emph{\textbf{i}}_5$  & $\emph{\textbf{i}}_4$
\\ \hline
$\emph{\textbf{i}}_{12}$ & $\emph{\textbf{i}}_{12}$  & $\emph{\textbf{i}}_{13}$ & $\emph{\textbf{i}}_{14}$ & $\emph{\textbf{i}}_{15}$ & $\emph{\textbf{i}}_8$ & $-\emph{\textbf{i}}_9$ & $-\emph{\textbf{i}}_{10}$  & $-\emph{\textbf{i}}_{11}$
& $-\emph{\textbf{i}}_4$ & $\emph{\textbf{i}}_5$ & $\emph{\textbf{i}}_6$ & $\emph{\textbf{i}}_7$ & $-1$ & $-\emph{\textbf{i}}_1$ & $-\emph{\textbf{i}}_2$  & $-\emph{\textbf{i}}_3$
\\
$\emph{\textbf{i}}_{13}$ & $\emph{\textbf{i}}_{13}$  & $-\emph{\textbf{i}}_{12}$ & $\emph{\textbf{i}}_{15}$ & $-\emph{\textbf{i}}_{14}$ & $\emph{\textbf{i}}_9$ & $\emph{\textbf{i}}_8$ & $\emph{\textbf{i}}_{11}$  & $-\emph{\textbf{i}}_{10}$
& $-\emph{\textbf{i}}_5$ & $-\emph{\textbf{i}}_4$ & $\emph{\textbf{i}}_7$ & $-\emph{\textbf{i}}_6$ & $\emph{\textbf{i}}_1$ &$-1$ &  $\emph{\textbf{i}}_3$  & $-\emph{\textbf{i}}_2$
\\
$\emph{\textbf{i}}_{14}$ & $\emph{\textbf{i}}_{14}$  & $-\emph{\textbf{i}}_{15}$ & $-\emph{\textbf{i}}_{12}$ & $\emph{\textbf{i}}_{13}$ & $\emph{\textbf{i}}_{10}$ & $-\emph{\textbf{i}}_{11}$ & $\emph{\textbf{i}}_8$  & $\emph{\textbf{i}}_9$
& $-\emph{\textbf{i}}_6$ & $-\emph{\textbf{i}}_7$ & $-\emph{\textbf{i}}_4$ & $\emph{\textbf{i}}_5$ & $\emph{\textbf{i}}_2$ & $-\emph{\textbf{i}}_3$  & $-1$ & $\emph{\textbf{i}}_1$
\\
$\emph{\textbf{i}}_{15}$ & $\emph{\textbf{i}}_{15}$  & $\emph{\textbf{i}}_{14}$ & $-\emph{\textbf{i}}_{13}$ & $-\emph{\textbf{i}}_{12}$ & $\emph{\textbf{i}}_{11}$ & $\emph{\textbf{i}}_{10}$ & $-\emph{\textbf{i}}_9$  & $\emph{\textbf{i}}_8$
& $-\emph{\textbf{i}}_7$ & $\emph{\textbf{i}}_6$ & $-\emph{\textbf{i}}_5$ & $-\emph{\textbf{i}}_4$ & $\emph{\textbf{i}}_3$ & $\emph{\textbf{i}}_2$  & $-\emph{\textbf{i}}_1$ & $-1$
\\
\hline\hline
\end{tabular}
\end{table}

\subsection{Field source}

In the four fields with the sedenion operator ($\lozenge_{16} + k_x \mathbb{\bar{X}} + k_a \mathbb{\bar{A}} + k_b \mathbb{\bar{B}} + k_s \mathbb{\bar{S}} + k_l \mathbb{\bar{L}} + k_w \mathbb{\bar{W}}$), the physical quantities will be expanded into the sedenion quantities, including the gravitational mass density, the mass continuity equation, the charge continuity equation, and the helicity etc. In this sedenion compounding space, the basis vector is $ \mathbb{E}_{16}$, the radius vector is $\mathbb{\bar{R}}$, and the velocity is $\mathbb{\bar{V}}$. The sedenion compounding field potential is $\mathbb{\bar{A}} = \mathbb{\bar{A}}_g + f_e \mathbb{\bar{A}}_e + f_w \mathbb{\bar{A}}_w + f_s \mathbb{\bar{A}}_s$, and is defined from the sedenion quantity $\mathbb{\bar{X}}$, that is $\mathbb{\bar{A}} = (\lozenge_{16} + k_x \mathbb{\bar{X}}) \circ \mathbb{\bar{X}}$ .

The sedenion compounding field potential $\mathbb{\bar{A}}$ encompasses,
\begin{align}
\mathbb{\bar{A}}_g = & ~ ( \lozenge_g \circ \mathbb{\bar{X}}_g + d_e f_e \lozenge_e \circ \mathbb{\bar{X}}_e  + d_w f_w \lozenge_w \circ \mathbb{\bar{X}}_w  + d_s f_s \lozenge_s \circ \mathbb{\bar{X}}_s )
\nonumber
\\
& + k_x ( \mathbb{\bar{X}}_g \circ \mathbb{\bar{X}}_g + f_e^2 \mathbb{\bar{X}}_e \circ \mathbb{\bar{X}}_e
+ f_w^2 \mathbb{\bar{X}}_w \circ \mathbb{\bar{X}}_w  + f_s^2 \mathbb{\bar{X}}_s \circ \mathbb{\bar{X}}_s )~,
\\
\mathbb{\bar{A}}_e = & ~ ( \lozenge_g \circ \mathbb{\bar{X}}_e + d_e f_e^{-1} \lozenge_e \circ \mathbb{\bar{X}}_g  + d_s f_w f_e^{-1} \lozenge_s \circ \mathbb{\bar{X}}_w  + d_w f_s f_e^{-1} \lozenge_w \circ \mathbb{\bar{X}}_s )
\nonumber
\\
& + k_x ( \mathbb{\bar{X}}_e \circ \mathbb{\bar{X}}_g + \mathbb{\bar{X}}_g \circ \mathbb{\bar{X}}_e) + k_x f_s f_w f_e^{-1} ( \mathbb{\bar{X}}_s \circ \mathbb{\bar{X}}_w + \mathbb{\bar{X}}_w \circ \mathbb{\bar{X}}_s)~,
\\
\mathbb{\bar{A}}_w = & ~ ( \lozenge_g \circ \mathbb{\bar{X}}_w + d_e f_s f_w^{-1} \lozenge_e \circ \mathbb{\bar{X}}_s  + d_w f_w^{-1} \lozenge_w \circ \mathbb{\bar{X}}_g + d_s f_e f_w^{-1} \lozenge_s \circ \mathbb{\bar{X}}_e)
\nonumber
\\
& + k_x ( \mathbb{\bar{X}}_g \circ \mathbb{\bar{X}}_w  + \mathbb{\bar{X}}_w \circ \mathbb{\bar{X}}_g ) + k_x f_e f_s f_w^{-1} ( \mathbb{\bar{X}}_e \circ \mathbb{\bar{X}}_s + \mathbb{\bar{X}}_s \circ \mathbb{\bar{X}}_e)~,
\\
\mathbb{\bar{A}}_s = & ~ ( \lozenge_g \circ \mathbb{\bar{X}}_s + d_e f_w f_s^{-1} \lozenge_e \circ \mathbb{\bar{X}}_w  + d_w f_e f_s^{-1} \lozenge_w \circ \mathbb{\bar{X}}_e  + d_s f_s^{-1} \lozenge_s \circ \mathbb{\bar{X}}_g )
\nonumber
\\
& + k_x ( \mathbb{\bar{X}}_g \circ \mathbb{\bar{X}}_s  + \mathbb{\bar{X}}_s \circ \mathbb{\bar{X}}_g ) + k_x f_e f_w f_s^{-1} ( \mathbb{\bar{X}}_e \circ \mathbb{\bar{X}}_w + \mathbb{\bar{X}}_w \circ \mathbb{\bar{X}}_e)~,
\end{align}
where the symbol $\circ$ denotes the sedenion multiplication. And the sedenion physical quantity $\mathbb{\bar{X}}$ covers $\mathbb{\bar{X}}_g = \Sigma ( \bar{x}_i \emph{\textbf{i}}_i) $, $\mathbb{\bar{X}}_e = \Sigma (\bar{x}_{(i+4)} \emph{\textbf{i}}_{(i+4)})$, $\mathbb{\bar{X}}_w = \Sigma ( \bar{x}_{(i+8)} \emph{\textbf{i}}_{(i+8)}) $, and $\mathbb{\bar{X}}_s = \Sigma (\bar{x}_{(i+12)} \emph{\textbf{i}}_{(i+12)})$. The $\mathbb{\bar{A}}_g$, $\mathbb{\bar{A}}_e$, $\mathbb{\bar{A}}_w$, and $\mathbb{\bar{A}}_s$ are the gravitational potential, electromagnetic potential, weak nuclear field potential, and strong nuclear field potential respectively.

The sedenion compounding field strength is $\mathbb{\bar{B}} = \mathbb{\bar{B}}_g + f_e \mathbb{\bar{B}}_e + f_w \mathbb{\bar{B}}_w + f_s \mathbb{\bar{B}}_s$, and is defined from the sedenion field potential $\mathbb{\bar{A}}$, that is $\mathbb{\bar{B}} = (\lozenge_{16} + k_x \mathbb{\bar{X}} + k_a \mathbb{\bar{A}}) \circ \mathbb{\bar{A}}$ .

The sedenion compounding field strength $\mathbb{\bar{B}}$ involves,
\begin{align}
\mathbb{\bar{B}}_g = & ~ ( \lozenge_g \circ \mathbb{\bar{A}}_g + d_e f_e \lozenge_e \circ \mathbb{\bar{A}}_e  + d_w f_w \lozenge_w \circ \mathbb{\bar{A}}_w  + d_s f_s \lozenge_s \circ \mathbb{\bar{A}}_s )
\nonumber
\\
&
+ k_x ( \mathbb{\bar{X}}_g \circ \mathbb{\bar{A}}_g + f_e^2 \mathbb{\bar{X}}_e \circ \mathbb{\bar{A}}_e )
+ k_x ( f_w^2 \mathbb{\bar{X}}_w \circ \mathbb{\bar{A}}_w  + f_s^2 \mathbb{\bar{X}}_s \circ \mathbb{\bar{A}}_s )
\nonumber
\\
&
+ k_a ( \mathbb{\bar{A}}_g \circ \mathbb{\bar{A}}_g + f_e^2 \mathbb{\bar{A}}_e \circ \mathbb{\bar{A}}_e
+ f_w^2 \mathbb{\bar{A}}_w \circ \mathbb{\bar{A}}_w  + f_s^2 \mathbb{\bar{A}}_s \circ \mathbb{\bar{A}}_s )~,
\\
\mathbb{\bar{B}}_e = & ~ ( \lozenge_g \circ \mathbb{\bar{A}}_e + d_e f_e^{-1} \lozenge_e \circ \mathbb{\bar{A}}_g  + d_s f_w f_e^{-1} \lozenge_s \circ \mathbb{\bar{A}}_w  + d_w f_s f_e^{-1} \lozenge_w \circ \mathbb{\bar{A}}_s )
\nonumber
\\
&
+ k_x ( \mathbb{\bar{X}}_e \circ \mathbb{\bar{A}}_g + \mathbb{\bar{X}}_g \circ \mathbb{\bar{A}}_e)
+ k_x f_s f_w f_e^{-1} ( \mathbb{\bar{X}}_s \circ \mathbb{\bar{A}}_w + \mathbb{\bar{X}}_w \circ \mathbb{\bar{A}}_s)
\nonumber
\\
&
+ k_a ( \mathbb{\bar{A}}_e \circ \mathbb{\bar{A}}_g + \mathbb{\bar{A}}_g \circ \mathbb{\bar{A}}_e)
+ k_a f_s f_w f_e^{-1} ( \mathbb{\bar{A}}_s \circ \mathbb{\bar{A}}_w + \mathbb{\bar{A}}_w \circ \mathbb{\bar{A}}_s)~,
\\
\mathbb{\bar{B}}_w = & ~ ( \lozenge_g \circ \mathbb{\bar{A}}_w + d_e f_s f_w^{-1} \lozenge_e \circ \mathbb{\bar{A}}_s  + d_w f_w^{-1} \lozenge_w \circ \mathbb{\bar{A}}_g + d_s f_e f_w^{-1} \lozenge_s \circ \mathbb{\bar{A}}_e)
\nonumber
\\
&
+ k_x ( \mathbb{\bar{X}}_g \circ \mathbb{\bar{A}}_w  + \mathbb{\bar{X}}_w \circ \mathbb{\bar{A}}_g )
+ k_x f_e f_s f_w^{-1} ( \mathbb{\bar{X}}_e \circ \mathbb{\bar{A}}_s + \mathbb{\bar{X}}_s \circ \mathbb{\bar{A}}_e)
\nonumber
\\
&
+ k_a ( \mathbb{\bar{A}}_g \circ \mathbb{\bar{A}}_w  + \mathbb{\bar{A}}_w \circ \mathbb{\bar{A}}_g )
+ k_a f_e f_s f_w^{-1} ( \mathbb{\bar{A}}_e \circ \mathbb{\bar{A}}_s + \mathbb{\bar{A}}_s \circ \mathbb{\bar{A}}_e)~,
\\
\mathbb{\bar{B}}_s = & ~ ( \lozenge_g \circ \mathbb{\bar{A}}_s + d_e f_w f_s^{-1} \lozenge_e \circ \mathbb{\bar{A}}_w  + d_w f_e f_s^{-1} \lozenge_w \circ \mathbb{\bar{A}}_e  + d_s f_s^{-1} \lozenge_s \circ \mathbb{\bar{A}}_g )
\nonumber
\\
&
+ k_x ( \mathbb{\bar{X}}_g \circ \mathbb{\bar{A}}_s  + \mathbb{\bar{X}}_s \circ \mathbb{\bar{A}}_g )
+ k_x f_e f_w f_s^{-1} ( \mathbb{\bar{X}}_e \circ \mathbb{\bar{A}}_w + \mathbb{\bar{X}}_w \circ \mathbb{\bar{A}}_e)
\nonumber
\\
&
+ k_a ( \mathbb{\bar{A}}_g \circ \mathbb{\bar{A}}_s  + \mathbb{\bar{A}}_s \circ \mathbb{\bar{A}}_g )
+ k_a f_e f_w f_s^{-1} ( \mathbb{\bar{A}}_e \circ \mathbb{\bar{A}}_w + \mathbb{\bar{A}}_w \circ \mathbb{\bar{A}}_e)~,
\end{align}
where $\mathbb{\bar{B}}_g = \Sigma ( \bar{h}_i \emph{\textbf{i}}_i)$, $\mathbb{\bar{B}}_e = \Sigma (\bar{h}_{(i+4)} \emph{\textbf{i}}_{(i+4)})$, $\mathbb{\bar{B}}_w = \Sigma ( \bar{h}_{(i+8)} \emph{\textbf{i}}_{(i+8)}) $, and $\mathbb{\bar{B}}_s = \Sigma (\bar{h}_{(i+12)} \emph{\textbf{i}}_{(i+12)})$. The gauge equations are $\bar{h}_0 = 0$, $\bar{h}_4 = 0$, $\bar{h}_8 = 0$, and $\bar{h}_{12} = 0$.

For the gravitational strength $\mathbb{\bar{B}}_g$, the linear momentum density $-\mu_{gg} \mathbb{\bar{S}}_{gg} = \lozenge_g^* \circ \mathbb{\bar{B}}_g$ is the source for the gravitational field, with three adjoint field sources for the gravitational field, that is $-\mu_{ge} \mathbb{\bar{S}}_{ge} = \lozenge_e^* \circ \mathbb{\bar{B}}_g$, $-\mu_{gw} \mathbb{\bar{S}}_{gw} = \lozenge_w^* \circ \mathbb{\bar{B}}_g$, and $-\mu_{gs} \mathbb{\bar{S}}_{gs} = \lozenge_s^* \circ \mathbb{\bar{B}}_g$. For the electromagnetic strength $\mathbb{\bar{B}}_e$, the electric current density $-\mu_{eg} \mathbb{\bar{S}}_{eg} = \lozenge_g^* \circ \mathbb{\bar{B}}_e$ is the source for electromagnetic field, with three adjoint field sources for the electromagnetic field, that is $-\mu_{ee} \mathbb{\bar{S}}_{ee} = \lozenge_e^* \circ \mathbb{\bar{B}}_e$, $-\mu_{ew} \mathbb{\bar{S}}_{ew} = \lozenge_w^* \circ \mathbb{\bar{B}}_e$, and $-\mu_{es} \mathbb{\bar{S}}_{es} = \lozenge_s^* \circ \mathbb{\bar{B}}_e$. For the weak nuclear field strength $\mathbb{\bar{B}}_w$, the $-\mu_{wg} \mathbb{\bar{S}}_{wg} = \lozenge_g^* \circ \mathbb{\bar{B}}_w$ is the source for weak nuclear field, with three kinds of adjoint field sources for the weak nuclear field, that is $-\mu_{we} \mathbb{\bar{S}}_{we} = \lozenge_e^* \circ \mathbb{\bar{B}}_w$, $-\mu_{ww} \mathbb{\bar{S}}_{ww} = \lozenge_w^* \circ \mathbb{\bar{B}}_w$, and $-\mu_{ws} \mathbb{\bar{S}}_{ws} = \lozenge_s^* \circ \mathbb{\bar{B}}_w$. For the strong nuclear field strength $\mathbb{\bar{B}}_s$, the $-\mu_{sg} \mathbb{\bar{S}}_{sg} = \lozenge_g^* \circ \mathbb{\bar{B}}_s$ is the source for strong nuclear field, with three kinds of adjoint field sources for the strong nuclear field, that is $-\mu_{se} \mathbb{\bar{S}}_{se} = \lozenge_e^* \circ \mathbb{\bar{B}}_s$, $-\mu_{sw} \mathbb{\bar{S}}_{sw} = \lozenge_w^* \circ \mathbb{\bar{B}}_s$, and $-\mu_{ss} \mathbb{\bar{S}}_{ss} = \lozenge_s^* \circ \mathbb{\bar{B}}_s$. The three adjoint field sources, $\mathbb{\bar{S}}_{ee}$, $\mathbb{\bar{S}}_{ww}$, and $\mathbb{\bar{S}}_{ss}$, possess the feature of the gravitational field, and are considered as the candidate of the dark matters.

The sedenion field source $\mathbb{\bar{S}}$ satisfies
\begin{align}
\mu \mathbb{\bar{S}} = & ~ - (\lozenge_{16} + k_x \mathbb{\bar{X}} + k_a \mathbb{\bar{A}} + k_b \mathbb{\bar{B}})^* \circ \mathbb{\bar{B}}
\nonumber\\
= & ~ \mu_{gg} \mathbb{\bar{S}}_{gg} + f_e \mu_{ge} \mathbb{\bar{S}}_{ge} + f_w \mu_{gw} \mathbb{\bar{S}}_{gw} + f_s \mu_{gs} \mathbb{\bar{S}}_{gs}
\nonumber\\
&
+ d_e ( \mu_{eg} \mathbb{\bar{S}}_{eg} + f_e \mu_{ee} \mathbb{\bar{S}}_{ee} + f_w \mu_{ew} \mathbb{\bar{S}}_{ew} + f_s \mu_{es} \mathbb{\bar{S}}_{es} )
\nonumber\\
&
+ d_w ( \mu_{wg} \mathbb{\bar{S}}_{wg} + f_e \mu_{we} \mathbb{\bar{S}}_{we} + f_w \mu_{ww} \mathbb{\bar{S}}_{ww} + f_s \mu_{ws} \mathbb{\bar{S}}_{ws} )
\nonumber\\
&
+ d_s ( \mu_{sg} \mathbb{\bar{S}}_{sg} + f_e \mu_{se} \mathbb{\bar{S}}_{se} + f_w \mu_{sw} \mathbb{\bar{S}}_{sw} + f_s \mu_{ss} \mathbb{\bar{S}}_{ss} )
\nonumber\\
&
- k_x \mathbb{\bar{X}}^* \circ \mathbb{\bar{B}} - k_a \mathbb{\bar{A}}^* \circ \mathbb{\bar{B}} - k_b \mathbb{\bar{B}}^* \circ \mathbb{\bar{B}}
~,
\end{align}
where $\mathbb{\bar{S}} = \Sigma (\bar{s}_i \emph{\textbf{i}}_i + f_e \bar{s}_{(i+4)} \emph{\textbf{i}}_{(i+4)} + f_w \bar{s}_{(i+8)} \emph{\textbf{i}}_{(i+8)} + f_s \bar{s}_{(i+12)} \emph{\textbf{i}}_{(i+12)})$. $\mu$, $\mu_{gg}$, $\mu_{ge}$, $\mu_{gw}$, $\mu_{gs}$, $\mu_{eg}$, $\mu_{ee}$, $\mu_{ew}$, $\mu_{es}$, $\mu_{wg}$, $\mu_{we}$, $\mu_{ww}$, $\mu_{ws}$, $\mu_{sg}$, $\mu_{se}$, $\mu_{sw}$, and $\mu_{ss}$ are the coefficients for the dimensional homogeneity. $\mu_{gg}$ and $\mu_{eg}$ are the gravitational constant and electromagnetic constant respectively. $(k_x \mathbb{\bar{X}}^* \cdot \mathbb{\bar{B}} + k_a \mathbb{\bar{A}}^* \cdot \mathbb{\bar{B}} + k_b \mathbb{\bar{B}}^* \cdot \mathbb{\bar{B}})$ is the field strength helicity. And $\mathbb{\bar{X}}^* \cdot \mathbb{\bar{B}}$ is the scalar part of the $\mathbb{\bar{X}}^* \circ \mathbb{\bar{B}}$. The $*$ denotes the sedenion conjugate \cite{culbert, dickson}.

\begin{table}[h]
\caption{The field equations of the four fields with their adjoint fields.}
\label{tab:table3}
\centering
\begin{tabular}{lll}
\hline\hline
$ fundamental~field $              & $ field~source $                           & $ field~equation $                    \\

\hline
gravitational~field                & gravitational~field~source                 & $-\mu_{gg} \mathbb{\bar{S}}_{gg} = \lozenge_g^* \circ \mathbb{\bar{B}}_g$    \\
                                   & adjoint~gravitational~field~source         & $-\mu_{ge} \mathbb{\bar{S}}_{ge} = \lozenge_e^* \circ \mathbb{\bar{B}}_g$    \\
                                   & adjoint~gravitational~field~source         & $-\mu_{gw} \mathbb{\bar{S}}_{gw} = \lozenge_w^* \circ \mathbb{\bar{B}}_g$    \\
                                   & adjoint~gravitational~field~source         & $-\mu_{gs} \mathbb{\bar{S}}_{gs} = \lozenge_s^* \circ \mathbb{\bar{B}}_g$    \\
\hline

electromagnetic~field              & electromagnetic~field~source               & $-\mu_{eg} \mathbb{\bar{S}}_{eg} = \lozenge_g^* \circ \mathbb{\bar{B}}_e$    \\
                                   & adjoint~electromagnetic~field~source       & $-\mu_{ee} \mathbb{\bar{S}}_{ee} = \lozenge_e^* \circ \mathbb{\bar{B}}_e$    \\
                                   & adjoint~electromagnetic~field~source       & $-\mu_{ew} \mathbb{\bar{S}}_{ew} = \lozenge_w^* \circ \mathbb{\bar{B}}_e$    \\
                                   & adjoint~electromagnetic~field~source       & $-\mu_{es} \mathbb{\bar{S}}_{es} = \lozenge_s^* \circ \mathbb{\bar{B}}_e$    \\
\hline

weak~nuclear~field                 & weak~nuclear~field~source                  & $-\mu_{wg} \mathbb{\bar{S}}_{wg} = \lozenge_g^* \circ \mathbb{\bar{B}}_w$    \\
                                   & adjoint~weak~nuclear~field~source          & $-\mu_{we} \mathbb{\bar{S}}_{we} = \lozenge_e^* \circ \mathbb{\bar{B}}_w$    \\
                                   & adjoint~weak~nuclear~field~source          & $-\mu_{ww} \mathbb{\bar{S}}_{ww} = \lozenge_w^* \circ \mathbb{\bar{B}}_w$    \\
                                   & adjoint~weak~nuclear~field~source          & $-\mu_{ws} \mathbb{\bar{S}}_{ws} = \lozenge_s^* \circ \mathbb{\bar{B}}_w$    \\
\hline

strong~nuclear~field               & strong~nuclear~field~source                & $-\mu_{sg} \mathbb{\bar{S}}_{sg} = \lozenge_g^* \circ \mathbb{\bar{B}}_s$    \\
                                   & adjoint~strong~nuclear~field~source        & $-\mu_{se} \mathbb{\bar{S}}_{se} = \lozenge_e^* \circ \mathbb{\bar{B}}_s$    \\
                                   & adjoint~strong~nuclear~field~source        & $-\mu_{sw} \mathbb{\bar{S}}_{sw} = \lozenge_w^* \circ \mathbb{\bar{B}}_s$    \\
                                   & adjoint~strong~nuclear~field~source        & $-\mu_{ss} \mathbb{\bar{S}}_{ss} = \lozenge_s^* \circ \mathbb{\bar{B}}_s$    \\

\hline\hline
\end{tabular}
\end{table}

\subsection{Torque}

The sedenion linear momentum density is $\mathbb{\bar{P}} = \mu \mathbb{\bar{S}} / \mu_{gg} = \Sigma (\bar{p}_i \emph{\textbf{i}}_i + f_e \bar{p}_{(i+4)} \emph{\textbf{i}}_{(i+4)} + f_w \bar{p}_{(i+8)} \emph{\textbf{i}}_{(i+8)} + f_s \bar{p}_{(i+12)} \emph{\textbf{i}}_{(i+12)})$ in the sedenion compounding space, and the sedenion angular momentum density is $\mathbb{\bar{L}} = \mathbb{\bar{R}} \circ \mathbb{\bar{P}} = \Sigma (\bar{l}_i \emph{\textbf{i}}_i + f_e \bar{l}_{(i+4)} \emph{\textbf{i}}_{(i+4)} + f_w \bar{l}_{(i+8)} \emph{\textbf{i}}_{(i+8)} + f_s \bar{l}_{(i+12)} \emph{\textbf{i}}_{(i+12)})$. And the sedenion torque-energy density $\mathbb{\bar{W}}$ is defined from the angular momentum density $\mathbb{\bar{L}}$, the field strength $\mathbb{\bar{B}}$, and the field source $\mathbb{\bar{S}}$ etc,
\begin{eqnarray}
\mathbb{\bar{W}} = v_0 (\lozenge_{16} + k_x \mathbb{\bar{X}} + k_a \mathbb{\bar{A}} + k_b \mathbb{\bar{B}} + k_s \mathbb{\bar{S}} + k_l \mathbb{\bar{L}}) \circ \mathbb{\bar{L}}~,
\end{eqnarray}
where $-\bar{w}_0/2$ is the energy density, $\bar{\textbf{w}}/2 = \Sigma (\bar{w}_j \emph{\textbf{i}}_j )/2$ is the torque density.

The scalar $\bar{w}_0$ of $\mathbb{\bar{W}} = \Sigma (\bar{w}_i \emph{\textbf{i}}_i + f_e \bar{w}_{(i+4)} \emph{\textbf{i}}_{(i+4)} + f_w \bar{w}_{(i+8)} \emph{\textbf{i}}_{(i+8)} + f_s \bar{w}_{(i+12)} \emph{\textbf{i}}_{(i+12)})$ is written as,
\begin{align}
\bar{w}_0 / v_0  = & ~
k_a \left\{ \bar{a}_0 \bar{l}_0 + \bar{\textbf{a}}_g \cdot \bar{\textbf{l}}_g + f_e^2 ( \bar{\textbf{a}}_4 \circ \bar{\textbf{l}}_4 + \bar{\textbf{a}}_e \cdot \bar{\textbf{l}}_e ) \right\}
\nonumber\\
&
+ k_a \left\{ f_w^2 ( \bar{\textbf{a}}_8 \circ \bar{\textbf{l}}_8 + \bar{\textbf{a}}_w \cdot \bar{\textbf{l}}_w ) + f_s^2 ( \bar{\textbf{a}}_{12} \circ \bar{\textbf{l}}_{12} + \bar{\textbf{a}}_s \cdot \bar{\textbf{l}}_s) \right\}
\nonumber\\
&
+ k_x \left\{ \bar{x}_0 \bar{l}_0 + \bar{\textbf{x}}_g \cdot \bar{\textbf{l}}_g + f_e^2 ( \bar{\textbf{x}}_4 \circ \bar{\textbf{l}}_4 + \bar{\textbf{x}}_e \cdot \bar{\textbf{l}}_e )  \right\}
\nonumber\\
&
+ k_x \left\{ f_w^2 ( \bar{\textbf{x}}_8 \circ \bar{\textbf{l}}_8 + \bar{\textbf{x}}_w \cdot \bar{\textbf{l}}_w )+ f_s^2 ( \bar{\textbf{x}}_{12} \circ \bar{\textbf{l}}_{12} + \bar{\textbf{x}}_s \cdot \bar{\textbf{l}}_s) \right\}
\nonumber\\
&
+ k_s \left\{ \bar{s}_0 \bar{l}_0 + \bar{\textbf{s}}_g \cdot \bar{\textbf{l}}_g + f_e^2 ( \bar{\textbf{s}}_4 \circ \bar{\textbf{l}}_4 + \bar{\textbf{s}}_e \cdot \bar{\textbf{l}}_e ) \right\}
\nonumber\\
&
+ k_s \left\{  f_w^2 ( \bar{\textbf{s}}_8 \circ \bar{\textbf{l}}_8 + \bar{\textbf{s}}_w \cdot \bar{\textbf{l}}_w )+ f_s^2 ( \bar{\textbf{s}}_{12} \circ \bar{\textbf{l}}_{12} + \bar{\textbf{s}}_s \cdot \bar{\textbf{l}}_s) \right\}
\nonumber\\
&
+ k_l \left\{ \bar{l}_0 \bar{l}_0 + \bar{\textbf{l}}_g \cdot \bar{\textbf{l}}_g + f_e^2 ( \bar{\textbf{l}}_4 \circ \bar{\textbf{l}}_4 + \bar{\textbf{l}}_e \cdot \bar{\textbf{l}}_e ) \right\}
\nonumber\\
&
+ k_l \left\{ f_w^2 ( \bar{\textbf{l}}_8 \circ \bar{\textbf{l}}_8 + \bar{\textbf{l}}_w \cdot \bar{\textbf{l}}_w ) + f_s^2 ( \bar{\textbf{l}}_{12} \circ \bar{\textbf{l}}_{12} + \bar{\textbf{l}}_s \cdot \bar{\textbf{l}}_s) \right\}
\nonumber\\
&
+ k_b ( \bar{\textbf{h}}_g \cdot \bar{\textbf{l}}_g + f_e^2 \bar{\textbf{h}}_e \cdot \bar{\textbf{l}}_e  + f_w^2 \bar{\textbf{h}}_w \cdot \bar{\textbf{l}}_w + f_s^2 \bar{\textbf{h}}_s \cdot \bar{\textbf{l}}_s )
\nonumber\\
&
+ \partial_0 \bar{l}_0 + \nabla_g \cdot \bar{\textbf{l}}_g + d_e f_e (\emph{\textbf{i}}_4 \circ \partial_4 \bar{\textbf{l}}_4 + \nabla_e \cdot \bar{\textbf{l}}_e)
\nonumber\\
&
+ d_w f_w (\emph{\textbf{i}}_8 \circ \partial_8 \bar{\textbf{l}}_8 + \nabla_w \cdot \bar{\textbf{l}}_w) + d_s f_s (\emph{\textbf{i}}_{12} \circ \partial_{12} \bar{\textbf{l}}_{12} + \nabla_s \cdot \bar{\textbf{l}}_s)~,
\end{align}
where $-\bar{w}_0/2$ includes the kinetic energy, the gravitational potential energy, the electric potential energy, the magnetic potential energy, the field energy, the work, the interacting energy between the dipole moment with the fields, and some new energy terms. $\bar{\textbf{l}}_g = \Sigma ( \emph{\textbf{i}}_j \bar{l}_j)$, $\bar{\textbf{l}}_e = \Sigma ( \emph{\textbf{i}}_{(j+4)} \bar{l}_{(j+4)})$, $\bar{\textbf{l}}_w = \Sigma ( \emph{\textbf{i}}_{(j+8)} \bar{l}_{(j+8)})$, $\bar{\textbf{l}}_s = \Sigma ( \emph{\textbf{i}}_{(j+12)} \bar{l}_{(j+12)})$; $\bar{\textbf{l}}_4 = \emph{\textbf{i}}_4 \bar{l}_4$, $\bar{\textbf{l}}_8 = \emph{\textbf{i}}_8 \bar{l}_8$, $\bar{\textbf{l}}_{12} = \emph{\textbf{i}}_{12} \bar{l}_{12}$ . $\bar{\textbf{x}}_g = \Sigma ( \emph{\textbf{i}}_j \bar{x}_j)$, $\bar{\textbf{x}}_e = \Sigma ( \emph{\textbf{i}}_{(j+4)} \bar{x}_{(j+4)})$, $\bar{\textbf{x}}_w = \Sigma ( \emph{\textbf{i}}_{(j+8)} \bar{x}_{(j+8)})$, $\bar{\textbf{x}}_s = \Sigma ( \emph{\textbf{i}}_{(j+12)} \bar{x}_{(j+12)})$; $\bar{\textbf{x}}_4 = \emph{\textbf{i}}_4 \bar{x}_4$, $\bar{\textbf{x}}_8 = \emph{\textbf{i}}_8 \bar{x}_8$, $\bar{\textbf{x}}_{12} = \emph{\textbf{i}}_{12} \bar{x}_{12}$ . $\bar{\textbf{a}}_g = \Sigma ( \emph{\textbf{i}}_j \bar{a}_j)$, $\bar{\textbf{a}}_e = \Sigma ( \emph{\textbf{i}}_{(j+4)} \bar{a}_{(j+4)})$, $\bar{\textbf{a}}_w = \Sigma ( \emph{\textbf{i}}_{(j+8)} \bar{a}_{(j+8)})$, $\bar{\textbf{a}}_s = \Sigma ( \emph{\textbf{i}}_{(j+12)} \bar{a}_{(j+12)})$; $\bar{\textbf{a}}_4 = \emph{\textbf{i}}_4 \bar{a}_4$, $\bar{\textbf{a}}_8 = \emph{\textbf{i}}_8 \bar{a}_8$, $\bar{\textbf{a}}_{12} = \emph{\textbf{i}}_{12} \bar{a}_{12}$ . $\bar{\textbf{s}}_g = \Sigma ( \emph{\textbf{i}}_j \bar{s}_j)$, $\bar{\textbf{s}}_e = \Sigma ( \emph{\textbf{i}}_{(j+4)} \bar{s}_{(j+4)})$, $\bar{\textbf{s}}_w = \Sigma ( \emph{\textbf{i}}_{(j+8)} \bar{s}_{(j+8)})$, $\bar{\textbf{s}}_s = \Sigma ( \emph{\textbf{i}}_{(j+12)} \bar{s}_{(j+12)})$. $\bar{\textbf{s}}_4 = \emph{\textbf{i}}_4 \bar{s}_4$, $\bar{\textbf{s}}_8 = \emph{\textbf{i}}_8 \bar{s}_8$, $\bar{\textbf{s}}_{12} = \emph{\textbf{i}}_{12} \bar{s}_{12}$ . $\bar{\textbf{h}}_g = \Sigma ( \emph{\textbf{i}}_j \bar{h}_j)$, $\bar{\textbf{h}}_e = \Sigma ( \emph{\textbf{i}}_{(j+4)} \bar{h}_{(j+4)})$, $\bar{\textbf{h}}_w = \Sigma ( \emph{\textbf{i}}_{(j+8)} \bar{h}_{(j+8)})$, $\bar{\textbf{h}}_s = \Sigma ( \emph{\textbf{i}}_{(j+12)} \bar{h}_{(j+12)})$. $\nabla_g = \Sigma ( \emph{\textbf{i}}_j \partial_j)$, $\nabla_e = \Sigma ( \emph{\textbf{i}}_{(j+4)} \partial_{(j+4)})$, $\nabla_w = \Sigma ( \emph{\textbf{i}}_{(j+8)} \partial_{(j+8)})$, $\nabla_s = \Sigma ( \emph{\textbf{i}}_{(j+12)} \partial_{(j+12)})$.

In a similar way, expressing the torque density $\bar{\textbf{w}}$ of $\mathbb{\bar{W}}$ as
\begin{align}
\bar{\textbf{w}} / v_0 = & ~ ( \partial_0 \bar{\textbf{l}}_g + \nabla_g \bar{l}_0 + \nabla_g \times \bar{\textbf{l}}_g )
+ d_e f_e ( \emph{\textbf{i}}_4 \circ \partial_4 \bar{\textbf{l}}_e + \nabla_e \circ \bar{\textbf{l}}_4 + \nabla_e \times \bar{\textbf{l}}_e )
\nonumber
\\
& + d_w f_w ( \emph{\textbf{i}}_8 \circ \partial_8 \bar{\textbf{l}}_w + \nabla_w \circ \bar{\textbf{l}}_8 + \nabla_w \times \bar{\textbf{l}}_w )
+ d_s f_s ( \emph{\textbf{i}}_{12} \circ \partial_{12} \bar{\textbf{l}}_s + \nabla_s \circ \bar{\textbf{l}}_{12} + \nabla_s \times \bar{\textbf{l}}_s )
\nonumber
\\
& + k_a ( \bar{\textbf{a}}_g \times \bar{\textbf{l}}_g + \bar{\textbf{a}}_g \bar{l}_0 + \bar{a}_0 \bar{\textbf{l}}_g ) + k_a f_e^2 ( \bar{\textbf{a}}_e \times \bar{\textbf{l}}_e + \bar{\textbf{a}}_4 \circ \bar{\textbf{l}}_e + \bar{\textbf{a}}_e \circ \bar{\textbf{l}}_4)
\nonumber
\\
& + k_a f_w^2 ( \bar{\textbf{a}}_w \times \bar{\textbf{l}}_w + \bar{\textbf{a}}_8 \circ \bar{\textbf{l}}_w + \bar{\textbf{a}}_w \circ \bar{\textbf{l}}_8)
+ k_a f_s^2 ( \bar{\textbf{a}}_s \times \bar{\textbf{l}}_s + \bar{\textbf{a}}_{12} \circ \bar{\textbf{l}}_s + \bar{\textbf{a}}_s \circ \bar{\textbf{l}}_{12})
\nonumber
\\
& + k_x ( \bar{\textbf{x}}_g \times \bar{\textbf{l}}_g + \bar{\textbf{x}}_g \bar{l}_0 + \bar{x}_0 \bar{\textbf{l}}_g ) + k_x f_e^2 ( \bar{\textbf{x}}_e \times \bar{\textbf{l}}_e + \bar{\textbf{x}}_4 \circ \bar{\textbf{l}}_e + \bar{\textbf{x}}_e \circ \bar{\textbf{l}}_4)
\nonumber
\\
& + k_x f_w^2 ( \bar{\textbf{x}}_w \times \bar{\textbf{l}}_w + \bar{\textbf{x}}_8 \circ \bar{\textbf{l}}_w + \bar{\textbf{x}}_w \circ \bar{\textbf{l}}_8)
+ k_x f_s^2 ( \bar{\textbf{x}}_s \times \bar{\textbf{l}}_s + \bar{\textbf{x}}_{12} \circ \bar{\textbf{l}}_s + \bar{\textbf{x}}_s \circ \bar{\textbf{l}}_{12})
\nonumber
\\
& + k_s ( \bar{\textbf{s}}_g \times \bar{\textbf{l}}_g + \bar{\textbf{s}}_g \bar{l}_0 + \bar{s}_0 \bar{\textbf{l}}_g ) + k_s f_e^2 ( \bar{\textbf{s}}_e \times \bar{\textbf{l}}_e + \bar{\textbf{s}}_4 \circ \bar{\textbf{l}}_e + \bar{\textbf{s}}_e \circ \bar{\textbf{l}}_4)
\nonumber
\\
& + k_s f_w^2 ( \bar{\textbf{s}}_w \times \bar{\textbf{l}}_w + \bar{\textbf{s}}_8 \circ \bar{\textbf{l}}_w + \bar{\textbf{s}}_w \circ \bar{\textbf{l}}_8)
+ k_s f_s^2 ( \bar{\textbf{s}}_s \times \bar{\textbf{l}}_s + \bar{\textbf{s}}_{12} \circ \bar{\textbf{l}}_s + \bar{\textbf{s}}_s \circ \bar{\textbf{l}}_{12})
\nonumber
\\
& + k_b ( \bar{\textbf{h}}_g \times \bar{\textbf{l}}_g + \bar{\textbf{h}}_g \bar{l}_0 ) + k_b f_e^2 ( \bar{\textbf{h}}_e \times \bar{\textbf{l}}_e + \bar{\textbf{h}}_e \circ \bar{\textbf{l}}_4)
\nonumber
\\
& + k_b f_w^2 ( \bar{\textbf{h}}_w \times \bar{\textbf{l}}_w + \bar{\textbf{h}}_w \circ \bar{\textbf{l}}_8)
+ k_b f_s^2 ( \bar{\textbf{h}}_s \times \bar{\textbf{l}}_s + \bar{\textbf{h}}_s \circ \bar{\textbf{l}}_{12}) + 2 k_l \bar{l}_0 \bar{\textbf{l}}_g
~,
\end{align}
where the above includes some new terms of the torque density.

\begin{table}[b]
\caption{The field sources of the four fields with their adjoint fields.}
\label{tab:table3}
\centering
\begin{tabular}{llll}
\hline\hline
$ sources $                              & $ fields $                                 & $ descriptions $                  & $ characterictics $    \\
\hline
$\mathbb{\bar{S}}_{gg}$                  & gravitational~field                        & linear~momentum                   & gravitation            \\
$\mathbb{\bar{S}}_{ee}$                  & electromagnetic~adjoint~field              & adjoint~electric~current          & gravitation            \\
$\mathbb{\bar{S}}_{ww}$                  & weak~nuclear~adjoint~field                 & adjoint~weak~nuclear~current      & gravitation            \\
$\mathbb{\bar{S}}_{ss}$                  & strong~nuclear~adjoint~field               & adjoint~strong~nuclear~current    & gravitation            \\
\hline
$\mathbb{\bar{S}}_{ge}$                  & gravitational~adjoint~field                & adjoint~linear~momentum           & electromagnetism       \\
$\mathbb{\bar{S}}_{eg}$                  & electromagnetic~field                      & electric~current                  & electromagnetism       \\
$\mathbb{\bar{S}}_{ws}$                  & weak~nuclear~adjoint~field                 & adjoint~weak~nuclear~current      & electromagnetism       \\
$\mathbb{\bar{S}}_{sw}$                  & strong~nuclear~adjoint~field               & adjoint~strong~nuclear~current    & electromagnetism       \\
\hline
$\mathbb{\bar{S}}_{gw}$                  & gravitational~adjoint~field                & adjoint~linear~momentum           & weak~nuclear~force     \\
$\mathbb{\bar{S}}_{es}$                  & electromagnetic~adjoint~field              & adjoint~electric~current          & weak~nuclear~force     \\
$\mathbb{\bar{S}}_{wg}$                  & weak~nuclear~field                         & weak~nuclear~current              & weak~nuclear~force     \\
$\mathbb{\bar{S}}_{se}$                  & strong~nuclear~adjoint~field               & adjoint~strong~nuclear~current    & weak~nuclear~force     \\
\hline
$\mathbb{\bar{S}}_{gs}$                  & gravitational~adjoint~field                & adjoint~linear~momentum           & strong~nuclear~force   \\
$\mathbb{\bar{S}}_{ew}$                  & electromagnetic~adjoint~field              & adjoint~electric~current          & strong~nuclear~force   \\
$\mathbb{\bar{S}}_{we}$                  & weak~nuclear~adjoint~field                 & adjoint~weak~nuclear~current      & strong~nuclear~force   \\
$\mathbb{\bar{S}}_{sg}$                  & strong~nuclear~field                       & strong~nuclear~current            & strong~nuclear~force   \\
\hline\hline
\end{tabular}
\end{table}

\subsection{Force}

In the sedenion compounding space with the operator $(\lozenge_{16} + k_x \mathbb{\bar{X}} + k_a \mathbb{\bar{A}} + k_b \mathbb{\bar{B}} + k_s \mathbb{\bar{S}} + k_l \mathbb{\bar{L}} + k_w \mathbb{\bar{W}})$, the sedenion force-power density $\mathbb{\bar{N}}$ is defined as follows,
\begin{eqnarray}
\mathbb{\bar{N}} = v_0 (\lozenge_{16} + k_x \mathbb{\bar{X}} + k_a \mathbb{\bar{A}} + k_b \mathbb{\bar{B}} + k_s \mathbb{\bar{S}} + k_l \mathbb{\bar{L}} + k_w \mathbb{\bar{W}})^* \circ \mathbb{\bar{W}}~,
\end{eqnarray}
where $\mathbb{\bar{N}} = \Sigma (\bar{n}_i \emph{\textbf{i}}_i) + f_e \Sigma (\bar{n}_{(i+4)} \emph{\textbf{i}}_{(i+4)}i) + f_w \Sigma (\bar{n}_{(i+8)} \emph{\textbf{i}}_{(i+8)}) + f_s \Sigma (\bar{n}_{(i+12)} \emph{\textbf{i}}_{(i+12)})$. The power density is $\bar{f}_0 = - \bar{n}_0/(2 v_0)$, and the vectorial part is $\bar{\textbf{n}} = \Sigma (\bar{n}_j \emph{\textbf{i}}_j )$.

Further expressing the scalar $\bar{n}_0$ of the sedenion $\mathbb{\bar{N}}$ as
\begin{align}
\bar{n}_0 / v_0 = & ~ k_a \left\{ \bar{a}_0 \bar{w}_0 + \bar{\textbf{a}}_g^* \cdot \bar{\textbf{w}}_g + f_e^2 ( \bar{\textbf{a}}_4^* \circ \bar{\textbf{w}}_4 + \bar{\textbf{a}}_e^* \cdot \bar{\textbf{w}}_e )  \right\}
\nonumber\\
&
+ k_a \left\{ f_w^2 ( \bar{\textbf{a}}_8^* \circ \bar{\textbf{w}}_8 + \bar{\textbf{a}}_w^* \cdot \bar{\textbf{w}}_w )
+ f_s^2 ( \bar{\textbf{a}}_{12}^* \circ \bar{\textbf{w}}_{12} + \bar{\textbf{a}}_s^* \cdot \bar{\textbf{w}}_s) \right\}
\nonumber\\
&
+ k_x \left\{ \bar{x}_0 \bar{w}_0 + \bar{\textbf{x}}_g^* \cdot \bar{\textbf{w}}_g + f_e^2 ( \bar{\textbf{x}}_4^* \circ \bar{\textbf{w}}_4 + \bar{\textbf{x}}_e^* \cdot \bar{\textbf{w}}_e )  \right\}
\nonumber\\
&
+ k_x \left\{ f_w^2 ( \bar{\textbf{x}}_8^* \circ \bar{\textbf{w}}_8 + \bar{\textbf{x}}_w^* \cdot \bar{\textbf{w}}_w ) + f_s^2 ( \bar{\textbf{x}}_{12}^* \circ \bar{\textbf{w}}_{12} + \bar{\textbf{x}}_s^* \cdot \bar{\textbf{w}}_s) \right\}
\nonumber\\
&
+ k_s \left\{ \bar{s}_0 \bar{w}_0 + \bar{\textbf{s}}_g^* \cdot \bar{\textbf{w}}_g +  f_e^2 ( \bar{\textbf{s}}_4^* \circ \bar{\textbf{w}}_4 + \bar{\textbf{s}}_e^* \cdot \bar{\textbf{w}}_e ) \right\}
\nonumber\\
&
+ k_s \left\{ f_w^2 ( \bar{\textbf{s}}_8^* \circ \bar{\textbf{w}}_8 + \bar{\textbf{s}}_w^* \cdot \bar{\textbf{w}}_w ) + f_s^2 ( \bar{\textbf{s}}_{12}^* \circ \bar{\textbf{w}}_{12} + \bar{\textbf{s}}_s^* \cdot \bar{\textbf{w}}_s) \right\}
\nonumber\\
&
+ k_l \left\{ \bar{l}_0 \bar{w}_0 + \bar{\textbf{l}}_g^* \cdot \bar{\textbf{w}}_g + f_e^2 ( \bar{\textbf{l}}_4^* \circ \bar{\textbf{w}}_4 + \bar{\textbf{l}}_e^* \cdot \bar{\textbf{w}}_e ) \right\}
\nonumber\\
&
+ k_l \left\{ f_w^2 ( \bar{\textbf{l}}_8^* \circ \bar{\textbf{w}}_8 + \bar{\textbf{l}}_w^* \cdot \bar{\textbf{w}}_w ) + f_s^2 ( \bar{\textbf{l}}_{12}^* \circ \bar{\textbf{w}}_{12} + \bar{\textbf{l}}_s^* \cdot \bar{\textbf{w}}_s) \right\}
\nonumber\\
&
+ k_w \left\{ \bar{w}_0 \bar{w}_0 + \bar{\textbf{w}}_g^* \cdot \bar{\textbf{w}}_g + f_e^2 ( \bar{\textbf{w}}_4^* \circ \bar{\textbf{w}}_4 + \bar{\textbf{w}}_e^* \cdot \bar{\textbf{w}}_e ) \right\}
\nonumber\\
&
+ k_w \left\{ f_w^2 ( \bar{\textbf{w}}_8^* \circ \bar{\textbf{w}}_8 + \bar{\textbf{w}}_w^* \cdot \bar{\textbf{w}}_w ) + f_s^2 ( \bar{\textbf{w}}_{12}^* \circ \bar{\textbf{w}}_{12} + \bar{\textbf{w}}_s^* \cdot \bar{\textbf{w}}_s) \right\}
\nonumber\\
&
+ k_b ( \bar{\textbf{h}}_g^* \cdot \bar{\textbf{w}}_g + f_e^2 \bar{\textbf{h}}_e^* \cdot \bar{\textbf{w}}_e  + f_w^2 \bar{\textbf{h}}_w^* \cdot \bar{\textbf{w}}_w + f_s^2 \bar{\textbf{h}}_s^* \cdot \bar{\textbf{w}}_s )
\nonumber\\
&
+ \partial_0 \bar{w}_0 + \nabla_g^* \cdot \bar{\textbf{w}}_g + d_e f_e (\emph{\textbf{i}}_4^* \circ \partial_4 \bar{\textbf{w}}_4 + \nabla_e^* \cdot \bar{\textbf{w}}_e)
\nonumber\\
&
+ d_w f_w (\emph{\textbf{i}}_8^* \circ \partial_8 \bar{\textbf{w}}_8 + \nabla_w^* \cdot \bar{\textbf{w}}_w) + d_s f_s (\emph{\textbf{i}}_{12}^* \circ \partial_{12} \bar{\textbf{w}}_{12} + \nabla_s^* \cdot \bar{\textbf{w}}_s)
~.
\end{align}

In the above four fields with the sedenion operator $(\lozenge_{16} + k_x \mathbb{\bar{X}} + k_a \mathbb{\bar{A}} + k_b \mathbb{\bar{B}} + k_s \mathbb{\bar{S}} + k_l \mathbb{\bar{L}} + k_w \mathbb{\bar{W}})$, the force density $\bar{\textbf{f}} = - \bar{\textbf{n}} / (2 v_0)$ can be defined from the vectorial part $\bar{\textbf{n}}$ of $\mathbb{\bar{N}}$ ,
\begin{align}
\bar{\textbf{n}} / v_0 = & ~ ( \partial_0 \bar{\textbf{w}}_g + \nabla_g^* \bar{w}_0 + \nabla_g^* \times \bar{\textbf{w}}_g )
+ d_e f_e ( \emph{\textbf{i}}_4^* \circ \partial_4 \bar{\textbf{w}}_e + \nabla_e^* \circ \bar{\textbf{w}}_4 + \nabla_e^* \times \bar{\textbf{w}}_e )
\nonumber
\\
& + d_w f_w ( \emph{\textbf{i}}_8^* \circ \partial_8 \bar{\textbf{w}}_w + \nabla_w^* \circ \bar{\textbf{w}}_8 + \nabla_w^* \times \bar{\textbf{w}}_w )
\nonumber
\\
&
+ d_s f_s ( \emph{\textbf{i}}_{12}^* \circ \partial_{12} \bar{\textbf{w}}_s + \nabla_s^* \circ \bar{\textbf{w}}_{12} + \nabla_s^* \times \bar{\textbf{w}}_s )
\nonumber
\\
& + k_a ( \bar{\textbf{a}}_g^* \times \bar{\textbf{w}}_g + \bar{\textbf{a}}_g^* \bar{w}_0 + \bar{a}_0 \bar{\textbf{w}}_g ) + k_a f_e^2 ( \bar{\textbf{a}}_e^* \times \bar{\textbf{w}}_e + \bar{\textbf{a}}_4^* \circ \bar{\textbf{w}}_e + \bar{\textbf{a}}_e^* \circ \bar{\textbf{w}}_4)
\nonumber
\\
& + k_a f_w^2 ( \bar{\textbf{a}}_w^* \times \bar{\textbf{w}}_w + \bar{\textbf{a}}_8^* \circ \bar{\textbf{w}}_w + \bar{\textbf{a}}_w^* \circ \bar{\textbf{w}}_8)
+ k_a f_s^2 ( \bar{\textbf{a}}_s^* \times \bar{\textbf{w}}_s + \bar{\textbf{a}}_{12}^* \circ \bar{\textbf{w}}_s + \bar{\textbf{a}}_s^* \circ \bar{\textbf{w}}_{12})
\nonumber
\\
& + k_x ( \bar{\textbf{x}}_g^* \times \bar{\textbf{w}}_g + \bar{\textbf{x}}_g^* \bar{w}_0 + \bar{x}_0 \bar{\textbf{w}}_g ) + k_x f_e^2 ( \bar{\textbf{x}}_e^* \times \bar{\textbf{w}}_e + \bar{\textbf{x}}_4^* \circ \bar{\textbf{w}}_e + \bar{\textbf{x}}_e^* \circ \bar{\textbf{w}}_4)
\nonumber
\\
& + k_x f_w^2 ( \bar{\textbf{x}}_w^* \times \bar{\textbf{w}}_w + \bar{\textbf{x}}_8^* \circ \bar{\textbf{w}}_w + \bar{\textbf{x}}_w^* \circ \bar{\textbf{w}}_8)
+ k_x f_s^2 ( \bar{\textbf{x}}_s^* \times \bar{\textbf{w}}_s + \bar{\textbf{x}}_{12}^* \circ \bar{\textbf{w}}_s + \bar{\textbf{x}}_s^* \circ \bar{\textbf{w}}_{12})
\nonumber
\\
& + k_s ( \bar{\textbf{s}}_g^* \times \bar{\textbf{w}}_g + \bar{\textbf{s}}_g^* \bar{w}_0 + \bar{s}_0 \bar{\textbf{w}}_g ) + k_s f_e^2 ( \bar{\textbf{s}}_e^* \times \bar{\textbf{w}}_e + \bar{\textbf{s}}_4^* \circ \bar{\textbf{w}}_e + \bar{\textbf{s}}_e^* \circ \bar{\textbf{w}}_4)
\nonumber
\\
& + k_s f_w^2 ( \bar{\textbf{s}}_w^* \times \bar{\textbf{w}}_w + \bar{\textbf{s}}_8^* \circ \bar{\textbf{w}}_w + \bar{\textbf{s}}_w^* \circ \bar{\textbf{w}}_8)
+ k_s f_s^2 ( \bar{\textbf{s}}_s^* \times \bar{\textbf{w}}_s + \bar{\textbf{s}}_{12}^* \circ \bar{\textbf{w}}_s + \bar{\textbf{s}}_s^* \circ \bar{\textbf{w}}_{12})
\nonumber
\\
& + k_l ( \bar{\textbf{l}}_g^* \times \bar{\textbf{w}}_g + \bar{\textbf{l}}_g^* \bar{w}_0 + \bar{l}_0 \bar{\textbf{w}}_g ) + k_l f_e^2 ( \bar{\textbf{l}}_e^* \times \bar{\textbf{w}}_e + \bar{\textbf{l}}_4^* \circ \bar{\textbf{w}}_e + \bar{\textbf{l}}_e^* \circ \bar{\textbf{w}}_4)
\nonumber
\\
& + k_l f_w^2 ( \bar{\textbf{l}}_w^* \times \bar{\textbf{w}}_w + \bar{\textbf{l}}_8^* \circ \bar{\textbf{w}}_w + \bar{\textbf{l}}_w^* \circ \bar{\textbf{w}}_8)
+ k_l f_s^2 ( \bar{\textbf{l}}_s^* \times \bar{\textbf{w}}_s + \bar{\textbf{l}}_{12}^* \circ \bar{\textbf{w}}_s + \bar{\textbf{l}}_s^* \circ \bar{\textbf{w}}_{12})
\nonumber
\\
& + k_b ( \bar{\textbf{h}}_g^* \times \bar{\textbf{w}}_g + \bar{\textbf{h}}_g^* \bar{w}_0 ) + k_b f_e^2 ( \bar{\textbf{h}}_e^* \times \bar{\textbf{w}}_e + \bar{\textbf{h}}_e^* \circ \bar{\textbf{w}}_4)
\nonumber
\\
& + k_b f_w^2 ( \bar{\textbf{h}}_w^* \times \bar{\textbf{w}}_w + \bar{\textbf{h}}_w^* \circ \bar{\textbf{w}}_8)
+ k_b f_s^2 ( \bar{\textbf{h}}_s^* \times \bar{\textbf{w}}_s + \bar{\textbf{h}}_s^* \circ \bar{\textbf{w}}_{12})
~,
\end{align}
where the force density $\bar{\textbf{f}}$ includes that of the inertial force, the gravitational force, the gradient of energy, Lorentz force, and the interacting force between the dipole moment with the fields, as well as the weak nuclear force and the strong nuclear force with their adjoint fields etc. The above force is much longer than that in the classical field theory, and includes more new force terms related to the gradient of energy, the field potential, and the angular velocity etc in the four fields with their related adjoint fields.

\begin{table}[h]
\caption{Some physical quantities of the four fields with their adjoint fields in the sedenion space.}
\label{tab:table3}
\centering
\begin{tabular}{ll}
\hline\hline
$ definitions $                                                                                                                            & $ meanings $ \\
\hline
$\mathbb{\bar{X}}$                                                                                                                         & field quantity \\
$\mathbb{\bar{A}} = (\lozenge_{16} + k_x \mathbb{\bar{X}}) \circ \mathbb{\bar{X}}$                                                         & field potential \\
$\mathbb{\bar{B}} = (\lozenge_{16} + k_x \mathbb{\bar{X}} + k_a \mathbb{\bar{A}}) \circ \mathbb{\bar{A}}$                                  & field strength \\
$\mathbb{\bar{R}}$                                                                                                                         & radius vector \\
$\mathbb{\bar{V}} = v_0 \lozenge_{16} \circ \mathbb{\bar{R}}$                                                                              & velocity \\
$\mathbb{\bar{U}} = \lozenge_{16} \circ \mathbb{\bar{V}}$                                                                                  & velocity curl \\
$\mu \mathbb{\bar{S}} = - ( \lozenge_{16} + k_x \mathbb{\bar{X}} + k_a \mathbb{\bar{A}} + k_b \mathbb{\bar{B}} )^* \circ \mathbb{\bar{B}}$ & field source \\
$\mathbb{\bar{H}}_b = (k_x \mathbb{\bar{X}} + k_a \mathbb{\bar{A}} + k_b \mathbb{\bar{B}})^* \cdot \mathbb{\bar{B}}$                       & field strength helicity\\
$\mathbb{\bar{P}} = \mu \mathbb{\bar{S}} / \mu_{gg}$                                                                                       & linear momentum density \\
$\mathbb{\bar{R}} = \mathbb{R} + k_{rx} \mathbb{X}$                                                                                        & compounding radius vector \\
$\mathbb{\bar{L}} = \mathbb{\bar{R}} \circ \mathbb{\bar{P}}$                                                                               & angular momentum density \\
$\mathbb{\bar{W}} = v_0 ( \lozenge_{16} + k_x \mathbb{\bar{X}} + k_a \mathbb{\bar{A}} + k_b \mathbb{\bar{B}} + k_s \mathbb{\bar{S}} + k_l \mathbb{\bar{L}} ) \circ \mathbb{\bar{L}}$                                                                                                                          & torque-energy densities \\
$\mathbb{\bar{N}} = v_0 ( \lozenge_{16} + k_x \mathbb{\bar{X}} + k_a \mathbb{\bar{A}} + k_b \mathbb{\bar{B}} + k_s \mathbb{\bar{S}} + k_l \mathbb{\bar{L}} + k_w \mathbb{\bar{W}})^* \circ \mathbb{\bar{W}}$                                                                                                & force-power density \\
$\mathbb{\bar{F}} = - \mathbb{\bar{N}} / (2v_0)$                                                                                           & force density \\
$\mathbb{\bar{H}}_s = ( k_x \mathbb{\bar{X}} + k_a \mathbb{\bar{A}} + k_b \mathbb{\bar{B}} + k_s \mathbb{\bar{S}} + k_l \mathbb{\bar{L}})^* \cdot \mathbb{\bar{P}}$
                                                                                                                                           & field source helicity\\
\hline\hline
\end{tabular}
\end{table}

\subsection{Helicity}

In the compounding space with the sedenion operator $(\lozenge_{16} + k_x \mathbb{\bar{X}} + k_a \mathbb{\bar{A}} + k_b \mathbb{\bar{B}} + k_s \mathbb{\bar{S}} + k_l \mathbb{\bar{L}} + k_w \mathbb{\bar{W}})$, some helicity terms \cite{guo1, guo2} will impact the mass continuity equation and the charge continuity equation etc.

\subsubsection{Field strength helicity}

The sedenion linear momentum density, $\mathbb{\bar{P}} = \mu \mathbb{\bar{S}} / \mu_{gg}$, is defined from the sedenion field source $\mathbb{\bar{S}}$ in the compounding space with the operator $(\lozenge_{16} + k_x \mathbb{\bar{X}} + k_a \mathbb{\bar{A}} + k_b \mathbb{\bar{B}} + k_s \mathbb{\bar{S}} + k_l \mathbb{\bar{L}} + k_w \mathbb{\bar{W}})$,
\begin{eqnarray}
\mathbb{\bar{P}} = \mathbb{\bar{P}}_g + f_e \mathbb{\bar{P}}_e + f_w \mathbb{\bar{P}}_w + f_s \mathbb{\bar{P}}_s~,
\end{eqnarray}
where $\mathbb{\bar{P}}_g = \Sigma (\bar{p}_i \emph{\textbf{i}}_i)$, $\mathbb{\bar{P}}_e = \Sigma (\bar{p}_{(i+4)} \emph{\textbf{i}}_{(i+4)})$, $\mathbb{\bar{P}}_w = \Sigma (\bar{p}_{(i+8)} \emph{\textbf{i}}_{(i+8)})$, $\mathbb{\bar{P}}_s = \Sigma (\bar{p}_{(i+12)} \emph{\textbf{i}}_{(i+12)})$. And $\lozenge_{16} \cdot \mathbb{\bar{R}}$ is the scalar part of the $\lozenge_{16} \circ \mathbb{\bar{R}}$. $\bar{p}_0 = \widehat{m} v_0 $, $\widehat{m} = m (\bar{v}_0^\delta / \bar{v}_0) + \bigtriangleup m$, $\bar{v}_0^\delta = v_0 \lozenge_{16} \cdot \mathbb{\bar{R}}$, $\bigtriangleup m = - (k_x \mathbb{\bar{X}}^* \cdot \mathbb{\bar{B}} + k_a \mathbb{\bar{A}}^* \cdot \mathbb{\bar{B}} + k_b \mathbb{\bar{B}}^* \cdot \mathbb{\bar{B}}) / ( \bar{v}_0 \mu_{gg} )$.

According to the sedenion features, the gravitational mass density $\widehat{m}$ is one reserved scalar, and is changed with the field strength ($\mathbb{\bar{B}}_g$, $\mathbb{\bar{B}}_e$, $\mathbb{\bar{B}}_w$, and $\mathbb{\bar{B}}_s$), the potential ($\mathbb{\bar{A}}_g$, $\mathbb{\bar{A}}_e$, $\mathbb{\bar{A}}_w$, and $\mathbb{\bar{A}}_s$), the helicity $(k_x \mathbb{\bar{X}}^* \cdot \mathbb{\bar{B}} + k_a \mathbb{\bar{A}}^* \cdot \mathbb{\bar{B}} + k_b \mathbb{\bar{B}}^* \cdot \mathbb{\bar{B}})$, and the sedenion ($\mathbb{\bar{X}}_g$, $\mathbb{\bar{X}}_e$, $\mathbb{\bar{X}}_w$, and $\mathbb{\bar{X}}_s$) etc in the sedenion compounding space. The helicity includes the magnetic helicity $ \bar{\textbf{a}}_e \cdot \bar{\textbf{h}}_e $, the kinetic helicity $ \bar{\textbf{v}}_g \cdot \bar{\textbf{u}}_g $, the cross helicity $ \bar{\textbf{v}}_e \cdot \bar{\textbf{h}}_e $, and the new helicity term $ \bar{\textbf{a}}_e \cdot \bar{\textbf{u}}_e $ etc.

Similarly to the physical quantities $\mathbb{\bar{S}}$ and $\mathbb{\bar{L}}$, the sedenion angular momentum $\mathbb{\bar{W}}$ has not an influence on the field strength helicity in the compounding space with the sedenion operator $(\lozenge_{16} + k_x \mathbb{\bar{X}} + k_a \mathbb{\bar{A}} + k_b \mathbb{\bar{B}} + k_s \mathbb{\bar{S}} + k_l \mathbb{\bar{L}} + k_w \mathbb{\bar{W}})$. Therefore we can not distinguish this field from some fields with other sedenion operators, according to the viewpoint of the field strength helicity.

\subsubsection{Field source helicity}

The part force density $\mathbb{\bar{F}}$ is one part of the sedenion force-power density $\mathbb{\bar{N}}$, and is defined from the sedenion linear momentum density $\mathbb{\bar{P}}$ ,
\begin{eqnarray}
\mathbb{\bar{F}} = v_0 (\lozenge_{16} + k_x \mathbb{\bar{X}} + k_a \mathbb{\bar{A}} + k_b \mathbb{\bar{B}} + k_s \mathbb{\bar{S}} + k_l \mathbb{\bar{L}} + k_w \mathbb{\bar{W}})^* \circ \mathbb{\bar{P}}~,
\end{eqnarray}
where the part force density includes that of the inertial force, gravitational force, Lorentz force, and the interacting force between the fields with the dipoles etc.

The scalar $\bar{f}_0$ of the part force density $\mathbb{\bar{F}}$ is written as,
\begin{align}
\bar{f}_0 / v_0 = & ~ k_a \left\{ \bar{a}_0 \bar{p}_0 + \bar{\textbf{a}}_g^* \cdot \bar{\textbf{p}}_g + f_e^2 ( \bar{\textbf{a}}_4^* \circ \bar{\textbf{p}}_4 + \bar{\textbf{a}}_e^* \cdot \bar{\textbf{p}}_e )  \right\}
\nonumber\\
&
+ k_a \left\{ f_w^2 ( \bar{\textbf{a}}_8^* \circ \bar{\textbf{p}}_8 + \bar{\textbf{a}}_w^* \cdot \bar{\textbf{p}}_w )  + f_s^2 ( \bar{\textbf{a}}_{12}^* \circ \bar{\textbf{p}}_{12} + \bar{\textbf{a}}_s^* \cdot \bar{\textbf{p}}_s) \right\}
\nonumber\\
&
+ k_x \left\{ \bar{x}_0 \bar{p}_0 + \bar{\textbf{x}}_g^* \cdot \bar{\textbf{p}}_g + f_e^2 ( \bar{\textbf{x}}_4^* \circ \bar{\textbf{p}}_4 + \bar{\textbf{x}}_e^* \cdot \bar{\textbf{p}}_e )  \right\}
\nonumber\\
&
+ k_x \left\{ f_w^2 ( \bar{\textbf{x}}_8^* \circ \bar{\textbf{p}}_8 + \bar{\textbf{x}}_w^* \cdot \bar{\textbf{p}}_w ) + f_s^2 ( \bar{\textbf{x}}_{12}^* \circ \bar{\textbf{p}}_{12} + \bar{\textbf{x}}_s^* \cdot \bar{\textbf{p}}_s) \right\}
\nonumber\\
&
+ k_s \left\{ \bar{s}_0 \bar{p}_0 + \bar{\textbf{s}}_g^* \cdot \bar{\textbf{p}}_g + f_e^2 ( \bar{\textbf{s}}_4^* \circ \bar{\textbf{p}}_4 + \bar{\textbf{s}}_e^* \cdot \bar{\textbf{p}}_e ) \right\}
\nonumber\\
&
+ k_s \left\{ f_w^2 ( \bar{\textbf{s}}_8^* \circ \bar{\textbf{p}}_8 + \bar{\textbf{s}}_w^* \cdot \bar{\textbf{p}}_w ) + f_s^2 ( \bar{\textbf{s}}_{12}^* \circ \bar{\textbf{p}}_{12} + \bar{\textbf{s}}_s^* \cdot \bar{\textbf{p}}_s) \right\}
\nonumber\\
&
+ k_l \left\{ \bar{l}_0 \bar{p}_0 + \bar{\textbf{l}}_g^* \cdot \bar{\textbf{p}}_g + f_e^2 ( \bar{\textbf{l}}_4^* \circ \bar{\textbf{p}}_4 + \bar{\textbf{l}}_e^* \cdot \bar{\textbf{p}}_e )  f_w^2 ( \bar{\textbf{l}}_8^* \circ \bar{\textbf{p}}_8 + \bar{\textbf{l}}_w^* \cdot \bar{\textbf{p}}_w ) \right\}
\nonumber\\
&
+ k_l \left\{  f_w^2 ( \bar{\textbf{l}}_8^* \circ \bar{\textbf{p}}_8 + \bar{\textbf{l}}_w^* \cdot \bar{\textbf{p}}_w )
+ f_s^2 ( \bar{\textbf{l}}_{12}^* \circ \bar{\textbf{p}}_{12} + \bar{\textbf{l}}_s^* \cdot \bar{\textbf{p}}_s) \right\}
\nonumber\\
&
+ k_w \left\{ \bar{w}_0 \bar{p}_0 + \bar{\textbf{w}}_g^* \cdot \bar{\textbf{p}}_g + f_e^2 ( \bar{\textbf{w}}_4^* \circ \bar{\textbf{p}}_4 + \bar{\textbf{w}}_e^* \cdot \bar{\textbf{p}}_e ) \right\}
\nonumber\\
&
+ k_w \left\{ f_w^2 ( \bar{\textbf{w}}_8^* \circ \bar{\textbf{p}}_8 + \bar{\textbf{w}}_w^* \cdot \bar{\textbf{p}}_w ) + f_s^2 ( \bar{\textbf{w}}_{12}^* \circ \bar{\textbf{p}}_{12} + \bar{\textbf{w}}_s^* \cdot \bar{\textbf{p}}_s) \right\}
\nonumber\\
&
+ k_b ( \bar{\textbf{h}}_g^* \cdot \bar{\textbf{p}}_g + f_e^2 \bar{\textbf{h}}_e^* \cdot \bar{\textbf{p}}_e  + f_w^2 \bar{\textbf{h}}_w^* \cdot \bar{\textbf{p}}_w + f_s^2 \bar{\textbf{h}}_s^* \cdot \bar{\textbf{p}}_s )
\nonumber\\
&
+ \partial_0 \bar{p}_0 + \nabla_g^* \cdot \bar{\textbf{p}}_g + d_e f_e (\emph{\textbf{i}}_4^* \circ \partial_4 \bar{\textbf{p}}_4 + \nabla_e^* \cdot \bar{\textbf{p}}_e)
\nonumber\\
&
+ d_w f_w (\emph{\textbf{i}}_8^* \circ \partial_8 \bar{\textbf{p}}_8 + \nabla_w^* \cdot \bar{\textbf{p}}_w) + d_s f_s (\emph{\textbf{i}}_{12}^* \circ \partial_{12} \bar{\textbf{p}}_{12} + \nabla_s^* \cdot \bar{\textbf{p}}_s)
~,
\end{align}
where the field source helicity in the fields with the sedenion operator $(\lozenge_{16} + k_x \mathbb{\bar{X}} + k_a \mathbb{\bar{A}} + k_b \mathbb{\bar{B}} + k_s \mathbb{\bar{S}} + k_l \mathbb{\bar{L}} + k_w \mathbb{\bar{W}})$ covers the helicity terms $( \bar{\textbf{a}}_g^* \cdot \bar{\textbf{p}}_g + \bar{a}_0 \bar{p}_0 )$, $( \bar{\textbf{a}}_e^* \cdot \bar{\textbf{p}}_e + \bar{\textbf{a}}_4^* \circ \bar{\textbf{p}}_4 )$, $( \bar{\textbf{h}}_g^* \cdot \bar{\textbf{p}}_g + f_e^2 \bar{\textbf{h}}_e^* \cdot \bar{\textbf{p}}_e )$, $(\bar{\textbf{x}}_g^* \cdot \bar{\textbf{p}}_g + \bar{x}_0 \bar{p}_0)$, $(\bar{\textbf{x}}_e^* \cdot \bar{\textbf{p}}_e + \bar{\textbf{x}}_4^* \circ \bar{\textbf{p}}_4)$, $( \bar{\textbf{s}}_g^* \cdot \bar{\textbf{p}}_g + \bar{s}_0 \bar{p}_0 )$, $( \bar{\textbf{s}}_e^* \cdot \bar{\textbf{p}}_e + \bar{\textbf{s}}_4^* \circ \bar{\textbf{p}}_4 )$, $( \bar{\textbf{l}}_g^* \cdot \bar{\textbf{p}}_g + \bar{l}_0 \bar{p}_0 )$, $( \bar{\textbf{l}}_e^* \cdot \bar{\textbf{p}}_e + \bar{\textbf{l}}_4^* \circ \bar{\textbf{p}}_4 )$, $( \bar{\textbf{w}}_g^* \cdot \bar{\textbf{p}}_g + \bar{w}_0 \bar{p}_0 )$, and $( \bar{\textbf{w}}_e^* \cdot \bar{\textbf{p}}_e + \bar{\textbf{w}}_4^* \circ \bar{\textbf{p}}_4 )$ etc. And that they include the magnetic helicity $ \textbf{a}_e \cdot \textbf{h}_e $, the kinetic helicity $ \textbf{v}_e \cdot \textbf{u}_e $, the cross helicity $ \textbf{v}_e \cdot \textbf{h}_e $, the current helicity $ \textbf{h}_e^* \cdot \textbf{p}_e $, and some other new helicity terms etc. $\bar{\textbf{p}}_g = \Sigma ( \emph{\textbf{i}}_j \bar{p}_j)$, $\bar{\textbf{p}}_e = \Sigma ( \emph{\textbf{i}}_{(j+4)} \bar{p}_{(j+4)})$, $\bar{\textbf{p}}_w = \Sigma ( \emph{\textbf{i}}_{(j+8)} \bar{p}_{(j+8)})$, $\bar{\textbf{p}}_s = \Sigma ( \emph{\textbf{i}}_{(j+12)} \bar{p}_{(j+12)})$; $\bar{\textbf{p}}_4 = \emph{\textbf{i}}_4 \bar{p}_4$, $\bar{\textbf{p}}_8 = \emph{\textbf{i}}_8 \bar{p}_8$, $\bar{\textbf{p}}_{12} = \emph{\textbf{i}}_{12} \bar{p}_{12}$ .

The above is the mass continuity equation in the four fields with the operator $(\lozenge_{16} + k_x \mathbb{\bar{X}} + k_a \mathbb{\bar{A}} + k_b \mathbb{\bar{B}} + k_s \mathbb{\bar{S}} + k_l \mathbb{\bar{L}} + k_w \mathbb{\bar{W}})$, and is effected by the speed of light $\bar{v}_0$ and some helicities of the rotational objects and of the spinning charged objects.

A new physical quantity $\mathbb{\bar{F}}_q$ can be defined from the part force density $\mathbb{\bar{F}}$,
\begin{eqnarray}
\mathbb{\bar{F}}_q = \mathbb{\bar{F}} \circ \emph{\textbf{i}}_4^*~.
\end{eqnarray}

The scalar part $\bar{F}_0$ of the $\mathbb{\bar{F}}_q$ is written as,
\begin{align}
\bar{F}_0 / v_0 = & ~
f_e (\partial_0 \bar{\textbf{p}}_4 + \nabla_g^* \cdot \bar{\textbf{p}}_e) \circ \emph{\textbf{i}}_4^*
+ d_e (\emph{\textbf{i}}_4^* \circ \partial_4 \bar{p}_0 + \nabla_e^* \cdot \bar{\textbf{p}}_g) \circ \emph{\textbf{i}}_4^*
\nonumber
\\
&
+ d_w f_s (\emph{\textbf{i}}_8^* \circ \partial_8 \bar{\textbf{p}}_{12} + \nabla_w^* \cdot \bar{\textbf{p}}_s) \circ \emph{\textbf{i}}_4^*
+ d_s f_w (\emph{\textbf{i}}_{12}^* \circ \partial_{12} \bar{\textbf{p}}_8 + \nabla_s^* \cdot \bar{\textbf{p}}_w) \circ \emph{\textbf{i}}_4^*
\nonumber
\\
&
+ k_b f_e ( \bar{\textbf{h}}_g^* \cdot \bar{\textbf{p}}_e + \bar{\textbf{h}}_e^* \cdot \bar{\textbf{p}}_g ) \circ \emph{\textbf{i}}_4^*
+ k_b f_w f_s ( \bar{\textbf{h}}_w^* \cdot \bar{\textbf{p}}_s + \bar{\textbf{h}}_s^* \cdot \bar{\textbf{p}}_w ) \circ \emph{\textbf{i}}_4^*
\nonumber
\\
&
+ k_a f_e ( \bar{\textbf{a}}_g^* \cdot \bar{\textbf{p}}_e + \bar{a}_0 \bar{\textbf{p}}_4 ) \circ \emph{\textbf{i}}_4^* + k_a f_e ( \bar{\textbf{a}}_e^* \cdot \bar{\textbf{p}}_g + \bar{p}_0 \bar{\textbf{a}}_4^* ) \circ \emph{\textbf{i}}_4^*
\nonumber
\\
&
+ k_a f_s f_w ( \bar{\textbf{a}}_w^* \cdot \bar{\textbf{p}}_s + \bar{\textbf{a}}_8^* \cdot \bar{\textbf{p}}_{12} ) \circ \emph{\textbf{i}}_4^* + k_a f_s f_w ( \bar{\textbf{a}}_s^* \cdot \bar{\textbf{p}}_w + \bar{\textbf{a}}_{12}^* \cdot \bar{\textbf{p}}_8 ) \circ \emph{\textbf{i}}_4^*
\nonumber
\\
&
+ k_x f_e ( \bar{\textbf{x}}_g^* \cdot \bar{\textbf{p}}_e + \bar{x}_0 \bar{\textbf{p}}_4 ) \circ \emph{\textbf{i}}_4^* + k_x f_e ( \bar{\textbf{x}}_e^* \cdot \bar{\textbf{p}}_g + \bar{p}_0 \bar{\textbf{x}}_4^* ) \circ \emph{\textbf{i}}_4^*
\nonumber
\\
&
+ k_x f_s f_w ( \bar{\textbf{x}}_w^* \cdot \bar{\textbf{p}}_s + \bar{\textbf{x}}_8^* \cdot \bar{\textbf{p}}_{12} ) \circ \emph{\textbf{i}}_4^* + k_x f_s f_w ( \bar{\textbf{x}}_s^* \cdot \bar{\textbf{p}}_w + \bar{\textbf{x}}_{12}^* \cdot \bar{\textbf{p}}_8 ) \circ \emph{\textbf{i}}_4^*
\nonumber
\\
&
+ k_s f_e ( \bar{\textbf{s}}_g^* \cdot \bar{\textbf{p}}_e + \bar{s}_0 \bar{\textbf{p}}_4 ) \circ \emph{\textbf{i}}_4^* + k_s f_e ( \bar{\textbf{s}}_e^* \cdot \bar{\textbf{p}}_g + \bar{p}_0 \bar{\textbf{s}}_4^* ) \circ \emph{\textbf{i}}_4^*
\nonumber
\\
&
+ k_s f_s f_w ( \bar{\textbf{s}}_w^* \cdot \bar{\textbf{p}}_s + \bar{\textbf{s}}_8^* \cdot \bar{\textbf{p}}_{12} ) \circ \emph{\textbf{i}}_4^* + k_s f_s f_w ( \bar{\textbf{s}}_s^* \cdot \bar{\textbf{p}}_w + \bar{\textbf{s}}_{12}^* \cdot \bar{\textbf{p}}_8 ) \circ \emph{\textbf{i}}_4^*
\nonumber
\\
&
+ k_l f_e ( \bar{\textbf{l}}_g^* \cdot \bar{\textbf{p}}_e + \bar{l}_0 \bar{\textbf{p}}_4 ) \circ \emph{\textbf{i}}_4^* + k_l f_e ( \bar{\textbf{l}}_e^* \cdot \bar{\textbf{p}}_g + \bar{p}_0 \bar{\textbf{l}}_4^* ) \circ \emph{\textbf{i}}_4^*
\nonumber
\\
&
+ k_l f_s f_w ( \bar{\textbf{l}}_w^* \cdot \bar{\textbf{p}}_s + \bar{\textbf{l}}_8^* \cdot \bar{\textbf{p}}_{12} ) \circ \emph{\textbf{i}}_4^* + k_l f_s f_w ( \bar{\textbf{l}}_s^* \cdot \bar{\textbf{p}}_w + \bar{\textbf{l}}_{12}^* \cdot \bar{\textbf{p}}_8 ) \circ \emph{\textbf{i}}_4^*
\nonumber
\\
&
+ k_w f_e ( \bar{\textbf{w}}_g^* \cdot \bar{\textbf{p}}_e + \bar{w}_0 \bar{\textbf{p}}_4 ) \circ \emph{\textbf{i}}_4^* + k_w f_e ( \bar{\textbf{w}}_e^* \cdot \bar{\textbf{p}}_g + \bar{p}_0 \bar{\textbf{w}}_4^* ) \circ \emph{\textbf{i}}_4^*
\nonumber
\\
&
+ k_w f_s f_w ( \bar{\textbf{w}}_w^* \cdot \bar{\textbf{p}}_s + \bar{\textbf{w}}_8^* \cdot \bar{\textbf{p}}_{12} ) \circ \emph{\textbf{i}}_4^* + k_w f_s f_w ( \bar{\textbf{w}}_s^* \cdot \bar{\textbf{p}}_w + \bar{\textbf{w}}_{12}^* \cdot \bar{\textbf{p}}_8 ) \circ \emph{\textbf{i}}_4^*
~,
\end{align}
where the above helicity terms include that in the gravitational field, the electromagnetic field, the weak nuclear field, and the strong nuclear field with their adjoint fields.

In the sedenion compounding space with the sedenion operator $(\lozenge_{16} + k_x \mathbb{\bar{X}} + k_a \mathbb{\bar{A}} + k_b \mathbb{\bar{B}} + k_s \mathbb{\bar{S}} + k_l \mathbb{\bar{L}} + k_w \mathbb{\bar{W}} )$, the above is the charge continuity equation in the presence of the gravitational field, the electromagnetic field, the weak nuclear field, and the strong nuclear field with their adjoint fields when the scalar part $F_0 = 0$. Similarly to the case with the octonion operator $(\lozenge_8 + k_x \mathbb{\bar{X}} + k_a \mathbb{\bar{A}} + k_b \mathbb{\bar{B}} + k_s \mathbb{\bar{S}} + k_l \mathbb{\bar{L}} + k_w \mathbb{\bar{W}} )$, this charge continuity equation is the invariant under the sedenion coordinate transformation, meanwhile the $\mathbb{\bar{X}}$, $\mathbb{\bar{A}}$, $\mathbb{\bar{B}}$, $\mathbb{\bar{S}}$, $\mathbb{\bar{L}}$, and $\mathbb{\bar{W}}$ of the above four fields have the influence on the charge continuity equation, although the impacts are usually paltry when the fields are puny.

In the same way, the above physical quantities will impact the continuity equation of the weak nuclear current in the definition $(\mathbb{\bar{F}} \circ \emph{\textbf{i}}_8^*)$, and the continuity equation of the strong nuclear current in the definition $(\mathbb{\bar{F}} \circ \emph{\textbf{i}}_{12}^*)$.

%--13--%

\section{The fields regarding the trigintaduonion operator $\lozenge_{32}$}

In the electromagnetic field, the gravitational field, the weak nuclear field, and the strong nuclear field, the sedenion operator ($\lozenge_{16} + k_x \mathbb{\bar{X}} + k_a \mathbb{\bar{A}} + k_b \mathbb{\bar{B}} + k_s \mathbb{\bar{S}} + k_l \mathbb{\bar{L}} + k_w \mathbb{\bar{W}}$) can deduce the sedenion physical properties of four fields with their related adjoint fields, including the sedenion linear momentum, the sedenion angular momentum, the energy, the sedenion torque, the power, the sedenion force, and some helicities of the rotational objects and the spinning charged objects etc. But there may exist other kinds of unknown fundamental fields in the nature theoretically, and then the above consequences can not cover the helicities of the these unknown fundamental fields with their related adjoint fields etc. In this section, the operator will substitute the trigintaduonion operator $\lozenge_{32}$ for the sedenion operator $\lozenge_{16}$ to encompass the physical properties of the above unknown fields simultaneously, besides that of the electromagnetic field, the gravitational field, the strong nuclear field, and the weak nuclear field.

\subsection{Trigintaduonion space}

The trigintaduonion space \cite{cawagas2} can encompass eight kinds of the independent and perpendicular quaternion spaces. Besides the four quaternion spaces for the gravitational field, the electromagnetic field, the weak nuclear field, and the strong nuclear field, we imagine that there exist other four kinds of quaternion spaces for the unknown fundamental fields. Those four unknown fields are called as the $\alpha$ field, the $\beta$ field, the $\gamma$ field, and the $\delta$ field.

In the quaternion space for the $\alpha$ field, the basis vector is $\mathbb{E}_\alpha$ = ($\emph{\textbf{i}}_{16}$,  $\emph{\textbf{i}}_{17}$, $\emph{\textbf{i}}_{18}$, $\emph{\textbf{i}}_{19}$), the radius vector is $\mathbb{R}_\alpha$ = ($r_{16}$, $r_{17}$, $r_{18}$, $r_{19}$), the velocity is $\mathbb{V}_\alpha$ = ($v_{16}$, $v_{17}$, $v_{18}$, $v_{19}$), and the $\alpha$ field potential is $\mathbb{A}_\alpha$ = ($a_{16}$, $a_{17}$, $a_{18}$, $a_{19}$), with the physical quantity $\mathbb{X}_\alpha$ = ($x_{16}$, $x_{17}$, $x_{18}$, $x_{19}$). In the quaternion space for the $\beta$ field, the basis vector is $\mathbb{E}_\beta$ = ($\emph{\textbf{i}}_{20}$, $\emph{\textbf{i}}_{21}$, $\emph{\textbf{i}}_{22}$, $\emph{\textbf{i}}_{23}$), the radius vector is $\mathbb{R}_\beta$ = ($r_{20}$, $r_{21}$, $r_{22}$, $r_{23}$), the velocity is $\mathbb{V}_\beta$ = ($v_{20}$, $v_{21}$, $v_{22}$, $v_{23}$), and the $\beta$ field potential is $\mathbb{A}_\beta$ = ($a_{20}$, $a_{21}$, $a_{22}$, $a_{23}$), with the physical quantity $\mathbb{X}_\beta$ = ($x_{20}$, $x_{21}$, $x_{22}$, $x_{23}$). In the quaternion space for the $\gamma$ field, the basis vector is $\mathbb{E}_\gamma$ = ($\emph{\textbf{i}}_{24}$,  $\emph{\textbf{i}}_{25}$, $\emph{\textbf{i}}_{26}$, $\emph{\textbf{i}}_{27}$), the radius vector is $\mathbb{R}_\gamma$ = ($r_{24}$, $r_{25}$, $r_{26}$, $r_{27}$), the velocity is $\mathbb{V}_\gamma$ = ($v_{24}$, $v_{25}$, $v_{26}$, $v_{27}$), and the $\gamma$ field potential is $\mathbb{A}_\gamma$ = ($a_{24}$, $a_{25}$, $a_{26}$, $a_{27}$), with the physical quantity $\mathbb{X}_\gamma$ = ($x_{24}$, $x_{25}$, $x_{26}$, $x_{27}$). In the quaternion space for the $\delta$ field, the basis vector is $\mathbb{E}_\delta$ = ($\emph{\textbf{i}}_{28}$, $\emph{\textbf{i}}_{29}$, $\emph{\textbf{i}}_{30}$, $\emph{\textbf{i}}_{31}$), the radius vector is $\mathbb{R}_\delta$ = ($r_{28}$, $r_{29}$, $r_{30}$, $r_{31}$), and the velocity is $\mathbb{V}_\delta$ = ($v_{28}$, $v_{29}$, $v_{30}$, $v_{31}$), and the $\delta$ field potential is $\mathbb{A}_\delta$ = ($a_{28}$, $a_{29}$, $a_{30}$, $a_{31}$), with the physical quantity $\mathbb{X}_\delta$ = ($x_{28}$, $x_{29}$, $x_{30}$, $x_{31}$).

The eight quaternion basis vectors, $\mathbb{E}_e$, $\mathbb{E}_g$, $\mathbb{E}_w$, $\mathbb{E}_s$, $\mathbb{E}_\alpha$, $\mathbb{E}_\beta$, $\mathbb{E}_\gamma$, and $\mathbb{E}_\delta$, are independent to each other, and that they can combine together to become the basis vector of the trigintaduonion space, $\mathbb{E}_{32}$ = ($\emph{\textbf{i}}_0$, $\emph{\textbf{i}}_1$, $\emph{\textbf{i}}_2$, $\emph{\textbf{i}}_3$, $\emph{\textbf{i}}_4$, $\emph{\textbf{i}}_5$, $\emph{\textbf{i}}_6$, $\emph{\textbf{i}}_7$, $\emph{\textbf{i}}_8$, $\emph{\textbf{i}}_9$, $\emph{\textbf{i}}_{10}$, $\emph{\textbf{i}}_{11}$, $\emph{\textbf{i}}_{12}$, $\emph{\textbf{i}}_{13}$, $\emph{\textbf{i}}_{14}$, $\emph{\textbf{i}}_{15}$, $\emph{\textbf{i}}_{16}$, $\emph{\textbf{i}}_{17}$, $\emph{\textbf{i}}_{18}$, $\emph{\textbf{i}}_{19}$, $\emph{\textbf{i}}_{20}$, $\emph{\textbf{i}}_{21}$, $\emph{\textbf{i}}_{22}$, $\emph{\textbf{i}}_{23}$, $\emph{\textbf{i}}_{24}$, $\emph{\textbf{i}}_{25}$, $\emph{\textbf{i}}_{26}$, $\emph{\textbf{i}}_{27}$, $\emph{\textbf{i}}_{28}$, $\emph{\textbf{i}}_{29}$, $\emph{\textbf{i}}_{30}$, $\emph{\textbf{i}}_{31}$). The trigintaduonion radius vector is $\mathbb{R}$ = $\mathbb{R}_g$ + $f_e \mathbb{R}_e$ + $f_w \mathbb{R}_w$ + $f_s \mathbb{R}_s$ + $f_\alpha \mathbb{R}_\alpha$ + $f_\beta \mathbb{R}_\beta$ + $f_\gamma \mathbb{R}_\gamma$ + $f_\delta \mathbb{R}_\delta$, the trigintaduonion velocity is $\mathbb{V}$ = $\mathbb{V}_g$ + $f_e \mathbb{V}_e$ + $f_w \mathbb{V}_w$ + $f_s \mathbb{V}_s$ + $f_\alpha \mathbb{V}_\alpha$ + $f_\beta \mathbb{V}_\beta$ + $f_\gamma \mathbb{V}_\gamma$ + $f_\delta \mathbb{V}_\delta$, and the trigintaduonion field potential is $\mathbb{A}$ = $\mathbb{A}_g$ + $f_e \mathbb{A}_e$ + $f_w \mathbb{A}_w$ + $f_s \mathbb{A}_s$ + $f_\alpha \mathbb{A}_\alpha$ + $f_\beta \mathbb{A}_\beta$ + $f_\gamma \mathbb{A}_\gamma$ + $f_\delta \mathbb{A}_\delta$, with the trigintaduonion physical quantity $\mathbb{X}$ = $\mathbb{X}_g$ + $f_e \mathbb{X}_e$ + $f_w \mathbb{X}_w$ + $f_s \mathbb{X}_s$ + $f_\alpha \mathbb{X}_\alpha$ + $f_\beta \mathbb{X}_\beta$ + $f_\gamma \mathbb{X}_\gamma$ + $f_\delta \mathbb{X}_\delta$. Herein $f_\alpha$, $f_\beta$, $f_\gamma$, and $f_\delta$ are the coefficients for the dimensional homogeneity. $\emph{\textbf{i}}_0 = 1$. $i = 0, 1, 2, 3$.

The trigintaduonion operator possesses eight kinds of independent and perpendicular quaternion operators in the trigintaduonion space. The $\alpha$ field operator is $\lozenge_\alpha = \Sigma ( \emph{\textbf{i}}_{(i+16)} \partial_{(i+16)})$, the $\beta$ field operator is $\lozenge_\beta = \Sigma ( \emph{\textbf{i}}_{(i+20)} \partial_{(i+20)})$, the $\gamma$ field operator is $\lozenge_\gamma = \Sigma ( \emph{\textbf{i}}_{(i+24)} \partial_{(i+24)})$, and the $\delta$ field operator is $\lozenge_\delta = \Sigma ( \emph{\textbf{i}}_{(i+28)} \partial_{(i+28)})$. Those eight operators can constitute the trigintaduonion operator $\lozenge_{32}$ = $\lozenge_g$ + $d_e \lozenge_e$ + $d_w \lozenge_w$ + $d_s \lozenge_s$ + $d_\alpha \lozenge_\alpha$ + $d_\beta \lozenge_\beta$ + $d_\gamma \lozenge_\gamma$ + $d_\delta \lozenge_\delta$ . Herein $d_\alpha$, $d_\beta$, $d_\gamma$, and $d_\delta$ are the coefficients for the dimensional homogeneity.

The trigintaduonion field strength is $\mathbb{B}$ = $\mathbb{B}_g$ + $f_e \mathbb{B}_e$ + $f_w \mathbb{B}_w$ + $f_s \mathbb{B}_s$ + $f_\alpha \mathbb{B}_\alpha$ + $f_\beta \mathbb{B}_\beta$ + $f_\gamma \mathbb{B}_\gamma$ + $f_\delta \mathbb{B}_\delta$, and it is defined from the trigintaduonion field potential $\mathbb{A}$ = $\mathbb{A}_g$ + $f_e \mathbb{A}_e$ + $f_w \mathbb{A}_w$ + $f_s \mathbb{A}_s$ + $f_\alpha \mathbb{A}_\alpha$ + $f_\beta \mathbb{A}_\beta$ + $f_\gamma \mathbb{A}_\gamma$ + $f_\delta \mathbb{A}_\delta$. And the trigintaduonion field strength $\mathbb{B}$ includes the $\alpha$ field strength $\lozenge_\alpha = \Sigma ( \emph{\textbf{i}}_{(i+16)} h_{(i+16)})$, the $\beta$ field strength $\lozenge_\beta = \Sigma ( \emph{\textbf{i}}_{(i+20)} h_{(i+20)})$, the $\gamma$ field strength $\lozenge_\gamma = \Sigma ( \emph{\textbf{i}}_{(i+24)} h_{(i+24)})$, and the $\delta$ field strength $\lozenge_\delta = \Sigma ( \emph{\textbf{i}}_{(i+28)} h_{(i+28)})$, besides the gravitational strength $\mathbb{B}_g$, the electromagnetic strength $\mathbb{B}_e$, the weak nuclear strength $\mathbb{B}_w$, and the strong nuclear strength $\mathbb{B}_s$. The eight gauge equations are $h_0 = 0$, $h_4 = 0$, $h_8 = 0$, $h_{12} = 0$, $h_{16} = 0$, $h_{20} = 0$, $h_{24} = 0$, and $h_{28} = 0$ respectively in the trigintaduonion space.

The radius vector $\mathbb{R}$ and the trigintaduonion quantity $\mathbb{X}$ can be combined together to become the compounding radius vector $\mathbb{\bar{R}} = \mathbb{R} + k_{rx} \mathbb{X}$, and the compounding quantity $\mathbb{\bar{X}} = \mathbb{X} + K_{rx} \mathbb{R}$. The related space is called as the trigintaduonion compounding space, which is one kind of function space also. In this space, the compounding field potential is $\mathbb{\bar{A}} = (\lozenge_{32} + k_x \mathbb{\bar{X}}) \circ \mathbb{\bar{X}}$, the compounding field strength is $\mathbb{\bar{B}} = (\lozenge_{32} + k_x \mathbb{\bar{X}} + k_a \mathbb{\bar{A}}) \circ \mathbb{\bar{A}}$, the compounding velocity is $\mathbb{\bar{V}} = \mathbb{V} + v_0 k_{rx} \mathbb{A}$, and the compounding velocity curl is $\mathbb{\bar{U}} = \mathbb{U} + v_0 k_{rx} \mathbb{B}$.

In terms of the trigintaduonion operator $\lozenge_{32}$ in the above eight fields, we can represent the field potential, the field strength, the field source, the linear momentum, the angular momentum, the energy, the torque, the force, and some helicities, including the influence of the adjoint fields on the torque, the force, and the helicity etc.

\subsection{Field source}

In the fields with the operator ($\lozenge_{32} + k_x \mathbb{\bar{X}} + k_a \mathbb{\bar{A}} + k_b \mathbb{\bar{B}} + k_s \mathbb{\bar{S}} + k_l \mathbb{\bar{L}} + k_w \mathbb{\bar{W}}$), the physical quantities will be expanded into the trigintaduonion quantity, including the gravitational mass density and the helicity etc. In this compounding space, the trigintaduonion basis vector is $ \mathbb{E}_{32}$, the trigintaduonion radius vector is $\mathbb{\bar{R}}$, and the trigintaduonion velocity is $\mathbb{\bar{V}}$. The trigintaduonion compounding field potential is $\mathbb{\bar{A}} = \mathbb{\bar{A}}_g + f_e \mathbb{\bar{A}}_e + f_w \mathbb{\bar{A}}_w + f_s \mathbb{\bar{A}}_s + f_\alpha \mathbb{\bar{A}}_\alpha + f_\beta \mathbb{\bar{A}}_\beta + f_\gamma \mathbb{\bar{A}}_\gamma + f_\delta \mathbb{\bar{A}}_\delta$, and is defined from the trigintaduonion quantity $\mathbb{\bar{X}}$, that is $\mathbb{\bar{A}} = (\lozenge_{32} + k_x \mathbb{\bar{X}}) \circ \mathbb{\bar{X}}$ .

The trigintaduonion compounding field potential $\mathbb{\bar{A}}$ encompasses,
\begin{align}
\mathbb{\bar{A}}_g = & ~ ( \lozenge_g \circ \mathbb{\bar{X}}_g + d_e f_e \lozenge_e \circ \mathbb{\bar{X}}_e  + d_w f_w \lozenge_w \circ \mathbb{\bar{X}}_w  + d_s f_s \lozenge_s \circ \mathbb{\bar{X}}_s )
\nonumber
\\
&
+ ( d_\alpha f_\alpha \lozenge_\alpha \circ \mathbb{\bar{X}}_\alpha + d_\beta f_\beta \lozenge_\beta \circ \mathbb{\bar{X}}_\beta  + d_\gamma f_\gamma \lozenge_\gamma \circ \mathbb{\bar{X}}_\gamma  + d_\delta f_\delta \lozenge_\delta \circ \mathbb{\bar{X}}_\delta )
\nonumber
\\
&
+ k_x ( \mathbb{\bar{X}}_g \circ \mathbb{\bar{X}}_g + f_e^2 \mathbb{\bar{X}}_e \circ \mathbb{\bar{X}}_e
+ f_w^2 \mathbb{\bar{X}}_w \circ \mathbb{\bar{X}}_w  + f_s^2 \mathbb{\bar{X}}_s \circ \mathbb{\bar{X}}_s )
\nonumber
\\
&
+ k_x ( f_\alpha^2 \mathbb{\bar{X}}_\alpha \circ \mathbb{\bar{X}}_\alpha + f_\beta^2 \mathbb{\bar{X}}_\beta \circ \mathbb{\bar{X}}_\beta
+ f_\gamma^2 \mathbb{\bar{X}}_\gamma \circ \mathbb{\bar{X}}_\gamma  + f_\delta^2 \mathbb{\bar{X}}_\delta \circ \mathbb{\bar{X}}_\delta ) ~,
\\
\mathbb{\bar{A}}_e = & ~ ( \lozenge_g \circ \mathbb{\bar{X}}_e + d_e f_e^{-1} \lozenge_e \circ \mathbb{\bar{X}}_g  + d_s f_w f_e^{-1} \lozenge_s \circ \mathbb{\bar{X}}_w  + d_w f_s f_e^{-1} \lozenge_w \circ \mathbb{\bar{X}}_s )
\nonumber
\\
&
+ f_e^{-1} ( d_\alpha f_\beta \lozenge_\alpha \circ \mathbb{\bar{X}}_\beta + d_\beta f_\alpha \lozenge_\beta \circ \mathbb{\bar{X}}_\alpha  + d_\gamma f_\delta  \lozenge_\gamma \circ \mathbb{\bar{X}}_\delta  + d_\delta f_\gamma \lozenge_\delta \circ \mathbb{\bar{X}}_\gamma )
\nonumber
\\
&
+ k_x ( \mathbb{\bar{X}}_e \circ \mathbb{\bar{X}}_g + \mathbb{\bar{X}}_g \circ \mathbb{\bar{X}}_e) + k_x f_s f_w f_e^{-1} ( \mathbb{\bar{X}}_s \circ \mathbb{\bar{X}}_w + \mathbb{\bar{X}}_w \circ \mathbb{\bar{X}}_s)
\nonumber
\\
&
+ k_x f_\alpha f_\beta f_e^{-1} ( \mathbb{\bar{X}}_\alpha \circ \mathbb{\bar{X}}_\beta + \mathbb{\bar{X}}_\beta \circ \mathbb{\bar{X}}_\alpha) + k_x f_\gamma f_\delta f_e^{-1} ( \mathbb{\bar{X}}_\gamma \circ \mathbb{\bar{X}}_\delta + \mathbb{\bar{X}}_\delta \circ \mathbb{\bar{X}}_\gamma ) ~,
\\
\mathbb{\bar{A}}_w = & ~ ( \lozenge_g \circ \mathbb{\bar{X}}_w + d_e f_s f_w^{-1} \lozenge_e \circ \mathbb{\bar{X}}_s  + d_w f_w^{-1} \lozenge_w \circ \mathbb{\bar{X}}_g + d_s f_e f_w^{-1} \lozenge_s \circ \mathbb{\bar{X}}_e)
\nonumber
\\
&
+ f_w^{-1} ( d_\alpha f_\gamma \lozenge_\alpha \circ \mathbb{\bar{X}}_\gamma + d_\beta f_\delta \lozenge_\beta \circ \mathbb{\bar{X}}_\delta  + d_\gamma f_\alpha \lozenge_\gamma \circ \mathbb{\bar{X}}_\alpha + d_\delta f_\beta \lozenge_\delta \circ \mathbb{\bar{X}}_\beta)
\nonumber
\\
&
+ k_x ( \mathbb{\bar{X}}_g \circ \mathbb{\bar{X}}_w  + \mathbb{\bar{X}}_w \circ \mathbb{\bar{X}}_g ) + k_x f_e f_s f_w^{-1} ( \mathbb{\bar{X}}_e \circ \mathbb{\bar{X}}_s + \mathbb{\bar{X}}_s \circ \mathbb{\bar{X}}_e)
\nonumber
\\
&
+ k_x f_\alpha f_\gamma f_w^{-1} ( \mathbb{\bar{X}}_\alpha \circ \mathbb{\bar{X}}_\gamma  + \mathbb{\bar{X}}_\gamma \circ \mathbb{\bar{X}}_\alpha ) + k_x f_\beta f_\delta f_w^{-1} ( \mathbb{\bar{X}}_\beta \circ \mathbb{\bar{X}}_\delta + \mathbb{\bar{X}}_\delta \circ \mathbb{\bar{X}}_\beta)
~,
\\
\mathbb{\bar{A}}_s = & ~ ( \lozenge_g \circ \mathbb{\bar{X}}_s + d_e f_w f_s^{-1} \lozenge_e \circ \mathbb{\bar{X}}_w  + d_w f_e f_s^{-1} \lozenge_w \circ \mathbb{\bar{X}}_e  + d_s f_s^{-1} \lozenge_s \circ \mathbb{\bar{X}}_g )
\nonumber
\\
&
+ f_s^{-1} ( d_\alpha f_\delta \lozenge_\alpha \circ \mathbb{\bar{X}}_\delta + d_\beta f_\gamma \lozenge_\beta \circ \mathbb{\bar{X}}_\gamma  + d_\gamma f_\beta  \lozenge_\gamma \circ \mathbb{\bar{X}}_\beta  + d_\delta f_\alpha \lozenge_\delta \circ \mathbb{\bar{X}}_\alpha )
\nonumber
\\
&
+ k_x ( \mathbb{\bar{X}}_g \circ \mathbb{\bar{X}}_s  + \mathbb{\bar{X}}_s \circ \mathbb{\bar{X}}_g ) + k_x f_e f_w f_s^{-1} ( \mathbb{\bar{X}}_e \circ \mathbb{\bar{X}}_w + \mathbb{\bar{X}}_w \circ \mathbb{\bar{X}}_e)
\nonumber
\\
&
+ k_x f_\alpha f_\delta f_s^{-1} ( \mathbb{\bar{X}}_\alpha \circ \mathbb{\bar{X}}_\delta  + \mathbb{\bar{X}}_\delta \circ \mathbb{\bar{X}}_\alpha ) + k_x f_\beta f_\gamma f_s^{-1} ( \mathbb{\bar{X}}_\beta \circ \mathbb{\bar{X}}_\gamma + \mathbb{\bar{X}}_\gamma \circ \mathbb{\bar{X}}_\beta)
~,
\\
\mathbb{\bar{A}}_\alpha = & ( \lozenge_g \circ \mathbb{\bar{X}}_\alpha + d_e f_\beta f_\alpha^{-1} \lozenge_e \circ \mathbb{\bar{X}}_\beta  + d_w f_\gamma  f_\alpha^{-1} \lozenge_w \circ \mathbb{\bar{X}}_\gamma  + d_s f_\delta  f_\alpha^{-1} \lozenge_s \circ \mathbb{\bar{X}}_\delta )
\nonumber
\\
&
+  f_\alpha^{-1} ( d_\alpha \lozenge_\alpha \circ \mathbb{\bar{X}}_g + d_\beta f_e \lozenge_\beta \circ \mathbb{\bar{X}}_e  + d_\gamma f_w \lozenge_\gamma \circ \mathbb{\bar{X}}_w  + d_\delta f_s \lozenge_\delta \circ \mathbb{\bar{X}}_s )
\nonumber
\\
&
+ k_x (  \mathbb{\bar{X}}_g \circ \mathbb{\bar{X}}_\alpha + \mathbb{\bar{X}}_\alpha \circ \mathbb{\bar{X}}_g )
+ k_x f_e f_\beta f_\alpha^{-1} ( \mathbb{\bar{X}}_e \circ \mathbb{\bar{X}}_\beta + \mathbb{\bar{X}}_\beta \circ \mathbb{\bar{X}}_e )
\nonumber
\\
&
+ k_x f_w f_\gamma f_\alpha^{-1} ( \mathbb{\bar{X}}_w \circ \mathbb{\bar{X}}_\gamma + \mathbb{\bar{X}}_\gamma \circ \mathbb{\bar{X}}_w )
+ k_x f_s f_\delta f_\alpha^{-1} ( \mathbb{\bar{X}}_s \circ \mathbb{\bar{X}}_\delta + \mathbb{\bar{X}}_\delta \circ \mathbb{\bar{X}}_s ) ~,
\\
\mathbb{\bar{A}}_\beta = & ~ ( \lozenge_g \circ \mathbb{\bar{X}}_\beta + d_e f_\alpha f_\beta^{-1} \lozenge_e \circ \mathbb{\bar{X}}_\alpha  + d_w f_\delta f_\beta^{-1} \lozenge_w \circ \mathbb{\bar{X}}_\delta  + d_s f_\gamma f_\beta^{-1} \lozenge_s \circ \mathbb{\bar{X}}_\gamma )
\nonumber
\\
&
+ f_\beta^{-1} ( d_\alpha f_e \lozenge_\alpha \circ \mathbb{\bar{X}}_e + d_\beta \lozenge_\beta \circ \mathbb{\bar{X}}_g  + d_\gamma f_s \lozenge_\gamma \circ \mathbb{\bar{X}}_s  + d_\delta f_w \lozenge_\delta \circ \mathbb{\bar{X}}_w )
\nonumber
\\
&
+ k_x ( \mathbb{\bar{X}}_\beta \circ \mathbb{\bar{X}}_g + \mathbb{\bar{X}}_g \circ \mathbb{\bar{X}}_\beta) + k_x f_e f_\alpha f_\beta^{-1} ( \mathbb{\bar{X}}_e \circ \mathbb{\bar{X}}_\alpha + \mathbb{\bar{X}}_\alpha \circ \mathbb{\bar{X}}_e )
\nonumber
\\
&
+ k_x f_w f_\delta f_\beta^{-1} ( \mathbb{\bar{X}}_w \circ \mathbb{\bar{X}}_\delta + \mathbb{\bar{X}}_\delta \circ \mathbb{\bar{X}}_w) + k_x f_\gamma f_s f_\beta^{-1} ( \mathbb{\bar{X}}_\gamma \circ \mathbb{\bar{X}}_s + \mathbb{\bar{X}}_s \circ \mathbb{\bar{X}}_\gamma ) ~,
\\
\mathbb{\bar{A}}_\gamma = & ~ ( \lozenge_g \circ \mathbb{\bar{X}}_\gamma + d_e f_\delta f_\gamma^{-1} \lozenge_e \circ \mathbb{\bar{X}}_\delta  + d_w f_\alpha f_\gamma^{-1} \lozenge_w \circ \mathbb{\bar{X}}_\alpha + d_s f_\beta f_\gamma^{-1} \lozenge_s \circ \mathbb{\bar{X}}_\beta )
\nonumber
\\
&
+ f_\gamma^{-1} ( d_\alpha f_w \lozenge_\alpha \circ \mathbb{\bar{X}}_w + d_\beta f_s \lozenge_\beta \circ \mathbb{\bar{X}}_s  + d_\gamma \lozenge_\gamma \circ \mathbb{\bar{X}}_g + d_\delta f_e \lozenge_\delta \circ \mathbb{\bar{X}}_e )
\nonumber
\\
&
+ k_x ( \mathbb{\bar{X}}_g \circ \mathbb{\bar{X}}_\gamma  + \mathbb{\bar{X}}_\gamma \circ \mathbb{\bar{X}}_g ) + k_x f_e f_\delta f_\gamma^{-1} ( \mathbb{\bar{X}}_e \circ \mathbb{\bar{X}}_\delta + \mathbb{\bar{X}}_\delta \circ \mathbb{\bar{X}}_e)
\nonumber
\\
&
+ k_x f_\alpha f_w f_\gamma^{-1} ( \mathbb{\bar{X}}_\alpha \circ \mathbb{\bar{X}}_w  + \mathbb{\bar{X}}_w \circ \mathbb{\bar{X}}_\alpha ) + k_x f_\beta f_s f_\gamma^{-1} ( \mathbb{\bar{X}}_\beta \circ \mathbb{\bar{X}}_s + \mathbb{\bar{X}}_s \circ \mathbb{\bar{X}}_\beta)
~,
\\
\mathbb{\bar{A}}_\delta = & ~ ( \lozenge_g \circ \mathbb{\bar{X}}_\delta + d_e f_\gamma f_\delta^{-1} \lozenge_e \circ \mathbb{\bar{X}}_\gamma  + d_w f_\beta f_\delta^{-1} \lozenge_w \circ \mathbb{\bar{X}}_\beta  + d_s f_\alpha f_\delta^{-1} \lozenge_s \circ \mathbb{\bar{X}}_\alpha )
\nonumber
\\
&
+ f_\delta^{-1} ( d_\alpha f_s \lozenge_\alpha \circ \mathbb{\bar{X}}_s + d_\beta f_w \lozenge_\beta \circ \mathbb{\bar{X}}_w  + d_\gamma f_e \lozenge_\gamma \circ \mathbb{\bar{X}}_e  + d_\delta \lozenge_\delta \circ \mathbb{\bar{X}}_g )
\nonumber
\\
&
+ k_x ( \mathbb{\bar{X}}_g \circ \mathbb{\bar{X}}_\delta  + \mathbb{\bar{X}}_\delta \circ \mathbb{\bar{X}}_g ) + k_x f_e f_\gamma f_\delta^{-1} ( \mathbb{\bar{X}}_e \circ \mathbb{\bar{X}}_\gamma + \mathbb{\bar{X}}_\gamma \circ \mathbb{\bar{X}}_e)
\nonumber
\\
&
+ k_x f_\alpha f_s f_\delta^{-1} ( \mathbb{\bar{X}}_\alpha \circ \mathbb{\bar{X}}_s  + \mathbb{\bar{X}}_s \circ \mathbb{\bar{X}}_\alpha ) + k_x f_\beta f_w f_\delta^{-1} ( \mathbb{\bar{X}}_\beta \circ \mathbb{\bar{X}}_w + \mathbb{\bar{X}}_w \circ \mathbb{\bar{X}}_\beta)
~,
\end{align}
where the symbol $\circ$ denotes the trigintaduonion multiplication. The trigintaduonion physical quantity $\mathbb{\bar{X}}$ includes the $\mathbb{\bar{X}}_\alpha = \Sigma ( \bar{x}_{(i+16)} \emph{\textbf{i}}_{(i+16)}) $, $\mathbb{\bar{X}}_\beta = \Sigma (\bar{x}_{(i+20)} \emph{\textbf{i}}_{(i+20)})$, $\mathbb{\bar{X}}_\gamma = \Sigma ( \bar{x}_{(i+24)} \emph{\textbf{i}}_{(i+24)}) $, and $\mathbb{\bar{X}}_\delta = \Sigma (\bar{x}_{(i+28)} \emph{\textbf{i}}_{(i+28)})$, besides the $\mathbb{\bar{X}}_g$, $\mathbb{\bar{X}}_e$, $\mathbb{\bar{X}}_w$, and $\mathbb{\bar{X}}_s$. The $\mathbb{\bar{A}}_\alpha$, $\mathbb{\bar{A}}_\beta$, $\mathbb{\bar{A}}_\gamma$, and $\mathbb{\bar{A}}_\delta$ are the $\alpha$ field potential, the $\beta$ field potential, the $\gamma$ field potential, and the $\delta$ field potential respectively.

The trigintaduonion compounding field strength is $\mathbb{\bar{B}} = \mathbb{\bar{B}}_g + f_e \mathbb{\bar{B}}_e + f_w \mathbb{\bar{B}}_w + f_s \mathbb{\bar{B}}_s + f_\alpha \mathbb{\bar{B}}_\alpha + f_\beta \mathbb{\bar{B}}_\beta + f_\gamma \mathbb{\bar{B}}_\gamma + f_\delta \mathbb{\bar{B}}_\delta$, and is defined from the trigintaduonion field potential $\mathbb{\bar{A}}$, that is $\mathbb{\bar{B}} = (\lozenge_{32} + k_x \mathbb{\bar{X}} + k_a \mathbb{\bar{A}}) \circ \mathbb{\bar{A}}$ .

The trigintaduonion compounding field strength $\mathbb{\bar{B}}$ involves,
\begin{align}
\mathbb{\bar{B}}_g = & ~ ( \lozenge_g \circ \mathbb{\bar{A}}_g + d_e f_e \lozenge_e \circ \mathbb{\bar{A}}_e  + d_w f_w \lozenge_w \circ \mathbb{\bar{A}}_w  + d_s f_s \lozenge_s \circ \mathbb{\bar{A}}_s )
\nonumber
\\
&
+ ( d_\alpha f_\alpha \lozenge_\alpha \circ \mathbb{\bar{A}}_\alpha + d_\beta f_\beta \lozenge_\beta \circ \mathbb{\bar{A}}_\beta  + d_\gamma f_\gamma \lozenge_\gamma \circ \mathbb{\bar{A}}_\gamma  + d_\delta f_\delta \lozenge_\delta \circ \mathbb{\bar{A}}_\delta )
\nonumber
\\
&
+ k_x ( \mathbb{\bar{X}}_g \circ \mathbb{\bar{A}}_g + f_e^2 \mathbb{\bar{X}}_e \circ \mathbb{\bar{A}}_e
+ f_w^2 \mathbb{\bar{X}}_w \circ \mathbb{\bar{A}}_w  + f_s^2 \mathbb{\bar{X}}_s \circ \mathbb{\bar{A}}_s )
\nonumber
\\
&
+ k_x ( f_\alpha^2 \mathbb{\bar{X}}_\alpha \circ \mathbb{\bar{A}}_\alpha + f_\beta^2 \mathbb{\bar{X}}_\beta \circ \mathbb{\bar{A}}_\beta
+ f_\gamma^2 \mathbb{\bar{X}}_\gamma \circ \mathbb{\bar{A}}_\gamma  + f_\delta^2 \mathbb{\bar{X}}_\delta \circ \mathbb{\bar{A}}_\delta )
\nonumber
\\
&
+ k_a ( \mathbb{\bar{A}}_g \circ \mathbb{\bar{A}}_g + f_e^2 \mathbb{\bar{A}}_e \circ \mathbb{\bar{A}}_e
+ f_w^2 \mathbb{\bar{A}}_w \circ \mathbb{\bar{A}}_w  + f_s^2 \mathbb{\bar{A}}_s \circ \mathbb{\bar{A}}_s )
\nonumber
\\
&
+ k_a ( f_\alpha^2 \mathbb{\bar{A}}_\alpha \circ \mathbb{\bar{A}}_\alpha + f_\beta^2 \mathbb{\bar{A}}_\beta \circ \mathbb{\bar{A}}_\beta
+ f_\gamma^2 \mathbb{\bar{A}}_\gamma \circ \mathbb{\bar{A}}_\gamma  + f_\delta^2 \mathbb{\bar{A}}_\delta \circ \mathbb{\bar{A}}_\delta ) ~,
\\
\mathbb{\bar{B}}_e = & ~ ( \lozenge_g \circ \mathbb{\bar{A}}_e + d_e f_e^{-1} \lozenge_e \circ \mathbb{\bar{A}}_g  + d_s f_w f_e^{-1} \lozenge_s \circ \mathbb{\bar{A}}_w  + d_w f_s f_e^{-1} \lozenge_w \circ \mathbb{\bar{A}}_s )
\nonumber
\\
&
+ f_e^{-1} ( d_\alpha f_\beta \lozenge_\alpha \circ \mathbb{\bar{A}}_\beta + d_\beta f_\alpha \lozenge_\beta \circ \mathbb{\bar{A}}_\alpha  + d_\gamma f_\delta \lozenge_\gamma \circ \mathbb{\bar{A}}_\delta  + d_\delta f_\gamma \lozenge_\delta \circ \mathbb{\bar{A}}_\gamma )
\nonumber
\\
&
+ k_x ( \mathbb{\bar{X}}_e \circ \mathbb{\bar{A}}_g + \mathbb{\bar{X}}_g \circ \mathbb{\bar{A}}_e) + k_x f_s f_w f_e^{-1} ( \mathbb{\bar{X}}_s \circ \mathbb{\bar{A}}_w + \mathbb{\bar{X}}_w \circ \mathbb{\bar{A}}_s)
\nonumber
\\
&
+ k_x f_\alpha f_\beta f_e^{-1} ( \mathbb{\bar{X}}_\alpha \circ \mathbb{\bar{A}}_\beta + \mathbb{\bar{X}}_\beta \circ \mathbb{\bar{A}}_\alpha) + k_x f_\gamma f_\delta f_e^{-1} ( \mathbb{\bar{X}}_\gamma \circ \mathbb{\bar{A}}_\delta + \mathbb{\bar{X}}_\delta \circ \mathbb{\bar{A}}_\gamma )
\nonumber
\\
&
+ k_a ( \mathbb{\bar{A}}_e \circ \mathbb{\bar{A}}_g + \mathbb{\bar{A}}_g \circ \mathbb{\bar{A}}_e) + k_a f_s f_w f_e^{-1} ( \mathbb{\bar{A}}_s \circ \mathbb{\bar{A}}_w + \mathbb{\bar{A}}_w \circ \mathbb{\bar{A}}_s)
\nonumber
\\
&
+ k_a f_\alpha f_\beta f_e^{-1} ( \mathbb{\bar{A}}_\alpha \circ \mathbb{\bar{A}}_\beta + \mathbb{\bar{A}}_\beta \circ \mathbb{\bar{A}}_\alpha) + k_a f_\gamma f_\delta f_e^{-1} ( \mathbb{\bar{A}}_\gamma \circ \mathbb{\bar{A}}_\delta + \mathbb{\bar{A}}_\delta \circ \mathbb{\bar{A}}_\gamma ) ~,
\\
\mathbb{\bar{B}}_w = & ~ ( \lozenge_g \circ \mathbb{\bar{A}}_w + d_e f_s f_w^{-1} \lozenge_e \circ \mathbb{\bar{A}}_s  + d_w f_w^{-1} \lozenge_w \circ \mathbb{\bar{A}}_g + d_s f_e f_w^{-1} \lozenge_s \circ \mathbb{\bar{A}}_e)
\nonumber
\\
&
+ f_w^{-1} ( d_\alpha f_\gamma \lozenge_\alpha \circ \mathbb{\bar{A}}_\gamma + d_\beta f_\delta \lozenge_\beta \circ \mathbb{\bar{A}}_\delta  + d_\gamma f_\alpha \lozenge_\gamma \circ \mathbb{\bar{A}}_\alpha + d_\delta f_\beta \lozenge_\delta \circ \mathbb{\bar{A}}_\beta)
\nonumber
\\
&
+ k_x ( \mathbb{\bar{X}}_g \circ \mathbb{\bar{A}}_w  + \mathbb{\bar{X}}_w \circ \mathbb{\bar{A}}_g ) + k_x f_e f_s f_w^{-1} ( \mathbb{\bar{X}}_e \circ \mathbb{\bar{A}}_s + \mathbb{\bar{X}}_s \circ \mathbb{\bar{A}}_e)
\nonumber
\\
&
+ k_x f_\alpha f_\gamma f_w^{-1} ( \mathbb{\bar{X}}_\alpha \circ \mathbb{\bar{A}}_\gamma  + \mathbb{\bar{X}}_\gamma \circ \mathbb{\bar{A}}_\alpha ) + k_x f_\beta f_\delta f_w^{-1} ( \mathbb{\bar{X}}_\beta \circ \mathbb{\bar{A}}_\delta + \mathbb{\bar{X}}_\delta \circ \mathbb{\bar{A}}_\beta)
\nonumber
\\
&
+ k_a ( \mathbb{\bar{A}}_g \circ \mathbb{\bar{A}}_w  + \mathbb{\bar{A}}_w \circ \mathbb{\bar{A}}_g ) + k_a f_e f_s f_w^{-1} ( \mathbb{\bar{A}}_e \circ \mathbb{\bar{A}}_s + \mathbb{\bar{A}}_s \circ \mathbb{\bar{A}}_e)
\nonumber
\\
&
+ k_a f_\alpha f_\gamma f_w^{-1} ( \mathbb{\bar{A}}_\alpha \circ \mathbb{\bar{A}}_\gamma  + \mathbb{\bar{A}}_\gamma \circ \mathbb{\bar{A}}_\alpha ) + k_a f_\beta f_\delta f_w^{-1} ( \mathbb{\bar{A}}_\beta \circ \mathbb{\bar{A}}_\delta + \mathbb{\bar{A}}_\delta \circ \mathbb{\bar{A}}_\beta) ~,
\\
\mathbb{\bar{B}}_s = & ~ ( \lozenge_g \circ \mathbb{\bar{A}}_s + d_e f_w f_s^{-1} \lozenge_e \circ \mathbb{\bar{A}}_w  + d_w f_e f_s^{-1} \lozenge_w \circ \mathbb{\bar{A}}_e  + d_s f_s^{-1} \lozenge_s \circ \mathbb{\bar{A}}_g )
\nonumber
\\
&
+ f_s^{-1} ( d_\alpha f_\delta \lozenge_\alpha \circ \mathbb{\bar{A}}_\delta + d_\beta f_\gamma \lozenge_\beta \circ \mathbb{\bar{A}}_\gamma  + d_\gamma f_\beta \lozenge_\gamma \circ \mathbb{\bar{A}}_\beta  + d_\delta f_\alpha \lozenge_\delta \circ \mathbb{\bar{A}}_\alpha )
\nonumber
\\
&
+ k_x ( \mathbb{\bar{X}}_g \circ \mathbb{\bar{A}}_s  + \mathbb{\bar{X}}_s \circ \mathbb{\bar{A}}_g ) + k_x f_e f_w f_s^{-1} ( \mathbb{\bar{X}}_e \circ \mathbb{\bar{A}}_w + \mathbb{\bar{X}}_w \circ \mathbb{\bar{A}}_e)
\nonumber
\\
&
+ k_x f_\alpha f_\delta f_s^{-1} ( \mathbb{\bar{X}}_\alpha \circ \mathbb{\bar{A}}_\delta  + \mathbb{\bar{X}}_\delta \circ \mathbb{\bar{A}}_\alpha ) + k_x f_\beta f_\gamma f_s^{-1} ( \mathbb{\bar{X}}_\beta \circ \mathbb{\bar{A}}_\gamma + \mathbb{\bar{X}}_\gamma \circ \mathbb{\bar{A}}_\beta)
\nonumber
\\
&
+ k_a ( \mathbb{\bar{A}}_g \circ \mathbb{\bar{A}}_s  + \mathbb{\bar{A}}_s \circ \mathbb{\bar{A}}_g ) + k_a f_e f_w f_s^{-1} ( \mathbb{\bar{A}}_e \circ \mathbb{\bar{A}}_w + \mathbb{\bar{A}}_w \circ \mathbb{\bar{A}}_e)
\nonumber
\\
&
+ k_a f_\alpha f_\delta f_s^{-1} ( \mathbb{\bar{A}}_\alpha \circ \mathbb{\bar{A}}_\delta  + \mathbb{\bar{A}}_\delta \circ \mathbb{\bar{A}}_\alpha ) + k_a f_\beta f_\gamma f_s^{-1} ( \mathbb{\bar{A}}_\beta \circ \mathbb{\bar{A}}_\gamma + \mathbb{\bar{A}}_\gamma \circ \mathbb{\bar{A}}_\beta) ~,
\\
\mathbb{\bar{B}}_\alpha = & ~ ( \lozenge_g \circ \mathbb{\bar{A}}_\alpha + d_e f_\beta f_\alpha^{-1} \lozenge_e \circ \mathbb{\bar{A}}_\beta  + d_w f_\gamma  f_\alpha^{-1} \lozenge_w \circ \mathbb{\bar{A}}_\gamma  + d_s f_\delta  f_\alpha^{-1} \lozenge_s \circ \mathbb{\bar{A}}_\delta )
\nonumber
\\
&
+  f_\alpha^{-1} ( d_\alpha \lozenge_\alpha \circ \mathbb{\bar{A}}_g + d_\beta f_e \lozenge_\beta \circ \mathbb{\bar{A}}_e  + d_\gamma f_w \lozenge_\gamma \circ \mathbb{\bar{A}}_w  + d_\delta f_s \lozenge_\delta \circ \mathbb{\bar{A}}_s )
\nonumber
\\
&
+ k_x (  \mathbb{\bar{X}}_g \circ \mathbb{\bar{A}}_\alpha + \mathbb{\bar{X}}_\alpha \circ \mathbb{\bar{A}}_g )
+ k_x f_e f_\beta f_\alpha^{-1} ( \mathbb{\bar{X}}_e \circ \mathbb{\bar{A}}_\beta + \mathbb{\bar{X}}_\beta \circ \mathbb{\bar{A}}_e )
\nonumber
\\
&
+ k_x f_w f_\gamma f_\alpha^{-1} ( \mathbb{\bar{X}}_w \circ \mathbb{\bar{A}}_\gamma + \mathbb{\bar{X}}_\gamma \circ \mathbb{\bar{A}}_w )
+ k_x f_s f_\delta f_\alpha^{-1} ( \mathbb{\bar{X}}_s \circ \mathbb{\bar{A}}_\delta + \mathbb{\bar{X}}_\delta \circ \mathbb{\bar{A}}_s )
\nonumber
\\
&
+ k_a (  \mathbb{\bar{A}}_g \circ \mathbb{\bar{A}}_\alpha + \mathbb{\bar{A}}_\alpha \circ \mathbb{\bar{A}}_g )
+ k_a f_e f_\beta f_\alpha^{-1} ( \mathbb{\bar{A}}_e \circ \mathbb{\bar{A}}_\beta + \mathbb{\bar{A}}_\beta \circ \mathbb{\bar{A}}_e )
\nonumber
\\
&
+ k_a f_w f_\gamma f_\alpha^{-1} ( \mathbb{\bar{A}}_w \circ \mathbb{\bar{A}}_\gamma + \mathbb{\bar{A}}_\gamma \circ \mathbb{\bar{A}}_w )
+ k_a f_s f_\delta f_\alpha^{-1} ( \mathbb{\bar{A}}_s \circ \mathbb{\bar{A}}_\delta + \mathbb{\bar{A}}_\delta \circ \mathbb{\bar{A}}_s ) ~,
\\
\mathbb{\bar{B}}_\beta = & ~ ( \lozenge_g \circ \mathbb{\bar{A}}_\beta + d_e f_\alpha f_\beta^{-1} \lozenge_e \circ \mathbb{\bar{A}}_\alpha  + d_w f_\delta f_\beta^{-1} \lozenge_w \circ \mathbb{\bar{A}}_\delta  + d_s f_\gamma f_\beta^{-1} \lozenge_s \circ \mathbb{\bar{A}}_\gamma )
\nonumber
\\
&
+ f_\beta^{-1} ( d_\alpha f_e \lozenge_\alpha \circ \mathbb{\bar{A}}_e + d_\beta \lozenge_\beta \circ \mathbb{\bar{A}}_g  + d_\gamma f_s \lozenge_\gamma \circ \mathbb{\bar{A}}_s  + d_\delta f_w \lozenge_\delta \circ \mathbb{\bar{A}}_w )
\nonumber
\\
&
+ k_x ( \mathbb{\bar{X}}_\beta \circ \mathbb{\bar{A}}_g + \mathbb{\bar{X}}_g \circ \mathbb{\bar{A}}_\beta) + k_x f_e f_\alpha f_\beta^{-1} ( \mathbb{\bar{X}}_e \circ \mathbb{\bar{A}}_\alpha + \mathbb{\bar{X}}_\alpha \circ \mathbb{\bar{A}}_e )
\nonumber
\\
&
+ k_x f_w f_\delta f_\beta^{-1} ( \mathbb{\bar{X}}_w \circ \mathbb{\bar{A}}_\delta + \mathbb{\bar{X}}_\delta \circ \mathbb{\bar{A}}_w) + k_x f_\gamma f_s f_\beta^{-1} ( \mathbb{\bar{X}}_\gamma \circ \mathbb{\bar{A}}_s + \mathbb{\bar{X}}_s \circ \mathbb{\bar{A}}_\gamma )
\nonumber
\\
&
+ k_a ( \mathbb{\bar{A}}_\beta \circ \mathbb{\bar{A}}_g + \mathbb{\bar{A}}_g \circ \mathbb{\bar{A}}_\beta) + k_a f_e f_\alpha f_\beta^{-1} ( \mathbb{\bar{A}}_e \circ \mathbb{\bar{A}}_\alpha + \mathbb{\bar{A}}_\alpha \circ \mathbb{\bar{A}}_e )
\nonumber
\\
&
+ k_a f_w f_\delta f_\beta^{-1} ( \mathbb{\bar{A}}_w \circ \mathbb{\bar{A}}_\delta + \mathbb{\bar{A}}_\delta \circ \mathbb{\bar{A}}_w) + k_a f_\gamma f_s f_\beta^{-1} ( \mathbb{\bar{A}}_\gamma \circ \mathbb{\bar{A}}_s + \mathbb{\bar{A}}_s \circ \mathbb{\bar{A}}_\gamma ) ~,
\\
\mathbb{\bar{B}}_\gamma = & ~ ( \lozenge_g \circ \mathbb{\bar{A}}_\gamma + d_e f_\delta f_\gamma^{-1} \lozenge_e \circ \mathbb{\bar{A}}_\delta  + d_w f_\alpha f_\gamma^{-1} \lozenge_w \circ \mathbb{\bar{A}}_\alpha + d_s f_\beta f_\gamma^{-1} \lozenge_s \circ \mathbb{\bar{A}}_\beta )
\nonumber
\\
&
+ f_\gamma^{-1} ( d_\alpha f_w \lozenge_\alpha \circ \mathbb{\bar{A}}_w + d_\beta f_s \lozenge_\beta \circ \mathbb{\bar{A}}_s  + d_\gamma \lozenge_\gamma \circ \mathbb{\bar{A}}_g + d_\delta f_e \lozenge_\delta \circ \mathbb{\bar{A}}_e )
\nonumber
\\
&
+ k_x ( \mathbb{\bar{X}}_g \circ \mathbb{\bar{A}}_\gamma  + \mathbb{\bar{X}}_\gamma \circ \mathbb{\bar{A}}_g ) + k_x f_e f_\delta f_\gamma^{-1} ( \mathbb{\bar{X}}_e \circ \mathbb{\bar{A}}_\delta + \mathbb{\bar{X}}_\delta \circ \mathbb{\bar{A}}_e)
\nonumber
\\
&
+ k_x f_\alpha f_w f_\gamma^{-1} ( \mathbb{\bar{X}}_\alpha \circ \mathbb{\bar{A}}_w  + \mathbb{\bar{X}}_w \circ \mathbb{\bar{A}}_\alpha ) + k_x f_\beta f_s f_\gamma^{-1} ( \mathbb{\bar{X}}_\beta \circ \mathbb{\bar{A}}_s + \mathbb{\bar{X}}_s \circ \mathbb{\bar{A}}_\beta)
\nonumber
\\
&
+ k_a ( \mathbb{\bar{A}}_g \circ \mathbb{\bar{A}}_\gamma  + \mathbb{\bar{A}}_\gamma \circ \mathbb{\bar{A}}_g ) + k_a f_e f_\delta f_\gamma^{-1} ( \mathbb{\bar{A}}_e \circ \mathbb{\bar{A}}_\delta + \mathbb{\bar{A}}_\delta \circ \mathbb{\bar{A}}_e)
\nonumber
\\
&
+ k_a f_\alpha f_w f_\gamma^{-1} ( \mathbb{\bar{A}}_\alpha \circ \mathbb{\bar{A}}_w  + \mathbb{\bar{A}}_w \circ \mathbb{\bar{A}}_\alpha ) + k_a f_\beta f_s f_\gamma^{-1} ( \mathbb{\bar{A}}_\beta \circ \mathbb{\bar{A}}_s + \mathbb{\bar{A}}_s \circ \mathbb{\bar{A}}_\beta) ~,
\\
\mathbb{\bar{B}}_\delta = & ~ ( \lozenge_g \circ \mathbb{\bar{A}}_\delta + d_e f_\gamma f_\delta^{-1} \lozenge_e \circ \mathbb{\bar{A}}_\gamma  + d_w f_\beta f_\delta^{-1} \lozenge_w \circ \mathbb{\bar{A}}_\beta  + d_s f_\alpha f_\delta^{-1} \lozenge_s \circ \mathbb{\bar{A}}_\alpha )
\nonumber
\\
&
+ f_\delta^{-1} ( d_\alpha f_s \lozenge_\alpha \circ \mathbb{\bar{A}}_s + d_\beta f_w \lozenge_\beta \circ \mathbb{\bar{A}}_w  + d_\gamma f_e \lozenge_\gamma \circ \mathbb{\bar{A}}_e  + d_\delta \lozenge_\delta \circ \mathbb{\bar{A}}_g )
\nonumber
\\
&
+ k_x ( \mathbb{\bar{X}}_g \circ \mathbb{\bar{A}}_\delta  + \mathbb{\bar{X}}_\delta \circ \mathbb{\bar{A}}_g ) + k_x f_e f_\gamma f_\delta^{-1} ( \mathbb{\bar{X}}_e \circ \mathbb{\bar{A}}_\gamma + \mathbb{\bar{X}}_\gamma \circ \mathbb{\bar{A}}_e)
\nonumber
\\
&
+ k_x f_\alpha f_s f_\delta^{-1} ( \mathbb{\bar{X}}_\alpha \circ \mathbb{\bar{A}}_s  + \mathbb{\bar{X}}_s \circ \mathbb{\bar{A}}_\alpha ) + k_x f_\beta f_w f_\delta^{-1} ( \mathbb{\bar{X}}_\beta \circ \mathbb{\bar{A}}_w + \mathbb{\bar{X}}_w \circ \mathbb{\bar{A}}_\beta)
\nonumber
\\
&
+ k_a ( \mathbb{\bar{A}}_g \circ \mathbb{\bar{A}}_\delta  + \mathbb{\bar{A}}_\delta \circ \mathbb{\bar{A}}_g ) + k_a f_e f_\gamma f_\delta^{-1} ( \mathbb{\bar{A}}_e \circ \mathbb{\bar{A}}_\gamma + \mathbb{\bar{A}}_\gamma \circ \mathbb{\bar{A}}_e)
\nonumber
\\
&
+ k_a f_\alpha f_s f_\delta^{-1} ( \mathbb{\bar{A}}_\alpha \circ \mathbb{\bar{A}}_s  + \mathbb{\bar{A}}_s \circ \mathbb{\bar{A}}_\alpha ) + k_a f_\beta f_w f_\delta^{-1} ( \mathbb{\bar{A}}_\beta \circ \mathbb{\bar{A}}_w + \mathbb{\bar{A}}_w \circ \mathbb{\bar{A}}_\beta) ~,
\end{align}
where the trigintaduonion compounding field strength $\mathbb{\bar{B}}$ is composed of $\mathbb{\bar{B}}_\alpha = \Sigma ( \bar{h}_{(i+16)} \emph{\textbf{i}}_{(i+16)}) $, $\mathbb{\bar{B}}_\beta = \Sigma (\bar{h}_{(i+20)} \emph{\textbf{i}}_{(i+20)})$, $\mathbb{\bar{B}}_\gamma = \Sigma ( \bar{h}_{(i+24)} \emph{\textbf{i}}_{(i+24)}) $, and $\mathbb{\bar{B}}_\delta = \Sigma (\bar{h}_{(i+28)} \emph{\textbf{i}}_{(i+28)})$, besides the $\mathbb{\bar{B}}_g$, $\mathbb{\bar{B}}_e$, $\mathbb{\bar{B}}_w$, and $\mathbb{\bar{B}}_s$. The gauge equations are $\bar{h}_0 = 0$, $\bar{h}_4 = 0$, $\bar{h}_8 = 0$, $\bar{h}_{12} = 0$, $\bar{h}_{16} = 0$, $\bar{h}_{20} = 0$, $\bar{h}_{24} = 0$, and $\bar{h}_{28} = 0$ respectively.

The trigintaduonion compounding field is composed of eight fundamental fields, and each field is accompanied by its seven adjoint fields. Herein the $*$ denotes the trigintaduonion conjugate.

For the gravitational strength $\mathbb{\bar{B}}_g$, the linear momentum density $-\mu_{gg} \mathbb{\bar{S}}_{gg} = \lozenge_g^* \circ \mathbb{\bar{B}}_g$ is the source for the gravitational field, with seven adjoint field sources for the gravitational field, that is $-\mu_{ge} \mathbb{\bar{S}}_{ge} = \lozenge_e^* \circ \mathbb{\bar{B}}_g$, $-\mu_{gw} \mathbb{\bar{S}}_{gw} = \lozenge_w^* \circ \mathbb{\bar{B}}_g$, $-\mu_{gs} \mathbb{\bar{S}}_{gs} = \lozenge_s^* \circ \mathbb{\bar{B}}_g$,
$-\mu_{g\alpha} \mathbb{\bar{S}}_{g\alpha} = \lozenge_\alpha^* \circ \mathbb{\bar{B}}_g$,
$-\mu_{g\beta} \mathbb{\bar{S}}_{g\beta} = \lozenge_\beta^* \circ \mathbb{\bar{B}}_g$,
$-\mu_{g\gamma} \mathbb{\bar{S}}_{g\gamma} = \lozenge_\gamma^* \circ \mathbb{\bar{B}}_g$, and
$-\mu_{g\delta} \mathbb{\bar{S}}_{g\delta} = \lozenge_\delta^* \circ \mathbb{\bar{B}}_g$.

For the electromagnetic strength $\mathbb{\bar{B}}_e$, the electric current density $-\mu_{eg} \mathbb{\bar{S}}_{eg} = \lozenge_g^* \circ \mathbb{\bar{B}}_e$ is the source for electromagnetic field, with seven adjoint field sources for the electromagnetic field, that is $-\mu_{ee} \mathbb{\bar{S}}_{ee} = \lozenge_e^* \circ \mathbb{\bar{B}}_e$, $-\mu_{ew} \mathbb{\bar{S}}_{ew} = \lozenge_w^* \circ \mathbb{\bar{B}}_e$, $-\mu_{es} \mathbb{\bar{S}}_{es} = \lozenge_s^* \circ \mathbb{\bar{B}}_e$,
$-\mu_{e\alpha} \mathbb{\bar{S}}_{e\alpha} = \lozenge_\alpha^* \circ \mathbb{\bar{B}}_e$,
$-\mu_{e\beta} \mathbb{\bar{S}}_{e\beta} = \lozenge_\beta^* \circ \mathbb{\bar{B}}_e$,
$-\mu_{e\gamma} \mathbb{\bar{S}}_{e\gamma} = \lozenge_\gamma^* \circ \mathbb{\bar{B}}_e$, and
$-\mu_{e\delta} \mathbb{\bar{S}}_{e\delta} = \lozenge_\delta^* \circ \mathbb{\bar{B}}_e$.

For the weak nuclear field strength $\mathbb{\bar{B}}_w$, the $-\mu_{wg} \mathbb{\bar{S}}_{wg} = \lozenge_g^* \circ \mathbb{\bar{B}}_w$ is the source for weak nuclear field, with seven kinds of adjoint field sources for the weak nuclear field, that is $-\mu_{we} \mathbb{\bar{S}}_{we} = \lozenge_e^* \circ \mathbb{\bar{B}}_w$, $-\mu_{ww} \mathbb{\bar{S}}_{ww} = \lozenge_w^* \circ \mathbb{\bar{B}}_w$, $-\mu_{ws} \mathbb{\bar{S}}_{ws} = \lozenge_s^* \circ \mathbb{\bar{B}}_w$,
$-\mu_{w\alpha} \mathbb{\bar{S}}_{w\alpha} = \lozenge_\alpha^* \circ \mathbb{\bar{B}}_w$,
$-\mu_{w\beta} \mathbb{\bar{S}}_{w\beta} = \lozenge_\beta^* \circ \mathbb{\bar{B}}_w$,
$-\mu_{w\gamma} \mathbb{\bar{S}}_{w\gamma} = \lozenge_\gamma^* \circ \mathbb{\bar{B}}_w$, and
$-\mu_{w\delta} \mathbb{\bar{S}}_{w\delta} = \lozenge_\delta^* \circ \mathbb{\bar{B}}_w$.

For the strong nuclear field strength $\mathbb{\bar{B}}_s$, the $-\mu_{sg} \mathbb{\bar{S}}_{sg} = \lozenge_g^* \circ \mathbb{\bar{B}}_s$ is the source for strong nuclear field, with seven kinds of adjoint field sources for the strong nuclear field, that is $-\mu_{se} \mathbb{\bar{S}}_{se} = \lozenge_e^* \circ \mathbb{\bar{B}}_s$, $-\mu_{sw} \mathbb{\bar{S}}_{sw} = \lozenge_w^* \circ \mathbb{\bar{B}}_s$, $-\mu_{ss} \mathbb{\bar{S}}_{ss} = \lozenge_s^* \circ \mathbb{\bar{B}}_s$,
$-\mu_{s\alpha} \mathbb{\bar{S}}_{s\alpha} = \lozenge_\alpha^* \circ \mathbb{\bar{B}}_s$,
$-\mu_{s\beta} \mathbb{\bar{S}}_{s\beta} = \lozenge_\beta^* \circ \mathbb{\bar{B}}_s$,
$-\mu_{s\gamma} \mathbb{\bar{S}}_{s\gamma} = \lozenge_\gamma^* \circ \mathbb{\bar{B}}_s$, and
$-\mu_{s\delta} \mathbb{\bar{S}}_{s\delta} = \lozenge_\delta^* \circ \mathbb{\bar{B}}_s$.

For the $\alpha$ field strength $\mathbb{\bar{B}}_\alpha$, the $-\mu_{\alpha g} \mathbb{\bar{S}}_{\alpha g} = \lozenge_g^* \circ \mathbb{\bar{B}}_\alpha$ is the source for $\alpha$ field, with seven kinds of adjoint field sources for the $\alpha$ field, that is $-\mu_{\alpha e} \mathbb{\bar{S}}_{\alpha e} = \lozenge_e^* \circ \mathbb{\bar{B}}_\alpha$, $-\mu_{\alpha w} \mathbb{\bar{S}}_{\alpha w} = \lozenge_\alpha^* \circ \mathbb{\bar{B}}_\alpha$, $-\mu_{\alpha s} \mathbb{\bar{S}}_{\alpha s} = \lozenge_s^* \circ \mathbb{\bar{B}}_\alpha$,
$-\mu_{\alpha \alpha} \mathbb{\bar{S}}_{\alpha \alpha} = \lozenge_\alpha^* \circ \mathbb{\bar{B}}_\alpha$,
$-\mu_{\alpha \beta} \mathbb{\bar{S}}_{\alpha \beta} = \lozenge_\beta^* \circ \mathbb{\bar{B}}_\alpha$,
$-\mu_{\alpha \gamma} \mathbb{\bar{S}}_{\alpha \gamma} = \lozenge_\gamma^* \circ \mathbb{\bar{B}}_\alpha$, and
$-\mu_{\alpha \delta} \mathbb{\bar{S}}_{\alpha \delta} = \lozenge_\delta^* \circ \mathbb{\bar{B}}_\alpha$.

For the $\beta$ field strength $\mathbb{\bar{B}}_\beta$, the $-\mu_{\beta g} \mathbb{\bar{S}}_{\beta g} = \lozenge_g^* \circ \mathbb{\bar{B}}_\beta$ is the source for $\beta$ field, with seven kinds of adjoint field sources for the $\beta$ field, that is $-\mu_{\beta e} \mathbb{\bar{S}}_{\beta e} = \lozenge_e^* \circ \mathbb{\bar{B}}_\beta$, $-\mu_{\beta w} \mathbb{\bar{S}}_{\beta w} = \lozenge_\beta^* \circ \mathbb{\bar{B}}_\beta$, $-\mu_{\beta s} \mathbb{\bar{S}}_{\beta s} = \lozenge_s^* \circ \mathbb{\bar{B}}_\beta$,
$-\mu_{\beta \alpha} \mathbb{\bar{S}}_{\beta \alpha} = \lozenge_\beta^* \circ \mathbb{\bar{B}}_\alpha$,
$-\mu_{\beta \beta} \mathbb{\bar{S}}_{\beta \beta} = \lozenge_\beta^* \circ \mathbb{\bar{B}}_\beta$,
$-\mu_{\beta \gamma} \mathbb{\bar{S}}_{\beta \gamma} = \lozenge_\gamma^* \circ \mathbb{\bar{B}}_\beta$, and
$-\mu_{\beta \delta} \mathbb{\bar{S}}_{\beta \delta} = \lozenge_\delta^* \circ \mathbb{\bar{B}}_\beta$.

For the $\gamma$ field strength $\mathbb{\bar{B}}_\gamma$, the $-\mu_{\gamma g} \mathbb{\bar{S}}_{\gamma g} = \lozenge_g^* \circ \mathbb{\bar{B}}_\gamma$ is the source for $\gamma$ field, with seven kinds of adjoint field sources for the $\gamma$ field, that is $-\mu_{\gamma e} \mathbb{\bar{S}}_{\gamma e} = \lozenge_e^* \circ \mathbb{\bar{B}}_\gamma$, $-\mu_{\gamma w} \mathbb{\bar{S}}_{\gamma w} = \lozenge_\gamma^* \circ \mathbb{\bar{B}}_\gamma$, $-\mu_{\gamma s} \mathbb{\bar{S}}_{\gamma s} = \lozenge_s^* \circ \mathbb{\bar{B}}_\gamma$,
$-\mu_{\gamma \alpha} \mathbb{\bar{S}}_{\gamma \alpha} = \lozenge_\gamma^* \circ \mathbb{\bar{B}}_\alpha$,
$-\mu_{\gamma \beta} \mathbb{\bar{S}}_{\gamma \beta} = \lozenge_\gamma^* \circ \mathbb{\bar{B}}_\beta$,
$-\mu_{\gamma \gamma} \mathbb{\bar{S}}_{\gamma \gamma} = \lozenge_\gamma^* \circ \mathbb{\bar{B}}_\gamma$, and
$-\mu_{\gamma \delta} \mathbb{\bar{S}}_{\gamma \delta} = \lozenge_\delta^* \circ \mathbb{\bar{B}}_\gamma$.

For the $\delta$ field strength $\mathbb{\bar{B}}_\delta$, the $-\mu_{\delta g} \mathbb{\bar{S}}_{\delta g} = \lozenge_g^* \circ \mathbb{\bar{B}}_\delta$ is the source for $\delta$ field, with seven kinds of adjoint field sources for the $\delta$ field, that is $-\mu_{\delta e} \mathbb{\bar{S}}_{\delta e} = \lozenge_e^* \circ \mathbb{\bar{B}}_\delta$, $-\mu_{\delta w} \mathbb{\bar{S}}_{\delta w} = \lozenge_\delta^* \circ \mathbb{\bar{B}}_\delta$, $-\mu_{\delta s} \mathbb{\bar{S}}_{\delta s} = \lozenge_s^* \circ \mathbb{\bar{B}}_\delta$,
$-\mu_{\delta \alpha} \mathbb{\bar{S}}_{\delta \alpha} = \lozenge_\delta^* \circ \mathbb{\bar{B}}_\alpha$,
$-\mu_{\delta \beta} \mathbb{\bar{S}}_{\delta \beta} = \lozenge_\delta^* \circ \mathbb{\bar{B}}_\beta$,
$-\mu_{\delta \gamma} \mathbb{\bar{S}}_{\delta \gamma} = \lozenge_\delta^* \circ \mathbb{\bar{B}}_\gamma$, and
$-\mu_{\delta \delta} \mathbb{\bar{S}}_{\delta \delta} = \lozenge_\delta^* \circ \mathbb{\bar{B}}_\delta$.

The seven adjoint field sources, $\mathbb{\bar{S}}_{ee}$, $\mathbb{\bar{S}}_{ww}$, $\mathbb{\bar{S}}_{ss}$, $\mathbb{\bar{S}}_{\alpha \alpha}$, $\mathbb{\bar{S}}_{\beta \beta}$, $\mathbb{\bar{S}}_{\gamma \gamma}$, $\mathbb{\bar{S}}_{\delta \delta}$, possess the feature of the gravitational field, and can be considered as the candidate of the dark matters.

The trigintaduonion field source $\mathbb{\bar{S}}$ satisfies
\begin{align}
\mu \mathbb{\bar{S}} = &~ - (\lozenge_{32} + k_x \mathbb{\bar{X}} + k_a \mathbb{\bar{A}} + k_b \mathbb{\bar{B}})^* \circ \mathbb{\bar{B}}
\nonumber\\
= & ~ \mu_{gg} \mathbb{\bar{S}}_{gg} + f_e \mu_{ge} \mathbb{\bar{S}}_{ge} + f_w \mu_{gw} \mathbb{\bar{S}}_{gw} + f_s \mu_{gs} \mathbb{\bar{S}}_{gs}
\nonumber\\
&
+ f_\alpha \mu_{g \alpha} \mathbb{\bar{S}}_{g \alpha} + f_\beta \mu_{g \beta} \mathbb{\bar{S}}_{g \beta} + f_\gamma \mu_{g \gamma} \mathbb{\bar{S}}_{g \gamma}
+ f_\delta \mu_{g \delta} \mathbb{\bar{S}}_{g \delta}
\nonumber\\
&
+ d_e ( \mu_{eg} \mathbb{\bar{S}}_{eg} + f_e \mu_{ee} \mathbb{\bar{S}}_{ee} + f_w \mu_{ew} \mathbb{\bar{S}}_{ew} + f_s \mu_{es} \mathbb{\bar{S}}_{es} )
\nonumber\\
&
+ d_e ( f_\alpha \mu_{e \alpha} \mathbb{\bar{S}}_{e \alpha} + f_\beta \mu_{e \beta} \mathbb{\bar{S}}_{e \beta} + f_\gamma \mu_{e \gamma} \mathbb{\bar{S}}_{e \gamma}
+ f_\delta \mu_{e \delta} \mathbb{\bar{S}}_{e \delta} )
\nonumber\\
&
+ d_w ( \mu_{wg} \mathbb{\bar{S}}_{wg} + f_e \mu_{we} \mathbb{\bar{S}}_{we} + f_w \mu_{ww} \mathbb{\bar{S}}_{ww} + f_s \mu_{ws} \mathbb{\bar{S}}_{ws} )
\nonumber\\
&
+ d_w ( f_\alpha \mu_{w \alpha} \mathbb{\bar{S}}_{w \alpha} + f_\beta \mu_{w \beta} \mathbb{\bar{S}}_{w \beta} + f_\gamma \mu_{w \gamma} \mathbb{\bar{S}}_{w \gamma}
+ f_\delta \mu_{w \delta} \mathbb{\bar{S}}_{w \delta} )
\nonumber\\
&
+ d_s ( \mu_{sg} \mathbb{\bar{S}}_{sg} + f_e \mu_{se} \mathbb{\bar{S}}_{se} + f_w \mu_{sw} \mathbb{\bar{S}}_{sw} + f_s \mu_{ss} \mathbb{\bar{S}}_{ss} )
\nonumber\\
&
+ d_s ( f_\alpha \mu_{s \alpha} \mathbb{\bar{S}}_{s \alpha} + f_\beta \mu_{s \beta} \mathbb{\bar{S}}_{s \beta} + f_\gamma \mu_{s \gamma} \mathbb{\bar{S}}_{s \gamma}
+ f_\delta \mu_{s \delta} \mathbb{\bar{S}}_{s \delta} )
\nonumber\\
&
+ d_\alpha  ( \mu_{\alpha g} \mathbb{\bar{S}}_{\alpha g} + f_e \mu_{\alpha e} \mathbb{\bar{S}}_{\alpha e} + f_w \mu_{\alpha w} \mathbb{\bar{S}}_{\alpha w} + f_s \mu_{\alpha s} \mathbb{\bar{S}}_{\alpha s} )
\nonumber\\
&
+ d_\alpha  ( f_\alpha \mu_{\alpha \alpha} \mathbb{\bar{S}}_{\alpha \alpha} + f_\beta \mu_{\alpha \beta} \mathbb{\bar{S}}_{\alpha \beta} + f_\gamma \mu_{\alpha \gamma} \mathbb{\bar{S}}_{\alpha \gamma} + f_\delta \mu_{\alpha \delta} \mathbb{\bar{S}}_{\alpha \delta} )
\nonumber\\
&
+ d_\beta ( \mu_{\beta g} \mathbb{\bar{S}}_{\beta g} + f_e \mu_{\beta e} \mathbb{\bar{S}}_{\beta e} + f_w \mu_{\beta w} \mathbb{\bar{S}}_{\beta w} + f_s \mu_{\beta s} \mathbb{\bar{S}}_{\beta s} )
\nonumber\\
&
+ d_\beta ( f_\alpha \mu_{\beta \alpha} \mathbb{\bar{S}}_{\beta \alpha} + f_\beta \mu_{\beta \beta} \mathbb{\bar{S}}_{\beta \beta} + f_\gamma \mu_{\beta \gamma} \mathbb{\bar{S}}_{\beta \gamma} + f_\delta \mu_{\beta \delta} \mathbb{\bar{S}}_{\beta \delta} )
\nonumber\\
&
+ d_\gamma ( \mu_{\gamma g} \mathbb{\bar{S}}_{\gamma g} + f_e \mu_{\gamma e} \mathbb{\bar{S}}_{\gamma e} + f_w \mu_{\gamma w} \mathbb{\bar{S}}_{\gamma w} + f_s \mu_{\gamma s} \mathbb{\bar{S}}_{\gamma s} )
\nonumber\\
&
+ d_\gamma ( f_\alpha \mu_{\gamma \alpha} \mathbb{\bar{S}}_{\gamma \alpha} + f_\beta \mu_{\gamma \beta} \mathbb{\bar{S}}_{\gamma \beta} + f_\gamma \mu_{\gamma \gamma} \mathbb{\bar{S}}_{\gamma \gamma} + f_\delta \mu_{\gamma \delta} \mathbb{\bar{S}}_{\gamma \delta} )
\nonumber\\
&
+ d_\delta ( \mu_{\delta g} \mathbb{\bar{S}}_{\delta g} + f_e \mu_{\delta e} \mathbb{\bar{S}}_{\delta e} + f_w \mu_{\delta w} \mathbb{\bar{S}}_{\delta w} + f_s \mu_{\delta s} \mathbb{\bar{S}}_{\delta s} )
\nonumber\\
&
+ d_\delta ( f_\alpha \mu_{\delta \alpha} \mathbb{\bar{S}}_{\delta \alpha} + f_\beta \mu_{\delta \beta} \mathbb{\bar{S}}_{\delta \beta} + f_\gamma \mu_{\delta \gamma} \mathbb{\bar{S}}_{\delta \gamma} + f_\delta \mu_{\delta \delta} \mathbb{\bar{S}}_{\delta \delta} )
\nonumber\\
&
- k_x \mathbb{\bar{X}}^* \circ \mathbb{\bar{B}} - k_a \mathbb{\bar{A}}^* \circ \mathbb{\bar{B}} - k_b \mathbb{\bar{B}}^* \circ \mathbb{\bar{B}} ~,
\end{align}
where $(k_x \mathbb{\bar{X}}^* \cdot \mathbb{\bar{B}} + k_a \mathbb{\bar{A}}^* \cdot \mathbb{\bar{B}} + k_b \mathbb{\bar{B}}^* \cdot \mathbb{\bar{B}})$ is the field strength helicity. $\mu$,
$\mu_{g g}$, $\mu_{g e}$, $\mu_{g w}$, $\mu_{g s}$, $\mu_{g \alpha}$, $\mu_{g \beta}$, $\mu_{g \gamma}$, $\mu_{g \delta}$,
$\mu_{e g}$, $\mu_{e e}$, $\mu_{e w}$, $\mu_{e s}$, $\mu_{e \alpha}$, $\mu_{e \beta}$, $\mu_{e \gamma}$, $\mu_{e \delta}$,
$\mu_{w g}$, $\mu_{w e}$, $\mu_{w w}$, $\mu_{w s}$, $\mu_{w \alpha}$, $\mu_{w \beta}$, $\mu_{w \gamma}$, $\mu_{w \delta}$,
$\mu_{s g}$, $\mu_{s e}$, $\mu_{s w}$, $\mu_{s s}$, $\mu_{s \alpha}$, $\mu_{s \beta}$, $\mu_{s \gamma}$, $\mu_{s \delta}$,
$\mu_{\alpha g}$, $\mu_{\alpha e}$, $\mu_{\alpha w}$, $\mu_{\alpha s}$, $\mu_{\alpha \alpha}$, $\mu_{\alpha \beta}$, $\mu_{\alpha \gamma}$, $\mu_{\alpha \delta}$,
$\mu_{\beta g}$, $\mu_{\beta e}$, $\mu_{\beta w}$, $\mu_{\beta s}$, $\mu_{\beta \alpha}$, $\mu_{\beta \beta}$, $\mu_{\beta \gamma}$, $\mu_{\beta \delta}$,
$\mu_{\gamma g}$, $\mu_{\gamma e}$, $\mu_{\gamma w}$, $\mu_{\gamma s}$, $\mu_{\gamma \alpha}$, $\mu_{\gamma \beta}$, $\mu_{\gamma \gamma}$, $\mu_{\gamma \delta}$,
$\mu_{\delta g}$, $\mu_{\delta e}$, $\mu_{\delta w}$, $\mu_{\delta s}$, $\mu_{\delta \alpha}$, $\mu_{\delta \beta}$, $\mu_{\delta \gamma}$, and $\mu_{\delta \delta}$
are the coefficients. $\mu_{gg}$ and $\mu_{eg}$ are the gravitational and electromagnetic constants respectively.

\subsection{Torque}

In the trigintaduonion compounding space, the linear momentum density is $\mathbb{\bar{P}} = \mu \mathbb{\bar{S}} / \mu_{gg} = \Sigma (\bar{p}_i \emph{\textbf{i}}_i + f_e \bar{p}_{(i+4)} \emph{\textbf{i}}_{(i+4)} + f_w \bar{p}_{(i+8)} \emph{\textbf{i}}_{(i+8)} + f_s \bar{p}_{(i+12)} \emph{\textbf{i}}_{(i+12)}) + f_\alpha \bar{p}_{(i+16)} \emph{\textbf{i}}_{(i+16)}) + f_\beta \bar{p}_{(i+20)} \emph{\textbf{i}}_{(i+20)}) + f_\gamma \bar{p}_{(i+24)} \emph{\textbf{i}}_{(i+24)}) + f_\delta \bar{p}_{(i+28)} \emph{\textbf{i}}_{(i+28)}) $,
and the angular momentum density is $\mathbb{\bar{L}} = \mathbb{\bar{R}} \circ \mathbb{\bar{P}} = \Sigma (\bar{l}_i \emph{\textbf{i}}_i + f_e \bar{l}_{(i+4)} \emph{\textbf{i}}_{(i+4)} + f_w \bar{l}_{(i+8)} \emph{\textbf{i}}_{(i+8)} + f_s \bar{l}_{(i+12)} \emph{\textbf{i}}_{(i+12)}) + f_\alpha \bar{l}_{(i+16)} \emph{\textbf{i}}_{(i+16)}) + f_\beta \bar{l}_{(i+20)} \emph{\textbf{i}}_{(i+20)}) + f_\gamma \bar{l}_{(i+24)} \emph{\textbf{i}}_{(i+24)}) + f_\delta \bar{l}_{(i+28)} \emph{\textbf{i}}_{(i+28)}) $.
And the trigintaduonion torque-energy density $\mathbb{\bar{W}}$ is defined from the angular momentum density $\mathbb{\bar{L}}$, the field strength $\mathbb{\bar{B}}$, and the field source $\mathbb{\bar{S}}$ etc,
\begin{eqnarray}
\mathbb{\bar{W}} = v_0 (\lozenge_{32} + k_x \mathbb{\bar{X}} + k_a \mathbb{\bar{A}} + k_b \mathbb{\bar{B}} + k_s \mathbb{\bar{S}} + k_l \mathbb{\bar{L}}) \circ \mathbb{\bar{L}}~,
\end{eqnarray}
where $-\bar{w}_0/2$ is the energy density, $\bar{\textbf{w}}/2 = \Sigma (\bar{w}_j \emph{\textbf{i}}_j )/2$ is the torque density. $\mathbb{\bar{S}} = \Sigma (\bar{s}_i \emph{\textbf{i}}_i + f_e \bar{s}_{(i+4)} \emph{\textbf{i}}_{(i+4)} + f_w \bar{s}_{(i+8)} \emph{\textbf{i}}_{(i+8)} + f_s \bar{s}_{(i+12)} \emph{\textbf{i}}_{(i+12)})
+ f_\alpha \bar{s}_{(i+16)} \emph{\textbf{i}}_{(i+16)}) + f_\beta \bar{s}_{(i+20)} \emph{\textbf{i}}_{(i+20)}) + f_\gamma \bar{s}_{(i+24)} \emph{\textbf{i}}_{(i+24)}) + f_\delta \bar{s}_{(i+28)} \emph{\textbf{i}}_{(i+28)})$.

The scalar part $\bar{w}_0$ of the torque-energy density $\mathbb{\bar{W}} = \Sigma (\bar{w}_i \emph{\textbf{i}}_i + f_e \bar{w}_{(i+4)} \emph{\textbf{i}}_{(i+4)} + f_w \bar{w}_{(i+8)} \emph{\textbf{i}}_{(i+8)} + f_s \bar{w}_{(i+12)} \emph{\textbf{i}}_{(i+12)} + f_\alpha \bar{w}_{(i+16)} \emph{\textbf{i}}_{(i+16)}) + f_\beta \bar{w}_{(i+20)} \emph{\textbf{i}}_{(i+20)}) + f_\gamma \bar{w}_{(i+24)} \emph{\textbf{i}}_{(i+24)}) + f_\delta \bar{w}_{(i+28)} \emph{\textbf{i}}_{(i+28)})$ is written as,
\begin{align}
\bar{w}_0 / v_0 = & ~
k_a \left\{ \bar{a}_0 \bar{l}_0 + \bar{\textbf{a}}_g \cdot \bar{\textbf{l}}_g + f_e^2 ( \bar{\textbf{a}}_4 \circ \bar{\textbf{l}}_4 + \bar{\textbf{a}}_e \cdot \bar{\textbf{l}}_e )  \right\}
\nonumber\\
&
+ k_a \left\{ f_w^2 ( \bar{\textbf{a}}_8 \circ \bar{\textbf{l}}_8 + \bar{\textbf{a}}_w \cdot \bar{\textbf{l}}_w ) +  f_s^2 ( \bar{\textbf{a}}_{12} \circ \bar{\textbf{l}}_{12} + \bar{\textbf{a}}_s \cdot \bar{\textbf{l}}_s) \right\}
\nonumber\\
&
+ k_a \left\{ f_\alpha^2 ( \bar{\textbf{a}}_{16} \circ \bar{\textbf{l}}_{16} + \bar{\textbf{a}}_\alpha \cdot \bar{\textbf{l}}_\alpha )
+ f_\beta^2 ( \bar{\textbf{a}}_{20} \circ \bar{\textbf{l}}_{20} + \bar{\textbf{a}}_\beta \cdot \bar{\textbf{l}}_\beta) \right\}
\nonumber\\
&
+ k_a \left\{ f_\gamma^2 ( \bar{\textbf{a}}_{24} \circ \bar{\textbf{l}}_{24} + \bar{\textbf{a}}_\gamma \cdot \bar{\textbf{l}}_\gamma)
+ f_\delta^2 ( \bar{\textbf{a}}_{28} \circ \bar{\textbf{l}}_{28} + \bar{\textbf{a}}_\delta \cdot \bar{\textbf{l}}_\delta ) \right\}
\nonumber\\
&
+ k_x \left\{ \bar{x}_0 \bar{l}_0 + \bar{\textbf{x}}_g \cdot \bar{\textbf{l}}_g + f_e^2 ( \bar{\textbf{x}}_4 \circ \bar{\textbf{l}}_4 + \bar{\textbf{x}}_e \cdot \bar{\textbf{l}}_e ) \right\}
\nonumber\\
&
+ k_x \left\{ f_w^2 ( \bar{\textbf{x}}_8 \circ \bar{\textbf{l}}_8 + \bar{\textbf{x}}_w \cdot \bar{\textbf{l}}_w )+ f_s^2 ( \bar{\textbf{x}}_{12} \circ \bar{\textbf{l}}_{12} + \bar{\textbf{x}}_s \cdot \bar{\textbf{l}}_s) \right\}
\nonumber\\
&
+ k_x \left\{ f_\alpha^2 ( \bar{\textbf{x}}_{16} \circ \bar{\textbf{l}}_{16} + \bar{\textbf{x}}_\alpha \cdot \bar{\textbf{l}}_\alpha)
+ f_\beta^2 ( \bar{\textbf{x}}_{20} \circ \bar{\textbf{l}}_{20} + \bar{\textbf{x}}_\beta \cdot \bar{\textbf{l}}_\beta) \right\}
\nonumber\\
&
+ k_x \left\{ f_\gamma^2 ( \bar{\textbf{x}}_{24} \circ \bar{\textbf{l}}_{24} + \bar{\textbf{x}}_\gamma \cdot \bar{\textbf{l}}_\gamma)
+ f_\delta^2 ( \bar{\textbf{x}}_{28} \circ \bar{\textbf{l}}_{28} + \bar{\textbf{x}}_\delta \cdot \bar{\textbf{l}}_\delta) \right\}
\nonumber\\
&
+ k_s \left\{ \bar{s}_0 \bar{l}_0 + \bar{\textbf{s}}_g \cdot \bar{\textbf{l}}_g + f_e^2 ( \bar{\textbf{s}}_4 \circ \bar{\textbf{l}}_4 + \bar{\textbf{s}}_e \cdot \bar{\textbf{l}}_e ) \right\}
\nonumber\\
&
+ k_s \left\{ f_w^2 ( \bar{\textbf{s}}_8 \circ \bar{\textbf{l}}_8 + \bar{\textbf{s}}_w \cdot \bar{\textbf{l}}_w )+ f_s^2 ( \bar{\textbf{s}}_{12} \circ \bar{\textbf{l}}_{12} + \bar{\textbf{s}}_s \cdot \bar{\textbf{l}}_s) \right\}
\nonumber\\
&
+ k_s \left\{ f_\alpha^2 ( \bar{\textbf{s}}_{16} \circ \bar{\textbf{l}}_{16} + \bar{\textbf{s}}_\alpha \cdot \bar{\textbf{l}}_\alpha)
+ f_\beta^2 ( \bar{\textbf{s}}_{20} \circ \bar{\textbf{l}}_{20} + \bar{\textbf{s}}_\beta \cdot \bar{\textbf{l}}_\beta) \right\}
\nonumber\\
&
+ k_s \left\{ f_\gamma^2 ( \bar{\textbf{s}}_{24} \circ \bar{\textbf{l}}_{24} + \bar{\textbf{s}}_\gamma \cdot \bar{\textbf{l}}_\gamma)
+ f_\delta^2 ( \bar{\textbf{s}}_{28} \circ \bar{\textbf{l}}_{28} + \bar{\textbf{s}}_\delta \cdot \bar{\textbf{l}}_\delta) \right\}
\nonumber\\
&
+ k_l \left\{ \bar{l}_0 \bar{l}_0 + \bar{\textbf{l}}_g \cdot \bar{\textbf{l}}_g + f_e^2 ( \bar{\textbf{l}}_4 \circ \bar{\textbf{l}}_4 + \bar{\textbf{l}}_e \cdot \bar{\textbf{l}}_e )  \right\}
\nonumber\\
&
+ k_l \left\{ f_w^2 ( \bar{\textbf{l}}_8 \circ \bar{\textbf{l}}_8 + \bar{\textbf{l}}_w \cdot \bar{\textbf{l}}_w ) + f_s^2 ( \bar{\textbf{l}}_{12} \circ \bar{\textbf{l}}_{12} + \bar{\textbf{l}}_s \cdot \bar{\textbf{l}}_s) \right\}
\nonumber\\
&
+ k_l \left\{ f_\alpha^2 ( \bar{\textbf{l}}_{16} \circ \bar{\textbf{l}}_{16} + \bar{\textbf{l}}_\alpha \cdot \bar{\textbf{l}}_\alpha)
+ f_\beta^2 ( \bar{\textbf{l}}_{20} \circ \bar{\textbf{l}}_{20} + \bar{\textbf{l}}_\beta \cdot \bar{\textbf{l}}_\beta) \right\}
\nonumber\\
&
+ k_l \left\{ f_\gamma^2 ( \bar{\textbf{l}}_{24} \circ \bar{\textbf{l}}_{24} + \bar{\textbf{l}}_\gamma \cdot \bar{\textbf{l}}_\gamma)
+ f_\delta^2 ( \bar{\textbf{l}}_{28} \circ \bar{\textbf{l}}_{28} + \bar{\textbf{l}}_\delta \cdot \bar{\textbf{l}}_\delta) \right\}
\nonumber\\
&
+ k_b ( \bar{\textbf{h}}_g \cdot \bar{\textbf{l}}_g + f_e^2 \bar{\textbf{h}}_e \cdot \bar{\textbf{l}}_e  + f_w^2 \bar{\textbf{h}}_w \cdot \bar{\textbf{l}}_w + f_s^2 \bar{\textbf{h}}_s \cdot \bar{\textbf{l}}_s )
\nonumber\\
&
+ k_b ( f_\alpha^2 \bar{\textbf{h}}_\alpha \cdot \bar{\textbf{l}}_\alpha + f_\beta^2 \bar{\textbf{h}}_\beta \cdot \bar{\textbf{l}}_\beta + f_\gamma^2 \bar{\textbf{h}}_\gamma \cdot \bar{\textbf{l}}_\gamma + f_\delta^2 \bar{\textbf{h}}_\delta \cdot \bar{\textbf{l}}_\delta )
\nonumber\\
&
+ \partial_0 \bar{l}_0 + \nabla_g \cdot \bar{\textbf{l}}_g + d_e f_e (\emph{\textbf{i}}_4 \circ \partial_4 \bar{\textbf{l}}_4 + \nabla_e \cdot \bar{\textbf{l}}_e)
\nonumber\\
&
+ d_w f_w (\emph{\textbf{i}}_8 \circ \partial_8 \bar{\textbf{l}}_8 + \nabla_w \cdot \bar{\textbf{l}}_w) + d_s f_s (\emph{\textbf{i}}_{12} \circ \partial_{12} \bar{\textbf{l}}_{12} + \nabla_s \cdot \bar{\textbf{l}}_s)
\nonumber\\
&
+ d_\alpha f_\alpha (\emph{\textbf{i}}_{16} \circ \partial_{16} \bar{\textbf{l}}_{16} + \nabla_\alpha \cdot \bar{\textbf{l}}_\alpha)
+ d_\beta f_\beta (\emph{\textbf{i}}_{20} \circ \partial_{20} \bar{\textbf{l}}_{20} + \nabla_\beta \cdot \bar{\textbf{l}}_\beta)
\nonumber\\
&
+ d_\gamma f_\gamma (\emph{\textbf{i}}_{24} \circ \partial_{24} \bar{\textbf{l}}_{24} + \nabla_\gamma \cdot \bar{\textbf{l}}_\gamma)
+ d_\delta f_\delta (\emph{\textbf{i}}_{28} \circ \partial_{28} \bar{\textbf{l}}_{28} + \nabla_\delta \cdot \bar{\textbf{l}}_\delta) ~,
\end{align}
where $-\bar{w}_0/2$ includes the kinetic energy, the gravitational potential energy, the electric potential energy, the magnetic potential energy, the field energy, the work, the interacting energy between the dipole moment with the fields, and some new energy terms.
$\bar{\textbf{l}}_\alpha = \Sigma ( \emph{\textbf{i}}_{(j+16)} \bar{l}_{(j+16)})$,
$\bar{\textbf{l}}_\beta = \Sigma ( \emph{\textbf{i}}_{(j+20)} \bar{l}_{(j+20)})$,
$\bar{\textbf{l}}_\gamma = \Sigma ( \emph{\textbf{i}}_{(j+24)} \bar{l}_{(j+24)})$,
$\bar{\textbf{l}}_\delta = \Sigma ( \emph{\textbf{i}}_{(j+28)} \bar{l}_{(j+28)})$;
$\bar{\textbf{l}}_{(j+16)} = \emph{\textbf{i}}_{(j+16)} \bar{l}_{(j+16)}$,
$\bar{\textbf{l}}_{(j+20)} = \emph{\textbf{i}}_{(j+20)} \bar{l}_{(j+20)}$,
$\bar{\textbf{l}}_{(j+24)} = \emph{\textbf{i}}_{(j+24)} \bar{l}_{(j+24)}$,
$\bar{\textbf{l}}_{(j+28)} = \emph{\textbf{i}}_{(j+28)} \bar{l}_{(j+28)}$.
$\bar{\textbf{x}}_\alpha = \Sigma ( \emph{\textbf{i}}_{(j+16)} \bar{x}_{(j+16)})$,
$\bar{\textbf{x}}_\beta = \Sigma ( \emph{\textbf{i}}_{(j+20)} \bar{x}_{(j+20)})$,
$\bar{\textbf{x}}_\gamma = \Sigma ( \emph{\textbf{i}}_{(j+24)} \bar{x}_{(j+24)})$,
$\bar{\textbf{x}}_\delta = \Sigma ( \emph{\textbf{i}}_{(j+28)} \bar{x}_{(j+28)})$;
$\bar{\textbf{x}}_{(j+16)} = \emph{\textbf{i}}_{(j+16)} \bar{x}_{(j+16)}$,
$\bar{\textbf{x}}_{(j+20)} = \emph{\textbf{i}}_{(j+20)} \bar{x}_{(j+20)}$,
$\bar{\textbf{x}}_{(j+24)} = \emph{\textbf{i}}_{(j+24)} \bar{x}_{(j+24)}$,
$\bar{\textbf{x}}_{(j+28)} = \emph{\textbf{i}}_{(j+28)} \bar{x}_{(j+28)}$.
$\bar{\textbf{a}}_\alpha = \Sigma ( \emph{\textbf{i}}_{(j+16)} \bar{a}_{(j+16)})$,
$\bar{\textbf{a}}_\beta = \Sigma ( \emph{\textbf{i}}_{(j+20)} \bar{a}_{(j+20)})$,
$\bar{\textbf{a}}_\gamma = \Sigma ( \emph{\textbf{i}}_{(j+24)} \bar{a}_{(j+24)})$,
$\bar{\textbf{a}}_\delta = \Sigma ( \emph{\textbf{i}}_{(j+28)} \bar{a}_{(j+28)})$;
$\bar{\textbf{a}}_{(j+16)} = \emph{\textbf{i}}_{(j+16)} \bar{a}_{(j+16)}$,
$\bar{\textbf{a}}_{(j+20)} = \emph{\textbf{i}}_{(j+20)} \bar{a}_{(j+20)}$,
$\bar{\textbf{a}}_{(j+24)} = \emph{\textbf{i}}_{(j+24)} \bar{a}_{(j+24)}$,
$\bar{\textbf{a}}_{(j+28)} = \emph{\textbf{i}}_{(j+28)} \bar{a}_{(j+28)}$.
$\bar{\textbf{s}}_\alpha = \Sigma ( \emph{\textbf{i}}_{(j+16)} \bar{s}_{(j+16)})$,
$\bar{\textbf{s}}_\beta = \Sigma ( \emph{\textbf{i}}_{(j+20)} \bar{s}_{(j+20)})$,
$\bar{\textbf{s}}_\gamma = \Sigma ( \emph{\textbf{i}}_{(j+24)} \bar{s}_{(j+24)})$,
$\bar{\textbf{s}}_\delta = \Sigma ( \emph{\textbf{i}}_{(j+28)} \bar{s}_{(j+28)})$;
$\bar{\textbf{s}}_{(j+16)} = \emph{\textbf{i}}_{(j+16)} \bar{s}_{(j+16)}$,
$\bar{\textbf{s}}_{(j+20)} = \emph{\textbf{i}}_{(j+20)} \bar{s}_{(j+20)}$,
$\bar{\textbf{s}}_{(j+24)} = \emph{\textbf{i}}_{(j+24)} \bar{s}_{(j+24)}$,
$\bar{\textbf{s}}_{(j+28)} = \emph{\textbf{i}}_{(j+28)} \bar{s}_{(j+28)}$.
$\bar{\textbf{h}}_\alpha = \Sigma ( \emph{\textbf{i}}_{(j+16)} \bar{h}_{(j+16)})$,
$\bar{\textbf{h}}_\beta = \Sigma ( \emph{\textbf{i}}_{(j+20)} \bar{h}_{(j+20)})$,
$\bar{\textbf{h}}_\gamma = \Sigma ( \emph{\textbf{i}}_{(j+24)} \bar{h}_{(j+24)})$,
$\bar{\textbf{h}}_\delta = \Sigma ( \emph{\textbf{i}}_{(j+28)} \bar{h}_{(j+28)})$;
$\bar{\textbf{h}}_\alpha = \Sigma ( \emph{\textbf{i}}_{(j+16)} \bar{h}_{(j+16)})$,
$\bar{\textbf{h}}_\beta = \Sigma ( \emph{\textbf{i}}_{(j+20)} \bar{h}_{(j+20)})$,
$\bar{\textbf{h}}_\gamma = \Sigma ( \emph{\textbf{i}}_{(j+24)} \bar{h}_{(j+24)})$,
$\bar{\textbf{h}}_\delta = \Sigma ( \emph{\textbf{i}}_{(j+28)} \bar{h}_{(j+28)})$;
$\nabla_\alpha = \Sigma ( \emph{\textbf{i}}_{(j+16)} \partial_{(j+16)})$,
$\nabla_\beta = \Sigma ( \emph{\textbf{i}}_{(j+20)} \partial_{(j+20)})$,
$\nabla_\gamma = \Sigma ( \emph{\textbf{i}}_{(j+24)} \partial_{(j+24)})$,
$\nabla_\delta = \Sigma ( \emph{\textbf{i}}_{(j+28)} \partial_{(j+28)})$.

In a similar way, expressing the torque density $\bar{\textbf{w}}$ of $\mathbb{\bar{W}}$ as
\begin{align}
\bar{\textbf{w}} / v_0 = &~ ( \partial_0 \bar{\textbf{l}}_g + \nabla_g \bar{l}_0 + \nabla_g \times \bar{\textbf{l}}_g )
+ d_e f_e ( \emph{\textbf{i}}_4 \circ \partial_4 \bar{\textbf{l}}_e + \nabla_e \circ \bar{\textbf{l}}_4 + \nabla_e \times \bar{\textbf{l}}_e )
\nonumber
\\
& + d_w f_w ( \emph{\textbf{i}}_8 \circ \partial_8 \bar{\textbf{l}}_w + \nabla_w \circ \bar{\textbf{l}}_8 + \nabla_w \times \bar{\textbf{l}}_w )
+ d_s f_s ( \emph{\textbf{i}}_{12} \circ \partial_{12} \bar{\textbf{l}}_s + \nabla_s \circ \bar{\textbf{l}}_{12} + \nabla_s \times \bar{\textbf{l}}_s )
\nonumber
\\
& + d_\alpha f_\alpha ( \emph{\textbf{i}}_{16} \circ \partial_{16} \bar{\textbf{l}}_\alpha + \nabla_\alpha \circ \bar{\textbf{l}}_{16} + \nabla_\alpha \times \bar{\textbf{l}}_\alpha )
+ d_\beta f_\beta ( \emph{\textbf{i}}_{20} \circ \partial_{20} \bar{\textbf{l}}_\beta + \nabla_\beta \circ \bar{\textbf{l}}_{20} + \nabla_\beta \times \bar{\textbf{l}}_\beta )
\nonumber
\\
& + d_\gamma f_\gamma ( \emph{\textbf{i}}_{24} \circ \partial_{24} \bar{\textbf{l}}_\gamma + \nabla_\gamma \circ \bar{\textbf{l}}_{24} + \nabla_\gamma \times \bar{\textbf{l}}_\gamma )
+ d_\delta f_\delta ( \emph{\textbf{i}}_{28} \circ \partial_{28} \bar{\textbf{l}}_\delta + \nabla_\delta \circ \bar{\textbf{l}}_{28} + \nabla_\delta \times \bar{\textbf{l}}_\delta )
\nonumber
\\
& + k_a ( \bar{\textbf{a}}_g \times \bar{\textbf{l}}_g + \bar{\textbf{a}}_g \bar{l}_0 + \bar{a}_0 \bar{\textbf{l}}_g ) + k_a f_e^2 ( \bar{\textbf{a}}_e \times \bar{\textbf{l}}_e + \bar{\textbf{a}}_4 \circ \bar{\textbf{l}}_e + \bar{\textbf{a}}_e \circ \bar{\textbf{l}}_4)
\nonumber
\\
& + k_a f_w^2 ( \bar{\textbf{a}}_w \times \bar{\textbf{l}}_w + \bar{\textbf{a}}_8 \circ \bar{\textbf{l}}_w + \bar{\textbf{a}}_w \circ \bar{\textbf{l}}_8)
+ k_a f_s^2 ( \bar{\textbf{a}}_s \times \bar{\textbf{l}}_s + \bar{\textbf{a}}_{12} \circ \bar{\textbf{l}}_s + \bar{\textbf{a}}_s \circ \bar{\textbf{l}}_{12})
\nonumber
\\
& + k_a f_\alpha^2 ( \bar{\textbf{a}}_\alpha \times \bar{\textbf{l}}_\alpha + \bar{\textbf{a}}_{16} \circ \bar{\textbf{l}}_\alpha + \bar{\textbf{a}}_\alpha \circ \bar{\textbf{l}}_{16})
+ k_a f_\beta^2 ( \bar{\textbf{a}}_\beta \times \bar{\textbf{l}}_\beta + \bar{\textbf{a}}_{20} \circ \bar{\textbf{l}}_\beta + \bar{\textbf{a}}_\beta \circ \bar{\textbf{l}}_{20})
\nonumber
\\
& + k_a f_\gamma^2 ( \bar{\textbf{a}}_\gamma \times \bar{\textbf{l}}_\gamma + \bar{\textbf{a}}_{24} \circ \bar{\textbf{l}}_\gamma + \bar{\textbf{a}}_\gamma \circ \bar{\textbf{l}}_{24})
+ k_a f_\delta^2 ( \bar{\textbf{a}}_\delta \times \bar{\textbf{l}}_\delta + \bar{\textbf{a}}_{28} \circ \bar{\textbf{l}}_\delta + \bar{\textbf{a}}_\delta \circ \bar{\textbf{l}}_{28})
\nonumber
\\
& + k_x ( \bar{\textbf{x}}_g \times \bar{\textbf{l}}_g + \bar{\textbf{x}}_g \bar{l}_0 + \bar{x}_0 \bar{\textbf{l}}_g ) + k_x f_e^2 ( \bar{\textbf{x}}_e \times \bar{\textbf{l}}_e + \bar{\textbf{x}}_4 \circ \bar{\textbf{l}}_e + \bar{\textbf{x}}_e \circ \bar{\textbf{l}}_4)
\nonumber
\\
& + k_x f_w^2 ( \bar{\textbf{x}}_w \times \bar{\textbf{l}}_w + \bar{\textbf{x}}_8 \circ \bar{\textbf{l}}_w + \bar{\textbf{x}}_w \circ \bar{\textbf{l}}_8)
+ k_x f_s^2 ( \bar{\textbf{x}}_s \times \bar{\textbf{l}}_s + \bar{\textbf{x}}_{12} \circ \bar{\textbf{l}}_s + \bar{\textbf{x}}_s \circ \bar{\textbf{l}}_{12})
\nonumber
\\
& + k_x f_\alpha^2 ( \bar{\textbf{x}}_\alpha \times \bar{\textbf{l}}_\alpha + \bar{\textbf{x}}_{16} \circ \bar{\textbf{l}}_\alpha + \bar{\textbf{x}}_\alpha \circ \bar{\textbf{l}}_{16})
+ k_x f_\beta^2 ( \bar{\textbf{x}}_\beta \times \bar{\textbf{l}}_\beta + \bar{\textbf{x}}_{20} \circ \bar{\textbf{l}}_\beta + \bar{\textbf{x}}_\beta \circ \bar{\textbf{l}}_{20})
\nonumber
\\
& + k_x f_\gamma^2 ( \bar{\textbf{x}}_\gamma \times \bar{\textbf{l}}_\gamma + \bar{\textbf{x}}_{24} \circ \bar{\textbf{l}}_\gamma + \bar{\textbf{x}}_\gamma \circ \bar{\textbf{l}}_{24})
+ k_x f_\delta^2 ( \bar{\textbf{x}}_\delta \times \bar{\textbf{l}}_\delta + \bar{\textbf{x}}_{28} \circ \bar{\textbf{l}}_\delta + \bar{\textbf{x}}_\delta \circ \bar{\textbf{l}}_{28})
\nonumber
\\
& + k_s ( \bar{\textbf{s}}_g \times \bar{\textbf{l}}_g + \bar{\textbf{s}}_g \bar{l}_0 + \bar{s}_0 \bar{\textbf{l}}_g ) + k_s f_e^2 ( \bar{\textbf{s}}_e \times \bar{\textbf{l}}_e + \bar{\textbf{s}}_4 \circ \bar{\textbf{l}}_e + \bar{\textbf{s}}_e \circ \bar{\textbf{l}}_4)
\nonumber
\\
& + k_s f_w^2 ( \bar{\textbf{s}}_w \times \bar{\textbf{l}}_w + \bar{\textbf{s}}_8 \circ \bar{\textbf{l}}_w + \bar{\textbf{s}}_w \circ \bar{\textbf{l}}_8)
+ k_s f_s^2 ( \bar{\textbf{s}}_s \times \bar{\textbf{l}}_s + \bar{\textbf{s}}_{12} \circ \bar{\textbf{l}}_s + \bar{\textbf{s}}_s \circ \bar{\textbf{l}}_{12})
\nonumber
\\
& + k_s f_\alpha^2 ( \bar{\textbf{s}}_\alpha \times \bar{\textbf{l}}_\alpha + \bar{\textbf{s}}_{16} \circ \bar{\textbf{l}}_\alpha + \bar{\textbf{s}}_\alpha \circ \bar{\textbf{l}}_{16})
+ k_s f_\beta^2 ( \bar{\textbf{s}}_\beta \times \bar{\textbf{l}}_\beta + \bar{\textbf{s}}_{20} \circ \bar{\textbf{l}}_\beta + \bar{\textbf{s}}_\beta \circ \bar{\textbf{l}}_{20})
\nonumber
\\
& + k_s f_\gamma^2 ( \bar{\textbf{s}}_\gamma \times \bar{\textbf{l}}_\gamma + \bar{\textbf{s}}_{24} \circ \bar{\textbf{l}}_\gamma + \bar{\textbf{s}}_\gamma \circ \bar{\textbf{l}}_{24})
+ k_s f_\delta^2 ( \bar{\textbf{s}}_\delta \times \bar{\textbf{l}}_\delta + \bar{\textbf{s}}_{28} \circ \bar{\textbf{l}}_\delta + \bar{\textbf{s}}_\delta \circ \bar{\textbf{l}}_{28})
\nonumber
\\
& + k_b ( \bar{\textbf{h}}_g \times \bar{\textbf{l}}_g + \bar{\textbf{h}}_g \bar{l}_0 ) + k_b f_e^2 ( \bar{\textbf{h}}_e \times \bar{\textbf{l}}_e + \bar{\textbf{h}}_e \circ \bar{\textbf{l}}_4)
\nonumber
\\
& + k_b f_w^2 ( \bar{\textbf{h}}_w \times \bar{\textbf{l}}_w + \bar{\textbf{h}}_w \circ \bar{\textbf{l}}_8)
+ k_b f_s^2 ( \bar{\textbf{h}}_s \times \bar{\textbf{l}}_s + \bar{\textbf{h}}_s \circ \bar{\textbf{l}}_{12})
\nonumber
\\
& + k_b f_\alpha^2 ( \bar{\textbf{h}}_\alpha \times \bar{\textbf{l}}_\alpha + \bar{\textbf{h}}_\alpha \circ \bar{\textbf{l}}_{16})
+ k_b f_\beta^2 ( \bar{\textbf{h}}_\beta \times \bar{\textbf{l}}_\beta + \bar{\textbf{h}}_\beta \circ \bar{\textbf{l}}_{20})
\nonumber
\\
& + k_b f_\gamma^2 ( \bar{\textbf{h}}_\gamma \times \bar{\textbf{l}}_\gamma + \bar{\textbf{h}}_\gamma \circ \bar{\textbf{l}}_{24})
+ k_b f_\delta^2 ( \bar{\textbf{h}}_\delta \times \bar{\textbf{l}}_\delta + \bar{\textbf{h}}_\delta \circ \bar{\textbf{l}}_{28})
+  2 k_l \bar{l}_0 \bar{\textbf{l}}_g ~,
\end{align}
where the above includes some new terms of the torque density.

\subsection{Force}

In the trigintaduonion compounding space with the operator $(\lozenge_{32} + k_x \mathbb{\bar{X}} + k_a \mathbb{\bar{A}} + k_b \mathbb{\bar{B}} + k_s \mathbb{\bar{S}} + k_l \mathbb{\bar{L}} + k_w \mathbb{\bar{W}})$, the trigintaduonion force-power density $\mathbb{\bar{N}}$ is defined as follows,
\begin{eqnarray}
\mathbb{\bar{N}} = v_0 (\lozenge_{32} + k_x \mathbb{\bar{X}} + k_a \mathbb{\bar{A}} + k_b \mathbb{\bar{B}} + k_s \mathbb{\bar{S}} + k_l \mathbb{\bar{L}} + k_w \mathbb{\bar{W}})^* \circ \mathbb{\bar{W}}~,
\end{eqnarray}
where the power density is $\bar{f}_0 = - \bar{n}_0/(2 v_0)$, and the vectorial part is $\bar{\textbf{n}} = \Sigma (\bar{n}_j \emph{\textbf{i}}_j )$. $\mathbb{\bar{N}} = \Sigma (\bar{n}_i \emph{\textbf{i}}_i + f_e \bar{n}_{(i+4)} \emph{\textbf{i}}_{(i+4)} + f_w \bar{n}_{(i+8)} \emph{\textbf{i}}_{(i+8)} + f_s \bar{n}_{(i+12)} \emph{\textbf{i}}_{(i+12)}) + f_\alpha \bar{n}_{(i+16)} \emph{\textbf{i}}_{(i+16)}) + f_\beta \bar{n}_{(i+20)} \emph{\textbf{i}}_{(i+20)}) + f_\gamma \bar{n}_{(i+24)} \emph{\textbf{i}}_{(i+24)}) + f_\delta \bar{n}_{(i+28)} \emph{\textbf{i}}_{(i+28)}) $.

Further expressing the scalar $\bar{n}_0$ of the trigintaduonion force-power density $\mathbb{\bar{N}}$ as
\begin{align}
\bar{n}_0 / v_0 = &~
k_a \left\{ \bar{a}_0 \bar{w}_0 + \bar{\textbf{a}}^*_g \cdot \bar{\textbf{w}}_g + f_e^2 ( \bar{\textbf{a}}^*_4 \circ \bar{\textbf{w}}_4 + \bar{\textbf{a}}^*_e \cdot \bar{\textbf{w}}_e )  \right\}
\nonumber\\
&
+ k_a \left\{ f_w^2 ( \bar{\textbf{a}}^*_8 \circ \bar{\textbf{w}}_8 + \bar{\textbf{a}}^*_w \cdot \bar{\textbf{w}}_w ) +  f_s^2 ( \bar{\textbf{a}}^*_{12} \circ \bar{\textbf{w}}_{12} + \bar{\textbf{a}}^*_s \cdot \bar{\textbf{w}}_s) \right\}
\nonumber\\
&
+ k_a \left\{ f_\alpha^2 ( \bar{\textbf{a}}^*_{16} \circ \bar{\textbf{w}}_{16} + \bar{\textbf{a}}^*_\alpha \cdot \bar{\textbf{w}}_\alpha )
+ f_\beta^2 ( \bar{\textbf{a}}^*_{20} \circ \bar{\textbf{w}}_{20} + \bar{\textbf{a}}^*_\beta \cdot \bar{\textbf{w}}_\beta) \right\}
\nonumber\\
&
+ k_a \left\{ f_\gamma^2 ( \bar{\textbf{a}}^*_{24} \circ \bar{\textbf{w}}_{24} + \bar{\textbf{a}}^*_\gamma \cdot \bar{\textbf{w}}_\gamma)
+ f_\delta^2 ( \bar{\textbf{a}}^*_{28} \circ \bar{\textbf{w}}_{28} + \bar{\textbf{a}}^*_\delta \cdot \bar{\textbf{w}}_\delta ) \right\}
\nonumber\\
&
+ k_x \left\{ \bar{x}_0 \bar{w}_0 + \bar{\textbf{x}}^*_g \cdot \bar{\textbf{w}}_g + f_e^2 ( \bar{\textbf{x}}^*_4 \circ \bar{\textbf{w}}_4 + \bar{\textbf{x}}^*_e \cdot \bar{\textbf{w}}_e )  \right\}
\nonumber\\
&
+ k_x \left\{ f_w^2 ( \bar{\textbf{x}}^*_8 \circ \bar{\textbf{w}}_8 + \bar{\textbf{x}}^*_w \cdot \bar{\textbf{w}}_w )+ f_s^2 ( \bar{\textbf{x}}^*_{12} \circ \bar{\textbf{w}}_{12} + \bar{\textbf{x}}^*_s \cdot \bar{\textbf{w}}_s) \right\}
\nonumber\\
&
+ k_x \left\{ f_\alpha^2 ( \bar{\textbf{x}}^*_{16} \circ \bar{\textbf{w}}_{16} + \bar{\textbf{x}}^*_\alpha \cdot \bar{\textbf{w}}_\alpha)
+ f_\beta^2 ( \bar{\textbf{x}}^*_{20} \circ \bar{\textbf{w}}_{20} + \bar{\textbf{x}}^*_\beta \cdot \bar{\textbf{w}}_\beta) \right\}
\nonumber\\
&
+ k_x \left\{ f_\gamma^2 ( \bar{\textbf{x}}^*_{24} \circ \bar{\textbf{w}}_{24} + \bar{\textbf{x}}^*_\gamma \cdot \bar{\textbf{w}}_\gamma)
+ f_\delta^2 ( \bar{\textbf{x}}^*_{28} \circ \bar{\textbf{w}}_{28} + \bar{\textbf{x}}^*_\delta \cdot \bar{\textbf{w}}_\delta) \right\}
\nonumber\\
&
+ k_s \left\{ \bar{s}_0 \bar{w}_0 + \bar{\textbf{s}}^*_g \cdot \bar{\textbf{w}}_g + f_e^2 ( \bar{\textbf{s}}^*_4 \circ \bar{\textbf{w}}_4 + \bar{\textbf{s}}^*_e \cdot \bar{\textbf{w}}_e )  \right\}
\nonumber\\
&
+ k_s \left\{ f_w^2 ( \bar{\textbf{s}}^*_8 \circ \bar{\textbf{w}}_8 + \bar{\textbf{s}}^*_w \cdot \bar{\textbf{w}}_w )+ f_s^2 ( \bar{\textbf{s}}^*_{12} \circ
\bar{\textbf{w}}_{12} + \bar{\textbf{s}}^*_s \cdot \bar{\textbf{w}}_s) \right\}
\nonumber\\
&
+ k_s \left\{ f_\alpha^2 ( \bar{\textbf{s}}^*_{16} \circ \bar{\textbf{w}}_{16} + \bar{\textbf{s}}^*_\alpha \cdot \bar{\textbf{w}}_\alpha)
+ f_\beta^2 ( \bar{\textbf{s}}^*_{20} \circ \bar{\textbf{w}}_{20} + \bar{\textbf{s}}^*_\beta \cdot \bar{\textbf{w}}_\beta) \right\}
\nonumber\\
&
+ k_s \left\{ f_\gamma^2 ( \bar{\textbf{s}}^*_{24} \circ \bar{\textbf{w}}_{24} + \bar{\textbf{s}}^*_\gamma \cdot \bar{\textbf{w}}_\gamma)
+ f_\delta^2 ( \bar{\textbf{s}}^*_{28} \circ \bar{\textbf{w}}_{28} + \bar{\textbf{s}}^*_\delta \cdot \bar{\textbf{w}}_\delta) \right\}
\nonumber\\
&
+ k_l \left\{ \bar{l}_0 \bar{w}_0 + \bar{\textbf{l}}^*_g \cdot \bar{\textbf{w}}_g + f_e^2 ( \bar{\textbf{l}}^*_4 \circ \bar{\textbf{w}}_4 + \bar{\textbf{l}}^*_e \cdot \bar{\textbf{w}}_e )  \right\}
\nonumber\\
&
+ k_l \left\{ f_w^2 ( \bar{\textbf{l}}^*_8 \circ \bar{\textbf{w}}_8 + \bar{\textbf{l}}^*_w \cdot \bar{\textbf{w}}_w ) + f_s^2 ( \bar{\textbf{l}}^*_{12} \circ \bar{\textbf{w}}_{12} + \bar{\textbf{l}}^*_s \cdot \bar{\textbf{w}}_s) \right\}
\nonumber\\
&
+ k_l \left\{ f_\alpha^2 ( \bar{\textbf{l}}^*_{16} \circ \bar{\textbf{w}}_{16} + \bar{\textbf{l}}^*_\alpha \cdot \bar{\textbf{w}}_\alpha)
+ f_\beta^2 ( \bar{\textbf{l}}^*_{20} \circ \bar{\textbf{w}}_{20} + \bar{\textbf{l}}^*_\beta \cdot \bar{\textbf{w}}_\beta) \right\}
\nonumber\\
&
+ k_l \left\{ f_\gamma^2 ( \bar{\textbf{l}}^*_{24} \circ \bar{\textbf{w}}_{24} + \bar{\textbf{l}}^*_\gamma \cdot \bar{\textbf{w}}_\gamma)
+ f_\delta^2 ( \bar{\textbf{l}}^*_{28} \circ \bar{\textbf{w}}_{28} + \bar{\textbf{l}}^*_\delta \cdot \bar{\textbf{w}}_\delta) \right\}
\nonumber\\
&
+ k_w \left\{ \bar{w}_0 \bar{w}_0 + \bar{\textbf{w}}^*_g \cdot \bar{\textbf{w}}_g + f_e^2 ( \bar{\textbf{w}}^*_4 \circ \bar{\textbf{w}}_4 + \bar{\textbf{w}}^*_e \cdot \bar{\textbf{w}}_e )  \right\}
\nonumber\\
&
+ k_w \left\{ f_w^2 ( \bar{\textbf{w}}^*_8 \circ \bar{\textbf{w}}_8 + \bar{\textbf{w}}^*_w \cdot \bar{\textbf{w}}_w ) + f_s^2 ( \bar{\textbf{w}}^*_{12} \circ \bar{\textbf{w}}_{12} + \bar{\textbf{w}}^*_s \cdot \bar{\textbf{w}}_s) \right\}
\nonumber\\
&
+ k_w \left\{ f_\alpha^2 ( \bar{\textbf{w}}^*_{16} \circ \bar{\textbf{w}}_{16} + \bar{\textbf{w}}^*_\alpha \cdot \bar{\textbf{w}}_\alpha)
+ f_\beta^2 ( \bar{\textbf{w}}^*_{20} \circ \bar{\textbf{w}}_{20} + \bar{\textbf{w}}^*_\beta \cdot \bar{\textbf{w}}_\beta) \right\}
\nonumber\\
&
+ k_w \left\{ f_\gamma^2 ( \bar{\textbf{w}}^*_{24} \circ \bar{\textbf{w}}_{24} + \bar{\textbf{w}}^*_\gamma \cdot \bar{\textbf{w}}_\gamma)
+ f_\delta^2 ( \bar{\textbf{w}}^*_{28} \circ \bar{\textbf{w}}_{28} + \bar{\textbf{w}}^*_\delta \cdot \bar{\textbf{w}}_\delta) \right\}
\nonumber\\
&
+ k_b ( \bar{\textbf{h}}^*_g \cdot \bar{\textbf{w}}_g + f_e^2 \bar{\textbf{h}}^*_e \cdot \bar{\textbf{w}}_e  + f_w^2 \bar{\textbf{h}}^*_w \cdot \bar{\textbf{w}}_w + f_s^2 \bar{\textbf{h}}^*_s \cdot \bar{\textbf{w}}_s )
\nonumber\\
&
+ k_b ( f_\alpha^2 \bar{\textbf{h}}^*_\alpha \cdot \bar{\textbf{w}}_\alpha + f_\beta^2 \bar{\textbf{h}}^*_\beta \cdot \bar{\textbf{w}}_\beta + f_\gamma^2 \bar{\textbf{h}}^*_\gamma \cdot \bar{\textbf{w}}_\gamma + f_\delta^2 \bar{\textbf{h}}^*_\delta \cdot \bar{\textbf{w}}_\delta )
\nonumber\\
&
+ \partial_0 \bar{w}_0 + \nabla^*_g \cdot \bar{\textbf{w}}_g + d_e f_e (\emph{\textbf{i}}^*_4 \circ \partial_4 \bar{\textbf{w}}_4 + \nabla^*_e \cdot \bar{\textbf{w}}_e)
\nonumber\\
&
+ d_w f_w (\emph{\textbf{i}}^*_8 \circ \partial_8 \bar{\textbf{w}}_8 + \nabla^*_w \cdot \bar{\textbf{w}}_w) + d_s f_s (\emph{\textbf{i}}^*_{12} \circ \partial_{12} \bar{\textbf{w}}_{12} + \nabla^*_s \cdot \bar{\textbf{w}}_s)
\nonumber\\
&
+ d_\alpha f_\alpha (\emph{\textbf{i}}^*_{16} \circ \partial_{16} \bar{\textbf{w}}_{16} + \nabla^*_\alpha \cdot \bar{\textbf{w}}_\alpha)
+ d_\beta f_\beta (\emph{\textbf{i}}^*_{20} \circ \partial_{20} \bar{\textbf{w}}_{20} + \nabla^*_\beta \cdot \bar{\textbf{w}}_\beta)
\nonumber\\
&
+ d_\gamma f_\gamma (\emph{\textbf{i}}^*_{24} \circ \partial_{24} \bar{\textbf{w}}_{24} + \nabla^*_\gamma \cdot \bar{\textbf{w}}_\gamma)
+ d_\delta f_\delta (\emph{\textbf{i}}^*_{28} \circ \partial_{28} \bar{\textbf{w}}_{28} + \nabla^*_\delta \cdot \bar{\textbf{w}}_\delta) ~.
\end{align}

In the eight fields with the trigintaduonion operator $(\lozenge_{16} + k_x \mathbb{\bar{X}} + k_a \mathbb{\bar{A}} + k_b \mathbb{\bar{B}} + k_s \mathbb{\bar{S}} + k_l \mathbb{\bar{L}} + k_w \mathbb{\bar{W}})$, the force density $\bar{\textbf{f}} = - \bar{\textbf{n}} / (2 v_0)$ can be defined from the vectorial part $\bar{\textbf{n}}$ of the trigintaduonion force-power density $\mathbb{\bar{N}}$ ,
\begin{align}
\bar{\textbf{w}} / v_0 = &~ ( \partial_0 \bar{\textbf{w}}_g + \nabla^*_g \bar{w}_0 + \nabla^*_g \times \bar{\textbf{w}}_g )
+ d_e f_e ( \emph{\textbf{i}}^*_4 \circ \partial_4 \bar{\textbf{w}}_e + \nabla^*_e \circ \bar{\textbf{w}}_4 + \nabla^*_e \times \bar{\textbf{w}}_e )
\nonumber
\\
& + d_w f_w ( \emph{\textbf{i}}^*_8 \circ \partial_8 \bar{\textbf{w}}_w + \nabla^*_w \circ \bar{\textbf{w}}_8 + \nabla^*_w \times \bar{\textbf{w}}_w )
+ d_s f_s ( \emph{\textbf{i}}^*_{12} \circ \partial_{12} \bar{\textbf{w}}_s + \nabla^*_s \circ \bar{\textbf{w}}_{12} + \nabla^*_s \times \bar{\textbf{w}}_s )
\nonumber
\\
& + d_\alpha f_\alpha ( \emph{\textbf{i}}^*_{16} \circ \partial_{16} \bar{\textbf{w}}_\alpha + \nabla^*_\alpha \circ \bar{\textbf{w}}_{16} + \nabla^*_\alpha \times \bar{\textbf{w}}_\alpha )
+ d_\beta f_\beta ( \emph{\textbf{i}}^*_{20} \circ \partial_{20} \bar{\textbf{w}}_\beta + \nabla^*_\beta \circ \bar{\textbf{w}}_{20} + \nabla^*_\beta \times \bar{\textbf{w}}_\beta )
\nonumber
\\
& + d_\gamma f_\gamma ( \emph{\textbf{i}}^*_{24} \circ \partial_{24} \bar{\textbf{w}}_\gamma + \nabla^*_\gamma \circ \bar{\textbf{w}}_{24} + \nabla^*_\gamma \times \bar{\textbf{w}}_\gamma )
+ d_\delta f_\delta ( \emph{\textbf{i}}^*_{28} \circ \partial_{28} \bar{\textbf{w}}_\delta + \nabla^*_\delta \circ \bar{\textbf{w}}_{28} + \nabla^*_\delta \times \bar{\textbf{w}}_\delta )
\nonumber
\\
& + k_a ( \bar{\textbf{a}}^*_g \times \bar{\textbf{w}}_g + \bar{\textbf{a}}^*_g \bar{w}_0 + \bar{a}_0 \bar{\textbf{w}}_g ) + k_a f_e^2 ( \bar{\textbf{a}}^*_e \times \bar{\textbf{w}}_e + \bar{\textbf{a}}^*_4 \circ \bar{\textbf{w}}_e + \bar{\textbf{a}}^*_e \circ \bar{\textbf{w}}_4)
\nonumber
\\
& + k_a f_w^2 ( \bar{\textbf{a}}^*_w \times \bar{\textbf{w}}_w + \bar{\textbf{a}}^*_8 \circ \bar{\textbf{w}}_w + \bar{\textbf{a}}^*_w \circ \bar{\textbf{w}}_8)
+ k_a f_s^2 ( \bar{\textbf{a}}^*_s \times \bar{\textbf{w}}_s + \bar{\textbf{a}}^*_{12} \circ \bar{\textbf{w}}_s + \bar{\textbf{a}}^*_s \circ \bar{\textbf{w}}_{12})
\nonumber
\\
& + k_a f_\alpha^2 ( \bar{\textbf{a}}^*_\alpha \times \bar{\textbf{w}}_\alpha + \bar{\textbf{a}}^*_{16} \circ \bar{\textbf{w}}_\alpha + \bar{\textbf{a}}^*_\alpha \circ \bar{\textbf{w}}_{16})
+ k_a f_\beta^2 ( \bar{\textbf{a}}^*_\beta \times \bar{\textbf{w}}_\beta + \bar{\textbf{a}}^*_{20} \circ \bar{\textbf{w}}_\beta + \bar{\textbf{a}}^*_\beta \circ \bar{\textbf{w}}_{20})
\nonumber
\\
& + k_a f_\gamma^2 ( \bar{\textbf{a}}^*_\gamma \times \bar{\textbf{w}}_\gamma + \bar{\textbf{a}}^*_{24} \circ \bar{\textbf{w}}_\gamma + \bar{\textbf{a}}^*_\gamma \circ \bar{\textbf{w}}_{24})
+ k_a f_\delta^2 ( \bar{\textbf{a}}^*_\delta \times \bar{\textbf{w}}_\delta + \bar{\textbf{a}}^*_{28} \circ \bar{\textbf{w}}_\delta + \bar{\textbf{a}}^*_\delta \circ \bar{\textbf{w}}_{28})
\nonumber
\\
& + k_x ( \bar{\textbf{x}}^*_g \times \bar{\textbf{w}}_g + \bar{\textbf{x}}^*_g \bar{w}_0 + \bar{x}_0 \bar{\textbf{w}}_g ) + k_x f_e^2 ( \bar{\textbf{x}}^*_e \times \bar{\textbf{w}}_e + \bar{\textbf{x}}^*_4 \circ \bar{\textbf{w}}_e + \bar{\textbf{x}}^*_e \circ \bar{\textbf{w}}_4)
\nonumber
\\
& + k_x f_w^2 ( \bar{\textbf{x}}^*_w \times \bar{\textbf{w}}_w + \bar{\textbf{x}}^*_8 \circ \bar{\textbf{w}}_w + \bar{\textbf{x}}^*_w \circ \bar{\textbf{w}}_8)
+ k_x f_s^2 ( \bar{\textbf{x}}^*_s \times \bar{\textbf{w}}_s + \bar{\textbf{x}}^*_{12} \circ \bar{\textbf{w}}_s + \bar{\textbf{x}}^*_s \circ \bar{\textbf{w}}_{12})
\nonumber
\\
& + k_x f_\alpha^2 ( \bar{\textbf{x}}^*_\alpha \times \bar{\textbf{w}}_\alpha + \bar{\textbf{x}}^*_{16} \circ \bar{\textbf{w}}_\alpha + \bar{\textbf{x}}^*_\alpha \circ \bar{\textbf{w}}_{16})
+ k_x f_\beta^2 ( \bar{\textbf{x}}^*_\beta \times \bar{\textbf{w}}_\beta + \bar{\textbf{x}}^*_{20} \circ \bar{\textbf{w}}_\beta + \bar{\textbf{x}}^*_\beta \circ \bar{\textbf{w}}_{20})
\nonumber
\\
& + k_x f_\gamma^2 ( \bar{\textbf{x}}^*_\gamma \times \bar{\textbf{w}}_\gamma + \bar{\textbf{x}}^*_{24} \circ \bar{\textbf{w}}_\gamma + \bar{\textbf{x}}^*_\gamma \circ \bar{\textbf{w}}_{24})
+ k_x f_\delta^2 ( \bar{\textbf{x}}^*_\delta \times \bar{\textbf{w}}_\delta + \bar{\textbf{x}}^*_{28} \circ \bar{\textbf{w}}_\delta + \bar{\textbf{x}}^*_\delta \circ \bar{\textbf{w}}_{28})
\nonumber
\\
& + k_s ( \bar{\textbf{s}}^*_g \times \bar{\textbf{w}}_g + \bar{\textbf{s}}^*_g \bar{w}_0 + \bar{s}_0 \bar{\textbf{w}}_g ) + k_s f_e^2 ( \bar{\textbf{s}}^*_e \times \bar{\textbf{w}}_e + \bar{\textbf{s}}^*_4 \circ \bar{\textbf{w}}_e + \bar{\textbf{s}}^*_e \circ \bar{\textbf{w}}_4)
\nonumber
\\
& + k_s f_w^2 ( \bar{\textbf{s}}^*_w \times \bar{\textbf{w}}_w + \bar{\textbf{s}}^*_8 \circ \bar{\textbf{w}}_w + \bar{\textbf{s}}^*_w \circ \bar{\textbf{w}}_8)
+ k_s f_s^2 ( \bar{\textbf{s}}^*_s \times \bar{\textbf{w}}_s + \bar{\textbf{s}}^*_{12} \circ \bar{\textbf{w}}_s + \bar{\textbf{s}}^*_s \circ \bar{\textbf{w}}_{12})
\nonumber
\\
& + k_s f_\alpha^2 ( \bar{\textbf{s}}^*_\alpha \times \bar{\textbf{w}}_\alpha + \bar{\textbf{s}}^*_{16} \circ \bar{\textbf{w}}_\alpha + \bar{\textbf{s}}^*_\alpha \circ \bar{\textbf{w}}_{16})
+ k_s f_\beta^2 ( \bar{\textbf{s}}^*_\beta \times \bar{\textbf{w}}_\beta + \bar{\textbf{s}}^*_{20} \circ \bar{\textbf{w}}_\beta + \bar{\textbf{s}}^*_\beta \circ \bar{\textbf{w}}_{20})
\nonumber
\\
& + k_s f_\gamma^2 ( \bar{\textbf{s}}^*_\gamma \times \bar{\textbf{w}}_\gamma + \bar{\textbf{s}}^*_{24} \circ \bar{\textbf{w}}_\gamma + \bar{\textbf{s}}^*_\gamma \circ \bar{\textbf{w}}_{24})
+ k_s f_\delta^2 ( \bar{\textbf{s}}^*_\delta \times \bar{\textbf{w}}_\delta + \bar{\textbf{s}}^*_{28} \circ \bar{\textbf{w}}_\delta + \bar{\textbf{s}}^*_\delta \circ \bar{\textbf{w}}_{28})
\nonumber
\\
& + k_l ( \bar{\textbf{l}}^*_g \times \bar{\textbf{w}}_g + \bar{\textbf{l}}^*_g \bar{w}_0 + \bar{s}_0 \bar{\textbf{w}}_g ) + k_s f_e^2 ( \bar{\textbf{l}}^*_e \times \bar{\textbf{w}}_e + \bar{\textbf{l}}^*_4 \circ \bar{\textbf{w}}_e + \bar{\textbf{l}}^*_e \circ \bar{\textbf{w}}_4)
\nonumber
\\
& + k_l f_w^2 ( \bar{\textbf{l}}^*_w \times \bar{\textbf{w}}_w + \bar{\textbf{l}}^*_8 \circ \bar{\textbf{w}}_w + \bar{\textbf{l}}^*_w \circ \bar{\textbf{w}}_8)
+ k_s f_s^2 ( \bar{\textbf{l}}^*_s \times \bar{\textbf{w}}_s + \bar{\textbf{l}}^*_{12} \circ \bar{\textbf{w}}_s + \bar{\textbf{l}}^*_s \circ \bar{\textbf{w}}_{12})
\nonumber
\\
& + k_l f_\alpha^2 ( \bar{\textbf{l}}^*_\alpha \times \bar{\textbf{w}}_\alpha + \bar{\textbf{l}}^*_{16} \circ \bar{\textbf{w}}_\alpha + \bar{\textbf{l}}^*_\alpha \circ \bar{\textbf{w}}_{16})
+ k_s f_\beta^2 ( \bar{\textbf{l}}^*_\beta \times \bar{\textbf{w}}_\beta + \bar{\textbf{l}}^*_{20} \circ \bar{\textbf{w}}_\beta + \bar{\textbf{l}}^*_\beta \circ \bar{\textbf{w}}_{20})
\nonumber
\\
& + k_l f_\gamma^2 ( \bar{\textbf{l}}^*_\gamma \times \bar{\textbf{w}}_\gamma + \bar{\textbf{l}}^*_{24} \circ \bar{\textbf{w}}_\gamma + \bar{\textbf{l}}^*_\gamma \circ \bar{\textbf{w}}_{24})
+ k_s f_\delta^2 ( \bar{\textbf{l}}^*_\delta \times \bar{\textbf{w}}_\delta + \bar{\textbf{l}}^*_{28} \circ \bar{\textbf{w}}_\delta + \bar{\textbf{l}}^*_\delta \circ \bar{\textbf{w}}_{28})
\nonumber
\\
& + k_b ( \bar{\textbf{h}}^*_g \times \bar{\textbf{w}}_g + \bar{\textbf{h}}^*_g \bar{w}_0 ) + k_b f_e^2 ( \bar{\textbf{h}}^*_e \times \bar{\textbf{w}}_e + \bar{\textbf{h}}^*_e \circ \bar{\textbf{w}}_4)
\nonumber
\\
& + k_b f_w^2 ( \bar{\textbf{h}}^*_w \times \bar{\textbf{w}}_w + \bar{\textbf{h}}^*_w \circ \bar{\textbf{w}}_8)
+ k_b f_s^2 ( \bar{\textbf{h}}^*_s \times \bar{\textbf{w}}_s + \bar{\textbf{h}}^*_s \circ \bar{\textbf{w}}_{12})
\nonumber
\\
& + k_b f_\alpha^2 ( \bar{\textbf{h}}^*_\alpha \times \bar{\textbf{w}}_\alpha + \bar{\textbf{h}}^*_\alpha \circ \bar{\textbf{w}}_{16})
+ k_b f_\beta^2 ( \bar{\textbf{h}}^*_\beta \times \bar{\textbf{w}}_\beta + \bar{\textbf{h}}^*_\beta \circ \bar{\textbf{w}}_{20})
\nonumber
\\
& + k_b f_\gamma^2 ( \bar{\textbf{h}}^*_\gamma \times \bar{\textbf{w}}_\gamma + \bar{\textbf{h}}^*_\gamma \circ \bar{\textbf{w}}_{24})
+ k_b f_\delta^2 ( \bar{\textbf{h}}^*_\delta \times \bar{\textbf{w}}_\delta + \bar{\textbf{h}}^*_\delta \circ \bar{\textbf{w}}_{28}) ~,
\end{align}
where the force density $\bar{\textbf{f}}$ includes that of the inertial force, gravitational force, gradient of energy, Lorentz force, and the interacting force between the dipole moment with the fields, as well as the unknown forces regarding the $\alpha$, $\beta$, $\gamma$, and $\delta$ fields etc. The above force is much more complex than that in the classical field theory, and includes more new force terms related to the gradient of energy, the field potential, and the angular velocity etc.

\begin{table}[h]
\caption{Some field sources of the eight fields with their adjoint fields.}
\label{tab:table3}
\centering
\begin{tabular}{llll}
\hline\hline
$ sources $                              & $ fields $                                 & $ descriptions $                  & $ characterictics $    \\
\hline

$\mathbb{\bar{S}}_{gg}$                  & gravitational~field                        & linear~momentum                   & gravitation            \\
$\mathbb{\bar{S}}_{ee}$                  & electromagnetic~adjoint~field              & adjoint~electric~current          & gravitation            \\
$\mathbb{\bar{S}}_{ww}$                  & weak~nuclear~adjoint~field                 & adjoint~weak~nuclear~current      & gravitation            \\
$\mathbb{\bar{S}}_{ss}$                  & strong~nuclear~adjoint~field               & adjoint~strong~nuclear~current    & gravitation            \\

$\mathbb{\bar{S}}_{\alpha \alpha}$       & $\alpha$~adjoint~field                     & adjoint~$\alpha$~current          & gravitation            \\
$\mathbb{\bar{S}}_{\beta \beta}$         & $\beta$~adjoint~field                      & adjoint~$\beta$~current           & gravitation            \\
$\mathbb{\bar{S}}_{\gamma \gamma}$       & $\gamma$~adjoint~field                     & adjoint~$\gamma$~current          & gravitation            \\
$\mathbb{\bar{S}}_{\delta \delta}$       & $\delta$~adjoint~field                     & adjoint~$\delta$~current          & gravitation            \\

\hline

$\mathbb{\bar{S}}_{ge}$                  & gravitational~adjoint~field                & adjoint~linear~momentum           & electromagnetism       \\
$\mathbb{\bar{S}}_{eg}$                  & electromagnetic~field                      & electric~current                  & electromagnetism       \\
$\mathbb{\bar{S}}_{ws}$                  & weak~nuclear~adjoint~field                 & adjoint~weak~nuclear~current      & electromagnetism       \\
$\mathbb{\bar{S}}_{sw}$                  & strong~nuclear~adjoint~field               & adjoint~strong~nuclear~current    & electromagnetism       \\

$\mathbb{\bar{S}}_{\alpha \beta}$        & $\alpha$~adjoint~field                     & adjoint~$\alpha$~current          & electromagnetism       \\
$\mathbb{\bar{S}}_{\beta \alpha}$        & $\beta$~adjoint~field                      & adjoint~$\beta$~current           & electromagnetism       \\
$\mathbb{\bar{S}}_{\gamma \delta}$       & $\gamma$~adjoint~field                     & adjoint~$\gamma$~current          & electromagnetism       \\
$\mathbb{\bar{S}}_{\delta \gamma}$       & $\delta$~adjoint~field                     & adjoint~$\delta$~current          & electromagnetism       \\

\hline

$\mathbb{\bar{S}}_{gw}$                  & gravitational~adjoint~field                & adjoint~linear~momentum           & weak~nuclear~force     \\
$\mathbb{\bar{S}}_{es}$                  & electromagnetic~adjoint~field              & adjoint~electric~current          & weak~nuclear~force     \\
$\mathbb{\bar{S}}_{wg}$                  & weak~nuclear~field                         & weak~nuclear~current              & weak~nuclear~force     \\
$\mathbb{\bar{S}}_{se}$                  & strong~nuclear~adjoint~field               & adjoint~strong~nuclear~current    & weak~nuclear~force     \\

$\mathbb{\bar{S}}_{\alpha \gamma}$       & $\alpha$~adjoint~field                     & adjoint~$\alpha$~current          & weak~nuclear~force     \\
$\mathbb{\bar{S}}_{\beta \delta}$        & $\beta$~adjoint~field                      & adjoint~$\beta$~current           & weak~nuclear~force     \\
$\mathbb{\bar{S}}_{\gamma \alpha}$       & $\gamma$~adjoint~field                     & adjoint~$\gamma$~current          & weak~nuclear~force     \\
$\mathbb{\bar{S}}_{\delta \beta}$        & $\delta$~adjoint~field                     & adjoint~$\delta$~current          & weak~nuclear~force     \\

\hline

$\mathbb{\bar{S}}_{gs}$                  & gravitational~adjoint~field                & adjoint~linear~momentum           & strong~nuclear~force   \\
$\mathbb{\bar{S}}_{ew}$                  & electromagnetic~adjoint~field              & adjoint~electric~current          & strong~nuclear~force   \\
$\mathbb{\bar{S}}_{we}$                  & weak~nuclear~adjoint~field                 & adjoint~weak~nuclear~current      & strong~nuclear~force   \\
$\mathbb{\bar{S}}_{sg}$                  & strong~nuclear~field                       & strong~nuclear~current            & strong~nuclear~force   \\

$\mathbb{\bar{S}}_{\alpha \delta}$       & $\alpha$~adjoint~field                     & adjoint~$\alpha$~current          & strong~nuclear~force   \\
$\mathbb{\bar{S}}_{\beta \gamma}$        & $\beta$~adjoint~field                      & adjoint~$\beta$~current           & strong~nuclear~force   \\
$\mathbb{\bar{S}}_{\gamma \beta}$        & $\gamma$~adjoint~field                     & adjoint~$\gamma$~current          & strong~nuclear~force   \\
$\mathbb{\bar{S}}_{\delta \alpha}$       & $\delta$~adjoint~field                     & adjoint~$\delta$~current          & strong~nuclear~force   \\

\hline\hline
\end{tabular}
\end{table}

\subsection{Helicity}

In the compounding space with the trigintaduonion operator $(\lozenge_{32} + k_x \mathbb{\bar{X}} + k_a \mathbb{\bar{A}} + k_b \mathbb{\bar{B}} + k_s \mathbb{\bar{S}} + k_l \mathbb{\bar{L}} + k_w \mathbb{\bar{W}})$, some helicity terms \cite{guo3} will impact the mass continuity equation and the charge continuity equation etc.

\subsubsection{Field strength helicity}

The trigintaduonion linear momentum density, $\mathbb{\bar{P}} = \mu \mathbb{\bar{S}} / \mu_{gg}$, is defined from the trigintaduonion field source $\mathbb{\bar{S}}$ in the compounding space with the operator $(\lozenge_{32} + k_x \mathbb{\bar{X}} + k_a \mathbb{\bar{A}} + k_b \mathbb{\bar{B}} + k_s \mathbb{\bar{S}} + k_l \mathbb{\bar{L}} + k_w \mathbb{\bar{W}})$,
\begin{eqnarray}
\mathbb{\bar{P}} = \mathbb{\bar{P}}_g + f_e \mathbb{\bar{P}}_e + f_w \mathbb{\bar{P}}_w + f_s \mathbb{\bar{P}}_s
+ f_\alpha \mathbb{\bar{P}}_\alpha + f_\beta \mathbb{\bar{P}}_\beta + f_\gamma \mathbb{\bar{P}}_\gamma + f_\delta \mathbb{\bar{P}}_\delta
~,
\end{eqnarray}
where $\mathbb{\bar{P}}_\alpha = \Sigma (\bar{p}_{(i+16)} \emph{\textbf{i}}_{(i+16)})$, $\mathbb{\bar{P}}_\beta = \Sigma (\bar{p}_{(i+20)} \emph{\textbf{i}}_{(i+20)})$, $\mathbb{\bar{P}}_\gamma = \Sigma (\bar{p}_{(i+24)} \emph{\textbf{i}}_{(i+24)})$, $\mathbb{\bar{P}}_\delta = \Sigma (\bar{p}_{(i+28)} \emph{\textbf{i}}_{(i+28)})$. And $\lozenge_{32} \cdot \mathbb{\bar{R}}$ is the scalar part of $\lozenge_{32} \circ \mathbb{\bar{R}}$. $\bar{p}_0 = \widehat{m} v_0 $, $\widehat{m} = m (\bar{v}_0^\delta / \bar{v}_0) + \bigtriangleup m$, $\bar{v}_0^\delta = v_0 \lozenge_{32} \cdot \mathbb{\bar{R}}$, $\bigtriangleup m = - (k_x \mathbb{\bar{X}}^* \cdot \mathbb{\bar{B}} + k_a \mathbb{\bar{A}}^* \cdot \mathbb{\bar{B}} + k_b \mathbb{\bar{B}}^* \cdot \mathbb{\bar{B}}) / ( \bar{v}_0 \mu_{gg} )$.

According to the trigintaduonion features, the gravitational mass density $\widehat{m}$ is one reserved scalar in the above, and is changed with the field strength ($\mathbb{\bar{B}}_g$, $\mathbb{\bar{B}}_e$, $\mathbb{\bar{B}}_w$, $\mathbb{\bar{B}}_s$, $\mathbb{\bar{B}}_\alpha$, $\mathbb{\bar{B}}_\beta$, $\mathbb{\bar{B}}_\gamma$, and $\mathbb{\bar{B}}_\delta$), the field potential ($\mathbb{\bar{A}}_g$, $\mathbb{\bar{A}}_e$, $\mathbb{\bar{A}}_w$, $\mathbb{\bar{A}}_s$£¬ $\mathbb{\bar{A}}_\alpha$, $\mathbb{\bar{A}}_\beta$, $\mathbb{\bar{A}}_\gamma$, and $\mathbb{\bar{A}}_\delta$), the helicity $(k_x \mathbb{\bar{X}}^* \cdot \mathbb{\bar{B}} + k_a \mathbb{\bar{A}}^* \cdot \mathbb{\bar{B}} + k_b \mathbb{\bar{B}}^* \cdot \mathbb{\bar{B}})$, and the trigintaduonion ($\mathbb{\bar{X}}_g$, $\mathbb{\bar{X}}_e$, $\mathbb{\bar{X}}_w$, $\mathbb{\bar{X}}_s$, $\mathbb{\bar{X}}_\alpha$, $\mathbb{\bar{X}}_\beta$, $\mathbb{\bar{X}}_\gamma$, $\mathbb{\bar{X}}_\delta$) etc in the trigintaduonion compounding space. The helicity includes the magnetic helicity $ \bar{\textbf{a}}_e \cdot \bar{\textbf{h}}_e $, the kinetic helicity $ \bar{\textbf{v}}_g \cdot \bar{\textbf{u}}_g $, the cross helicity $ \bar{\textbf{v}}_e \cdot \bar{\textbf{h}}_e $, and the new helicity term $ \bar{\textbf{a}}_e \cdot \bar{\textbf{u}}_e $ etc.

Similarly to the $\mathbb{\bar{S}}$ and the $\mathbb{\bar{L}}$, the trigintaduonion angular momentum $\mathbb{\bar{W}}$ has not an influence on the field strength helicity in the trigintaduonion compounding space with the operator $(\lozenge_{32} + k_x \mathbb{\bar{X}} + k_a \mathbb{\bar{A}} + k_b \mathbb{\bar{B}} + k_s \mathbb{\bar{S}} + k_l \mathbb{\bar{L}} + k_w \mathbb{\bar{W}})$. Therefore we can not distinguish this field from some fields with other trigintaduonion operators, according to the viewpoint of the field strength helicity.

\subsubsection{Field source helicity}

The part force density $\mathbb{\bar{F}}$ is one part of the trigintaduonion compounding force-power density $\mathbb{\bar{N}}$, and is defined from the trigintaduonion linear momentum density $\mathbb{\bar{P}}$ ,
\begin{eqnarray}
\mathbb{\bar{F}} = v_0 (\lozenge_{32} + k_x \mathbb{\bar{X}} + k_a \mathbb{\bar{A}} + k_b \mathbb{\bar{B}} + k_s \mathbb{\bar{S}} + k_l \mathbb{\bar{L}} + k_w \mathbb{\bar{W}})^* \circ \mathbb{\bar{P}}~,
\end{eqnarray}
where the part force density includes that of the inertial force, gravitational force, Lorentz force, and the interacting force between the fields with the dipoles etc.

The scalar $\bar{f}_0$ of $\mathbb{\bar{F}}$ is written as,
\begin{align}
\bar{n}_0 / v_0 = &~
k_a \left\{ \bar{a}_0 \bar{p}_0 + \bar{\textbf{a}}^*_g \cdot \bar{\textbf{p}}_g + f_e^2 ( \bar{\textbf{a}}^*_4 \circ \bar{\textbf{p}}_4 + \bar{\textbf{a}}^*_e \cdot \bar{\textbf{p}}_e )  \right\}
\nonumber\\
&
+ k_a \left\{ f_w^2 ( \bar{\textbf{a}}^*_8 \circ \bar{\textbf{p}}_8 + \bar{\textbf{a}}^*_w \cdot \bar{\textbf{p}}_w ) +  f_s^2 ( \bar{\textbf{a}}^*_{12} \circ \bar{\textbf{p}}_{12} + \bar{\textbf{a}}^*_s \cdot \bar{\textbf{p}}_s) \right\}
\nonumber\\
&
+ k_a \left\{ f_\alpha^2 ( \bar{\textbf{a}}^*_{16} \circ \bar{\textbf{p}}_{16} + \bar{\textbf{a}}^*_\alpha \cdot \bar{\textbf{p}}_\alpha )
+ f_\beta^2 ( \bar{\textbf{a}}^*_{20} \circ \bar{\textbf{p}}_{20} + \bar{\textbf{a}}^*_\beta \cdot \bar{\textbf{p}}_\beta) \right\}
\nonumber\\
&
+ k_a \left\{ f_\gamma^2 ( \bar{\textbf{a}}^*_{24} \circ \bar{\textbf{p}}_{24} + \bar{\textbf{a}}^*_\gamma \cdot \bar{\textbf{p}}_\gamma)
+ f_\delta^2 ( \bar{\textbf{a}}^*_{28} \circ \bar{\textbf{p}}_{28} + \bar{\textbf{a}}^*_\delta \cdot \bar{\textbf{p}}_\delta ) \right\}
\nonumber\\
&
+ k_x \left\{ \bar{x}_0 \bar{p}_0 + \bar{\textbf{x}}^*_g \cdot \bar{\textbf{p}}_g + f_e^2 ( \bar{\textbf{x}}^*_4 \circ \bar{\textbf{p}}_4 + \bar{\textbf{x}}^*_e \cdot \bar{\textbf{p}}_e )  \right\}
\nonumber\\
&
+ k_x \left\{ f_w^2 ( \bar{\textbf{x}}^*_8 \circ \bar{\textbf{p}}_8 + \bar{\textbf{x}}^*_w \cdot \bar{\textbf{p}}_w )+ f_s^2 ( \bar{\textbf{x}}^*_{12} \circ \bar{\textbf{p}}_{12} + \bar{\textbf{x}}^*_s \cdot \bar{\textbf{p}}_s) \right\}
\nonumber\\
&
+ k_x \left\{ f_\alpha^2 ( \bar{\textbf{x}}^*_{16} \circ \bar{\textbf{p}}_{16} + \bar{\textbf{x}}^*_\alpha \cdot \bar{\textbf{p}}_\alpha)
+ f_\beta^2 ( \bar{\textbf{x}}^*_{20} \circ \bar{\textbf{p}}_{20} + \bar{\textbf{x}}^*_\beta \cdot \bar{\textbf{p}}_\beta) \right\}
\nonumber\\
&
+ k_x \left\{ f_\gamma^2 ( \bar{\textbf{x}}^*_{24} \circ \bar{\textbf{p}}_{24} + \bar{\textbf{x}}^*_\gamma \cdot \bar{\textbf{p}}_\gamma)
+ f_\delta^2 ( \bar{\textbf{x}}^*_{28} \circ \bar{\textbf{p}}_{28} + \bar{\textbf{x}}^*_\delta \cdot \bar{\textbf{p}}_\delta) \right\}
\nonumber\\
&
+ k_s \left\{ \bar{s}_0 \bar{p}_0 + \bar{\textbf{s}}^*_g \cdot \bar{\textbf{p}}_g + f_e^2 ( \bar{\textbf{s}}^*_4 \circ \bar{\textbf{p}}_4 + \bar{\textbf{s}}^*_e \cdot \bar{\textbf{p}}_e )  \right\}
\nonumber\\
&
+ k_s \left\{ f_w^2 ( \bar{\textbf{s}}^*_8 \circ \bar{\textbf{p}}_8 + \bar{\textbf{s}}^*_w \cdot \bar{\textbf{p}}_w )+ f_s^2 ( \bar{\textbf{s}}^*_{12} \circ
\bar{\textbf{p}}_{12} + \bar{\textbf{s}}^*_s \cdot \bar{\textbf{p}}_s) \right\}
\nonumber\\
&
+ k_s \left\{ f_\alpha^2 ( \bar{\textbf{s}}^*_{16} \circ \bar{\textbf{p}}_{16} + \bar{\textbf{s}}^*_\alpha \cdot \bar{\textbf{p}}_\alpha)
+ f_\beta^2 ( \bar{\textbf{s}}^*_{20} \circ \bar{\textbf{p}}_{20} + \bar{\textbf{s}}^*_\beta \cdot \bar{\textbf{p}}_\beta) \right\}
\nonumber\\
&
+ k_s \left\{ f_\gamma^2 ( \bar{\textbf{s}}^*_{24} \circ \bar{\textbf{p}}_{24} + \bar{\textbf{s}}^*_\gamma \cdot \bar{\textbf{p}}_\gamma)
+ f_\delta^2 ( \bar{\textbf{s}}^*_{28} \circ \bar{\textbf{p}}_{28} + \bar{\textbf{s}}^*_\delta \cdot \bar{\textbf{p}}_\delta) \right\}
\nonumber\\
&
+ k_l \left\{ \bar{l}_0 \bar{p}_0 + \bar{\textbf{l}}^*_g \cdot \bar{\textbf{p}}_g + f_e^2 ( \bar{\textbf{l}}^*_4 \circ \bar{\textbf{p}}_4 + \bar{\textbf{l}}^*_e \cdot \bar{\textbf{p}}_e )  \right\}
\nonumber\\
&
+ k_l \left\{ f_w^2 ( \bar{\textbf{l}}^*_8 \circ \bar{\textbf{p}}_8 + \bar{\textbf{l}}^*_w \cdot \bar{\textbf{p}}_w ) + f_s^2 ( \bar{\textbf{l}}^*_{12} \circ \bar{\textbf{p}}_{12} + \bar{\textbf{l}}^*_s \cdot \bar{\textbf{p}}_s) \right\}
\nonumber\\
&
+ k_l \left\{ f_\alpha^2 ( \bar{\textbf{l}}^*_{16} \circ \bar{\textbf{p}}_{16} + \bar{\textbf{l}}^*_\alpha \cdot \bar{\textbf{p}}_\alpha)
+ f_\beta^2 ( \bar{\textbf{l}}^*_{20} \circ \bar{\textbf{p}}_{20} + \bar{\textbf{l}}^*_\beta \cdot \bar{\textbf{p}}_\beta) \right\}
\nonumber\\
&
+ k_l \left\{ f_\gamma^2 ( \bar{\textbf{l}}^*_{24} \circ \bar{\textbf{p}}_{24} + \bar{\textbf{l}}^*_\gamma \cdot \bar{\textbf{p}}_\gamma)
+ f_\delta^2 ( \bar{\textbf{l}}^*_{28} \circ \bar{\textbf{p}}_{28} + \bar{\textbf{l}}^*_\delta \cdot \bar{\textbf{p}}_\delta) \right\}
\nonumber\\
&
+ k_w \left\{ \bar{w}_0 \bar{p}_0 + \bar{\textbf{w}}^*_g \cdot \bar{\textbf{p}}_g + f_e^2 ( \bar{\textbf{w}}^*_4 \circ \bar{\textbf{p}}_4 + \bar{\textbf{w}}^*_e \cdot \bar{\textbf{p}}_e )  \right\}
\nonumber\\
&
+ k_w \left\{ f_w^2 ( \bar{\textbf{w}}^*_8 \circ \bar{\textbf{p}}_8 + \bar{\textbf{w}}^*_w \cdot \bar{\textbf{p}}_w ) + f_s^2 ( \bar{\textbf{w}}^*_{12} \circ \bar{\textbf{p}}_{12} + \bar{\textbf{w}}^*_s \cdot \bar{\textbf{p}}_s) \right\}
\nonumber\\
&
+ k_w \left\{ f_\alpha^2 ( \bar{\textbf{w}}^*_{16} \circ \bar{\textbf{p}}_{16} + \bar{\textbf{w}}^*_\alpha \cdot \bar{\textbf{p}}_\alpha)
+ f_\beta^2 ( \bar{\textbf{w}}^*_{20} \circ \bar{\textbf{p}}_{20} + \bar{\textbf{w}}^*_\beta \cdot \bar{\textbf{p}}_\beta) \right\}
\nonumber\\
&
+ k_w \left\{ f_\gamma^2 ( \bar{\textbf{w}}^*_{24} \circ \bar{\textbf{p}}_{24} + \bar{\textbf{w}}^*_\gamma \cdot \bar{\textbf{p}}_\gamma)
+ f_\delta^2 ( \bar{\textbf{w}}^*_{28} \circ \bar{\textbf{p}}_{28} + \bar{\textbf{w}}^*_\delta \cdot \bar{\textbf{p}}_\delta) \right\}
\nonumber\\
&
+ k_b ( \bar{\textbf{h}}^*_g \cdot \bar{\textbf{p}}_g + f_e^2 \bar{\textbf{h}}^*_e \cdot \bar{\textbf{p}}_e  + f_w^2 \bar{\textbf{h}}^*_w \cdot \bar{\textbf{p}}_w + f_s^2 \bar{\textbf{h}}^*_s \cdot \bar{\textbf{p}}_s )
\nonumber\\
&
+ k_b ( f_\alpha^2 \bar{\textbf{h}}^*_\alpha \cdot \bar{\textbf{p}}_\alpha + f_\beta^2 \bar{\textbf{h}}^*_\beta \cdot \bar{\textbf{p}}_\beta + f_\gamma^2 \bar{\textbf{h}}^*_\gamma \cdot \bar{\textbf{p}}_\gamma + f_\delta^2 \bar{\textbf{h}}^*_\delta \cdot \bar{\textbf{p}}_\delta )
\nonumber\\
&
+ \partial_0 \bar{p}_0 + \nabla^*_g \cdot \bar{\textbf{p}}_g + d_e f_e (\emph{\textbf{i}}^*_4 \circ \partial_4 \bar{\textbf{p}}_4 + \nabla^*_e \cdot \bar{\textbf{p}}_e)
\nonumber\\
&
+ d_w f_w (\emph{\textbf{i}}^*_8 \circ \partial_8 \bar{\textbf{p}}_8 + \nabla^*_w \cdot \bar{\textbf{p}}_w) + d_s f_s (\emph{\textbf{i}}^*_{12} \circ \partial_{12} \bar{\textbf{p}}_{12} + \nabla^*_s \cdot \bar{\textbf{p}}_s)
\nonumber\\
&
+ d_\alpha f_\alpha (\emph{\textbf{i}}^*_{16} \circ \partial_{16} \bar{\textbf{p}}_{16} + \nabla^*_\alpha \cdot \bar{\textbf{p}}_\alpha)
+ d_\beta f_\beta (\emph{\textbf{i}}^*_{20} \circ \partial_{20} \bar{\textbf{p}}_{20} + \nabla^*_\beta \cdot \bar{\textbf{p}}_\beta)
\nonumber\\
&
+ d_\gamma f_\gamma (\emph{\textbf{i}}^*_{24} \circ \partial_{24} \bar{\textbf{p}}_{24} + \nabla^*_\gamma \cdot \bar{\textbf{p}}_\gamma)
+ d_\delta f_\delta (\emph{\textbf{i}}^*_{28} \circ \partial_{28} \bar{\textbf{p}}_{28} + \nabla^*_\delta \cdot \bar{\textbf{p}}_\delta) ~£¬
\end{align}
where the field source helicity in the fields with the trigintaduonion operator $(\lozenge_{32} + k_x \mathbb{\bar{X}} + k_a \mathbb{\bar{A}} + k_b \mathbb{\bar{B}} + k_s \mathbb{\bar{S}} + k_l \mathbb{\bar{L}} + k_w \mathbb{\bar{W}})$ covers the
$( \bar{\textbf{a}}^*_8 \circ \bar{\textbf{p}}_8 + \bar{\textbf{a}}^*_w \cdot \bar{\textbf{p}}_w )$, $( \bar{\textbf{a}}^*_{12} \circ \bar{\textbf{p}}_{12} + \bar{\textbf{a}}^*_s \cdot \bar{\textbf{p}}_s)$,
$( \bar{\textbf{x}}^*_{16} \circ \bar{\textbf{p}}_{16} + \bar{\textbf{x}}^*_\alpha \cdot \bar{\textbf{p}}_\alpha)$,
$( \bar{\textbf{x}}^*_{20} \circ \bar{\textbf{p}}_{20} + \bar{\textbf{x}}^*_\beta \cdot \bar{\textbf{p}}_\beta)$,
$( \bar{\textbf{s}}^*_{24} \circ \bar{\textbf{p}}_{24} + \bar{\textbf{s}}^*_\gamma \cdot \bar{\textbf{p}}_\gamma)$,
$( \bar{\textbf{s}}^*_{28} \circ \bar{\textbf{p}}_{28} + \bar{\textbf{s}}^*_\delta \cdot \bar{\textbf{p}}_\delta)$,
$( \bar{\textbf{l}}^*_{24} \circ \bar{\textbf{p}}_{24} + \bar{\textbf{l}}^*_\gamma \cdot \bar{\textbf{p}}_\gamma)$,
$( \bar{\textbf{l}}^*_{28} \circ \bar{\textbf{p}}_{28} + \bar{\textbf{l}}^*_\delta \cdot \bar{\textbf{p}}_\delta)$,
besides the $( \bar{\textbf{a}}_g^* \cdot \bar{\textbf{p}}_g + \bar{a}_0 \bar{p}_0 )$, $( \bar{\textbf{a}}_e^* \cdot \bar{\textbf{p}}_e + \bar{\textbf{a}}_4^* \circ \bar{\textbf{p}}_4 )$, $( \bar{\textbf{h}}_g^* \cdot \bar{\textbf{p}}_g + f_e^2 \bar{\textbf{h}}_e^* \cdot \bar{\textbf{p}}_e )$, $( \bar{\textbf{x}}_g^* \cdot \bar{\textbf{p}}_g + \bar{x}_0 \bar{p}_0)$, $( \bar{\textbf{x}}_e^* \cdot \bar{\textbf{p}}_e + \bar{\textbf{x}}_4^* \circ \bar{\textbf{p}}_4)$, $( \bar{\textbf{s}}_g^* \cdot \bar{\textbf{p}}_g + \bar{s}_0 \bar{p}_0 )$, $( \bar{\textbf{s}}_e^* \cdot \bar{\textbf{p}}_e + \bar{\textbf{s}}_4^* \circ \bar{\textbf{p}}_4 )$, $( \bar{\textbf{l}}_g^* \cdot \bar{\textbf{p}}_g + \bar{l}_0 \bar{p}_0 )$, $( \bar{\textbf{l}}_e^* \cdot \bar{\textbf{p}}_e + \bar{\textbf{l}}_4^* \circ \bar{\textbf{p}}_4 )$, and $( \bar{\textbf{w}}_e^* \cdot \bar{\textbf{p}}_e + \bar{\textbf{w}}_4^* \circ \bar{\textbf{p}}_4 )$ etc.
And that they include the magnetic helicity $ \textbf{a}_e \cdot \textbf{h}_e $, the kinetic helicity $ \textbf{v}_e \cdot \textbf{u}_e $, the cross helicity $ \textbf{v}_e \cdot \textbf{h}_e $, the current helicity $ \textbf{h}_e^* \cdot \textbf{p}_e $, and some other new helicity terms etc.
$\bar{\textbf{p}}_\alpha = \Sigma ( \emph{\textbf{i}}_{(j+16)} \bar{p}_{(j+16)})$,
$\bar{\textbf{p}}_\beta = \Sigma ( \emph{\textbf{i}}_{(j+20)} \bar{p}_{(j+20)})$,
$\bar{\textbf{p}}_\gamma = \Sigma ( \emph{\textbf{i}}_{(j+24)} \bar{p}_{(j+24)})$,
$\bar{\textbf{p}}_\delta = \Sigma ( \emph{\textbf{i}}_{(j+28)} \bar{p}_{(j+28)})$;
$\bar{\textbf{p}}_{16} = \emph{\textbf{i}}_{16} \bar{p}_{16}$,
$\bar{\textbf{p}}_{20} = \emph{\textbf{i}}_{20} \bar{p}_{20}$,
$\bar{\textbf{p}}_{24} = \emph{\textbf{i}}_{24} \bar{p}_{24}$,
$\bar{\textbf{p}}_{28} = \emph{\textbf{i}}_{28} \bar{p}_{28}$.

The above is the mass continuity equation in the eight fields with the trigintaduonion operator $(\lozenge_{32} + k_x \mathbb{\bar{X}} + k_a \mathbb{\bar{A}} + k_b \mathbb{\bar{B}} + k_s \mathbb{\bar{S}} + k_l \mathbb{\bar{L}} + k_w \mathbb{\bar{W}})$, and is effected by the speed of light $\bar{v}_0$ and some helicities of the rotational objects and of the spinning charged objects.

A new physical quantity $\mathbb{\bar{F}}_q$ can be defined from the part force density $\mathbb{\bar{F}}$,
\begin{eqnarray}
\mathbb{\bar{F}}_q = \mathbb{\bar{F}} \circ \emph{\textbf{i}}_4^*~.
\end{eqnarray}

The scalar part $\bar{F}_0$ of the $\mathbb{\bar{F}}_q$ is written as,
\begin{align}
\bar{F}_0 / v_0 = & ~
f_e (\partial_0 \bar{\textbf{p}}_4 + \nabla_g^* \cdot \bar{\textbf{p}}_e) \circ \emph{\textbf{i}}_4^*
+ d_e (\emph{\textbf{i}}_4^* \circ \partial_4 \bar{p}_0 + \nabla_e^* \cdot \bar{\textbf{p}}_g) \circ \emph{\textbf{i}}_4^*
\nonumber
\\
&
+ d_w f_s (\emph{\textbf{i}}_8^* \circ \partial_8 \bar{\textbf{p}}_{12} + \nabla_w^* \cdot \bar{\textbf{p}}_s) \circ \emph{\textbf{i}}_4^*
+ d_s f_w (\emph{\textbf{i}}_{12}^* \circ \partial_{12} \bar{\textbf{p}}_8 + \nabla_s^* \cdot \bar{\textbf{p}}_w) \circ \emph{\textbf{i}}_4^*
\nonumber
\\
&
+ d_\alpha f_\beta (\emph{\textbf{i}}_{16}^* \circ \partial_{16} \bar{\textbf{p}}_{20} + \nabla_\alpha^* \cdot \bar{\textbf{p}}_\beta) \circ \emph{\textbf{i}}_4^*
+ d_\beta f_\alpha (\emph{\textbf{i}}_{20}^* \circ \partial_{20} \bar{\textbf{p}}_{16} + \nabla_\beta^* \cdot \bar{\textbf{p}}_\alpha) \circ \emph{\textbf{i}}_4^*
\nonumber
\\
&
+ d_\gamma f_\delta (\emph{\textbf{i}}_{24}^* \circ \partial_{24} \bar{\textbf{p}}_{28} + \nabla_\gamma^* \cdot \bar{\textbf{p}}_\delta) \circ \emph{\textbf{i}}_4^*
+ d_\delta f_\gamma (\emph{\textbf{i}}_{28}^* \circ \partial_{28} \bar{\textbf{p}}_{24} + \nabla_\delta^* \cdot \bar{\textbf{p}}_\gamma) \circ \emph{\textbf{i}}_4^*
\nonumber
\\
&
+ k_b f_e ( \bar{\textbf{h}}_g^* \cdot \bar{\textbf{p}}_e + \bar{\textbf{h}}_e^* \cdot \bar{\textbf{p}}_g ) \circ \emph{\textbf{i}}_4^*
+ k_b f_w f_s ( \bar{\textbf{h}}_w^* \cdot \bar{\textbf{p}}_s + \bar{\textbf{h}}_s^* \cdot \bar{\textbf{p}}_w ) \circ \emph{\textbf{i}}_4^*
\nonumber
\\
&
+ k_b f_\alpha f_\beta ( \bar{\textbf{h}}_\alpha^* \cdot \bar{\textbf{p}}_\beta + \bar{\textbf{h}}_\beta^* \cdot \bar{\textbf{p}}_\alpha ) \circ \emph{\textbf{i}}_4^*
+ k_b f_\gamma f_\delta ( \bar{\textbf{h}}_\gamma^* \cdot \bar{\textbf{p}}_\delta + \bar{\textbf{h}}_\delta^* \cdot \bar{\textbf{p}}_\gamma ) \circ \emph{\textbf{i}}_4^*
\nonumber
\\
&
+ k_a f_e ( \bar{\textbf{a}}_g^* \cdot \bar{\textbf{p}}_e + \bar{a}_0 \bar{\textbf{p}}_4 ) \circ \emph{\textbf{i}}_4^* + k_a f_e ( \bar{\textbf{a}}_e^* \cdot \bar{\textbf{p}}_g + \bar{p}_0 \bar{\textbf{a}}_4^* ) \circ \emph{\textbf{i}}_4^*
\nonumber
\\
&
+ k_a f_s f_w ( \bar{\textbf{a}}_w^* \cdot \bar{\textbf{p}}_s + \bar{\textbf{a}}_8^* \cdot \bar{\textbf{p}}_{12} ) \circ \emph{\textbf{i}}_4^* + k_a f_s f_w ( \bar{\textbf{a}}_s^* \cdot \bar{\textbf{p}}_w + \bar{\textbf{a}}_{12}^* \cdot \bar{\textbf{p}}_8 ) \circ \emph{\textbf{i}}_4^*
\nonumber
\\
&
+ k_a f_\beta f_\alpha ( \bar{\textbf{a}}_\alpha^* \cdot \bar{\textbf{p}}_\beta + \bar{\textbf{a}}_{16}^* \cdot \bar{\textbf{p}}_{20} ) \circ \emph{\textbf{i}}_4^* + k_a f_\beta f_\alpha ( \bar{\textbf{a}}_\beta^* \cdot \bar{\textbf{p}}_\alpha + \bar{\textbf{a}}_{20}^* \cdot \bar{\textbf{p}}_{16} ) \circ \emph{\textbf{i}}_4^*
\nonumber
\\
&
+ k_a f_\delta f_\gamma ( \bar{\textbf{a}}_\gamma^* \cdot \bar{\textbf{p}}_\delta + \bar{\textbf{a}}_{24}^* \cdot \bar{\textbf{p}}_{28} ) \circ \emph{\textbf{i}}_4^* + k_a f_\delta f_\gamma ( \bar{\textbf{a}}_\delta^* \cdot \bar{\textbf{p}}_\gamma + \bar{\textbf{a}}_{28}^* \cdot \bar{\textbf{p}}_{24} ) \circ \emph{\textbf{i}}_4^*
\nonumber
\\
&
+ k_x f_e ( \bar{\textbf{x}}_g^* \cdot \bar{\textbf{p}}_e + \bar{x}_0 \bar{\textbf{p}}_4 ) \circ \emph{\textbf{i}}_4^* + k_x f_e ( \bar{\textbf{x}}_e^* \cdot \bar{\textbf{p}}_g + \bar{p}_0 \bar{\textbf{x}}_4^* ) \circ \emph{\textbf{i}}_4^*
\nonumber
\\
&
+ k_x f_s f_w ( \bar{\textbf{x}}_w^* \cdot \bar{\textbf{p}}_s + \bar{\textbf{x}}_8^* \cdot \bar{\textbf{p}}_{12} ) \circ \emph{\textbf{i}}_4^* + k_x f_s f_w ( \bar{\textbf{x}}_s^* \cdot \bar{\textbf{p}}_w + \bar{\textbf{x}}_{12}^* \cdot \bar{\textbf{p}}_8 ) \circ \emph{\textbf{i}}_4^*
\nonumber
\\
&
+ k_x f_\beta f_\alpha ( \bar{\textbf{x}}_\alpha^* \cdot \bar{\textbf{p}}_\beta + \bar{\textbf{x}}_{16}^* \cdot \bar{\textbf{p}}_{20} ) \circ \emph{\textbf{i}}_4^* + k_x f_\beta f_\alpha ( \bar{\textbf{x}}_\beta^* \cdot \bar{\textbf{p}}_\alpha + \bar{\textbf{x}}_{20}^* \cdot \bar{\textbf{p}}_{16} ) \circ \emph{\textbf{i}}_4^*
\nonumber
\\
&
+ k_x f_\delta f_\gamma ( \bar{\textbf{x}}_\gamma^* \cdot \bar{\textbf{p}}_\delta + \bar{\textbf{x}}_{24}^* \cdot \bar{\textbf{p}}_{28} ) \circ \emph{\textbf{i}}_4^* + k_x f_\delta f_\gamma ( \bar{\textbf{x}}_\delta^* \cdot \bar{\textbf{p}}_\gamma + \bar{\textbf{x}}_{28}^* \cdot \bar{\textbf{p}}_{24} ) \circ \emph{\textbf{i}}_4^*
\nonumber
\\
&
+ k_s f_e ( \bar{\textbf{s}}_g^* \cdot \bar{\textbf{p}}_e + \bar{s}_0 \bar{\textbf{p}}_4 ) \circ \emph{\textbf{i}}_4^* + k_s f_e ( \bar{\textbf{s}}_e^* \cdot \bar{\textbf{p}}_g + \bar{p}_0 \bar{\textbf{s}}_4^* ) \circ \emph{\textbf{i}}_4^*
\nonumber
\\
&
+ k_s f_s f_w ( \bar{\textbf{s}}_w^* \cdot \bar{\textbf{p}}_s + \bar{\textbf{s}}_8^* \cdot \bar{\textbf{p}}_{12} ) \circ \emph{\textbf{i}}_4^* + k_s f_s f_w ( \bar{\textbf{s}}_s^* \cdot \bar{\textbf{p}}_w + \bar{\textbf{s}}_{12}^* \cdot \bar{\textbf{p}}_8 ) \circ \emph{\textbf{i}}_4^*
\nonumber
\\
&
+ k_s f_\beta f_\alpha ( \bar{\textbf{s}}_\alpha^* \cdot \bar{\textbf{p}}_\beta + \bar{\textbf{s}}_{16}^* \cdot \bar{\textbf{p}}_{20} ) \circ \emph{\textbf{i}}_4^* + k_s f_\beta f_\alpha ( \bar{\textbf{s}}_\beta^* \cdot \bar{\textbf{p}}_\alpha + \bar{\textbf{s}}_{20}^* \cdot \bar{\textbf{p}}_{16} ) \circ \emph{\textbf{i}}_4^*
\nonumber
\\
&
+ k_s f_\delta f_\gamma ( \bar{\textbf{s}}_\gamma^* \cdot \bar{\textbf{p}}_\delta + \bar{\textbf{s}}_{24}^* \cdot \bar{\textbf{p}}_{28} ) \circ \emph{\textbf{i}}_4^* + k_s f_\delta f_\gamma ( \bar{\textbf{s}}_\delta^* \cdot \bar{\textbf{p}}_\gamma + \bar{\textbf{s}}_{28}^* \cdot \bar{\textbf{p}}_{24} ) \circ \emph{\textbf{i}}_4^*
\nonumber
\\
&
+ k_l f_e ( \bar{\textbf{l}}_g^* \cdot \bar{\textbf{p}}_e + \bar{l}_0 \bar{\textbf{p}}_4 ) \circ \emph{\textbf{i}}_4^* + k_l f_e ( \bar{\textbf{l}}_e^* \cdot \bar{\textbf{p}}_g + \bar{p}_0 \bar{\textbf{l}}_4^* ) \circ \emph{\textbf{i}}_4^*
\nonumber
\\
&
+ k_l f_s f_w ( \bar{\textbf{l}}_w^* \cdot \bar{\textbf{p}}_s + \bar{\textbf{l}}_8^* \cdot \bar{\textbf{p}}_{12} ) \circ \emph{\textbf{i}}_4^* + k_l f_s f_w ( \bar{\textbf{l}}_s^* \cdot \bar{\textbf{p}}_w + \bar{\textbf{l}}_{12}^* \cdot \bar{\textbf{p}}_8 ) \circ \emph{\textbf{i}}_4^*
\nonumber
\\
&
+ k_l f_\beta f_\alpha ( \bar{\textbf{l}}_\alpha^* \cdot \bar{\textbf{p}}_\beta + \bar{\textbf{l}}_{16}^* \cdot \bar{\textbf{p}}_{20} ) \circ \emph{\textbf{i}}_4^* + k_l f_\beta f_\alpha ( \bar{\textbf{l}}_\beta^* \cdot \bar{\textbf{p}}_\alpha + \bar{\textbf{l}}_{20}^* \cdot \bar{\textbf{p}}_{16} ) \circ \emph{\textbf{i}}_4^*
\nonumber
\\
&
+ k_l f_\delta f_\gamma ( \bar{\textbf{l}}_\gamma^* \cdot \bar{\textbf{p}}_\delta + \bar{\textbf{l}}_{24}^* \cdot \bar{\textbf{p}}_{28} ) \circ \emph{\textbf{i}}_4^* + k_l f_\delta f_\gamma ( \bar{\textbf{l}}_\delta^* \cdot \bar{\textbf{p}}_\gamma + \bar{\textbf{l}}_{28}^* \cdot \bar{\textbf{p}}_{24} ) \circ \emph{\textbf{i}}_4^*
\nonumber
\\
&
+ k_w f_e ( \bar{\textbf{w}}_g^* \cdot \bar{\textbf{p}}_e + \bar{w}_0 \bar{\textbf{p}}_4 ) \circ \emph{\textbf{i}}_4^* + k_w f_e ( \bar{\textbf{w}}_e^* \cdot \bar{\textbf{p}}_g + \bar{p}_0 \bar{\textbf{w}}_4^* ) \circ \emph{\textbf{i}}_4^*
\nonumber
\\
&
+ k_w f_s f_w ( \bar{\textbf{w}}_w^* \cdot \bar{\textbf{p}}_s + \bar{\textbf{w}}_8^* \cdot \bar{\textbf{p}}_{12} ) \circ \emph{\textbf{i}}_4^* + k_w f_s f_w ( \bar{\textbf{w}}_s^* \cdot \bar{\textbf{p}}_w + \bar{\textbf{w}}_{12}^* \cdot \bar{\textbf{p}}_8 ) \circ \emph{\textbf{i}}_4^*
\nonumber
\\
&
+ k_w f_\beta f_\alpha ( \bar{\textbf{w}}_\alpha^* \cdot \bar{\textbf{p}}_\beta + \bar{\textbf{w}}_{16}^* \cdot \bar{\textbf{p}}_{20} ) \circ \emph{\textbf{i}}_4^* + k_w f_\beta f_\alpha ( \bar{\textbf{w}}_\beta^* \cdot \bar{\textbf{p}}_\alpha + \bar{\textbf{w}}_{20}^* \cdot \bar{\textbf{p}}_{16} ) \circ \emph{\textbf{i}}_4^*
\nonumber
\\
&
+ k_w f_\delta f_\gamma ( \bar{\textbf{w}}_\gamma^* \cdot \bar{\textbf{p}}_\delta + \bar{\textbf{w}}_{24}^* \cdot \bar{\textbf{p}}_{28} ) \circ \emph{\textbf{i}}_4^* + k_w f_\delta f_\gamma ( \bar{\textbf{w}}_\delta^* \cdot \bar{\textbf{p}}_\gamma + \bar{\textbf{w}}_{28}^* \cdot \bar{\textbf{p}}_{24} ) \circ \emph{\textbf{i}}_4^*
~,
\end{align}
where the above helicity terms encompass that in the gravitational field, the electromagnetic field, the weak nuclear field, and the strong nuclear field with their adjoint fields, as well as that in the $\alpha$, $\beta$, $\gamma$, and $\delta$ fields.

In the compounding space with the trigintaduonion operator $(\lozenge_{32} + k_x \mathbb{\bar{X}} + k_a \mathbb{\bar{A}} + k_b \mathbb{\bar{B}} + k_s \mathbb{\bar{S}} + k_l \mathbb{\bar{L}} + k_w \mathbb{\bar{W}} )$, the above is the charge continuity equation in the presence of the gravitational field, the electromagnetic field, the weak nuclear field, and the strong nuclear field with their adjoint fields when the scalar part $F_0 = 0$. Similarly to the case with the sedenion operator $(\lozenge_{16} + k_x \mathbb{\bar{X}} + k_a \mathbb{\bar{A}} + k_b \mathbb{\bar{B}} + k_s \mathbb{\bar{S}} + k_l \mathbb{\bar{L}} + k_w \mathbb{\bar{W}} )$, this charge continuity equation is the invariant under the trigintaduonion coordinate transformation, meanwhile the $\mathbb{\bar{X}}$, $\mathbb{\bar{A}}$, $\mathbb{\bar{B}}$, $\mathbb{\bar{S}}$, $\mathbb{\bar{L}}$, and $\mathbb{\bar{W}}$ of the above eight fields have the influence on the charge continuity equation, although the impacts are usually tiny when the fields are overlooked.

In the same way, the above physical quantities will influence the continuity equations in the definitions $(\mathbb{\bar{F}} \circ \emph{\textbf{i}}_8^*)$ and $(\mathbb{\bar{F}} \circ \emph{\textbf{i}}_{12}^*)$, and more continuity equations in the $\alpha$, $\beta$, $\gamma$, and $\delta$ fields.

\begin{table}[h]
\caption{Some physical quantities of the eight fields with their adjoint fields in the trigintaduonion space.}
\label{tab:table3}
\centering
\begin{tabular}{ll}
\hline\hline
$ definitions $                                                                                                                            & $ meanings $ \\
\hline
$\mathbb{\bar{X}}$                                                                                                                         & field quantity \\
$\mathbb{\bar{A}} = (\lozenge_{32} + k_x \mathbb{\bar{X}}) \circ \mathbb{\bar{X}}$                                                         & field potential \\
$\mathbb{\bar{B}} = (\lozenge_{32} + k_x \mathbb{\bar{X}} + k_a \mathbb{\bar{A}}) \circ \mathbb{\bar{A}}$                                  & field strength \\
$\mathbb{\bar{R}}$                                                                                                                         & radius vector \\
$\mathbb{\bar{V}} = v_0 \lozenge_{32} \circ \mathbb{\bar{R}}$                                                                              & velocity \\
$\mathbb{\bar{U}} = \lozenge_{32} \circ \mathbb{\bar{V}}$                                                                                  & velocity curl \\
$\mu \mathbb{\bar{S}} = - ( \lozenge_{32} + k_x \mathbb{\bar{X}} + k_a \mathbb{\bar{A}} + k_b \mathbb{\bar{B}} )^* \circ \mathbb{\bar{B}}$ & field source \\
$\mathbb{\bar{H}}_b = (k_x \mathbb{\bar{X}} + k_a \mathbb{\bar{A}} + k_b \mathbb{\bar{B}})^* \cdot \mathbb{\bar{B}}$                       & field strength helicity\\
$\mathbb{\bar{P}} = \mu \mathbb{\bar{S}} / \mu_{gg}$                                                                                       & linear momentum density \\
$\mathbb{\bar{R}} = \mathbb{R} + k_{rx} \mathbb{X}$                                                                                        & compounding radius vector \\
$\mathbb{\bar{L}} = \mathbb{\bar{R}} \circ \mathbb{\bar{P}}$                                                                               & angular momentum density \\
$\mathbb{\bar{W}} = v_0 ( \lozenge_{32} + k_x \mathbb{\bar{X}} + k_a \mathbb{\bar{A}} + k_b \mathbb{\bar{B}} + k_s \mathbb{\bar{S}} + k_l \mathbb{\bar{L}} ) \circ \mathbb{\bar{L}}$                                                                                                                          & torque-energy densities \\
$\mathbb{\bar{N}} = v_0 ( \lozenge_{32} + k_x \mathbb{\bar{X}} + k_a \mathbb{\bar{A}} + k_b \mathbb{\bar{B}} + k_s \mathbb{\bar{S}} + k_l \mathbb{\bar{L}} + k_w \mathbb{\bar{W}})^* \circ \mathbb{\bar{W}}$                                                                                                & force-power density \\
$\mathbb{\bar{F}} = - \mathbb{\bar{N}} / (2v_0)$                                                                                           & force density \\
$\mathbb{\bar{H}}_s = ( k_x \mathbb{\bar{X}} + k_a \mathbb{\bar{A}} + k_b \mathbb{\bar{B}} + k_s \mathbb{\bar{S}} + k_l \mathbb{\bar{L}})^* \cdot \mathbb{\bar{P}}$
                                                                                                                                           & field source helicity\\
\hline\hline
\end{tabular}
\end{table}

%--14--%

\section{CONCLUSIONS}

In the octonion compounding space with the octonion operator $\lozenge$ and the compounding field strength, the features of the electromagnetic field and the gravitational field can be depicted by the algebra of octonions, including the field source, the gravitational mass density, and the mass continuity equation etc. The physical quantities are influenced by the current helicity, the field energy, and the enstrophy in the electromagnetic and the gravitational fields.

Similarly to the above compounding fields, there may exist other kinds of fields with different operators. The magnetic helicity, the cross helicity, and the kinetic helicity cause these fields with different kinds of operators combine together to become one compounding field. In the octonion compounding space with the operator $\lozenge$ and the octonion quantity $\mathbb{\bar{X}}$, the field potential $\mathbb{\bar{A}}$, the field strength $\mathbb{\bar{B}}$, and the field source $\mathbb{\bar{S}}$ etc, much more helicity terms can be concluded in the electromagnetic field and the gravitational field, including the magnetic helicity, the current helicity, the cross helicity, the kinetic helicity, the enstrophy, the field energy, and some other helicity terms. It is found that those helicity terms will effect the gravitational mass density, the charge continuity equation, and the mass continuity equation etc directly. The impact of these helicity terms may be significant in the strong fields of the electromagnetic field and the gravitational field with their related adjoint fields.

It should be noted that the study for the helicity terms with different kinds of operators examined only some simple cases in the electromagnetic field and the gravitational field. Despite its preliminary features, this study can clearly indicate the above helicity terms and the enstrophy are only some simple inferences of the field strength helicity and the field source helicity. They will impact the charge continuity equation and the mass continuity equation in the electromagnetic and gravitational fields. For the future studies, the research will concentrate on only the predictions about some new cross helicity terms related to different physical quantities in the case of the high velocity curl and the strong strength in the electromagnetic field and the gravitational field.

\begin{acknowledgments}
This project was supported partially by the National Natural Science Foundation of China under grant number 60677039.
\end{acknowledgments}

\clearpage

\begin{table}[b]
\centering
\hbox{
\rotcaption{The trigintaduonion multiplication table.}
\label{rotfloat3}%
\begin{sideways}
\begin{tabular}[b]{c|cccc|cccc|cccc|cccc||cccc|cccc|cccc|ccccc}

\hline\hline

$ $ & $1$ & $\emph{\textbf{i}}_1$  & $\emph{\textbf{i}}_2$ & $\emph{\textbf{i}}_3$  & $\emph{\textbf{i}}_4$ & $\emph{\textbf{i}}_5$ & $\emph{\textbf{i}}_6$  & $\emph{\textbf{i}}_7$
& $\emph{\textbf{i}}_8$ & $\emph{\textbf{i}}_9$ & $\emph{\textbf{i}}_{10}$  & $\emph{\textbf{i}}_{11}$ & $\emph{\textbf{i}}_{12}$  & $\emph{\textbf{i}}_{13}$ & $\emph{\textbf{i}}_{14}$ & $\emph{\textbf{i}}_{15}$
& $\emph{\textbf{i}}_{16}$ & $\emph{\textbf{i}}_{17}$  & $\emph{\textbf{i}}_{18}$ & $\emph{\textbf{i}}_{19}$  & $\emph{\textbf{i}}_{20}$ & $\emph{\textbf{i}}_{21}$ & $\emph{\textbf{i}}_{22}$  & $\emph{\textbf{i}}_{23}$
& $\emph{\textbf{i}}_{24}$ & $\emph{\textbf{i}}_{25}$ & $\emph{\textbf{i}}_{26}$  & $\emph{\textbf{i}}_{27}$ & $\emph{\textbf{i}}_{28}$  & $\emph{\textbf{i}}_{29}$ & $\emph{\textbf{i}}_{30}$ & $\emph{\textbf{i}}_{31}$
%0
\\
\hline
$1$ & $1$ & $\emph{\textbf{i}}_1$  & $\emph{\textbf{i}}_2$ & $\emph{\textbf{i}}_3$  & $\emph{\textbf{i}}_4$  & $\emph{\textbf{i}}_5$ & $\emph{\textbf{i}}_6$  & $\emph{\textbf{i}}_7$
& $\emph{\textbf{i}}_8$ & $\emph{\textbf{i}}_9$ & $\emph{\textbf{i}}_{10}$  & $\emph{\textbf{i}}_{11}$ & $\emph{\textbf{i}}_{12}$  & $\emph{\textbf{i}}_{13}$ & $\emph{\textbf{i}}_{14}$ & $\emph{\textbf{i}}_{15}$
& $\emph{\textbf{i}}_{16}$ & $\emph{\textbf{i}}_{17}$  & $\emph{\textbf{i}}_{18}$ & $\emph{\textbf{i}}_{19}$  & $\emph{\textbf{i}}_{20}$ & $\emph{\textbf{i}}_{21}$ & $\emph{\textbf{i}}_{22}$  & $\emph{\textbf{i}}_{23}$
& $\emph{\textbf{i}}_{24}$ & $\emph{\textbf{i}}_{25}$ & $\emph{\textbf{i}}_{26}$  & $\emph{\textbf{i}}_{27}$ & $\emph{\textbf{i}}_{28}$  & $\emph{\textbf{i}}_{29}$ & $\emph{\textbf{i}}_{30}$ & $\emph{\textbf{i}}_{31}$
%1
\\
$\emph{\textbf{i}}_1$ & $\emph{\textbf{i}}_1$ & $-1$ & $\emph{\textbf{i}}_3$  & $-\emph{\textbf{i}}_2$ & $\emph{\textbf{i}}_5$ & $-\emph{\textbf{i}}_4$ &  $-\emph{\textbf{i}}_7$ & $\emph{\textbf{i}}_6$
& $\emph{\textbf{i}}_9$ & $-\emph{\textbf{i}}_8$ & $-\emph{\textbf{i}}_{11}$  & $\emph{\textbf{i}}_{10}$ & $-\emph{\textbf{i}}_{13}$  & $\emph{\textbf{i}}_{12}$ & $\emph{\textbf{i}}_{15}$ & $-\emph{\textbf{i}}_{14}$
& $\emph{\textbf{i}}_{17}$ & $-\emph{\textbf{i}}_{16}$  & $-\emph{\textbf{i}}_{19}$ & $\emph{\textbf{i}}_{18}$  & $-\emph{\textbf{i}}_{21}$ & $\emph{\textbf{i}}_{20}$ & $\emph{\textbf{i}}_{23}$  & $-\emph{\textbf{i}}_{22}$
& $-\emph{\textbf{i}}_{25}$ & $\emph{\textbf{i}}_{24}$ & $\emph{\textbf{i}}_{27}$  & $-\emph{\textbf{i}}_{26}$ & $\emph{\textbf{i}}_{29}$  & $-\emph{\textbf{i}}_{28}$ & $-\emph{\textbf{i}}_{31}$ & $\emph{\textbf{i}}_{30}$
%2
\\
$\emph{\textbf{i}}_2$ & $\emph{\textbf{i}}_2$ & $-\emph{\textbf{i}}_3$ & $-1$ & $\emph{\textbf{i}}_1$  & $\emph{\textbf{i}}_6$  & $\emph{\textbf{i}}_7$ & $-\emph{\textbf{i}}_4$ & $-\emph{\textbf{i}}_5$
& $\emph{\textbf{i}}_{10}$ & $\emph{\textbf{i}}_{11}$ & $-\emph{\textbf{i}}_8$  & $-\emph{\textbf{i}}_9$ & $-\emph{\textbf{i}}_{14}$  & $-\emph{\textbf{i}}_{15}$ & $\emph{\textbf{i}}_{12}$ & $\emph{\textbf{i}}_{13}$
& $\emph{\textbf{i}}_{18}$ & $\emph{\textbf{i}}_{19}$  & $-\emph{\textbf{i}}_{16}$ & $-\emph{\textbf{i}}_{17}$  & $-\emph{\textbf{i}}_{22}$ & $-\emph{\textbf{i}}_{23}$ & $\emph{\textbf{i}}_{20}$  & $\emph{\textbf{i}}_{21}$
& $-\emph{\textbf{i}}_{26}$ & $-\emph{\textbf{i}}_{27}$ & $\emph{\textbf{i}}_{24}$  & $\emph{\textbf{i}}_{25}$ & $\emph{\textbf{i}}_{30}$  & $\emph{\textbf{i}}_{31}$ & $-\emph{\textbf{i}}_{28}$ & $-\emph{\textbf{i}}_{29}$
%3
\\
$\emph{\textbf{i}}_3$ & $\emph{\textbf{i}}_3$ & $\emph{\textbf{i}}_2$ & $-\emph{\textbf{i}}_1$ & $-1$ & $\emph{\textbf{i}}_7$  & $-\emph{\textbf{i}}_6$ & $\emph{\textbf{i}}_5$  & $-\emph{\textbf{i}}_4$
& $\emph{\textbf{i}}_{11}$ & $-\emph{\textbf{i}}_{10}$ & $\emph{\textbf{i}}_9$  & $-\emph{\textbf{i}}_8$ & $-\emph{\textbf{i}}_{15}$  & $\emph{\textbf{i}}_{14}$ & $-\emph{\textbf{i}}_{13}$ & $\emph{\textbf{i}}_{12}$
& $\emph{\textbf{i}}_{19}$ & $-\emph{\textbf{i}}_{18}$  & $\emph{\textbf{i}}_{17}$ & $-\emph{\textbf{i}}_{16}$  & $-\emph{\textbf{i}}_{23}$ & $\emph{\textbf{i}}_{22}$ & $-\emph{\textbf{i}}_{21}$  & $\emph{\textbf{i}}_{20}$
& $-\emph{\textbf{i}}_{27}$ & $\emph{\textbf{i}}_{26}$ & $-\emph{\textbf{i}}_{25}$  & $\emph{\textbf{i}}_{24}$ & $\emph{\textbf{i}}_{31}$  & $-\emph{\textbf{i}}_{30}$ & $\emph{\textbf{i}}_{29}$ & $-\emph{\textbf{i}}_{28}$
%4
\\
\hline
$\emph{\textbf{i}}_4$ & $\emph{\textbf{i}}_4$ & $-\emph{\textbf{i}}_5$ & $-\emph{\textbf{i}}_6$ & $-\emph{\textbf{i}}_7$ & $-1$ & $\emph{\textbf{i}}_1$ & $\emph{\textbf{i}}_2$  & $\emph{\textbf{i}}_3$
& $\emph{\textbf{i}}_{12}$ & $\emph{\textbf{i}}_{13}$ & $\emph{\textbf{i}}_{14}$  & $\emph{\textbf{i}}_{15}$ & $-\emph{\textbf{i}}_8$  & $-\emph{\textbf{i}}_9$ & $-\emph{\textbf{i}}_{10}$ & $-\emph{\textbf{i}}_{11}$
& $\emph{\textbf{i}}_{20}$ & $\emph{\textbf{i}}_{21}$  & $\emph{\textbf{i}}_{22}$ & $\emph{\textbf{i}}_{23}$  & $-\emph{\textbf{i}}_{16}$ & $-\emph{\textbf{i}}_{17}$ & $-\emph{\textbf{i}}_{18}$  & $-\emph{\textbf{i}}_{19}$
& $-\emph{\textbf{i}}_{28}$ & $-\emph{\textbf{i}}_{29}$ & $-\emph{\textbf{i}}_{30}$  & $-\emph{\textbf{i}}_{31}$ & $\emph{\textbf{i}}_{24}$  & $\emph{\textbf{i}}_{25}$ & $\emph{\textbf{i}}_{26}$ & $\emph{\textbf{i}}_{27}$
%5
\\
$\emph{\textbf{i}}_5$ & $\emph{\textbf{i}}_5$ & $\emph{\textbf{i}}_4$ & $-\emph{\textbf{i}}_7$ & $\emph{\textbf{i}}_6$  & $-\emph{\textbf{i}}_1$ & $-1$ & $-\emph{\textbf{i}}_3$ & $\emph{\textbf{i}}_2$
& $\emph{\textbf{i}}_{13}$ & $-\emph{\textbf{i}}_{12}$ & $\emph{\textbf{i}}_{15}$  & $-\emph{\textbf{i}}_{14}$ & $\emph{\textbf{i}}_9$  & $-\emph{\textbf{i}}_8$ & $\emph{\textbf{i}}_{11}$ & $-\emph{\textbf{i}}_{10}$
& $\emph{\textbf{i}}_{21}$ & $-\emph{\textbf{i}}_{20}$  & $\emph{\textbf{i}}_{23}$ & $-\emph{\textbf{i}}_{22}$  & $\emph{\textbf{i}}_{17}$ & $-\emph{\textbf{i}}_{16}$ & $\emph{\textbf{i}}_{19}$  & $-\emph{\textbf{i}}_{18}$
& $-\emph{\textbf{i}}_{29}$ & $\emph{\textbf{i}}_{28}$ & $-\emph{\textbf{i}}_{31}$  & $\emph{\textbf{i}}_{30}$ & $-\emph{\textbf{i}}_{25}$  & $\emph{\textbf{i}}_{24}$ & $-\emph{\textbf{i}}_{27}$ & $\emph{\textbf{i}}_{26}$
%6
\\
$\emph{\textbf{i}}_6$ & $\emph{\textbf{i}}_6$ & $\emph{\textbf{i}}_7$ & $\emph{\textbf{i}}_4$  & $-\emph{\textbf{i}}_5$ & $-\emph{\textbf{i}}_2$ & $\emph{\textbf{i}}_3$  & $-1$ & $-\emph{\textbf{i}}_1$
& $\emph{\textbf{i}}_{14}$ & $-\emph{\textbf{i}}_{15}$ & $-\emph{\textbf{i}}_{12}$  & $\emph{\textbf{i}}_{13}$ & $\emph{\textbf{i}}_{10}$  & $-\emph{\textbf{i}}_{11}$ & $-\emph{\textbf{i}}_8$ & $\emph{\textbf{i}}_9$
& $\emph{\textbf{i}}_{22}$ & $-\emph{\textbf{i}}_{23}$  & $-\emph{\textbf{i}}_{20}$ & $\emph{\textbf{i}}_{21}$  & $\emph{\textbf{i}}_{18}$ & $-\emph{\textbf{i}}_{19}$ & $-\emph{\textbf{i}}_{16}$  & $\emph{\textbf{i}}_{17}$
& $-\emph{\textbf{i}}_{30}$ & $\emph{\textbf{i}}_{31}$ & $\emph{\textbf{i}}_{28}$  & $-\emph{\textbf{i}}_{29}$ & $-\emph{\textbf{i}}_{26}$  & $\emph{\textbf{i}}_{27}$ & $\emph{\textbf{i}}_{24}$ & $-\emph{\textbf{i}}_{25}$
%7
\\
$\emph{\textbf{i}}_7$ & $\emph{\textbf{i}}_7$ & $-\emph{\textbf{i}}_6$ & $\emph{\textbf{i}}_5$  & $\emph{\textbf{i}}_4$  & $-\emph{\textbf{i}}_3$ & $-\emph{\textbf{i}}_2$ & $\emph{\textbf{i}}_1$  & $-1$
& $\emph{\textbf{i}}_{15}$ & $\emph{\textbf{i}}_{14}$ & $-\emph{\textbf{i}}_{13}$  & $-\emph{\textbf{i}}_{12}$ & $\emph{\textbf{i}}_{11}$  & $\emph{\textbf{i}}_{10}$ & $-\emph{\textbf{i}}_9$ & $-\emph{\textbf{i}}_8$
& $\emph{\textbf{i}}_{23}$ & $\emph{\textbf{i}}_{22}$  & $-\emph{\textbf{i}}_{21}$ & $-\emph{\textbf{i}}_{20}$  & $\emph{\textbf{i}}_{19}$ & $\emph{\textbf{i}}_{18}$ & $-\emph{\textbf{i}}_{17}$  & $-\emph{\textbf{i}}_{16}$
& $-\emph{\textbf{i}}_{31}$ & $-\emph{\textbf{i}}_{30}$ & $\emph{\textbf{i}}_{29}$  & $\emph{\textbf{i}}_{28}$ & $-\emph{\textbf{i}}_{27}$  & $-\emph{\textbf{i}}_{26}$ & $\emph{\textbf{i}}_{25}$ & $\emph{\textbf{i}}_{24}$
%8
%---%
\\
\hline
$\emph{\textbf{i}}_8$ & $\emph{\textbf{i}}_8$ & $-\emph{\textbf{i}}_9$ & $-\emph{\textbf{i}}_{10}$  & $-\emph{\textbf{i}}_{11}$  & $-\emph{\textbf{i}}_{12}$  & $-\emph{\textbf{i}}_{13}$ & $-\emph{\textbf{i}}_{14}$ & $-\emph{\textbf{i}}_{15}$
& $-1$ & $\emph{\textbf{i}}_1$  & $\emph{\textbf{i}}_2$ & $\emph{\textbf{i}}_3$ & $\emph{\textbf{i}}_4$  & $\emph{\textbf{i}}_5$ & $\emph{\textbf{i}}_6$  & $\emph{\textbf{i}}_7$
& $\emph{\textbf{i}}_{24}$ & $\emph{\textbf{i}}_{25}$ & $\emph{\textbf{i}}_{26}$  & $\emph{\textbf{i}}_{27}$ & $\emph{\textbf{i}}_{28}$  & $\emph{\textbf{i}}_{29}$ & $\emph{\textbf{i}}_{30}$ & $\emph{\textbf{i}}_{31}$
& $-\emph{\textbf{i}}_{16}$ & $-\emph{\textbf{i}}_{17}$ & $-\emph{\textbf{i}}_{18}$  & $-\emph{\textbf{i}}_{19}$ & $-\emph{\textbf{i}}_{20}$  & $-\emph{\textbf{i}}_{21}$ & $-\emph{\textbf{i}}_{22}$ & $-\emph{\textbf{i}}_{23}$
%9
\\
$\emph{\textbf{i}}_9$ & $\emph{\textbf{i}}_9$ & $\emph{\textbf{i}}_8$ & $-\emph{\textbf{i}}_{11}$  & $\emph{\textbf{i}}_{10}$  & $-\emph{\textbf{i}}_{13}$  & $\emph{\textbf{i}}_{12}$ & $\emph{\textbf{i}}_{15}$ & $-\emph{\textbf{i}}_{14}$
& $-\emph{\textbf{i}}_1$ & $-1$ & $-\emph{\textbf{i}}_3$ & $\emph{\textbf{i}}_2$ & $-\emph{\textbf{i}}_5$  & $\emph{\textbf{i}}_4$ & $\emph{\textbf{i}}_7$  & $-\emph{\textbf{i}}_6$
& $\emph{\textbf{i}}_{25}$ & $-\emph{\textbf{i}}_{24}$  & $\emph{\textbf{i}}_{27}$ & $-\emph{\textbf{i}}_{26}$  & $\emph{\textbf{i}}_{29}$ & $-\emph{\textbf{i}}_{28}$ & $-\emph{\textbf{i}}_{31}$  & $\emph{\textbf{i}}_{30}$
& $\emph{\textbf{i}}_{17}$ & $-\emph{\textbf{i}}_{16}$ & $\emph{\textbf{i}}_{19}$  & $-\emph{\textbf{i}}_{18}$ & $\emph{\textbf{i}}_{21}$  & $-\emph{\textbf{i}}_{20}$ & $-\emph{\textbf{i}}_{23}$ & $\emph{\textbf{i}}_{22}$
%10
\\
$\emph{\textbf{i}}_{10}$ & $\emph{\textbf{i}}_{10}$ & $\emph{\textbf{i}}_{11}$ & $\emph{\textbf{i}}_8$  & $-\emph{\textbf{i}}_9$  & $-\emph{\textbf{i}}_{14}$  & $-\emph{\textbf{i}}_{15}$ & $\emph{\textbf{i}}_{12}$ & $\emph{\textbf{i}}_{13}$
& $-\emph{\textbf{i}}_2$  & $\emph{\textbf{i}}_3$ & $-1$ & $-\emph{\textbf{i}}_1$ & $-\emph{\textbf{i}}_6$  & $-\emph{\textbf{i}}_7$ & $\emph{\textbf{i}}_4$  & $\emph{\textbf{i}}_5$
& $\emph{\textbf{i}}_{26}$ & $-\emph{\textbf{i}}_{27}$  & $-\emph{\textbf{i}}_{24}$ & $\emph{\textbf{i}}_{25}$  & $\emph{\textbf{i}}_{30}$ & $\emph{\textbf{i}}_{31}$ & $-\emph{\textbf{i}}_{28}$  & $-\emph{\textbf{i}}_{29}$
& $\emph{\textbf{i}}_{18}$ & $-\emph{\textbf{i}}_{19}$ & $-\emph{\textbf{i}}_{16}$  & $\emph{\textbf{i}}_{17}$ & $\emph{\textbf{i}}_{22}$  & $\emph{\textbf{i}}_{23}$ & $-\emph{\textbf{i}}_{20}$ & $-\emph{\textbf{i}}_{21}$
%11
\\
$\emph{\textbf{i}}_{11}$ & $\emph{\textbf{i}}_{11}$ & $-\emph{\textbf{i}}_{10}$ & $\emph{\textbf{i}}_9$  & $\emph{\textbf{i}}_8$  & $-\emph{\textbf{i}}_{15}$  & $\emph{\textbf{i}}_{14}$ & $-\emph{\textbf{i}}_{13}$ & $\emph{\textbf{i}}_{12}$
& $-\emph{\textbf{i}}_3$  & $-\emph{\textbf{i}}_2$ & $\emph{\textbf{i}}_1$ & $-1$ & $-\emph{\textbf{i}}_7$  & $\emph{\textbf{i}}_6$ & $-\emph{\textbf{i}}_5$  & $\emph{\textbf{i}}_4$
& $\emph{\textbf{i}}_{27}$ & $\emph{\textbf{i}}_{26}$  & $-\emph{\textbf{i}}_{25}$ & $-\emph{\textbf{i}}_{24}$  & $\emph{\textbf{i}}_{31}$ & $-\emph{\textbf{i}}_{30}$ & $\emph{\textbf{i}}_{29}$  & $-\emph{\textbf{i}}_{28}$
& $\emph{\textbf{i}}_{19}$ & $\emph{\textbf{i}}_{18}$ & $-\emph{\textbf{i}}_{17}$  & $-\emph{\textbf{i}}_{16}$ & $\emph{\textbf{i}}_{23}$  & $-\emph{\textbf{i}}_{22}$ & $\emph{\textbf{i}}_{21}$ & $-\emph{\textbf{i}}_{20}$
%12
\\
\hline
$\emph{\textbf{i}}_{12}$ & $\emph{\textbf{i}}_{12}$  & $\emph{\textbf{i}}_{13}$ & $\emph{\textbf{i}}_{14}$ & $\emph{\textbf{i}}_{15}$ & $\emph{\textbf{i}}_8$ & $-\emph{\textbf{i}}_9$ & $-\emph{\textbf{i}}_{10}$  & $-\emph{\textbf{i}}_{11}$
& $-\emph{\textbf{i}}_4$ & $\emph{\textbf{i}}_5$ & $\emph{\textbf{i}}_6$ & $\emph{\textbf{i}}_7$ & $-1$ & $-\emph{\textbf{i}}_1$ & $-\emph{\textbf{i}}_2$  & $-\emph{\textbf{i}}_3$
& $\emph{\textbf{i}}_{28}$  & $-\emph{\textbf{i}}_{29}$ & $-\emph{\textbf{i}}_{30}$ & $-\emph{\textbf{i}}_{31}$  & $-\emph{\textbf{i}}_{24}$ & $\emph{\textbf{i}}_{25}$ & $\emph{\textbf{i}}_{26}$  & $\emph{\textbf{i}}_{27}$
& $\emph{\textbf{i}}_{20}$ & $-\emph{\textbf{i}}_{21}$ & $-\emph{\textbf{i}}_{22}$  & $-\emph{\textbf{i}}_{23}$ & $-\emph{\textbf{i}}_{16}$  & $\emph{\textbf{i}}_{17}$ & $\emph{\textbf{i}}_{18}$ & $\emph{\textbf{i}}_{19}$
%13
\\
$\emph{\textbf{i}}_{13}$ & $\emph{\textbf{i}}_{13}$  & $-\emph{\textbf{i}}_{12}$ & $\emph{\textbf{i}}_{15}$ & $-\emph{\textbf{i}}_{14}$ & $\emph{\textbf{i}}_9$ & $\emph{\textbf{i}}_8$ & $\emph{\textbf{i}}_{11}$  & $-\emph{\textbf{i}}_{10}$
& $-\emph{\textbf{i}}_5$ & $-\emph{\textbf{i}}_4$ & $\emph{\textbf{i}}_7$ & $-\emph{\textbf{i}}_6$ & $\emph{\textbf{i}}_1$ &$-1$ &  $\emph{\textbf{i}}_3$  & $-\emph{\textbf{i}}_2$
& $\emph{\textbf{i}}_{29}$ & $\emph{\textbf{i}}_{28}$  & $-\emph{\textbf{i}}_{31}$ & $\emph{\textbf{i}}_{30}$  & $-\emph{\textbf{i}}_{25}$ & $-\emph{\textbf{i}}_{24}$ & $-\emph{\textbf{i}}_{27}$  & $\emph{\textbf{i}}_{26}$
& $\emph{\textbf{i}}_{21}$ & $\emph{\textbf{i}}_{20}$ & $-\emph{\textbf{i}}_{23}$  & $\emph{\textbf{i}}_{22}$ & $-\emph{\textbf{i}}_{17}$  & $-\emph{\textbf{i}}_{16}$ & $-\emph{\textbf{i}}_{19}$ & $\emph{\textbf{i}}_{18}$
%14
\\
$\emph{\textbf{i}}_{14}$ & $\emph{\textbf{i}}_{14}$  & $-\emph{\textbf{i}}_{15}$ & $-\emph{\textbf{i}}_{12}$ & $\emph{\textbf{i}}_{13}$ & $\emph{\textbf{i}}_{10}$ & $-\emph{\textbf{i}}_{11}$ & $\emph{\textbf{i}}_8$  & $\emph{\textbf{i}}_9$
& $-\emph{\textbf{i}}_6$ & $-\emph{\textbf{i}}_7$ & $-\emph{\textbf{i}}_4$ & $\emph{\textbf{i}}_5$ & $\emph{\textbf{i}}_2$ & $-\emph{\textbf{i}}_3$  & $-1$ & $\emph{\textbf{i}}_1$
& $\emph{\textbf{i}}_{30}$ & $\emph{\textbf{i}}_{31}$  & $\emph{\textbf{i}}_{28}$ & $-\emph{\textbf{i}}_{29}$  & $-\emph{\textbf{i}}_{26}$ & $\emph{\textbf{i}}_{27}$ & $-\emph{\textbf{i}}_{24}$  & $-\emph{\textbf{i}}_{25}$
& $\emph{\textbf{i}}_{22}$ & $\emph{\textbf{i}}_{23}$ & $\emph{\textbf{i}}_{20}$  & $-\emph{\textbf{i}}_{21}$ & $-\emph{\textbf{i}}_{18}$  & $\emph{\textbf{i}}_{19}$ & $-\emph{\textbf{i}}_{16}$ & $-\emph{\textbf{i}}_{17}$
%15
\\
$\emph{\textbf{i}}_{15}$ & $\emph{\textbf{i}}_{15}$  & $\emph{\textbf{i}}_{14}$ & $-\emph{\textbf{i}}_{13}$ & $-\emph{\textbf{i}}_{12}$ & $\emph{\textbf{i}}_{11}$ & $\emph{\textbf{i}}_{10}$ & $-\emph{\textbf{i}}_9$  & $\emph{\textbf{i}}_8$
& $-\emph{\textbf{i}}_7$ & $\emph{\textbf{i}}_6$ & $-\emph{\textbf{i}}_5$ & $-\emph{\textbf{i}}_4$ & $\emph{\textbf{i}}_3$ & $\emph{\textbf{i}}_2$  & $-\emph{\textbf{i}}_1$ & $-1$
& $\emph{\textbf{i}}_{31}$ & $-\emph{\textbf{i}}_{30}$  & $\emph{\textbf{i}}_{29}$ & $\emph{\textbf{i}}_{28}$  & $-\emph{\textbf{i}}_{27}$ & $-\emph{\textbf{i}}_{26}$ & $\emph{\textbf{i}}_{25}$  & $-\emph{\textbf{i}}_{24}$
& $\emph{\textbf{i}}_{23}$ & $-\emph{\textbf{i}}_{22}$ & $\emph{\textbf{i}}_{21}$  & $\emph{\textbf{i}}_{20}$ & $-\emph{\textbf{i}}_{19}$  & $-\emph{\textbf{i}}_{18}$ & $\emph{\textbf{i}}_{17}$ & $-\emph{\textbf{i}}_{16}$
%16

\\
\hline\hline
$\emph{\textbf{i}}_{16}$ & $\emph{\textbf{i}}_{16}$ & $-\emph{\textbf{i}}_{17}$  & $-\emph{\textbf{i}}_{18}$ & $-\emph{\textbf{i}}_{19}$  & $-\emph{\textbf{i}}_{20}$  & $-\emph{\textbf{i}}_{21}$ & $-\emph{\textbf{i}}_{22}$  & $-\emph{\textbf{i}}_{23}$
& $-\emph{\textbf{i}}_{24}$ & $-\emph{\textbf{i}}_{25}$ & $-\emph{\textbf{i}}_{26}$  & $-\emph{\textbf{i}}_{27}$ & $-\emph{\textbf{i}}_{28}$  & $-\emph{\textbf{i}}_{29}$ & $-\emph{\textbf{i}}_{30}$ & $-\emph{\textbf{i}}_{31}$
& $-1$ & $\emph{\textbf{i}}_1$  & $\emph{\textbf{i}}_2$ & $\emph{\textbf{i}}_3$  & $\emph{\textbf{i}}_4$ & $\emph{\textbf{i}}_5$ & $\emph{\textbf{i}}_6$  & $\emph{\textbf{i}}_7$
& $\emph{\textbf{i}}_8$ & $\emph{\textbf{i}}_9$ & $\emph{\textbf{i}}_{10}$  & $\emph{\textbf{i}}_{11}$ & $\emph{\textbf{i}}_{12}$  & $\emph{\textbf{i}}_{13}$ & $\emph{\textbf{i}}_{14}$ & $\emph{\textbf{i}}_{15}$
%17
\\
$\emph{\textbf{i}}_{17}$ & $\emph{\textbf{i}}_{17}$ & $\emph{\textbf{i}}_{16}$ & $-\emph{\textbf{i}}_{19}$  & $\emph{\textbf{i}}_{18}$ & $-\emph{\textbf{i}}_{21}$ & $\emph{\textbf{i}}_{20}$ &  $\emph{\textbf{i}}_{23}$ & $-\emph{\textbf{i}}_{22}$
& $-\emph{\textbf{i}}_{25}$ & $\emph{\textbf{i}}_{24}$ & $\emph{\textbf{i}}_{27}$  & $-\emph{\textbf{i}}_{26}$ & $\emph{\textbf{i}}_{29}$  & $-\emph{\textbf{i}}_{28}$ & $-\emph{\textbf{i}}_{31}$ & $\emph{\textbf{i}}_{30}$
& $-\emph{\textbf{i}}_1$ & $-1$  & $-\emph{\textbf{i}}_3$ & $\emph{\textbf{i}}_2$  & $-\emph{\textbf{i}}_5$ & $\emph{\textbf{i}}_4$ & $\emph{\textbf{i}}_7$  & $-\emph{\textbf{i}}_6$
& $-\emph{\textbf{i}}_9$ & $\emph{\textbf{i}}_8$ & $\emph{\textbf{i}}_{11}$  & $-\emph{\textbf{i}}_{10}$ & $\emph{\textbf{i}}_{13}$  & $-\emph{\textbf{i}}_{12}$ & $-\emph{\textbf{i}}_{15}$ & $\emph{\textbf{i}}_{14}$
%18
\\
$\emph{\textbf{i}}_{18}$ & $\emph{\textbf{i}}_{18}$ & $\emph{\textbf{i}}_{19}$ & $\emph{\textbf{i}}_{16}$ & $-\emph{\textbf{i}}_{17}$  & $-\emph{\textbf{i}}_{22}$  & $-\emph{\textbf{i}}_{23}$ & $\emph{\textbf{i}}_{20}$ & $\emph{\textbf{i}}_{21}$
& $-\emph{\textbf{i}}_{26}$ & $-\emph{\textbf{i}}_{27}$ & $\emph{\textbf{i}}_{24}$  & $\emph{\textbf{i}}_{25}$ & $\emph{\textbf{i}}_{30}$  & $\emph{\textbf{i}}_{31}$ & $-\emph{\textbf{i}}_{28}$ & $-\emph{\textbf{i}}_{29}$
& $-\emph{\textbf{i}}_2$ & $\emph{\textbf{i}}_3$  & $-1$ & $-\emph{\textbf{i}}_1$  & $-\emph{\textbf{i}}_6$ & $-\emph{\textbf{i}}_7$ & $\emph{\textbf{i}}_4$  & $\emph{\textbf{i}}_5$
& $-\emph{\textbf{i}}_{10}$ & $-\emph{\textbf{i}}_{11}$ & $\emph{\textbf{i}}_8$  & $\emph{\textbf{i}}_9$ & $\emph{\textbf{i}}_{14}$  & $\emph{\textbf{i}}_{15}$ & $-\emph{\textbf{i}}_{12}$ & $-\emph{\textbf{i}}_{13}$
%19
\\
$\emph{\textbf{i}}_{19}$ & $\emph{\textbf{i}}_{19}$ & $-\emph{\textbf{i}}_{18}$ & $\emph{\textbf{i}}_{17}$ & $\emph{\textbf{i}}_{16}$ & $-\emph{\textbf{i}}_{23}$ & $\emph{\textbf{i}}_{22}$ & $-\emph{\textbf{i}}_{21}$  & $\emph{\textbf{i}}_{20}$
& $-\emph{\textbf{i}}_{27}$ & $\emph{\textbf{i}}_{26}$ & $-\emph{\textbf{i}}_{25}$  & $\emph{\textbf{i}}_{24}$ & $\emph{\textbf{i}}_{31}$  & $-\emph{\textbf{i}}_{30}$ & $\emph{\textbf{i}}_{29}$ & $-\emph{\textbf{i}}_{28}$
& $-\emph{\textbf{i}}_3$ & $-\emph{\textbf{i}}_2$  & $\emph{\textbf{i}}_1$ & $-1$  & $-\emph{\textbf{i}}_7$ & $\emph{\textbf{i}}_6$ & $-\emph{\textbf{i}}_5$  & $\emph{\textbf{i}}_4$
& $-\emph{\textbf{i}}_{11}$ & $\emph{\textbf{i}}_{10}$ & $-\emph{\textbf{i}}_9$  & $\emph{\textbf{i}}_8$ & $\emph{\textbf{i}}_{15}$  & $-\emph{\textbf{i}}_{14}$ & $\emph{\textbf{i}}_{13}$ & $-\emph{\textbf{i}}_{12}$
%20
\\
\hline
$\emph{\textbf{i}}_{20}$ & $\emph{\textbf{i}}_{20}$ & $\emph{\textbf{i}}_{21}$ & $\emph{\textbf{i}}_{22}$ & $\emph{\textbf{i}}_{23}$ & $\emph{\textbf{i}}_{16}$ & $-\emph{\textbf{i}}_{17}$ & $-\emph{\textbf{i}}_{18}$  & $-\emph{\textbf{i}}_{19}$
& $-\emph{\textbf{i}}_{28}$ & $-\emph{\textbf{i}}_{29}$ & $-\emph{\textbf{i}}_{30}$  & $-\emph{\textbf{i}}_{31}$ & $\emph{\textbf{i}}_{24}$  & $\emph{\textbf{i}}_{25}$ & $\emph{\textbf{i}}_{26}$ & $\emph{\textbf{i}}_{27}$
& $-\emph{\textbf{i}}_4$ & $\emph{\textbf{i}}_5$  & $\emph{\textbf{i}}_6$ & $\emph{\textbf{i}}_7$  & $-1$ & $-\emph{\textbf{i}}_1$ & $-\emph{\textbf{i}}_2$  & $-\emph{\textbf{i}}_3$
& $-\emph{\textbf{i}}_{12}$ & $-\emph{\textbf{i}}_{13}$ & $-\emph{\textbf{i}}_{14}$  & $-\emph{\textbf{i}}_{15}$ & $\emph{\textbf{i}}_{8}$  & $\emph{\textbf{i}}_{9}$ & $\emph{\textbf{i}}_{10}$ & $\emph{\textbf{i}}_{11}$
%21
\\
$\emph{\textbf{i}}_{21}$ & $\emph{\textbf{i}}_{21}$  & $-\emph{\textbf{i}}_{20}$ & $\emph{\textbf{i}}_{23}$  & $-\emph{\textbf{i}}_{22}$ & $\emph{\textbf{i}}_{17}$ & $\emph{\textbf{i}}_{16}$ & $\emph{\textbf{i}}_{19}$ & $-\emph{\textbf{i}}_{18}$
& $-\emph{\textbf{i}}_{29}$ & $\emph{\textbf{i}}_{28}$ & $-\emph{\textbf{i}}_{31}$  & $\emph{\textbf{i}}_{30}$ & $-\emph{\textbf{i}}_{25}$  & $\emph{\textbf{i}}_{24}$ & $-\emph{\textbf{i}}_{27}$ & $\emph{\textbf{i}}_{26}$
& $-\emph{\textbf{i}}_5$ & $-\emph{\textbf{i}}_4$  & $\emph{\textbf{i}}_7$ & $-\emph{\textbf{i}}_6$  & $\emph{\textbf{i}}_1$ & $-1$ & $\emph{\textbf{i}}_3$  & $-\emph{\textbf{i}}_2$
& $-\emph{\textbf{i}}_{13}$ & $\emph{\textbf{i}}_{12}$ & $-\emph{\textbf{i}}_{15}$  & $\emph{\textbf{i}}_{14}$ & $-\emph{\textbf{i}}_9$  & $\emph{\textbf{i}}_8$ & $-\emph{\textbf{i}}_{11}$ & $\emph{\textbf{i}}_{10}$
%22
\\
$\emph{\textbf{i}}_{22}$ & $\emph{\textbf{i}}_{22}$  & $-\emph{\textbf{i}}_{23}$ & $-\emph{\textbf{i}}_{20}$  & $\emph{\textbf{i}}_{21}$ & $\emph{\textbf{i}}_{18}$ & $-\emph{\textbf{i}}_{19}$  & $\emph{\textbf{i}}_{16}$ & $\emph{\textbf{i}}_{17}$
& $-\emph{\textbf{i}}_{30}$ & $\emph{\textbf{i}}_{31}$ & $\emph{\textbf{i}}_{28}$  & $-\emph{\textbf{i}}_{29}$ & $-\emph{\textbf{i}}_{26}$  & $\emph{\textbf{i}}_{27}$ & $\emph{\textbf{i}}_{24}$ & $-\emph{\textbf{i}}_{25}$
& $-\emph{\textbf{i}}_6$ & $-\emph{\textbf{i}}_7$  & $-\emph{\textbf{i}}_4$ & $\emph{\textbf{i}}_5$  & $\emph{\textbf{i}}_2$ & $-\emph{\textbf{i}}_3$ & $-1$  & $\emph{\textbf{i}}_1$
& $-\emph{\textbf{i}}_{14}$ & $\emph{\textbf{i}}_{15}$ & $\emph{\textbf{i}}_{12}$  & $-\emph{\textbf{i}}_{13}$ & $-\emph{\textbf{i}}_{10}$  & $\emph{\textbf{i}}_{11}$ & $\emph{\textbf{i}}_8$ & $-\emph{\textbf{i}}_9$
%23
\\
$\emph{\textbf{i}}_{23}$ & $\emph{\textbf{i}}_{23}$  & $\emph{\textbf{i}}_{22}$ & $-\emph{\textbf{i}}_{21}$  & $-\emph{\textbf{i}}_{20}$  & $\emph{\textbf{i}}_{19}$ & $\emph{\textbf{i}}_{18}$ & $-\emph{\textbf{i}}_{17}$  & $\emph{\textbf{i}}_{16}$
& $-\emph{\textbf{i}}_{31}$ & $-\emph{\textbf{i}}_{30}$ & $\emph{\textbf{i}}_{29}$  & $\emph{\textbf{i}}_{28}$ & $-\emph{\textbf{i}}_{27}$  & $-\emph{\textbf{i}}_{26}$ & $\emph{\textbf{i}}_{25}$ & $\emph{\textbf{i}}_{24}$
& $-\emph{\textbf{i}}_7$ & $\emph{\textbf{i}}_6$  & $-\emph{\textbf{i}}_5$ & $-\emph{\textbf{i}}_4$  & $\emph{\textbf{i}}_3$ & $\emph{\textbf{i}}_2$ & $-\emph{\textbf{i}}_1$  & $-1$
& $-\emph{\textbf{i}}_{15}$ & $-\emph{\textbf{i}}_{14}$ & $\emph{\textbf{i}}_{13}$  & $\emph{\textbf{i}}_{12}$ & $-\emph{\textbf{i}}_{11}$  & $-\emph{\textbf{i}}_{10}$ & $\emph{\textbf{i}}_9$ & $\emph{\textbf{i}}_8$
%24
%---%
\\
\hline
$\emph{\textbf{i}}_{24}$ & $\emph{\textbf{i}}_{24}$  & $\emph{\textbf{i}}_{25}$ & $\emph{\textbf{i}}_{26}$  & $\emph{\textbf{i}}_{27}$  & $\emph{\textbf{i}}_{28}$  & $\emph{\textbf{i}}_{29}$ & $\emph{\textbf{i}}_{30}$ & $\emph{\textbf{i}}_{31}$
& $\emph{\textbf{i}}_{16}$ & $-\emph{\textbf{i}}_{17}$  & $-\emph{\textbf{i}}_{18}$ & $-\emph{\textbf{i}}_{19}$ & $-\emph{\textbf{i}}_{20}$  & $-\emph{\textbf{i}}_{21}$ & $-\emph{\textbf{i}}_{22}$  & $-\emph{\textbf{i}}_{23}$
& $-\emph{\textbf{i}}_8$ & $\emph{\textbf{i}}_9$  & $\emph{\textbf{i}}_{10}$ & $\emph{\textbf{i}}_{11}$  & $\emph{\textbf{i}}_{12}$ & $\emph{\textbf{i}}_{13}$ & $\emph{\textbf{i}}_{14}$  & $\emph{\textbf{i}}_{15}$
& $-1$ & $-\emph{\textbf{i}}_1$ & $-\emph{\textbf{i}}_2$  & $-\emph{\textbf{i}}_3$ & $-\emph{\textbf{i}}_4$  & $-\emph{\textbf{i}}_5$ & $-\emph{\textbf{i}}_6$ & $-\emph{\textbf{i}}_7$
%25
\\
$\emph{\textbf{i}}_{25}$ & $\emph{\textbf{i}}_{25}$  & $-\emph{\textbf{i}}_{24}$ & $\emph{\textbf{i}}_{27}$  & $-\emph{\textbf{i}}_{26}$  & $\emph{\textbf{i}}_{29}$  & $-\emph{\textbf{i}}_{28}$ & $-\emph{\textbf{i}}_{31}$ & $\emph{\textbf{i}}_{30}$
& $\emph{\textbf{i}}_{17}$ & $\emph{\textbf{i}}_{16}$ & $\emph{\textbf{i}}_{19}$ & $-\emph{\textbf{i}}_{18}$ & $\emph{\textbf{i}}_{21}$  & $-\emph{\textbf{i}}_{20}$ & $-\emph{\textbf{i}}_{23}$  & $\emph{\textbf{i}}_{22}$
& $-\emph{\textbf{i}}_9$ & $-\emph{\textbf{i}}_8$  & $\emph{\textbf{i}}_{11}$ & $-\emph{\textbf{i}}_{10}$  & $\emph{\textbf{i}}_{13}$ & $-\emph{\textbf{i}}_{12}$ & $-\emph{\textbf{i}}_{15}$  & $\emph{\textbf{i}}_{14}$
& $\emph{\textbf{i}}_1$ & $-1$ & $\emph{\textbf{i}}_3$  & $-\emph{\textbf{i}}_2$ & $\emph{\textbf{i}}_5$  & $-\emph{\textbf{i}}_4$ & $-\emph{\textbf{i}}_7$ & $\emph{\textbf{i}}_6$
%26
\\
$\emph{\textbf{i}}_{26}$ & $\emph{\textbf{i}}_{26}$  & $-\emph{\textbf{i}}_{27}$ & $-\emph{\textbf{i}}_{24}$  & $\emph{\textbf{i}}_{25}$  & $\emph{\textbf{i}}_{30}$  & $\emph{\textbf{i}}_{31}$ & $-\emph{\textbf{i}}_{28}$ & $-\emph{\textbf{i}}_{29}$
& $\emph{\textbf{i}}_{18}$  & $-\emph{\textbf{i}}_{19}$ & $\emph{\textbf{i}}_{16}$ & $\emph{\textbf{i}}_{17}$ & $\emph{\textbf{i}}_{22}$  & $\emph{\textbf{i}}_{23}$ & $-\emph{\textbf{i}}_{20}$  & $-\emph{\textbf{i}}_{21}$
& $-\emph{\textbf{i}}_{10}$ & $-\emph{\textbf{i}}_{11}$  & $-\emph{\textbf{i}}_8$ & $\emph{\textbf{i}}_9$  & $\emph{\textbf{i}}_{14}$ & $\emph{\textbf{i}}_{15}$ & $-\emph{\textbf{i}}_{12}$  & $-\emph{\textbf{i}}_{13}$
& $\emph{\textbf{i}}_2$ & $-\emph{\textbf{i}}_3$ & $-1$  & $\emph{\textbf{i}}_1$ & $\emph{\textbf{i}}_6$  & $\emph{\textbf{i}}_7$ & $-\emph{\textbf{i}}_4$ & $-\emph{\textbf{i}}_5$
%27
\\
$\emph{\textbf{i}}_{27}$ & $\emph{\textbf{i}}_{27}$  & $\emph{\textbf{i}}_{26}$ & $-\emph{\textbf{i}}_{25}$  & $-\emph{\textbf{i}}_{24}$  & $\emph{\textbf{i}}_{31}$  & $-\emph{\textbf{i}}_{30}$ & $\emph{\textbf{i}}_{29}$ & $-\emph{\textbf{i}}_{28}$
& $\emph{\textbf{i}}_{19}$  & $\emph{\textbf{i}}_{18}$ & $-\emph{\textbf{i}}_{17}$ & $\emph{\textbf{i}}_{16}$ & $\emph{\textbf{i}}_{23}$  & $-\emph{\textbf{i}}_{22}$ & $\emph{\textbf{i}}_{21}$  & $-\emph{\textbf{i}}_{20}$
& $-\emph{\textbf{i}}_{11}$ & $\emph{\textbf{i}}_{10}$  & $-\emph{\textbf{i}}_9$ & $-\emph{\textbf{i}}_8$  & $\emph{\textbf{i}}_{15}$ & $-\emph{\textbf{i}}_{14}$ & $\emph{\textbf{i}}_{13}$  & $-\emph{\textbf{i}}_{12}$
& $\emph{\textbf{i}}_3$ & $\emph{\textbf{i}}_2$ & $-\emph{\textbf{i}}_1$  & $-1$ & $\emph{\textbf{i}}_7$  & $-\emph{\textbf{i}}_6$ & $\emph{\textbf{i}}_5$ & $-\emph{\textbf{i}}_4$
%28
\\
\hline
$\emph{\textbf{i}}_{28}$ & $\emph{\textbf{i}}_{28}$   & $-\emph{\textbf{i}}_{29}$ & $-\emph{\textbf{i}}_{30}$ & $-\emph{\textbf{i}}_{31}$ & $-\emph{\textbf{i}}_{24}$ & $\emph{\textbf{i}}_{25}$ & $\emph{\textbf{i}}_{26}$  & $\emph{\textbf{i}}_{27}$
& $\emph{\textbf{i}}_{20}$ & $-\emph{\textbf{i}}_{21}$ & $-\emph{\textbf{i}}_{22}$ & $-\emph{\textbf{i}}_{23}$ & $\emph{\textbf{i}}_{16}$ & $\emph{\textbf{i}}_{17}$ & $\emph{\textbf{i}}_{18}$  & $\emph{\textbf{i}}_{19}$
& $-\emph{\textbf{i}}_{12}$ & $-\emph{\textbf{i}}_{13}$  & $-\emph{\textbf{i}}_{14}$ & $-\emph{\textbf{i}}_{15}$  & $-\emph{\textbf{i}}_8$ & $\emph{\textbf{i}}_9$ & $\emph{\textbf{i}}_{10}$  & $\emph{\textbf{i}}_{11}$
& $\emph{\textbf{i}}_4$ & $-\emph{\textbf{i}}_5$ & $-\emph{\textbf{i}}_6$  & $-\emph{\textbf{i}}_7$ & $-1$  & $\emph{\textbf{i}}_1$ & $\emph{\textbf{i}}_2$ & $\emph{\textbf{i}}_3$
%29
\\
$\emph{\textbf{i}}_{29}$ & $\emph{\textbf{i}}_{29}$   & $\emph{\textbf{i}}_{28}$ & $-\emph{\textbf{i}}_{31}$ & $\emph{\textbf{i}}_{30}$ & $-\emph{\textbf{i}}_{25}$ & $-\emph{\textbf{i}}_{24}$ & $-\emph{\textbf{i}}_{27}$  & $\emph{\textbf{i}}_{26}$
& $\emph{\textbf{i}}_{21}$ & $\emph{\textbf{i}}_{20}$ & $-\emph{\textbf{i}}_{23}$ & $\emph{\textbf{i}}_{22}$ & $-\emph{\textbf{i}}_{17}$ & $\emph{\textbf{i}}_{16}$ &  $-\emph{\textbf{i}}_{19}$  & $\emph{\textbf{i}}_{18}$
& $-\emph{\textbf{i}}_{13}$ & $\emph{\textbf{i}}_{12}$  & $-\emph{\textbf{i}}_{15}$ & $\emph{\textbf{i}}_{14}$  & $-\emph{\textbf{i}}_9$ & $-\emph{\textbf{i}}_8$ & $-\emph{\textbf{i}}_{11}$  & $\emph{\textbf{i}}_{10}$
& $\emph{\textbf{i}}_5$ & $\emph{\textbf{i}}_4$ & $-\emph{\textbf{i}}_7$  & $\emph{\textbf{i}}_6$ & $-\emph{\textbf{i}}_1$  & $-1$ & $-\emph{\textbf{i}}_3$ & $\emph{\textbf{i}}_2$
%30
\\
$\emph{\textbf{i}}_{30}$ & $\emph{\textbf{i}}_{30}$   & $\emph{\textbf{i}}_{31}$ & $\emph{\textbf{i}}_{28}$ & $-\emph{\textbf{i}}_{29}$ & $-\emph{\textbf{i}}_{26}$ & $\emph{\textbf{i}}_{27}$ & $-\emph{\textbf{i}}_{24}$  & $-\emph{\textbf{i}}_{25}$
& $\emph{\textbf{i}}_{22}$ & $\emph{\textbf{i}}_{23}$ & $\emph{\textbf{i}}_{20}$ & $-\emph{\textbf{i}}_{21}$ & $-\emph{\textbf{i}}_{18}$ & $\emph{\textbf{i}}_{19}$  & $\emph{\textbf{i}}_{16}$ & $-\emph{\textbf{i}}_{17}$
& $-\emph{\textbf{i}}_{14}$ & $\emph{\textbf{i}}_{15}$  & $\emph{\textbf{i}}_{12}$ & $-\emph{\textbf{i}}_{13}$  & $-\emph{\textbf{i}}_{10}$ & $\emph{\textbf{i}}_{11}$ & $-\emph{\textbf{i}}_8$  & $-\emph{\textbf{i}}_9$
& $\emph{\textbf{i}}_6$ & $\emph{\textbf{i}}_7$ & $\emph{\textbf{i}}_4$  & $-\emph{\textbf{i}}_5$ & $-\emph{\textbf{i}}_2$  & $\emph{\textbf{i}}_3$ & $-1$ & $-\emph{\textbf{i}}_1$
%31
\\
$\emph{\textbf{i}}_{31}$ & $\emph{\textbf{i}}_{31}$  & $-\emph{\textbf{i}}_{30}$ & $\emph{\textbf{i}}_{29}$ & $\emph{\textbf{i}}_{28}$ & $-\emph{\textbf{i}}_{27}$ & $-\emph{\textbf{i}}_{26}$ & $\emph{\textbf{i}}_{25}$  & $-\emph{\textbf{i}}_{24}$
& $\emph{\textbf{i}}_{23}$ & $-\emph{\textbf{i}}_{22}$ & $\emph{\textbf{i}}_{21}$ & $\emph{\textbf{i}}_{20}$ & $-\emph{\textbf{i}}_{19}$ & $-\emph{\textbf{i}}_{18}$  & $\emph{\textbf{i}}_{17}$ & $\emph{\textbf{i}}_{16}$
& $-\emph{\textbf{i}}_{15}$ & $-\emph{\textbf{i}}_{14}$  & $\emph{\textbf{i}}_{13}$ & $\emph{\textbf{i}}_{12}$  & $-\emph{\textbf{i}}_{11}$ & $-\emph{\textbf{i}}_{10}$ & $\emph{\textbf{i}}_9$  & $-\emph{\textbf{i}}_8$
& $\emph{\textbf{i}}_7$ & $-\emph{\textbf{i}}_6$ & $\emph{\textbf{i}}_5$  & $\emph{\textbf{i}}_4$ & $-\emph{\textbf{i}}_3$  & $-\emph{\textbf{i}}_2$ & $\emph{\textbf{i}}_1$ & $-1$
%32
\\
\hline\hline
\end{tabular}
\end{sideways}
}
\end{table}

\end{document}